\documentclass[a4paper,fleqn,usenatbib,sorting=nyt]{mnras}

\usepackage[T1]{fontenc}
\usepackage{ae,aecompl}

\usepackage{graphicx}	% Including figure files
\usepackage{subcaption}
\usepackage{amssymb,amsmath}

\newcommand{\be}{\begin{eqnarray}}
\newcommand{\ee}{\end{eqnarray}}

\title[Evolution of radio continuum emission with SKA]{Evolution of thermal and nonthermal radio continuum emission on kpc scales--Predictions for SKA}

\author[M. Ghasemi-Nodehi et al.]{M. Ghasemi-Nodehi$^{1}$, %\thanks{E-mail: mghasemin@ipm.ir (MGN)}
Fatemeh S. Tabatabaei$^{1,2,3}$\thanks{Corresponding author E-mail: ftaba@ipm.ir}, 
Mark Sargent$^{4,5}$,
Eric J. Murphy$^{6}$,
\newauthor
Habib Khosroshahi$^{1}$,
Rob Beswick,$^{7}$,
Anna Bonaldi,$^{8}$
Eva Schinnerer$^{3}$
\\
$^{1}$School of Astronomy, Institute for Research in Fundamental Sciences, 19395-5531, Tehran, Iran\\
$^{2}$Instituto de Astrof\'{\i}sica de Canarias, E-38205 La Laguna, Tenerife, Spain\\
$^{3}$Max-Planck-Institut f{\"u}r Astronomie, K{\"o}nigstuhl 17, D-69117 Heidelberg, Germany\\
$^{4}$International Space Science Institute (ISSI), Hallerstrasse 6, CH-3012 Bern, Switzerland\\
$^{5}$Astronomy Centre, Department of Physics \& Astronomy, University of Sussex, Brighton, BN1 9QH, England\\
$^{6}$National Radio Astronomy Observatory, 520 Edgemont Road, Charlottesville, VA 22903, USA\\
$^{7}$Jodrell Bank Centre for Astrophysics/e-MERLIN, The University of Manchester, M13 9PL, UK\\
$^{8}$SKA Organisation, Jodrell Bank, Lower Withington, Macclesfield, Cheshire SK11 9FT, UK
}
\date{Accepted XXX. Received YYY; in original form ZZZ}

\pubyear{2020}

\begin{document}
\label{firstpage}
\pagerange{\pageref{firstpage}--\pageref{lastpage}}
\maketitle

\begin{abstract}
{ Resolved maps of the thermal and nonthermal radio continuum (RC) emission of distant galaxies are a powerful tool for understanding the role of the interstellar medium (ISM) in the evolution of galaxies. We simulate the RC surface brightness of present-day star forming galaxies in the past at $0.15<z<3$  considering two cases of radio size evolution: (1)~no evolution, and (2)~same evolution as in the optical. We aim to investigate the a)~structure of the thermal and nonthermal emission on kpc scales, b)~evolution of the thermal fraction and synchrotron spectrum at mid-radio frequencies ($\simeq$1-10\,GHz), and c)~capability of the proposed SKA1-MID reference surveys in detecting the RC emitting structures. The synchrotron spectrum flattens with $z$ causing curvature in the observed mid-radio SEDs of galaxies at higher $z$. The spectral index reported in recent observational studies agrees better with the no size evolution scenario. In this case, the mean thermal fraction observed at 1.4\,GHz increases with redshift by more than 30\% from $z=0.15$ to $z=2$ because of the drop of the synchrotron emission at higher rest-frame~frequencies. More massive galaxies have lower thermal fractions and experience a faster flattening of the nonthermal spectrum. The proposed SKA1-MID band~2 reference survey, unveils the ISM in M51- and NGC6946-like galaxies (with ${\rm M_{\star}}\simeq10^{10}\,{\rm M}_{\odot}$) up to $z=3$. This survey detects lower-mass galaxies like M33 (${\rm M_{\star}}\simeq10^{9}\,{\rm M}_{\odot}$) only at low redshifts $z\lesssim 0.5$. For a proper separation of the RC emitting processes at the peak of star formation, it is vital to include band~1 into the SKA1-MID reference surveys.}
\end{abstract}


%%%%%%%%%%%%%%%%%%%%%%%%%%%%%%%%%%%%%%%%%%%%%%%%%%

%%%%%%%%%%%%%%%%% BODY OF PAPER %%%%%%%%%%%%%%%%%%

\section{Introduction}\label{intro}

{The interstellar and intergalactic media play an important role in galaxy formation and evolution as they host the cool gas reservoir needed to form stars and fuel supermassive black hole accretion \cite[e.g. ][]{Daddi,Greve,Tacconi,Walter,Walter20,zovaro}. 
%Galaxies quench their massive star formation over 
{Galaxies experience star formation quenching over}
cosmic time that is linked to a lack of cool gas caused by different mechanisms such as feedback and starvation \cite[e.g. ][]{schaye,Gatto,Peng}. However, ample amounts of cool gas can be found in quenched systems which are, for an unknown reason, inefficient in forming massive stars \cite[][]{brown,Hunt,scovil,Tacconi18}.  Thus,  factors controlling gas against collapse and accretion  over cosmic time must be better understood. Observations show that nonthermal pressures inserted by cosmic rays and magnetic fields can be more significant than the thermal gas pressure in nearby galaxies \cite[][]{Beck_07,taba0708,Hassani}. % while only the thermal winds and radiation is often considered in models \cite[e.g.][]{schaye}. 
 The nonthermal processes can support cool gas against accretion from the IGM \citep{Gron,Mathews,Owen} and collapse/fragmentation in the ISM \cite[e.g.,][]{pillai,tab18}  decreasing the star formation rate (SFR). Therefore, it is very insightful to dissect the thermal and nonthermal processes over cosmic time. These processes can be very well studied through mapping the radio continuum (RC) emission in galaxies. The thermal free-free emission and the nonthermal synchrotron emission, the two main components of the RC emission, emerge from the thermal ionized gas and the nonthermal magnetized/relativistic ISM, respectively. %This paper presents decomposed maps of the thermal and nonthermal RC emission at different redshifts simulated for the upcoming SKA observations. 
The power-law index of the nonthermal synchrotron spectrum, that is related to the energy index of cosmic ray electrons, changes with location in a galaxy, being flatter in star-forming regions and steeper in inter-arm regions and outer disks \citep{tab7,tab_a,Hassani}. This indicates that cosmic rays are more energetic closer to their {birthplace} in star-forming regions (e.g., in supernova remnants) than in other media due to cooling mechanisms \cite[e.g.,][]{Longair}.  Global surveys such as KINGFISH \citep[Key Insights on Nearby Galaxies; a Far-Infrared Survey with Herschel, ][]{kenn} also find that the nonthermal spectrum in galaxies with higher star formation rate surface density is flatter. This shows that the bulk of the cosmic ray electron population is younger and more energetic due to massive star formation activities \citep{cal2}. 
Combined with polarization observations, the synchrotron emission reveals that the turbulent magnetic field becomes stronger with star formation rate as shown in both local and global studies \citep{nikandbeck,chy,hee2}. Strong turbulent magnetic fields and a large number of high energy particles can cause strong winds and outflows in starbursting galaxies. This implies an important role of magnetic fields and cosmic rays in driving so called 'feedback' in galaxies~\citep{pfm}.  

Deep and resolved RC observations at high-redshifts are needed to address the exact role of the thermal and nonthermal processes and the origin of feedback. Currently available radio deep field surveys have, however, {too} low angular resolution and sensitivity to allow for separation of the thermal and nonthermal emission in distant galaxies.
The advent of the Square Kilometer Array (SKA) and its instrumental capabilities combined with ground-breaking results from ALMA, VLT/MUSE (and others) which trace different phases of the gaseous ISM will shed light on this subject. %The SKA will map the synchrotron emission not only in star forming regions and AGNs, but also in more quiescent regions in molecular clouds and diffuse ISM in distant galaxies. 

The RC emission from star-forming galaxies has been often used as an extinction-free tracer of star formation rate (SFR) at all redshifts mainly because of its tight correlation with the infrared (IR) emission \citep[see e.g., ][and references therein]{cond_2,gran,Garrett,Appleton,Sajina,sarg_10,Leslie}. However, there are works questioning the RC--IR's direct connection to star formation suggesting a conspiracy of several factors \cite[e.g.][]{bell,Lacki}.  Moreover, several studies show that the RC--IR correlation can deviate from linearity at high-z \cite[e.g.,][]{Magneli,Basu,Delhaize} and in nearby galaxies \cite[e.g.][]{nikandbeck,dumas,hee2}  depending on galaxy mass and type \cite[e.g.][]{Delvecchio,Read,molnar} and the ISM conditions and spatial scale \cite[e.g.][]{tab_a,tab_b}. The non-linearity of the global correlation occurs mainly at high SFRs \cite[][]{cal2} and this can be observed with redshift as the SFR increases with redshift \cite[e.g.,][]{molnar}.
%(e.g., being optically thick to both UV photons and cosmic ray electrons or if a coupling between magnetic field with gas occurs).

 \cite{cal1} presented the RC--SFR calibration relations for both the thermal and nonthermal emission which are independent from the RC--IR correlation and show that, in the HII complexes of NGC6946, the thermal RC  emission provides a more robust SFR tracer as it can be directly linked to the ionizing photon rate arising from newly formed massive stars. Using the same physically motivated relations and separating the thermal and nonthermal RC emission through a detailed study of the mid-radio (1-10\,GHz) spectral energy distribution (SED), \cite{cal2} showed that, on galaxy scales, the standard non-radio SFR tracers hold a tighter correlation with the nonthermal RC than with the thermal RC, although the linearity is better achieved with the thermal emission. These authors link the non-linearity, i.e, the excess of the nonthermal emission, to a stronger  magnetic field in galaxies with higher SFR.  
 
 In this paper, we study the cosmic evolution of the thermal and nonthermal RC emission from star-forming galaxies taking into account the evolution of SFR and using the thermal/nonthermal RC--SFR correlations \cite[given by][]{cal1,cal2}. %A similar calibration of SFR was also presented for high-redshift galaxies by \cite{Algera}.  
 The most fundamental questions are if the thermal fraction of the RC emission remains invariant with redshift and how the nonthermal spectrum at mid-radio frequencies of $\simeq$1-10 GHz evolve with cosmic time. %These are key to study the importance of the nonthermal pressure in the ISM energy balance in distant universe. 
 We perform simulations to investigate the detectability of the ISM structures by the proposed SKA surveys. { The} Galaxies selected are like those in the local universe as if they were at higher redshifts. Comparing these predictions with the upcoming real SKA observations will further help to infer any difference in the ISM properties of galaxies in the distant universe.               
The goals of this study are then to investigate 1) the structure of the thermal and nonthermal emission on 1 kpc scales at different redshifts, 2) possible evolution of the thermal fraction and the mid-radio synchrotron spectral index, and 3) the capability of the proposed SKA reference surveys {to detect} the RC emitting structures, {which is} important to address the role of the thermal/nonthermal processes in the evolution of galaxies. }

{Separating the thermal and nonthermal { components of the RC emission} has { historically} been a technical challenge. It basically needs multi-frequency radio observations with proper frequency sampling and consistent sensitivities at different frequencies. Insufficient frequency coverage { requires} pre-assumptions about the nonthermal spectral index \citep[e.g., ][]{klein84}.  Moreover,  because the sensitivity changes naturally with frequency, available separations are limited to only a few bright sources or very coarse angular resolutions in nearby galaxies \citep[e.g.,][]{Westcott}. We developed a separation technique for nearby galaxies that is based on a tracer of the free-free emission, { like} a de-reddened H$\alpha$ map \citep{tab7,tab_a,tab_b}. This method can { separate} the thermal and nonthermal components down to very low surface brightnesses enabling us to study different galactic structures. The present study uses the results of this separation method applied to local galaxies as { initial inputs} to further investigate the ability of the SKA to detect such structures at higher redshifts. }

The paper is organized as follows. After presenting the theoretical framework, we explain the different steps {taken to construct simulated maps} in section~\ref{theory}. The resulting maps of the thermal and nonthermal RC emission and their redshift profiles are presented  in Sect~\ref{res}. We further discuss the evolution of synchrotron spectral indices, the spectral energy distribution, and the compatibility with the SKA surveys in Sect.~\ref{discu} and summarize in Sect.~\ref{summary}.

%%%%%%%%%%%%%%%%%%%%%%%%%%%%%%%

\section{Methodology}\label{theory}

%Several studies show a tight and linear correlation between the thermal radio continuum and SFR and a tight, but slightly non-linear correlation between the nonthermal radio continuum and SFR in galaxies \citep{cal1,cal2,molnar}. On the other hand, 
{ Deep field observations show that galaxies' SFRs evolve with redshift $z$ depending on the stellar mass of galaxies.
We first review this SFR evolution and then obtain relations for the evolution of the thermal and nonthermal RC emission. %, $I^{\rm th}(z,M_{\star})$ and $I^{\rm nt}(z,M_{\star})$. 
These relations are then used to simulate the RC maps of three prototypical galaxies as observed by SKA. We use $H_0 = 67.4 \,\,\textrm{km/s/Mpc}, \Omega_m = 0.31$, and $\Omega_{\Lambda} =  0.68$~\citep{planck} throughout the paper.}   %maps of $I^{\rm th}$(z) and $I^{\rm nt}$(z) for galaxies resembling those in the local and nearby universe and study variations in the thermal fraction and the nonthermal spectral index with redshift. }

\subsection{Evolution of star formation rate}

{Evolution of SFR has been the topic of several observational studies \citep[e.g., ][]{sey,Driver,Madau,Speagle,Scoville}. Star-forming galaxies follow a tight SFR--stellar mass (M$_{\star}$) correlation, with a dispersion of about 0.3 dex forming the so called {\it main sequence} (MS) of galaxies in the SFR--M$_{\star}$ plane \citep{Noeske}. Although the log-log slope of this correlation and its normalization can depend on galaxy sample and the SFR tracer \citep[e.g., ][]{Speagle,Bisigello,Leslie}, a large number of studies find strong evolution out to at least $z=4$ with a roughly linear relation at low masses and an increasing tendency for bending or flattening at high mass and lower redshift~{ \citep[e.g., ][]{Whitaker,Lee,Schreiber,Tom,Leslie}}. A well-presented parametrization of the SFR evolution of the MS galaxies, including all the aforementioned features, was introduced by \cite{Schreiber}.
This was performed by combining deep far-infrared data taken with the Herschel Space Observatory with the {UV to near-infrared} data  of four major extragalactic fields GOODS-North, GOODS-South, UDS, and COSMOS for a mass complete sample of star-forming galaxies. They found that the SFR of galaxies  of all stellar masses (M$_{\star}$) at $z\leq 4$ follow a universal scaling law showing a close-to-linear slope of the log(SFR)-log(M$_{\star}$) relation: }

\be\label{sfrz}
\textrm{log}_{10}({\rm SFR_{MS}}[\rm M_\odot yr^{-1}]) &=& m - m_0 + a_0 r\\ 
&&- a_1 [max(0,m - m_1 - a_2 r) ]^2 \nonumber
\ee
with $r \equiv \textrm{log} (1+z)$ and $m \equiv \textrm{log}(\rm M_{\star}/10^9 M_{\odot})$. Other parameters in the above relation are $m_0 = 0.5 \pm 0.07, a_0 = 1.5 \pm 0.15, a_1 = 0.3 \pm 0.08, m_1 = 0.36 \pm 0.3$, and $a_2 = 2.5 \pm 0.6$.
Dispersion in this relation is reportedly about 0.3 dex, i.e., at a fixed redshift and stellar mass, about 68\% of star-forming galaxies form stars at a universal rate within a factor 2 \citep{Schreiber}. %This relation given in Eq.~1 is the main basis of the present work but the use of other SFR(z) models are also discussed in section~4.

%We use Eq. (1) to study the evolution of the thermal and the nonthermal radio continuum emission. The usage of other SFR(z) models are further discussed in section~4. 

\subsection{Evolution of thermal and nonthermal emission}\label{evothnon}
The  { RC emission} is often described by a power-law function of frequency $\nu$, { $S_{\nu}\propto \nu^{-\alpha}$, with $S_{\nu}$ the integrated flux density} and $\alpha$ the power-law spectral index. It originates mainly from two different emission mechanisms, the nonthermal synchrotron emission and the thermal free-free emission, { $S_{\nu}=S_{\nu}^{\rm th} + S_{\nu}^{\rm nt}$}. The nonthermal and thermal emission are also power-law functions of $\nu$ with different spectral indices $\alpha_{\rm nt}$ and $\alpha_{\rm th}$, respectively: { $S_{\nu}^{\rm nt} \propto \nu^{-\alpha_{\rm nt}}$ and  $S_{\nu}^{\rm th} \propto \nu^{-\alpha_{\rm th}}$}. 
{ For a galaxy at  {redshift $z \sim 0$ the} relation between the integrated flux density and luminosity is: $L_{\nu} =  S_{\nu}\, 4 \pi \,D^2$, }
%In the local universe, the radio luminosity surface density is related to the intensity by $\Sigma L_{\nu} =  I_{\nu}\, 4 \pi \,D^2$, 
with D the distance, while at high redshifts, taking into account the k-correction of $K(z)=(1+z)^{-(1-\alpha)}$, we { get}\footnote{$\alpha$ is equivalent to $\alpha_{\rm nt}$ for the nonthermal emission and $\alpha_{\rm th}$ for the thermal emission}
\be\label{kcor} 
L_{\nu_1} = \frac{4 \pi \,D_L^2(z)}{(1+z)^{1-\alpha}}\, \left(\frac{\nu_1}{\nu_2}\right)^{-\alpha} S_{\nu_2},
\ee
{ where $L_{\nu_1} $ is the radio luminosity at rest-frame requency $\nu_1$ and $S_{\nu_2}$ is the integrated flux density observed at frequency $\nu_2$} and $D_L(z)$ is the luminosity distance at redshift $z$.
%
%\be\label{kcor} 
%\Sigma L_{\nu_1} = \frac{4 \pi \,D_L^2(z)}{(1+z)^{1-\alpha}}\, \left(\frac{\nu_1}{\nu_2}\right)^{-\alpha} I_{\nu_2},
%\ee
%where $\Sigma L_{\nu_1} $ is the  rest-frame radio luminosity density at frequency $\nu_1$ and $I_{\nu_2}$ is the observed intensity at frequency $\nu_2$ and $D_L(z)$ is the luminosity distance at redshift $z$, 
%\be\label{ldis}
%D_L (z)= (1+z) \frac{c}{H_0} \int_0^z \frac{dz'}{\sqrt{\Omega_m (1+z')^3 + \Omega_{\Lambda}}},
%\ee
%with $c= 3 \times 10^5 \,\,\textrm{km/s}, H_0 = 67.4 \,\,\textrm{km/s/Mpc}, \Omega_m = 0.31$, and $\Omega_{\Lambda} =  0.68$~\citep{planck}.

\cite{cal1} obtained the following calibration relations between the SFR and the thermal and nonthermal radio luminosities, $L_{\nu}^{th}$ and $L_{\nu}^{nt}$, using a Kroupa IMF~{ \citep{imf}} and assuming a solar metallicity and continuous star formation over a timescale of $\sim$100 Myr ,
\be\label{calth}
\left( \frac{\rm SFR_{\nu_1}^{th}}{\rm M_{\odot} yr^{-1}} \right) &=& 4.6 \times 10^{-28} \left(\frac{T_e}{10^4 {\rm K}}\right)^{-0.45}  \\ 
&& \times \left(\frac{\nu_1}{\rm GHz}\right)^{\alpha_{\rm th}}  \left( \frac{L_{\nu_1}^{th}}{\rm erg\,\, s^{-1}\,Hz^{-1}} \right),\nonumber
\ee

\be\label{calnt}
\left( \frac{\rm SFR_{\nu_1}^{nt}}{\rm M_{\odot} yr^{-1}} \right) &=& 6.64 \times 10^{-29}  \left(\frac{\nu_1}{\rm GHz}\right)^{\alpha_{nt}} \\ 
&& \times \left(\frac{L_{\nu_1}^{nt}}{\rm erg\, s^{-1} Hz^{-1}} \right),\nonumber 
\ee
with $T_e$  the electron temperature. A calibration between the supernova rate and the SFR was applied for Eq.~\ref{calnt} using the empirical relations between supernova rate and nonthermal spectral luminosity of the Milky Way. { Hence, $S_{\nu_2}$ in Eq.~\ref{kcor} evolves as
\be\label{S}
S_{\nu_2}(z)= S(0)\,\, \frac{\rm SFR_{\nu_{1}}(z)}{\rm SFR(0)}  \,\, \frac{D^2}{D_L^2} \,\,(1+z)^{1-\alpha},\ee
with $S(0)$, $\rm SFR(0)$, and $D$ being the actual integrated flux density, star formation rate, and distance to the $z\sim0$ local Universe galaxies \footnote{Note that the rest-frame frequency is equal to the observed frequency at z=0 and $S_{\nu_{1}}(0)=S_{\nu_{2}}(0)=S(0)$.} used in our simulation ($S(0)=\frac{L_{\nu_1}(0)}{4\pi\,D^2}\propto \,{\rm SFR(0)}$) following Eq.~\ref{calth} for the thermal emission and Eq.~\ref{calnt} for the nonthermal emission\footnote{ It is clear that ${\rm SFR^{th}}={\rm SFR^{nt}}={\rm SFR}$ }.
 }

{ Similarly, to obtain the evolution of the surface brightness, $I_{\nu_2}(z)$, the evolution of the luminosity density at the rest-frame frequency, $\Sigma L_{\nu_1}$, must be known,
\be\label{kcor1} 
\Sigma L_{\nu_1} = \frac{4 \pi \,D_L^2(z)}{(1+z)^{1-\alpha}}\, \left(\frac{\nu_1}{\nu_2}\right)^{-\alpha} I_{\nu_2},
\ee
The same relations as in Eqs.~\ref{calth}~and~\ref{calnt} hold between  $\Sigma\,L_{\nu_1}$ and the surface density of the star formation rate, $\Sigma$\,SFR$_{\nu_1}$. However, it is not that straightforward to directly link the evolution of $\Sigma$\,SFR$_{\nu_1}$ to that of SFR$_{\nu_1}$. }  
 \begin{table*}
\begin{center}
\begin{tabular}{ l c c c c c c c c l}
  \hline 
  Galaxy & $D^{a}$& $M_{\star}$$^b$  & { SFR}$^{c}$ &   { SFR(0)$^{d}$}&   { SFR(0.5)$^{e}$}&   { SFR(1)$^{f}$}& { SFR(2)$^{g}$} & { SFR(3)$^{h}$} &$\alpha_{\rm nt}(0)^{j}$   \\
              & Mpc      &  $M_{\odot}$       & ${\rm M_{\odot}yr^{-1}}$ & ${\rm M_{\odot}yr^{-1}}$ & ${\rm M_{\odot}yr^{-1}}$ & ${\rm M_{\odot}yr^{-1}}$&  ${\rm M_{\odot}yr^{-1}}$ & ${\rm M_{\odot}yr^{-1}}$&                                           \\     
  \hline
% M31 & 0.76 & $10^{10.02}$ &$0.95$ &$0.30$& 2.45 &$3.8\times 10^{-4}$ & Tabatabaei et al. 2010 \\ %3.4 \times 10^{-3} 
 M51 & 7.60    & $10^{10.56}$ & 3.1 & 4.2 & 14.2 & 28.3 & 59.7 & 91.8 & $0.95\pm\,0.09$  \\
 N6946 &6.80 &$10^{9.96}$  & 3.2 & 2.3 & 5.2 & 8.2 & 15.0 & 23.1 &$0.80\pm\,0.06$ \\
 M33 & 0.84    & $10^{9.54}$ & 0.5 & 1.1& 2.0 & 3.1 & 5.7 & 8.8 &$0.86\pm\,0.08$\\ %0.74\times 10^{-3}
 \hline
\end{tabular}
\caption{Galaxy sample and their properties. $(a)$ Distance to M51: \citet{Ciardullo}, NGC6946:  \citet{Karachentsev20}, and M33: \citet{Freedman}.$(b)$ Stellar mass of M51:  \citet{Karachentsev}, NGC6946: \citet{kenn}, and M33: \citet{Corbelli}.  $(c)$ SFR estimates from literature for M51 \& NGC6946: \citet{leroy_08}, M33: \citet{verl}. $(d)-(h)$ SFR of the main sequence galaxies with $M_{\star}$ of M51, NGC6946, and M33 given by Eq.~(\ref{sfrz}) at $z=0,~0.5,~1,~2$ and $3$ respectively.  $(j)$ Nonthermal spectral index of M51 \& NGC6946: \citet{cal2}, and M33: \citet{Berkhuijsen}.\label{table}}
\end{center}
\end{table*} 
{Several deep field optical studies find that the stellar size of galaxies evolves with redshift { \citep[e.g., ][]{Ferguson,Mosleh,Ono,van,Shibuya,sue}}. These studies also show that the evolution becomes shallower at longer wavelengths. In the radio domain, this subject awaits multi-wavelength observations and it is still under debate (see Sect.\ref{struc}). Therefore, { accounting} for different possibilities, we consider two extreme scenarios}: (1) The SFR and its surface density evolve similarly with $z$, i.e., 
\be\label{c1}
\frac{\Sigma\,\textrm{SFR}(z)}{\Sigma \, \textrm{SFR}(0)}= \frac{\rm SFR(z)}{ \rm SFR(0)},
\ee
with $\rm SFR(0)$ and $\Sigma \, \textrm{SFR}(0)$ the corresponding rates at z=0. In case (2), the surface density of star formation rate changes additionally due to the evolution of the galaxies' effective radii\footnote{The effective radius $r_{e}$ is defined as the semi-major axis of the ellipse that contains half of the total flux of the best-fitting S\'ersic model.}, $r_e$ as  $\Sigma\, \textrm{SFR}\propto\, \rm SFR/r_{e}^2$.   
The size evolution of galaxies is given by $r_{e}~\propto~(1+z)^{-0.75}$  \citep{van,straat}, thus for case (2) we obtain
\be\label{c2}
\frac{\Sigma \, \textrm{SFR}(z)}{\Sigma \, \textrm{SFR}(0)} &=&  \frac{\textrm{SFR}(z)}{ \textrm{SFR}(0)}\, \,\,(1+z)^{1.5}
\ee
The ${\rm SFR}_{\nu_1}-L_{\nu_1}$ relations given in Eqs.~\ref{calth}~and~\ref{calnt} can be converted to a $\Sigma\,{\rm SFR}_{\nu_1}-I_{\nu_2}$ expression using Eq.~\ref{kcor1}. Then, the evolution of the { surface brightness} with $z$ is obtained for case (1) using Eq.~\ref{c1},
%e
%Using equations \ref{sfrz} to \ref{kcor}, the change in the thermal and nonthemal intensities with $z$ is derived as follows,
\be\label{intensityth}
I_{\nu_2}(z)= I(0)\,\, \frac{\rm SFR_{\nu_{1}}(z)}{\rm SFR(0)}  \,\, \frac{D^2}{D_L^2} \,\,(1+z)^{1-\alpha},\ee
and for case (2) using Eq.~\ref{c2},

\be\label{intensitynth}
I_{\nu_2}(z) = I(0)\,\, \frac{\rm SFR_{\nu_{1}}(z)}{\rm SFR(0)} \,\, \frac{D^2}{D_L^2}\,\,(1+z)^{2.5-\alpha}, 
\ee
{ with $I(0)$ the surface brightness at $z=0$ and ${\rm SFR(0)}$ the model normalization ({obtained from Eq.~\ref{sfrz} by assuming $z=0$}). For the thermal emission $\alpha=\alpha_{\rm th}$ and for the nonthermal emission  $\alpha=\alpha_{\rm nt}$. We note that the surface brightness of the thermal emission given by Eqs.~\ref{intensityth}~and~\ref{intensitynth} does not depend on $T_e$ provided that  $T_e$ does not vary with redshift.}

The spectral index of the thermal emission is given by the Planck function that is $\alpha_{\rm th}=0.1$ in the optically thin condition, as expected for the ISM at $\geq$ 1 kpc scales.   
The nonthermal spectral index $\alpha_{\rm nt}$ is proportional to the energy index of cosmic ray electrons. It can change depending on the cooling mechanism and interaction of cosmic ray electrons with the ISM~\citep{Longair}. As mentioned in Sect.~\ref{intro}, resolved studies show that $\alpha_{\rm nt}$ is flatter in star-forming regions compared to more quiescent regions of the ISM \citep{tab7,tab_a}. { This is expected as cosmic rays have higher energies closer to star-forming regions and supernova remnants where they are injected}. Intense star formation feedback in the form of relativistic shocks can also re-accelerate cosmic ray electrons, causing a { flatter} energy index. For the KINGFISH galaxy sample,  \citet{cal2} showed that $\alpha_{\rm nt}$ changes with $\Sigma \,\textrm{SFR}$ as $\alpha_{\rm nt}\sim\,(-0.41\pm 0.05)\times ({\rm log}\,\Sigma\,{\rm SFR})$.  Hence, it is expected that $\alpha_{\rm nt}$ changes with redshift due to the { increase in} $\Sigma$\,SFR:
%where, from~\cite{cal2}, we have relation for $\alpha_{nt}$ vs $\Sigma \,\textrm{SFR}$ that is 
\be\label{alphant}
\alpha_{\rm nt}(z) = \alpha_{\rm nt}(0) - (0.41\pm 0.05)\,\, \textrm{log} \,\,\frac{\Sigma \, \textrm{SFR}(z)}{\Sigma \, \textrm{SFR}(0)}, %- 0.0134263 
\ee
with $\alpha_{\rm nt}(0)$ the nonthermal spectral index at $z=0$ and $\Sigma \,\textrm{SFR}$ in ${\rm M}_{\odot} \textrm{yr}^{-1} \textrm{kpc}^{-2}$.

\begin{figure*}
\vspace{0.4cm}
\centering
\textbf{case 1}
\hspace{6cm}
\textbf{case 2}\par\medskip
\includegraphics[width=7.5cm]{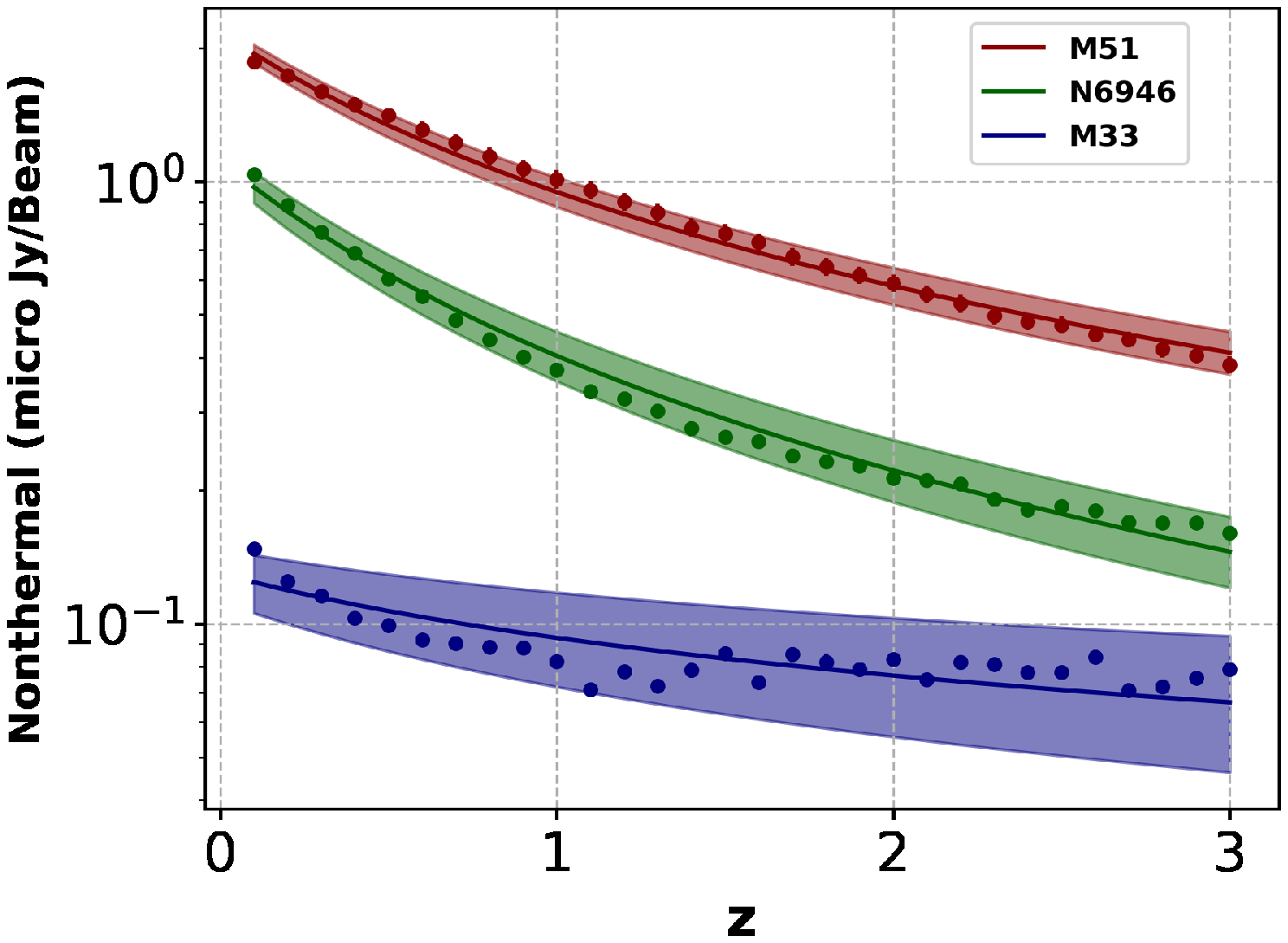}
\hspace{0cm}
\includegraphics[width=7.5cm]{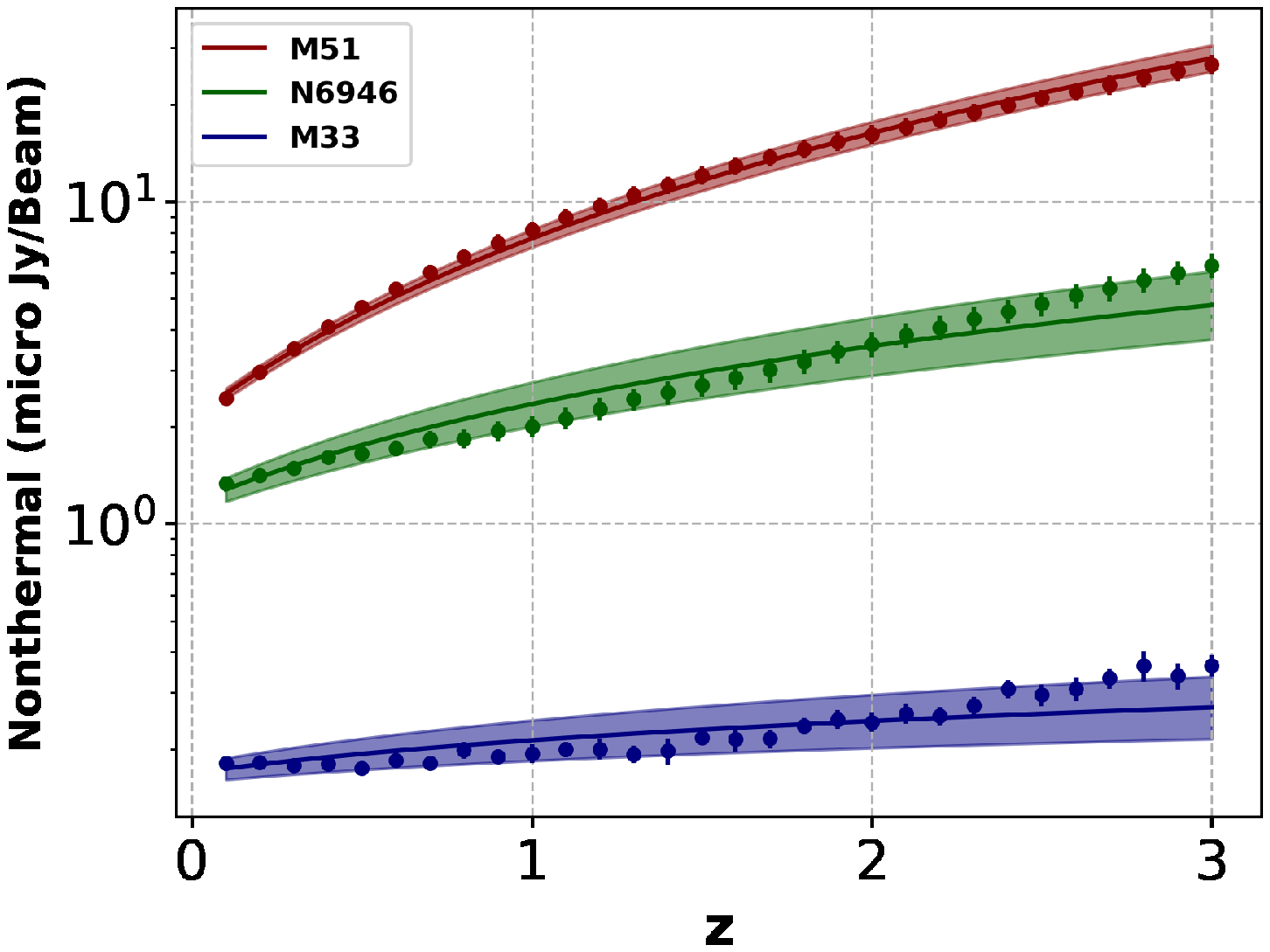}\\
\includegraphics[width=7.5cm]{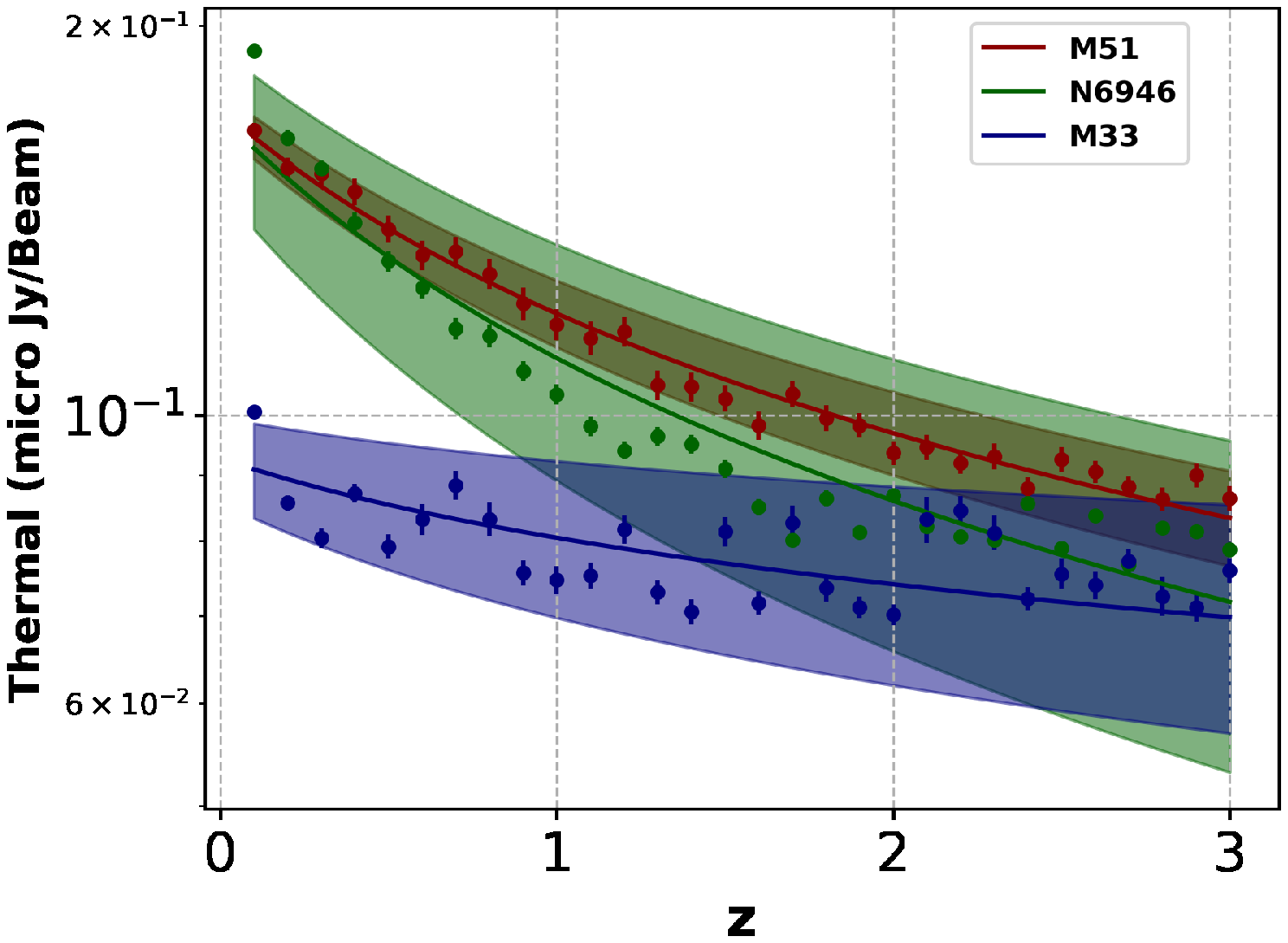}
\hspace{0cm}
\includegraphics[width=7.5cm]{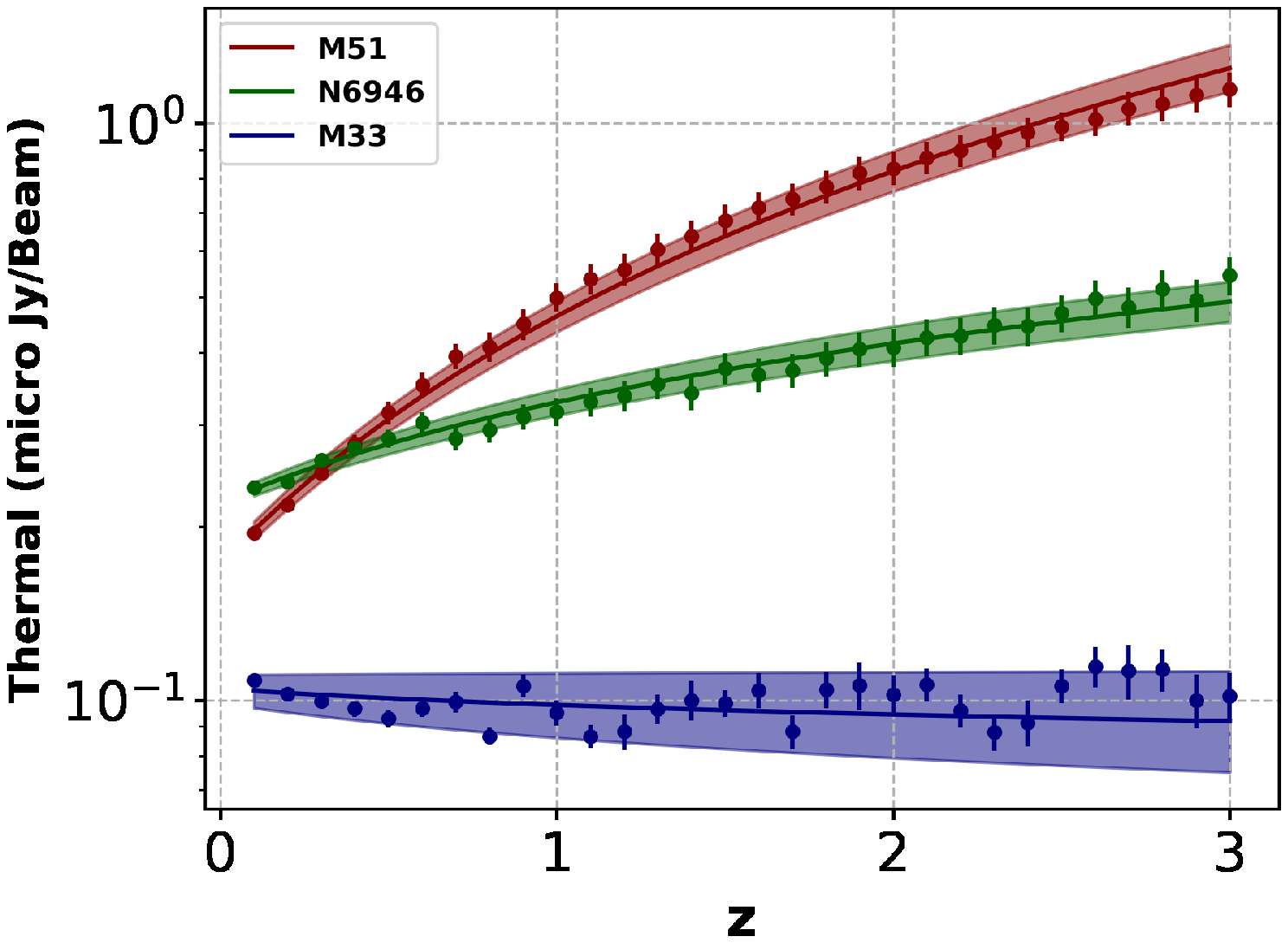}\\
\includegraphics[width=7.5cm]{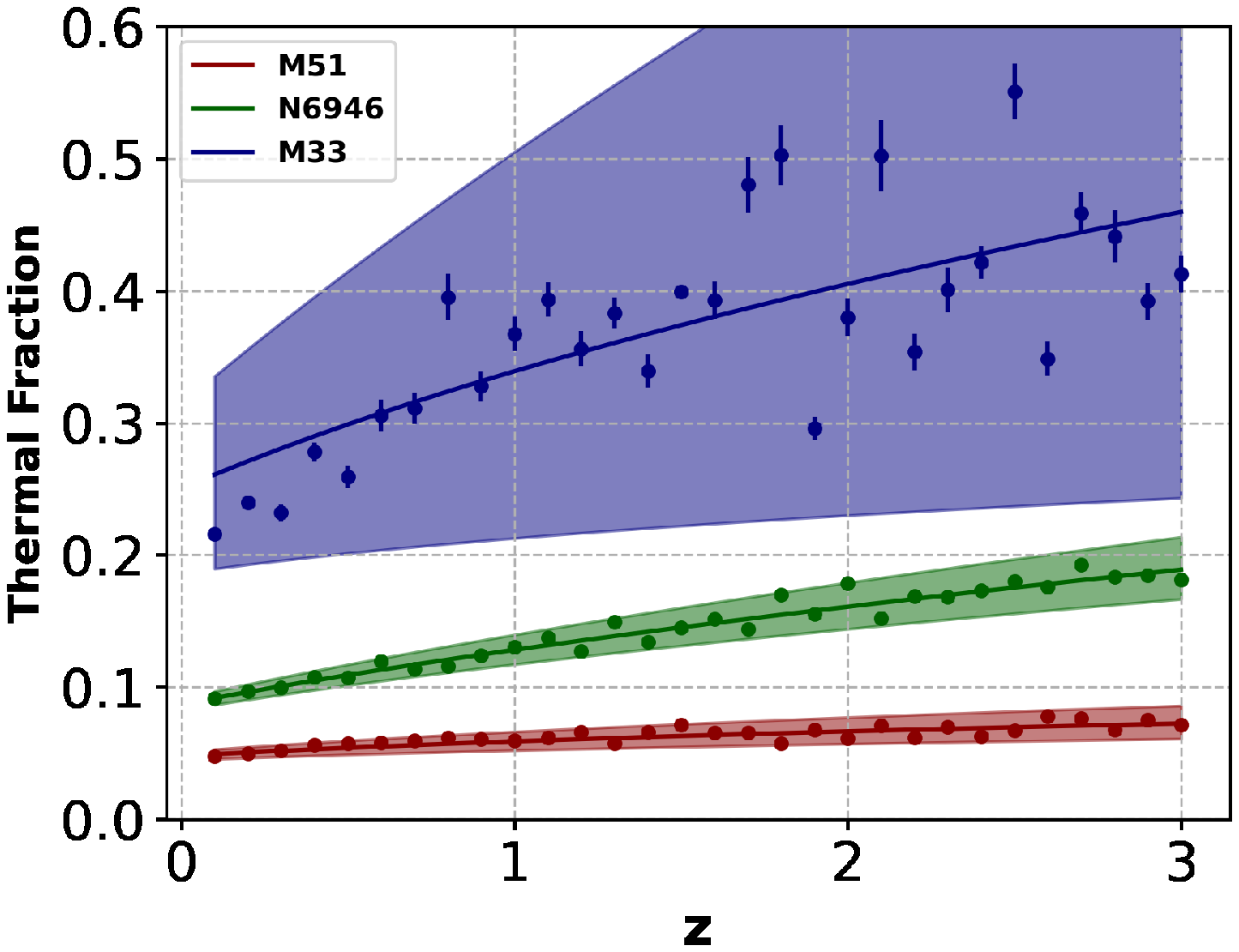}
\hspace{0cm}
\includegraphics[width=7.5cm]{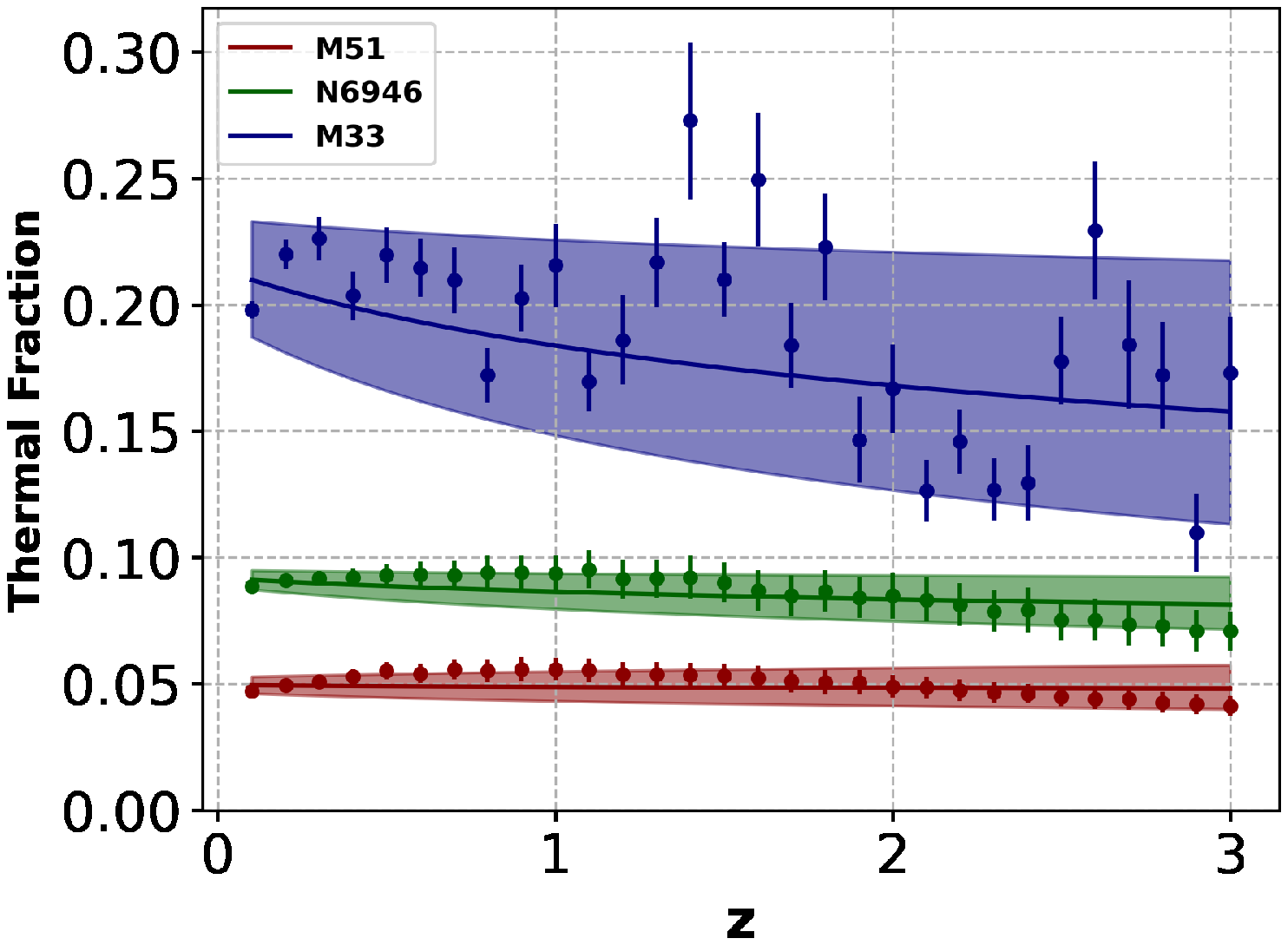}
\vspace{-0.3cm}
\caption{ Redshift evolution of the nonthermal $\langle {\rm I_{\nu_2}^{\rm nt}}\rangle$ ({\it top}) and thermal $\langle {\rm I_{\nu_2}^{\rm th}}\rangle$ ({\it middle}) mean surface brightness and the corresponding thermal fraction, $\langle {\rm f_{\nu_2}^{\rm th}}\rangle$, ({\it bottom}) for case~1 ({\it left}) and case~2 ({\it right}) at the observed frequency of $\nu_2=1.4$\,GHz (equivalent to a rest-frame frequency of 1.6--5.6\,GHz at $0.15\leq z \leq3$). {The mean values of the maps simulated with the UDT-noise sky level ({\it points}) are shown with their standard errors $\sigma/\sqrt n$, where $\sigma$ is the standard deviation and $n$ the number of map pixels}. Curves are fitted functions of the form $a (1+z)^b$ with $a$ and $b$ listed in Table~\ref{param}. {Shaded regions} show the $3\sigma$ uncertainties of the fits.  \label{thernon} }
\end{figure*}

{

\subsection{Input data for the simulations}\label{ini}
Eqs.~\ref{intensityth}  and~\ref{intensitynth} allow us to predict the radio {  surface brightness}
%morphology 
of present-day galaxies when placed at higher redshift, provided their $z\sim0$ RC surface brightness distribution is known.  %The nearby galaxies selected and their data used as input data at $z=0$ ($I(0)$ and $\alpha_{\rm nt}(0)$ in Eqs.~8~to~10).
Therefore, we select nearby star-forming galaxies whose RC properties are well-studied with current radio telescopes. These are M51, NGC6946, and M33 which lie within 0.4 dex of the main sequence (Table~\ref{table}). {Their recent SFR traced by a combination of the far-UV and infrared emission are comparable with those obtained for the main sequence galaxies (Eq.~\ref{sfrz}) at $z=0$ taking into account their uncertainties  \citep[about 50\% for hybrid SFR calibrations,][]{leroy_08}}. The galaxies selected also have decomposed maps of their thermal and nonthermal RC emission sensitive to all ISM structures down to sub-kpc scales.  The data of these prototypical star-forming galaxies used to obtain $I^{\rm th}(0)$ and $I^{\rm nt}(0)$ are described in the following. %\citep[][]{tab_a,tab_b,tab7}

{\large{ NGC\,5194 (M\,51)--}} is a classical grand-design spiral galaxy of Hubble type Sbc. It represents the most massive galaxy in our sample. Its iconic spiral arms are also bright in the radio. %Having a molecular gas dominated ISM \citep[][in their inner 10\,kpc radius]{Scov},
M51 emits stronger RC in the north than in the south {particularly} at lower frequencies \citep{Mulcahy}, indicating a nonthermal origin. M51 was observed with the VLA at 1.4~GHz with C-array \citep{nein} and D-array \citep{ho}. These data were re-reduced and combined by \cite{flet_11} resulting in a map sensitive to emission on all scales down to a beam size of 15\arcsec~equivalent to a physical scale of $\simeq$0.55\,kpc. This map was decomposed into the thermal and nonthermal components using the Thermal Radio Template technique {\citep[TRT, ][see also Appendix]{tab7,tab_a}}. This method uses no prior assumptions about the RC spectrum
 as mentioned in Sect.~1. 
	
{\large{ NGC\,6946--}} is a late-type Scd spiral galaxy.  It is less massive than M51 but has a higher specific SFR.  The distribution of star formation is asymmetric with one prominent spiral arm in the north-east. Its spiral arms are generally not as prominent as those of M51 indicating weaker density waves~\citep{cedr}.   NGC\,6946 has been extensively studied in the radio \citep[see][and references therein]{Beck_07}. %Apart from star forming regions, the radio continuum emission is bright in the inner disk and also in its magnetized inter arms \citep{Beck_96a}. 
This study uses the combined VLA C- and D- array observations at 1.4~GHz \citep{Beck_91} at an angular resolution of 15\arcsec~(equivalent to a physical scale of $\simeq$0.5\,kpc at its 6.8\,Mpc distance). The thermal/nonthermal decomposition of this map was presented by  \cite{tab_a} based on the TRT method at $\simeq$0.6\,kpc linear resolution.
	
{\large{ NGC\,598 (M\,33)--}} {is a low-surface brightness Scd spiral galaxy with a flocculent structure. Residing between the dwarfs and normal spirals \citep[e.g., ][]{shao,muller}, galaxies like M33 are important to assess} capabilities of the SKA surveys in detecting the ISM in low-mass, low-surface brightness galaxies at high-redshifts\footnote{Lower mass dwarf galaxies are not considered in this study as they are hardly detectable at high-z with the proposed SKA1-MID surveys (see Sect.~4.3).}. M33 was observed with the VLA at 1.4~GHz in D-array \citep{tab7a}. As the VLA primary beam of 30\arcmin~(at 1.4~GHz) is smaller than the apparent galaxy size ($\simeq$70\arcmin), a mosaic of 12 pointings  was used to cover the entire galaxy. The final VLA map has an angular resolution of 51\arcsec~equivalent to a physical scale of $\simeq$0.2\,kpc at its 840\,kpc distance. To  recover the extended emission or correct for missing short spacing, the VLA map was combined with the Effelsberg 100-m observations at the same frequency~\citep{tab7a}. We presented the TRT thermal/nonthermal separation in \cite{tab_b}. 
	
The thermal and nonthermal emission obtained for these galaxies\footnote{The maps are shown in the first rows of Figs.~\ref{mapM51c12}, \ref{mapN6946c12} and \ref{mapM33c12}.} serve as the input data $I^{\rm th}(0)$ and $I^{\rm nt}(0)$ in Eqs,~(\ref{intensityth})~and~(\ref{intensitynth}) to obtain the surface brightness maps at other redshifts taking into account the geometrical effects.  
\begin{table*}
	\begin{center}
		\begin{tabular}{ l c c c c c c c c}
			\hline			
			Galaxy& $z$ & $\langle {\rm I_{\nu_2}^{\rm nt}}\rangle$ &$\langle {\rm I_{\nu_2}^{\rm th}}\rangle$ &$\langle {\rm I_{\nu_2}^{\rm tot}}\rangle$ & $\langle {\rm f_{\nu_2}^{\rm th}}\rangle$  & $\langle{{\rm S/N}_{\rm UDT}}\rangle$  & $\langle{{\rm S/N}_{\rm DT}}\rangle$& $\langle{{\rm S/N}_{\rm WT}}\rangle$\\
			\hline
			M\,51& & && &&  &&\\
			& 0.15& 1790$\pm$39  & 155$\pm$3 & 1945$\pm$42 & 0.048$\pm$0.001 &  43 &11 &2\\
			& 0.3&   1600$\pm$51  & 154$\pm$3 & 1754$\pm$54 & 0.051$\pm$0.002 &  39 &10& 2\\
			& 0.5&   1409$\pm$55  & 139$\pm$3 & 1548$\pm$58 & 0.057$\pm$0.002 &  34 & 8&2\\
			& 1&      1011$\pm$47  & 117$\pm$3 & 1058$\pm$50 & 0.059$\pm$0.003 & 24 & 6& 1\\
			& 2&      589$\pm$27  & 93$\pm$2 & 682$\pm$29 & 0.061$\pm$0.002 &  15 &4& $<1$\\
			\hline
			NGC\,6946&& && && & & \\
			& 0.15& 957$\pm$18  & 177$\pm$2 & 1134$\pm$20 & 0.095$\pm$0.002& 23 &6&$1$\\
			& 0.3&   769$\pm$21  & 155$\pm$3 & 924$\pm$24 & 0.099$\pm$0.002 &  18 &4&$<1$\\
			& 0.5&   603$\pm$21  & 131$\pm$2 &  734$\pm$23 & 0.106$\pm$0.003 & 14 & 4&$<1$\\
			& 1&      374$\pm$14  & 104$\pm$2 & 478$\pm$16 & 0.130$\pm$0.004 & 9 &2&$<1$\\
			& 2&      214$\pm$7  & 87$\pm$1    & 301$\pm$8    & 0.178$\pm$0.004 & 5 &1&$<1$\\ 
			\hline
			M\,33& & & && & && \\
			& 0.15&  133$\pm$2  &   89$\pm$1   & 222$\pm$3 & 0.224$\pm$0.004  &  4 &$<1$&$<1$ \\
			& 0.3&    115$\pm$2 &    80$\pm$1   & 195$\pm$3 & 0.231$\pm$0.006 &  3 &$<1$&$<1$ \\
			& 0.5&    99$\pm$3 &    79$\pm$2   & 178$\pm$5 & 0.259$\pm$0.009 &  2 &$<1$ & $<1$\\
			& 1&       82$\pm$2 &    74$\pm$2   & 156$\pm$4  & 0.367$\pm$0.013  &  2&$<1$& $<1$\\
			& 2&       83$\pm$2 &    70$\pm$1   & 153$\pm$3 & 0.380$\pm$0.014  &  1&$<1$&$<1$ \\ 
			\hline
		\end{tabular}
		\caption{Mean  surface brightness of the nonthermal ($\langle {\rm I_{\nu_2}^{\rm nt}}\rangle$), thermal ($\langle {\rm I_{\nu_2}^{\rm th}}\rangle$), and total ($\langle {\rm I_{\nu_2}^{\rm tot}}\rangle$) radio continuum  maps simulated with the UDT-noise sky level in unites of nJy ($10^{-9}$Jy)/beam  at 5 selected redshifts for case (1) (no radio size evolution) at the observed frequency of 1.4\,GHz. Also listed are the mean thermal fraction ($\langle {\rm f_{\nu_2}^{\rm th}}\rangle$) and the signal-to-noise ratio (see Sect.~\ref{ska}) for the SKAI-MID band2 UDT  ($\langle{{\rm S/N}_{\rm UDT}}\rangle$), DT ($\langle{{\rm S/N}_{\rm DT}}\rangle$), and WT ($\langle{{\rm S/N}_{\rm WT}}\rangle$).}\label{table2} %$\textrm{SFR}_0$ from model normalization, }
	\end{center}
\end{table*}

\begin{table*}
	\begin{center}
		\begin{tabular}{ l c c c c c c c c}
			\hline			
			Galaxy&  $z$ & $\langle{I_{\nu_2}^{\rm nt}}\rangle$ & $\langle{I_{\nu_2}^{\rm th}}\rangle$ &$\langle {\rm I_{\nu_2}^{\rm tot}}\rangle$ & $\langle{f_{\nu_2}^{\rm th}}\rangle$  & $\langle{{\rm S/N}_{\rm UDT}}\rangle$  & $\langle{{\rm S/N}_{\rm DT}}\rangle$& $\langle{{\rm S/N}_{\rm WT}}\rangle$\\
			\hline
			M\,51& & && && & &\\
			&0.15& 2707$\pm$64  & 206$\pm$4 & 2913$\pm$68      & 0.048$\pm$ 0.001 &  64 &16&3\\
			& 0.3&   3497$\pm$131  & 248$\pm$7 & 3745$\pm$138  & 0.050$\pm$0.002 &  80 &20& 4\\
			& 0.5&   4704$\pm$231  & 315$\pm$14 & 5019$\pm$245 & 0.055$\pm$0.003 &  103& 26 &5\\
			& 1&      8194$\pm$479  & 499$\pm$28 & 8693$\pm$507 & 0.056$\pm$0.005 & 174 &43& 9\\
			&&      16202$\pm$1063  & 835$\pm$54 & 17037$\pm$1117 & 0.049$\pm$0.005 &  341 &85& 17\\
			\hline
			NGC\,6946&& && && & & \\
			&0.15&  1372$\pm$29  &  240$\pm$4 & 1612$\pm$33       & 0.089$\pm$0.002  & 34 &8&1\\
			&0.3&   1485$\pm$54  & 261$\pm$6 & 1746$\pm$60     & 0.092$\pm$0.003 &  37 &9&2\\
			&0.5&   1650$\pm$86  & 284$\pm$10 & 1934$\pm$96 & 0.093$\pm$0.004 & 40 & 10 &2\\
			&1&      2006$\pm$152  & 316$\pm$17 & 2322$\pm$169 & 0.094$\pm$0.07 & 50 &12&3\\
			&2&      3601$\pm$320 & 409$\pm$31 & 4010$\pm$351 & 0.085$\pm$0.009 & 83 &21&4\\  
			\hline
			M\,33&& && &  & & &\\
			& 0.15&  179$\pm$3  &   106$\pm$1   & 285$\pm$4 & 0.195$\pm$ 0.004   &  5 &1& $<1$\\
			& 0.3&    178$\pm$6 &    100$\pm$3   & 278$\pm$9 & 0.226$\pm$0.009 &  5 &1&$<1$\\
			& 0.5&    174$\pm$8 &    93$\pm$3   & 267$\pm$11 & 0.220$\pm$0.011 &  5 &1 & $<1$\\
			&1&       193$\pm$13 &    95$\pm$4   & 288$\pm$17  & 0.216$\pm$0.005  &  6 &1&$<1$ \\
			&2&       242$\pm$16 &    102$\pm$5   & 344$\pm$21 & 0.167$\pm$0.005  &  7 &2&$<1$ \\
			\hline
		\end{tabular}
		\caption{ Same as Table~\ref{table2} for case (2) (radio size evolution with $z$). }\label{table3} %$\textrm{SFR}_0$ from model normalization, }
	\end{center}
\end{table*}

\begin{table*}
	\begin{center}
		\begin{tabular}{ l c c c c c}
			\hline
			Galaxy           &  Component           &  $a^{(1)}$        & $b^{(1)}$  & $a^{(2)}$ & $b^{(2)}$\\
			\hline
			&     $\langle {\rm I_{\nu_2}^{\rm nt}}\rangle$       &      2.18$\pm$0.09           &     -1.21$\pm$0.05           &      2.12 $\pm$0.07         &      1.86$\pm$0.04          \\
			M\,51           &    $\langle {\rm I_{\nu_2}^{\rm th}}\rangle$        &      0.17$\pm$0.006          &    -0.52$\pm$0.04           &       0.17$\pm$0.005       &       1.43$\pm$0.05       \\
			&     $\langle{f_{\nu_2}^{\rm th}}\rangle$       &      0.05$\pm$0.003         &          0.30$\pm$ 0.08        &        0.05 $\pm$0.003        &          -0.02$\pm$0.09       \\
			\hline 
			&     $\langle {\rm I_{\nu_2}^{\rm nt}}\rangle$       &      1.12$\pm$0.08         &     -1.47$\pm$0.08          &       1.16$\pm$0.08          &      1.02$\pm$0.12            \\
			NGC\,6946   &     $\langle {\rm I_{\nu_2}^{\rm th}}\rangle$       &      0.17$\pm$0.02         &        -0.63$\pm$ 0.12       &      0.22$\pm$0.005          &      0.58$\pm$0.04           \\
			&     $\langle{f_{\nu_2}^{\rm th}}\rangle$       &      0.09$\pm$0.004          &          0.56$\pm$0.05         &        0.09$\pm$0.003         &        -0.09$\pm$0.07         \\
			\hline
			&    $\langle {\rm I_{\nu_2}^{\rm nt}}\rangle$        &     0.13$\pm$0.02            &      -0.48$\pm$0.16            &     0.17 $\pm$0.01           &      0.34$\pm$0.11           \\	
			M\,33           &     $\langle {\rm I_{\nu_2}^{\rm th}}\rangle$       &     0.09$\pm$0.007          &      -0.20$\pm$ 0.09        &       0.10$\pm$0.006        &         -0.09$\pm$0.10        \\
			&     $\langle {\rm f_{\nu_2}^{\rm th}}\rangle$       &     0.25$\pm$0.06           &         0.44$\pm$0.25          &        0.21$\pm$0.02        &        -0.22$\pm$0.17         \\
			\hline
		\end{tabular}
		\caption{  Fit parameters $a$ and $b$ of the functional form $a (1+z)^b$ shown in Fig.~\ref{thernon}. The parameters are obtained using a least-square regression. Indices (1) and (2) represent case~1 and case~2, respectively. }\label{param} 
	\end{center}
\end{table*}

\subsection{SKA reference surveys}	\label{srs}
A main goal of this study is to predict the structure of the RC emitting ISM in distant normal star-forming galaxies as observed with the proposed SKA reference surveys. {In phase 1, SKA will offer observations for three of the five bands covering the frequency range 0.35-13.8 GHz (SKA1-MID), specifically band 1 (350-1000 MHz), band 2 (950-1760 MHz) and band 5 (4.6-13.8 GHz).  However, SKA1-MID continuum reference surveys are proposed only for band~2 and band~5  to achieve high-angular resolutions resolving high-redshift galaxies at 0.05\arcsec--0.5\arcsec \citep[][hereafter PS15]{pran}\footnote{Band~1 provides lower angular resolutions of about 0.7\arcsec}.} We investigate the compatibility with the band~2 survey centered at $\simeq$1.4\,GHz, because our input data are mostly available at this frequency.  The band~2 survey is a {\it Three-tiered survey} aiming to observe a set of fields of views (FoVs) on the sky with different depths. The FoVs are about 1000~deg$^2$, 10~deg$^2$, and 1~deg$^2$ reaching a sky level (rms sensitivity) of $1  \mu$Jy,  $0.2 \mu$Jy, and $0.05 \mu$Jy in the  wide tier (WT),  deep tier (DT), and ultra-deep tier (UDT),  respectively. {PS15 calculated the time needed to achieve these depths using the full-bandwidth available. They} assumed an angular resolution of 0.5\arcsec.  However, we note that it will be possible to sample angular scales down to $\sim0.35\arcsec$ in band 2 given the 150\,km maximum baseline of SKA1-MID (see Fig.~10 in the System Baseline description published by the SKA office\footnote{$\rm  http://skacontinuum.pbworks.com/w/file/fetch/119268525/SKA-TEL-SKO-0000308\_ SKA1\_System\_Baseline\_v2\_DescriptionRev01-part-1-signed.pdf$}). At this maximum resolution, the sensitivity is, however, reduced compared to lower resolutions. We also note that the final image resolution depends both on frequency and also the weighting method used by the imaging pipeline.  In this paper, we adopt the smallest angular scale at which the MID array provides maximal sensitivity in band 2, that is $\theta_{\rm SKA}=0.6\arcsec$. At this resolution, the above sky levels will be achieved earlier than the observing time proposed by PS15 (e.g., 3300 hours per pointing for UDT).

\subsubsection{Thermal/nonthermal separation methods with SKA1-MID} \label{trtska}
As mentioned in Sect.~\ref{intro}, mapping the thermal and nonthermal components of the RC emission requires a proper frequency sampling, particularly at the low and high frequency ends, with consistent sensitivities. This becomes more vital for resolved studies addressing the evolution of the thermal and nonthermal surface brightness in distant galaxies. Particularly, band~1 provides an important constrain on the mid-radio SEDs when studying the epoch of maximum star formation rate at $z \approx 2$ (see Sect.~\ref{SED}). Therefore, the currently proposed Band~2~\&~5 surveys will not be sufficient for a radio-based separation in distant galaxies. 
Avoiding any assumption about the radio SED, the TRT technique~\citep{tab7,tab_a} (see Appendix) can be ideally used at a single radio frequency. This will be feasible using deep narrow-band or spectroscopy imaging surveys of recombination lines tracing the free-free emission at resolutions comparable with those of the SKA. %Moreover, the forthcoming observations with the James Webb Telescope along with other facilities (see e.g., PS15) justifies the use of the TRT technique to decompose the components of the RC maps taken with the SKAI-MID reference surveys. 
%The resolution planned in these 

%In terms of resolution, as mentioned earlier, the simulations are performed to match the SKA1-MID design antenna distribution, with a maximum baseline of Bmax=150\,km resulting in a synthesized beam of FWHM$\sim 0.6$''. This configuration is optimized to reach maximum sensitivities in { Band 2}. The $1 \sigma$ rms sensitivities of the {\it Three-tiered survey} are about $0.05 \mu$Jy,  $0.2 \mu$Jy, and $1  \mu$Jy for the ultra deep,  deep, and wide tiers, respectively. 

\subsection{Simulating maps at high redshifts}\label{simu}
The maps of the thermal and nonthermal surface brightness $I_{\nu_2}^{th}(z)$ and $I_{\nu_2}^{nt}(z)$ of M51--, NGC6946--, and M33--like galaxies are simulated {over the redshifts $0<z\leq3$ where the main sequence typically spans the mass range of  galaxies selected} \cite[$\simeq 10^{9.5}-10^{10.6} M_{\sun},$][]{Schreiber}. Depending on redshift, the change in apparent size of a galaxy observed in a fixed FoV of the telescope, i.e.,  the redshift dependence of angular scales, geometrical effects, and instrumental resolution must be taken into account. The simulated maps should also contain noise to mimic real observations.  How we account for these effects is explained in the following.

}
\subsubsection{Geometrical effects}\label{geo}
{ The radio continuum  maps simulated at high redshifts should resemble those that will be observed by SKA with an angular resolution of $\theta_{\rm SKA}$.} Thus, the angular resolutions and pixel sizes of the $I(0)$ maps should be converted to those of the SKA observations. Because the radio measurements are per unit beam area, this is done by first convolving the maps to the angle subtended to redshift $z$ by the linear resolution element $x$ of the nearby galaxy at distance $D$ and { the resulting map is} then resampled.  Considering that { tangent}\,$(\theta) \simeq \theta$ for a small angle,  the  { linear scale element} $x$ for case (1) is given by
\be
x= \theta(z)\,\frac{D_L}{(1+z)^2}= \theta_0 \frac{D}{(1+z_0)^2},
\ee
with $\theta(z)$ and $\theta_0$ the angles subtended by $x$ at redshifts $z$ and $z_0\simeq 0$ in units of radian. 
%For this study,  we set $\theta(z)=\theta_{\rm SKA}$, and obtain,
Hence, the angle representing the physical resolution of the $I(0)$ maps at redshift $z$ is given by
\be
\theta(z)= \theta_0 (1+z)^2 \frac{D}{D_L}.
\ee
For case (2), { $x$ changes with redshift as a result of the galaxy size evolution, thus $x=x_0\, (1+z)^{-0.75}$ with $x_0$ the linear resolution element at $z=0$.} This leads to
 \be
\theta(z)= \theta_0 (1+z)^{1.25} \frac{D}{D_L}.
\ee

Given that $\theta(z)$ represents the original angular resolution at redshift $z$, the maps are then convolved using a Gaussian kernel with the width of $\sqrt{\theta_{\rm SKA}^2 - \theta(z)^2}$ for each case (1) and (2) resulting in maps at the SKA resolution\footnote{ This means that we can start the simulations from the redshifts at which $\theta(z)<\theta_{\rm SKA}$.}. 
%{ We take $\theta_{\rm SKA}=0.6\arcsec$  corresponding to the maximum baseline of B$_{\rm max}=$150\,km in Band~1. This resolution is the smallest angular scale at which the SKA array provides maximal sensitivity in Band~2.} 
%{The SKA1-MID telescope will have a maximal baseline length of 150km, sampling spatial scales of up to $\sim0.35\arcsec$. Here we adopt $\theta_{\rm SKA}=0.6\arcsec$, the smallest angular scale at which the MID array provides maximal sensitivity in band 2.}
These maps are then re-sampled to have 4 pixels per beam at each redshift. { As a result, the number of pixels reduces with redshift and the maps become more unresolved. To keep the observation FoV unchanged, we add more number of pixels as sky to the surrounding of the maps. Therefore, the final maps simulated have the same number of pixels at different redshifts.   }

%This corresponds to a linear resolution $x$ that varies with redshift as $x(z)~=~\theta_{\rm SKA}\,\, D_A(z)/206265$, with $D_A(z) = D_L/(1+z)^2$ the angular distance. Then to simulate the scales at different redshifts, our initial maps of $I_{th}(0)$, $I_{nt}(0)$ were first smoothed to angular scales corresponding to those physical scales at $z=0$, i.e., to $\theta(z)~=~206265\,\,x(z)/D$ in units of aradio continuumsec using Gaussian kernels. These maps were resampled to have 4 pixels per beam after smoothing.} 
%
\subsubsection{Adding noise}\label{noise}
The radio images simulated inherit the background noise from the VLA\footnote{VLA $\&$ Effelsberg for M33} observations. However, as { the noise values are} much smaller than the signal from the galaxy in the original maps, the simulated maps are practically noiseless. It is then important to add background noise due to the SKA observations. According to the anticipated SKA1 science performance \citep{brau}, the image noise depends on the sensitivity of the antennas, correlator efficiency, frequency bandwidth, and integration time. Moreover, as a source of correlated noise, the dirty synthesized beam pattern is determined by the array configuration, the (u,v) coverage of  observation and the data weighting scheme that is employed. The (u,v) coverage and data weighting scheme are given by the observing strategy and are { thus} up to the user.  { The sensitivity of the images is set in accordance with the SKA1-MID reference surveys outlined in Sect.~\ref{srs}.   
%For example, the 1\,$\sigma$ rms sky level in the images taken with the Band 2 ultra-deep survey is about $0.05\,\mu$Jy per $0.6\arcsec$ beam  at the central frequency of $\nu=$1.4$\,$GHz  and a bandwidth of BW = $0.3\nu=0.42\,$GHz after $\sim3300\,$hours observation time per pointing.?????????????? 
We inject the 1\,$\sigma$ rms sky levels of the {\it Three-tiered Band 2 Survey} into our simulated total RC (= thermal\,+\,nonthermal) maps after generating Gaussian noise of those  levels\footnote{using the task numpy.random.normal in Python version 3.0.}.} % to compare them with the  images of the SKA reference surveys (see Sect.\,\ref{ska}).  
A more realistic noise treatment can include other instrumental correlated noise that is unknown at this stage for the SKA. %This noise can affect quantitive structural studies at scales of the angular resolution but not the galaxy-averaged properties presented in this work.  
{ Sky levels injected are the same for the thermal and nonthermal radio maps that is expected if the radio SED method is used to separate these RC components. We note that  using the TRT separation method (Sect.~\ref{trtska}) instead, the sky level in the thermal radio map can differ depending on the map of the free-free emission template (e.g., H$\alpha$ emission).}  %As the original thermal radio maps are obtained independently using the H$\alpha$ observations as a thermal radio tracer \citep{tab7,tab_a,tab_b}, we determine the sky levels of the thermal maps independently. This is particularly useful assuming that the same method will be used to separate the thermal  changing in and nonthermal components of the emission observed with SKA. 
%If using H$\alpha$ emission as a tracer of the free-free emission, the background in the thermal radio maps will depend on the instrument that observed the rest-frame H$\alpha$ emission. We use  the Hubble space telescope Deep Field (HDF) survey\footnote{https://cosmos.astro.caltech.edu/page/hst} to simulate the sky level in these maps. Observations with the NICMOS F160W camera provide a 27.2 mag at $10\,\sigma$ level rms noise in this survey. Hence, the equivalent K-band H$\alpha$ observations at $z=2$ has a sensitivity ($1\sigma$ rms level) of $6.75 \times10^{-3} \mu$Jy per $0.6\arcsec$ beam of the SKA used in generating Gaussian noise and injecting into the simulated thermal maps. 

% Bars show the surface brightness in units of $\mu$Jy/beam. \cite{tab7}

%%%%%%%%%%%%%%%%%%%%%%%%%

\begin{figure*}
\centering
\textbf{case 1}
\hspace{6cm}
\textbf{case 2}\par\medskip
\begin{subfigure}[t]{0.20\textwidth}
    \makebox[0pt][r]{\makebox[30pt]{\raisebox{40pt}{\rotatebox[origin=c]{90}{Input}}}}%
    \includegraphics[width=3.2cm]
    {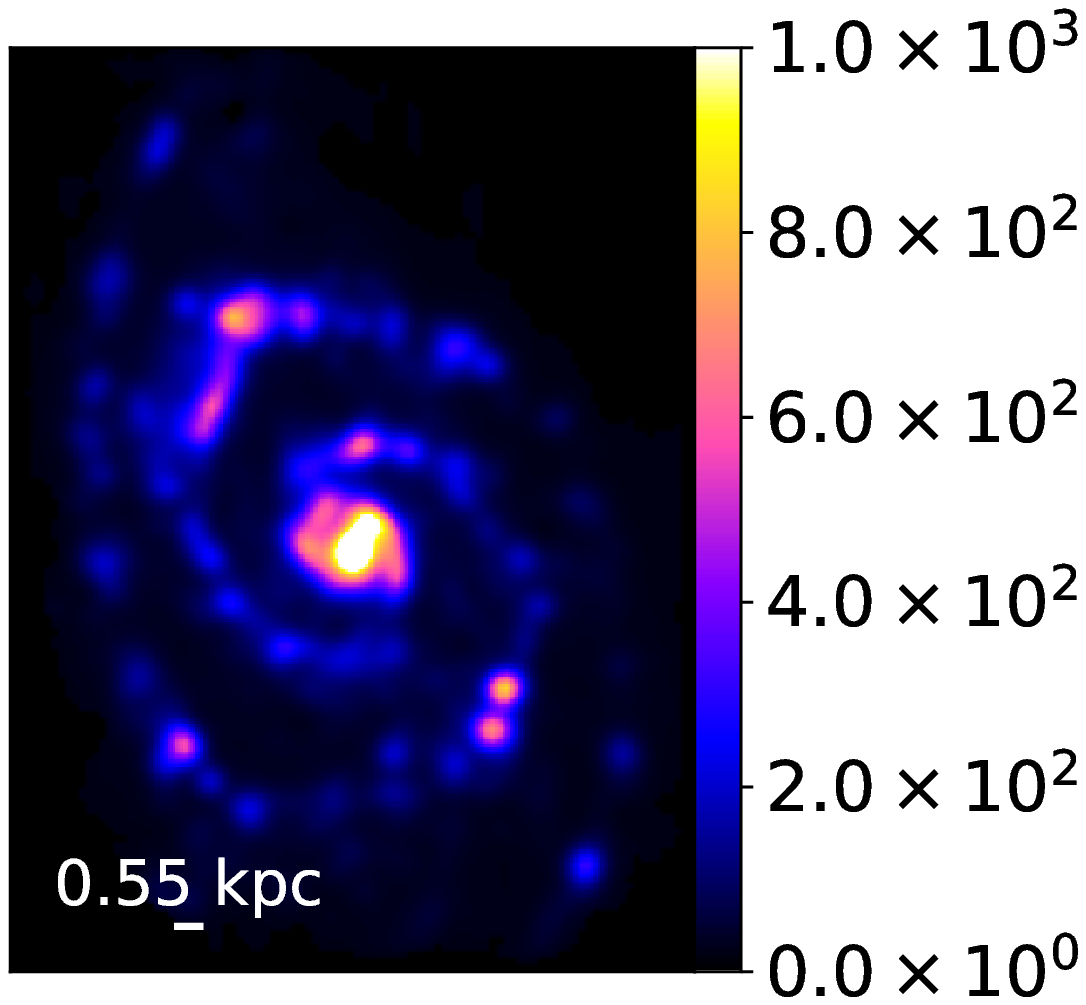}
    \makebox[0pt][r]{\makebox[30pt]{\raisebox{40pt}{\rotatebox[origin=c]{90}{$z=0.15$}}}}%
    \includegraphics[width=3.2cm]
    {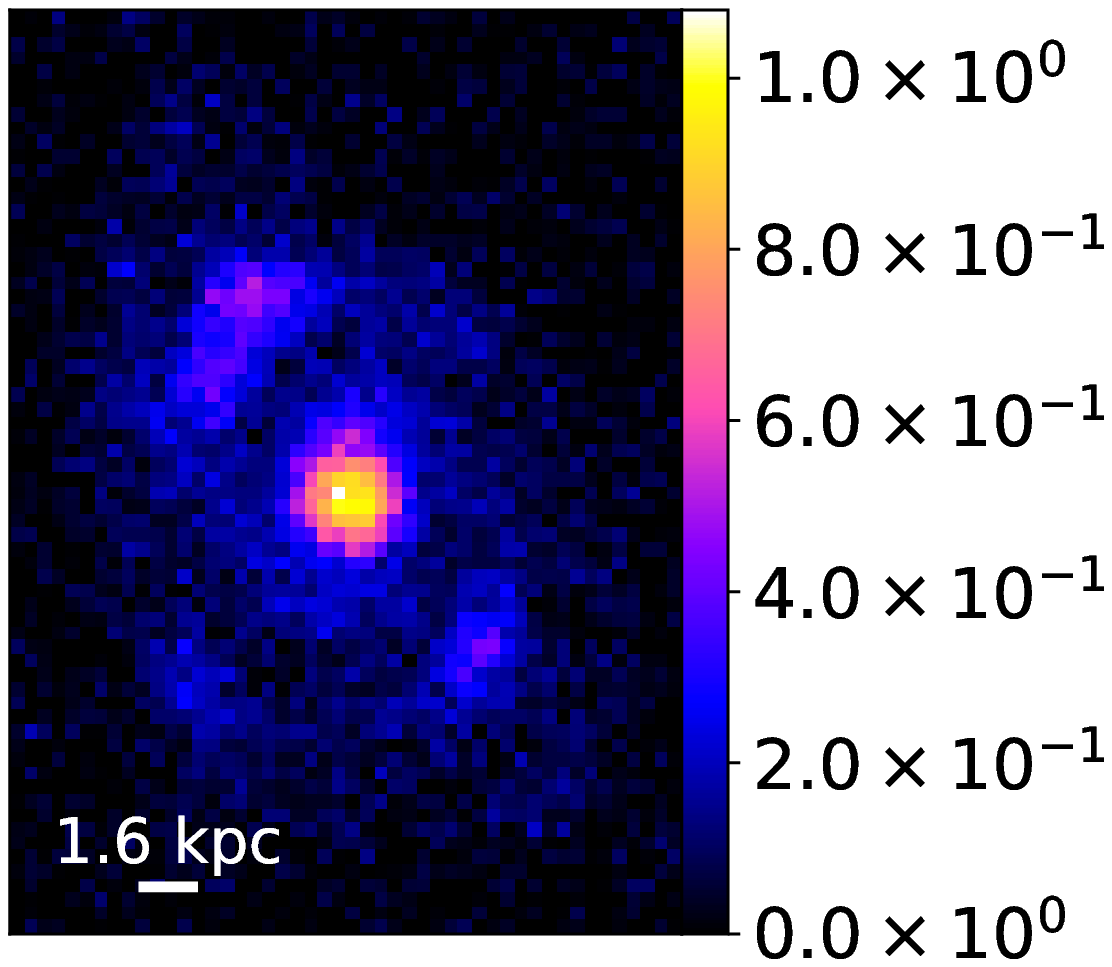}
    \makebox[0pt][r]{\makebox[30pt]{\raisebox{40pt}{\rotatebox[origin=c]{90}{$z=0.3$}}}}%
    \includegraphics[width=3.2cm]
    {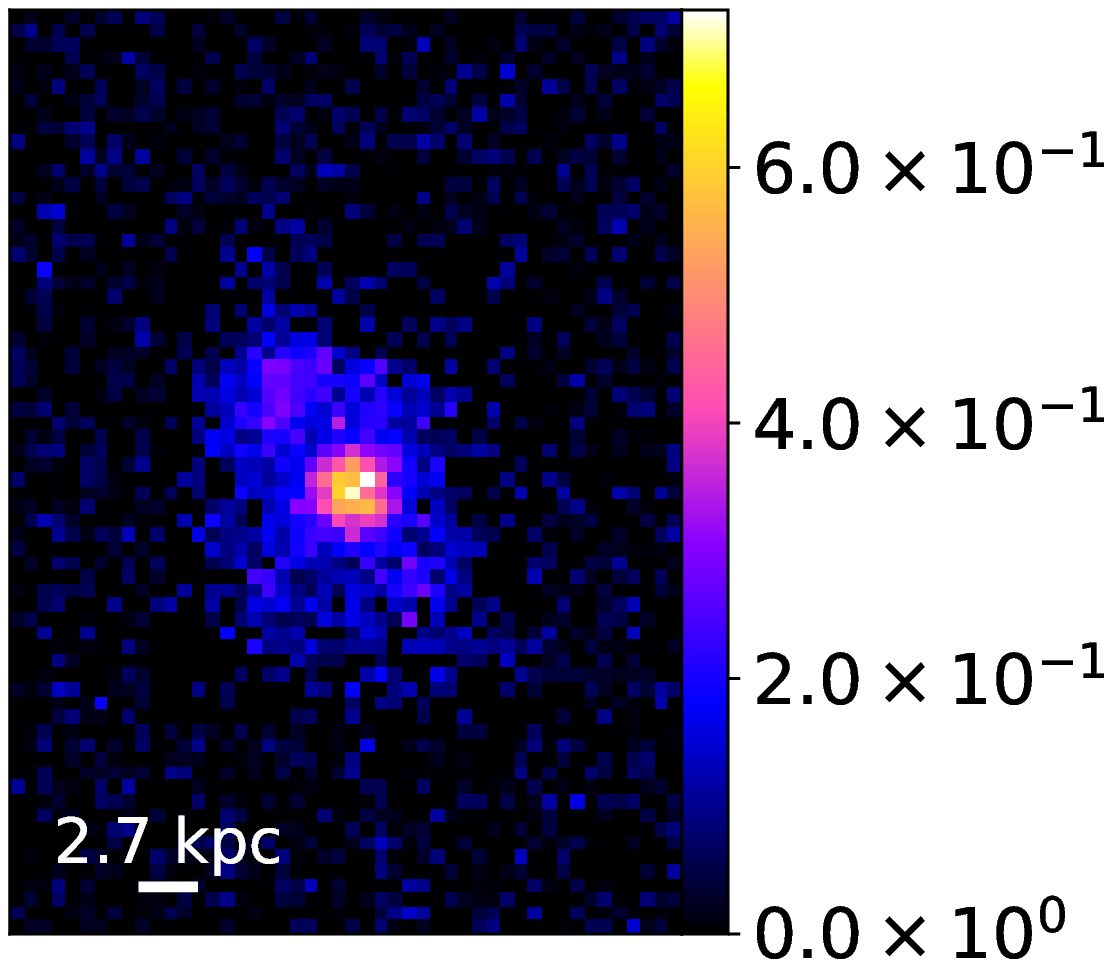}
    \makebox[0pt][r]{\makebox[30pt]{\raisebox{40pt}{\rotatebox[origin=c]{90}{$z=0.5$}}}}%
    \includegraphics[width=3.2cm]
    {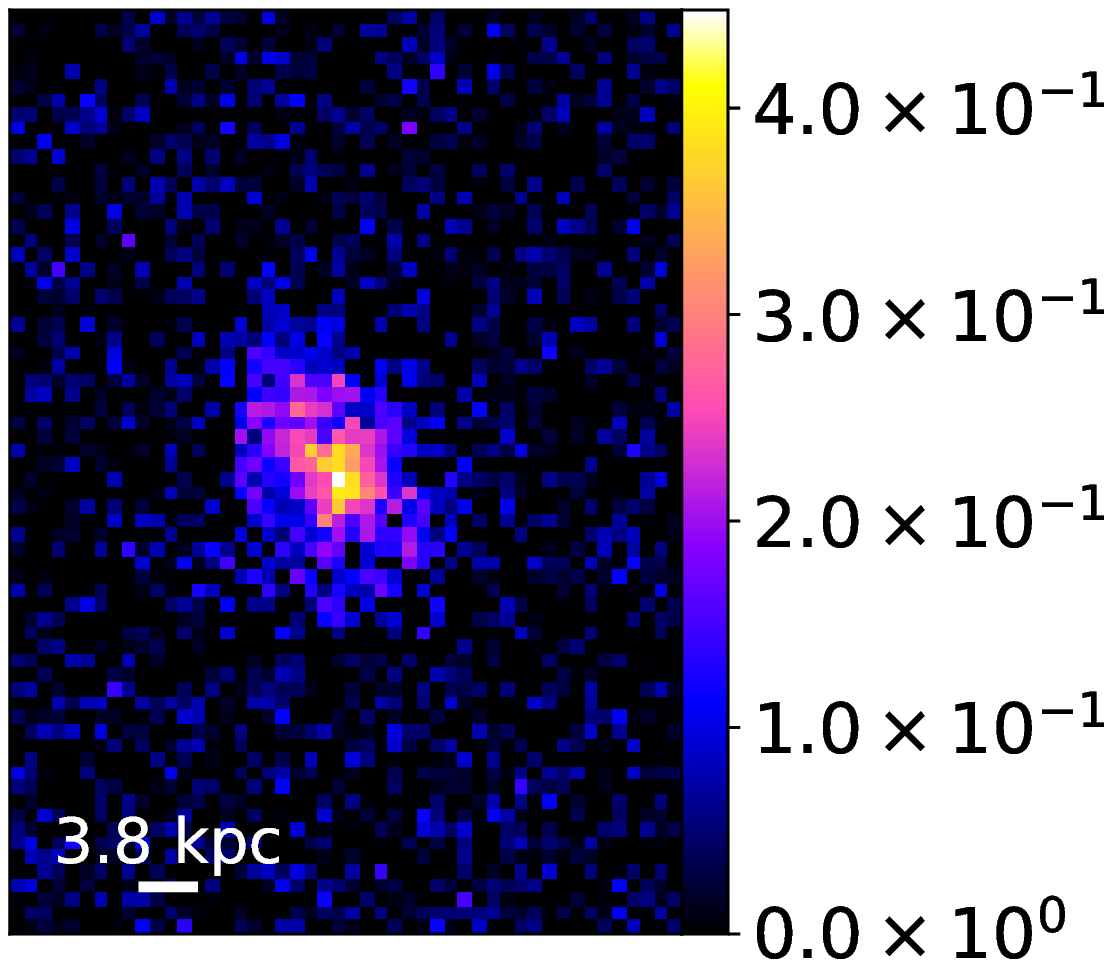}
    \makebox[0pt][r]{\makebox[30pt]{\raisebox{40pt}{\rotatebox[origin=c]{90}{$z=1$}}}}%
    \includegraphics[width=3.2cm]
    {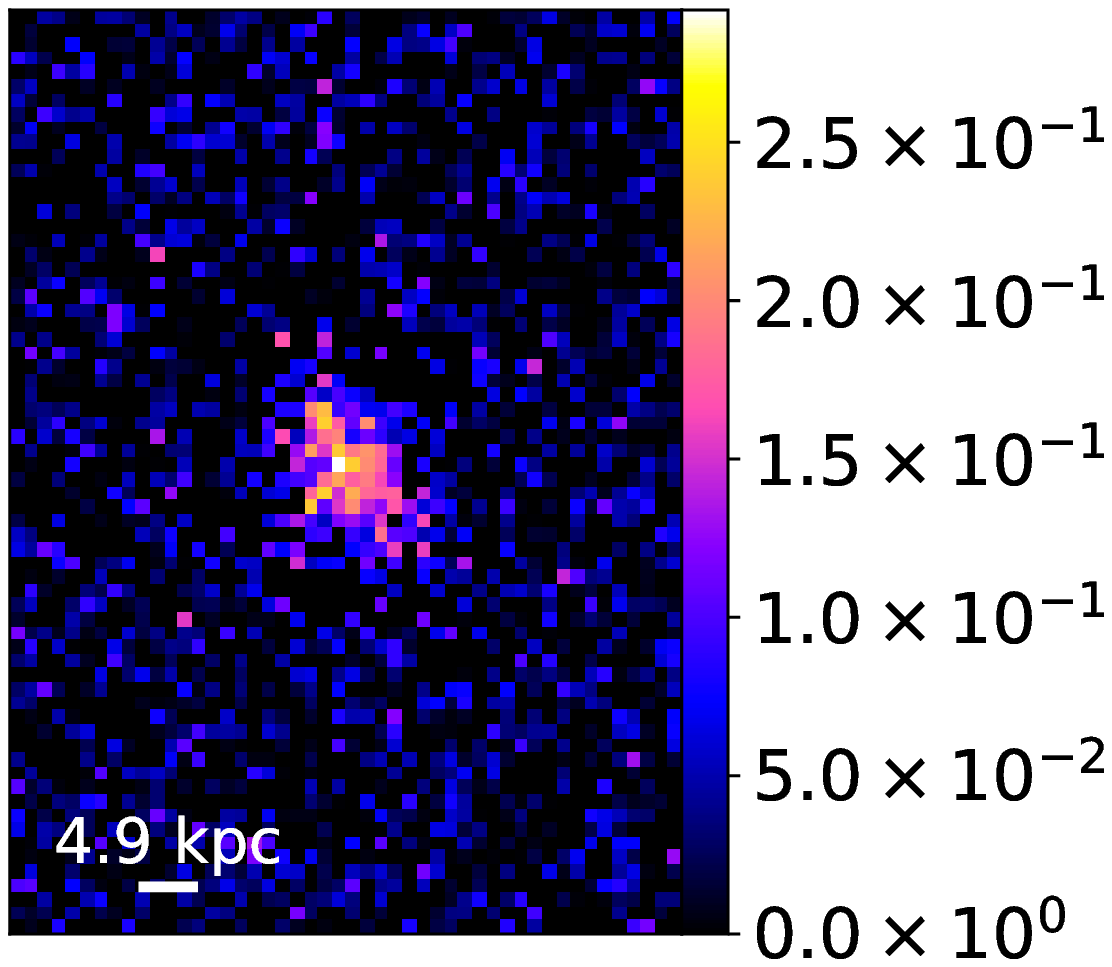}
     \makebox[0pt][r]{\makebox[30pt]{\raisebox{40pt}{\rotatebox[origin=c]{90}{$z=2$}}}}%
    \includegraphics[width=3.2cm]
    {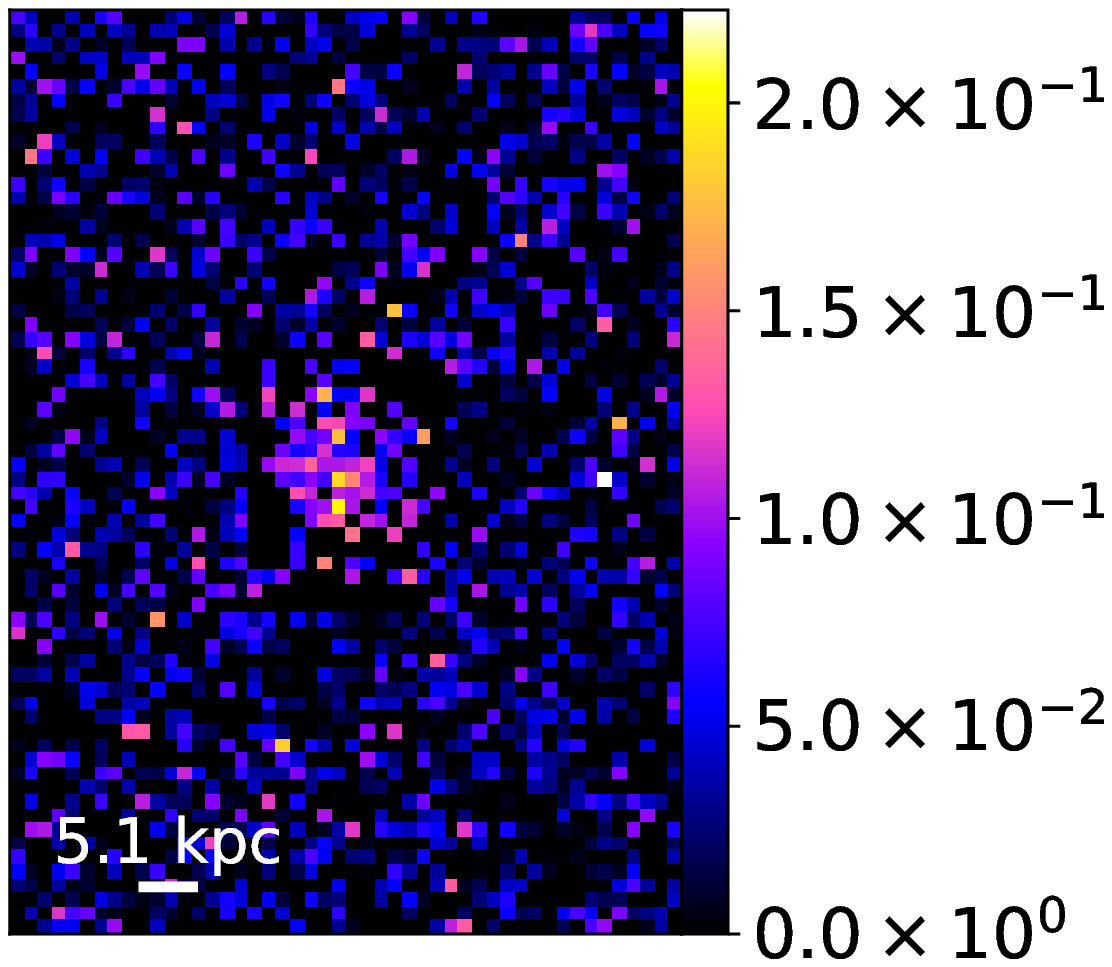}
    \caption{Thermal}
\end{subfigure}
\hspace{1em}
\begin{subfigure}[t]{0.20\textwidth}
    \includegraphics[width=3.1cm]  
    {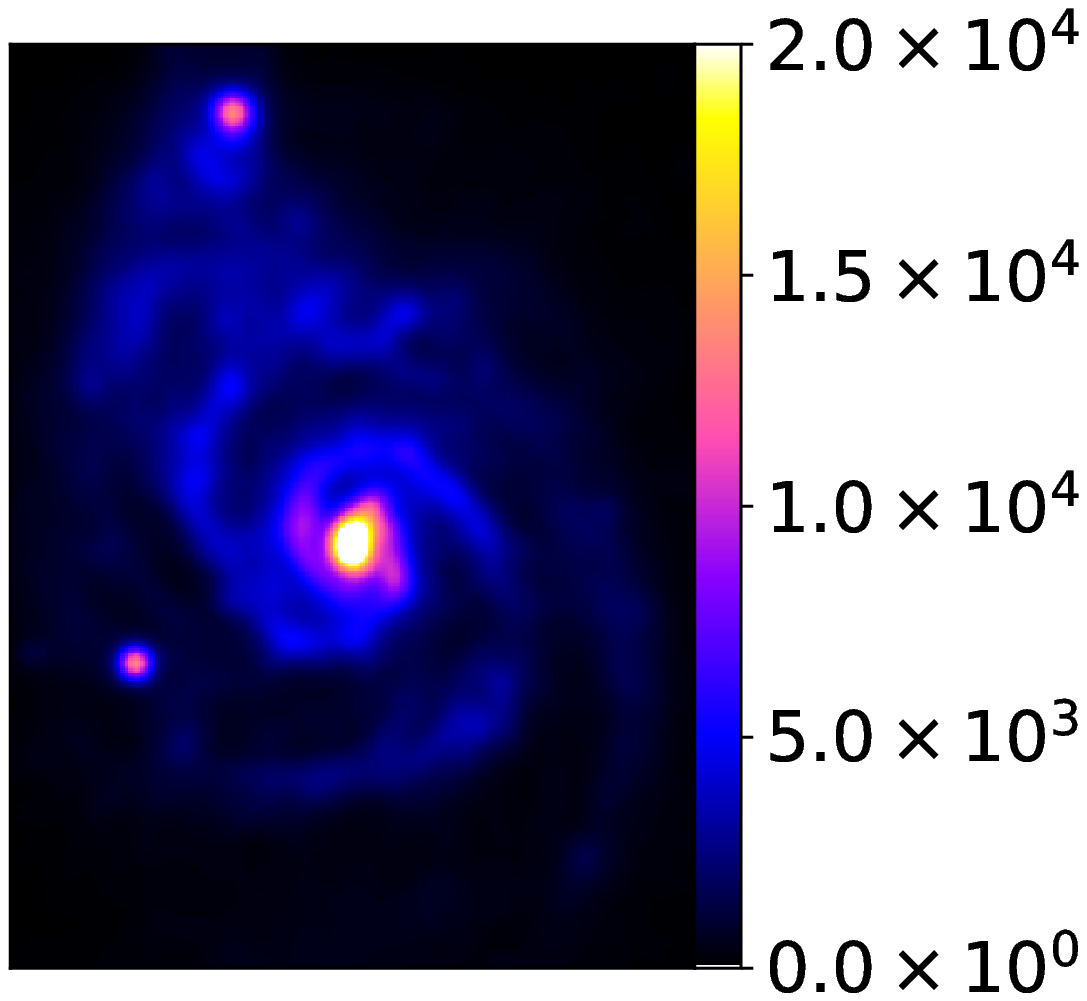}
    \includegraphics[width=3.1cm]
    {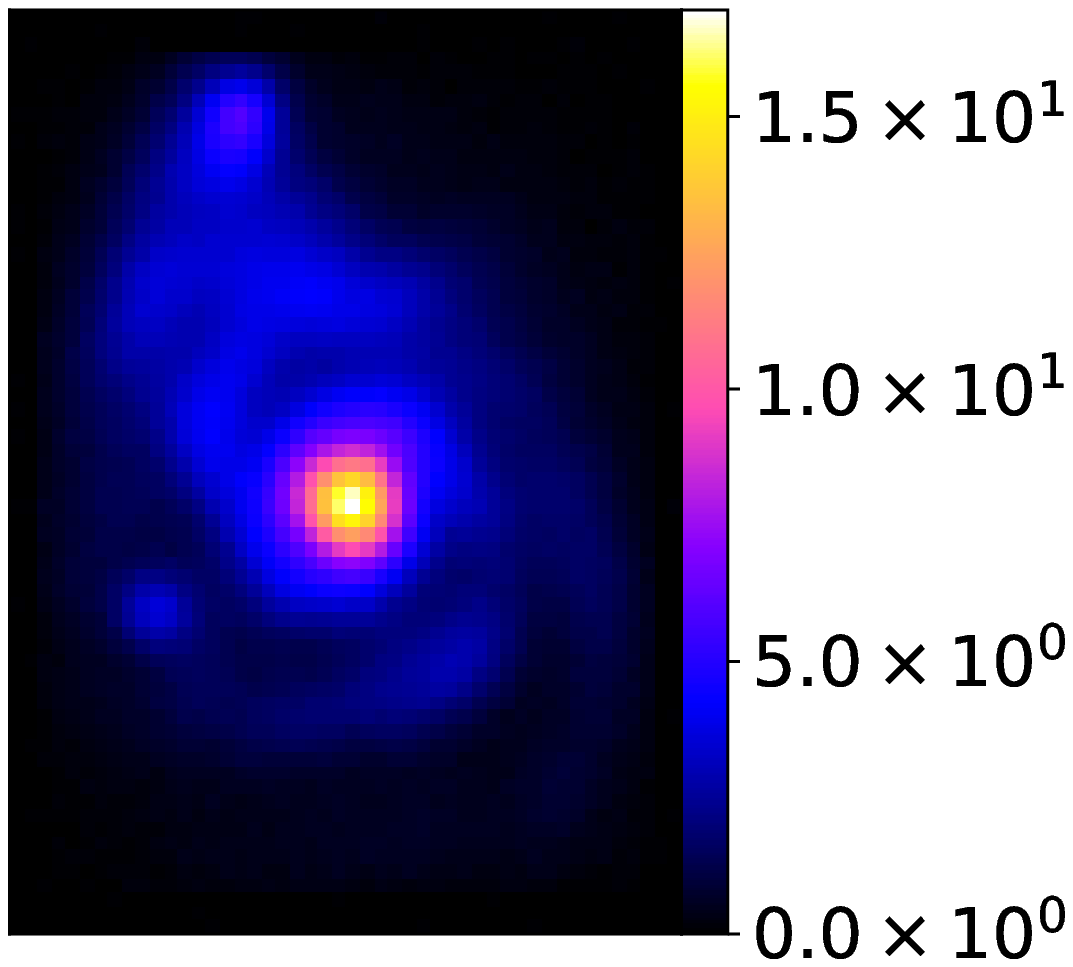}
    \includegraphics[width=3.1cm]
    {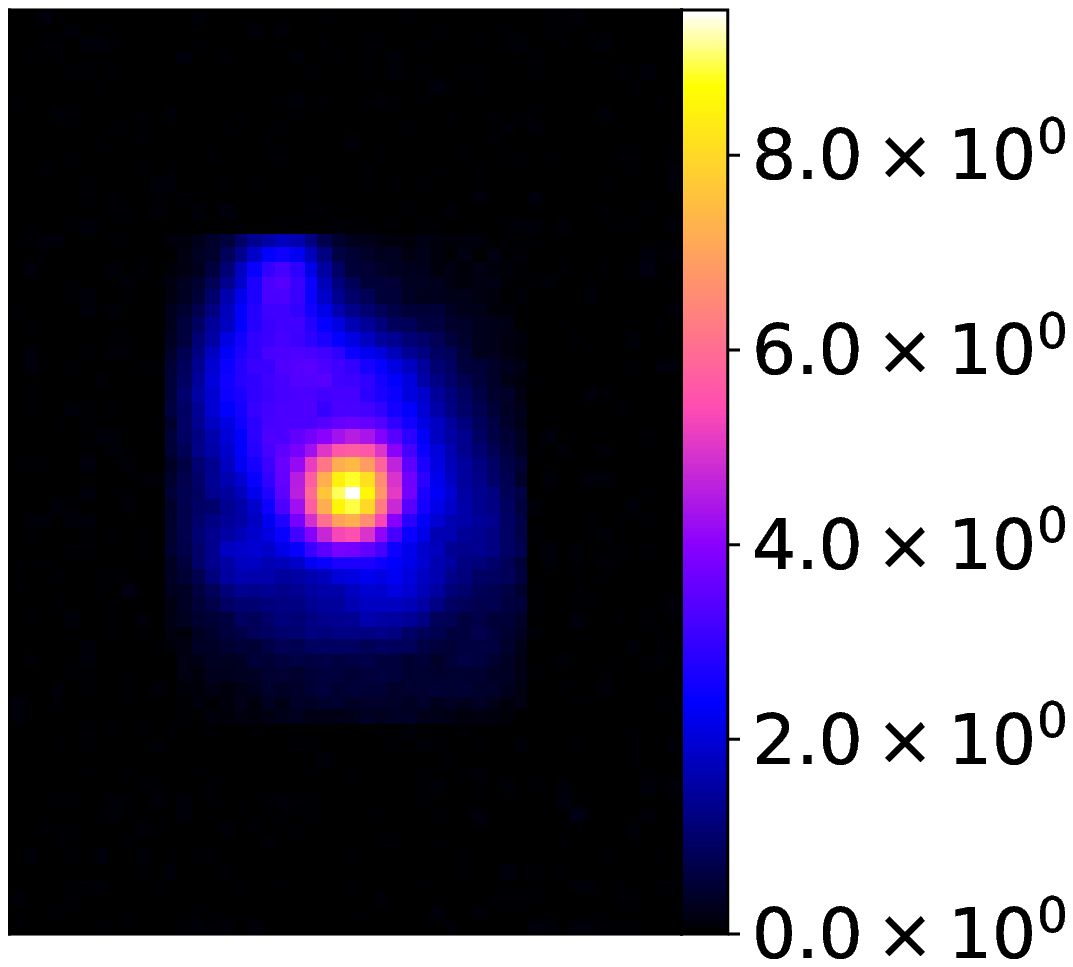}
    \includegraphics[width=3.1cm]
    {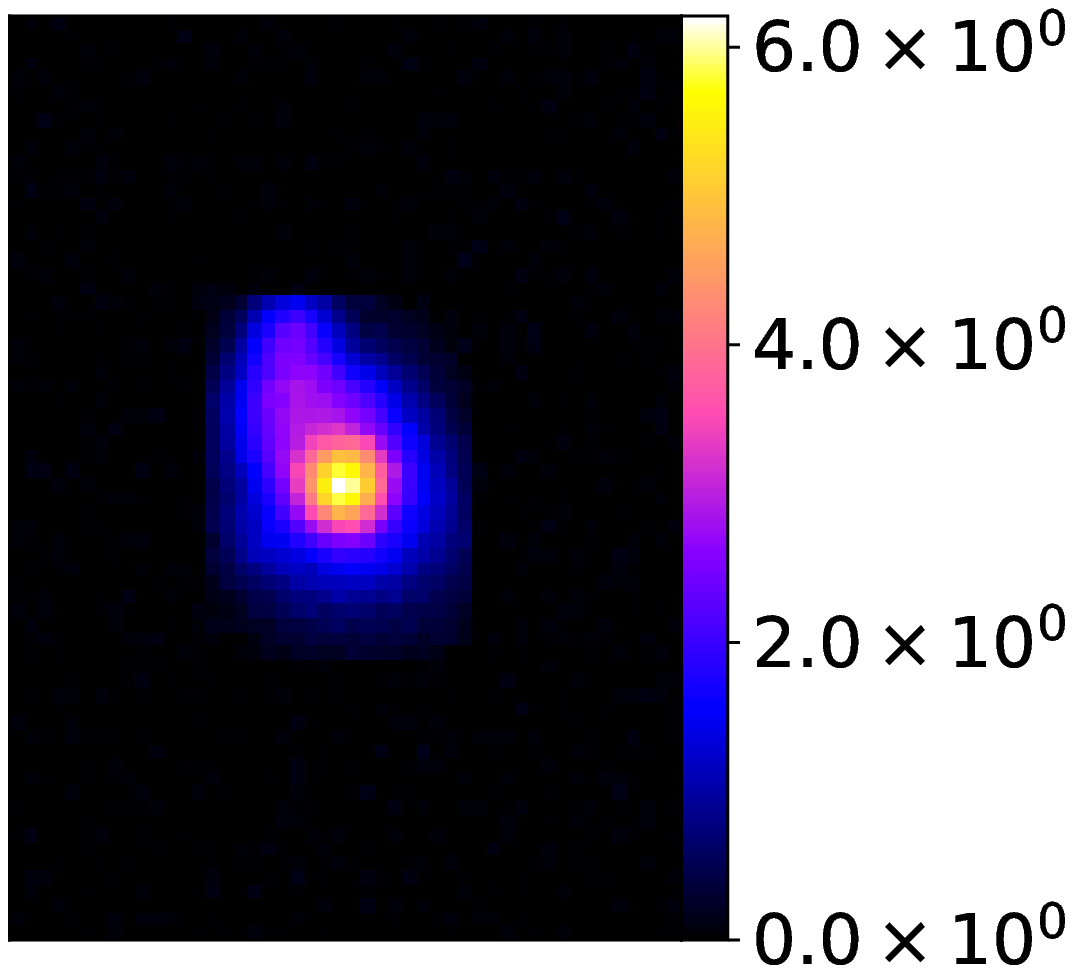}
    \includegraphics[width=3.1cm]
    {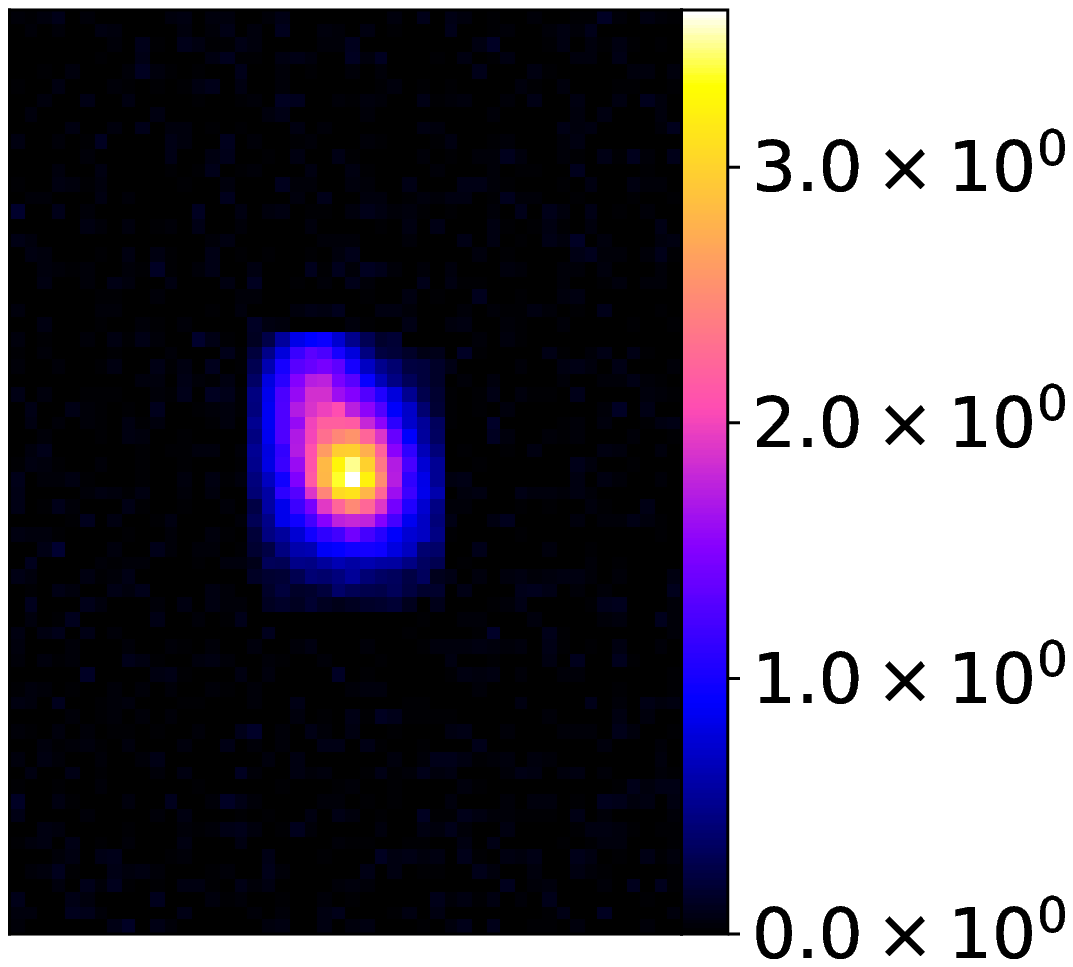}
     \includegraphics[width=3.1cm]
    {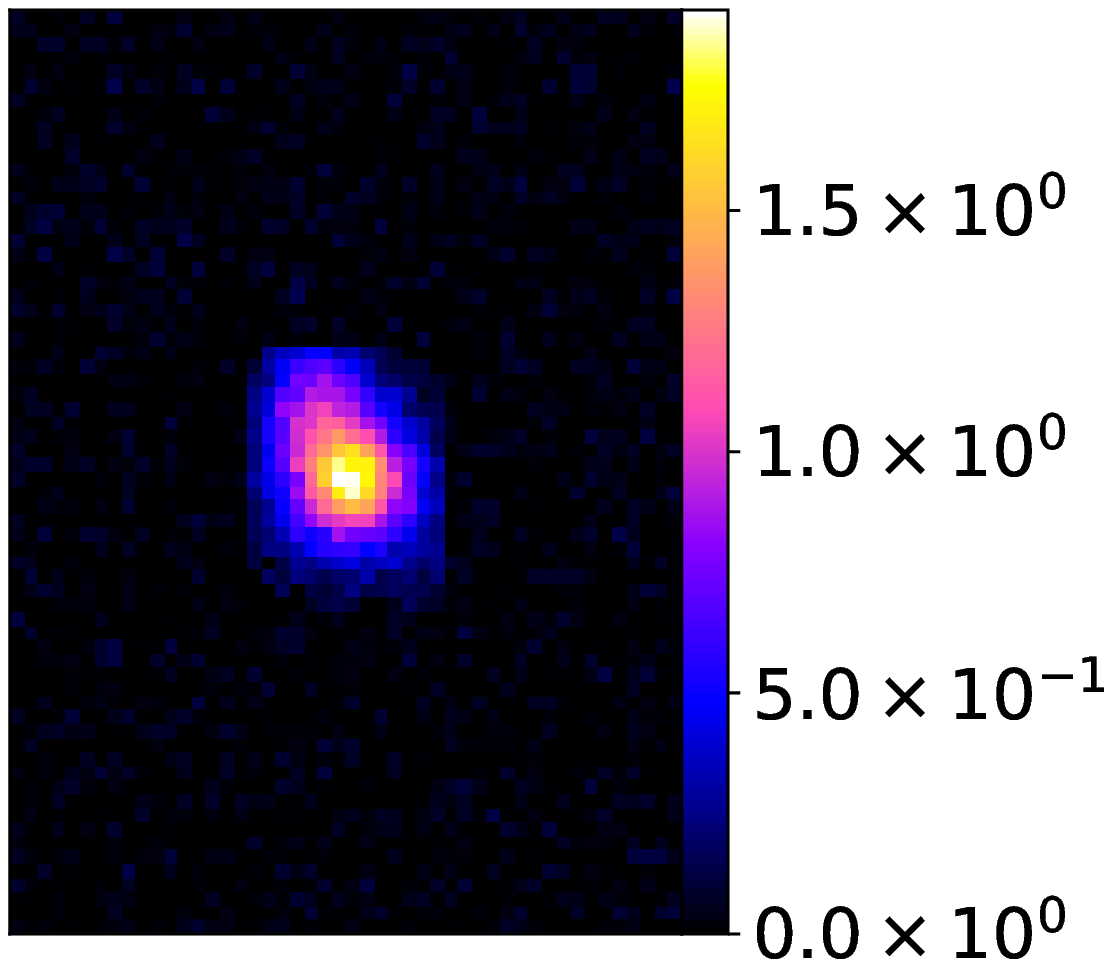}
    \caption{Nonthermal}
\end{subfigure}
\hspace{1em}
\begin{subfigure}[t]{0.20\textwidth}
    \includegraphics[width=3.2cm]  
    {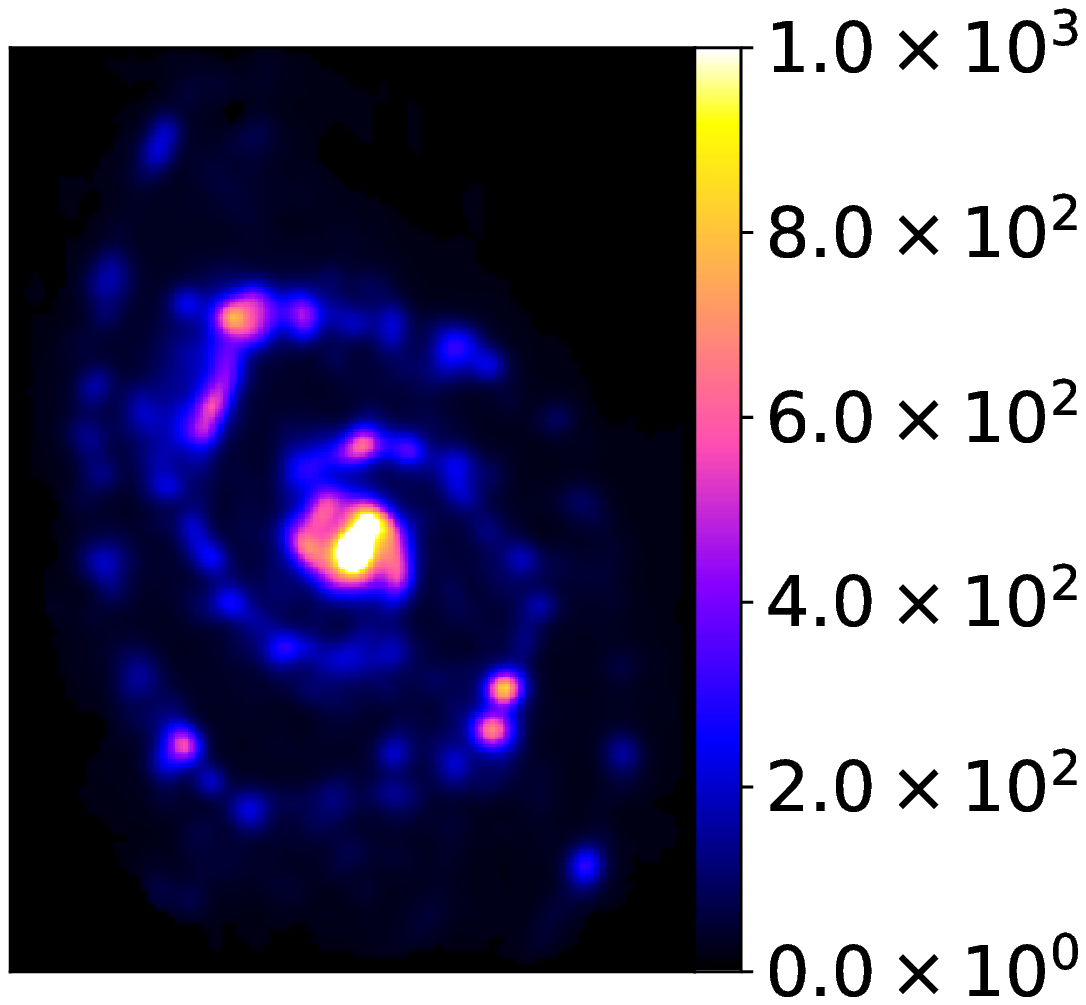}
    \includegraphics[width=3.2cm]
    {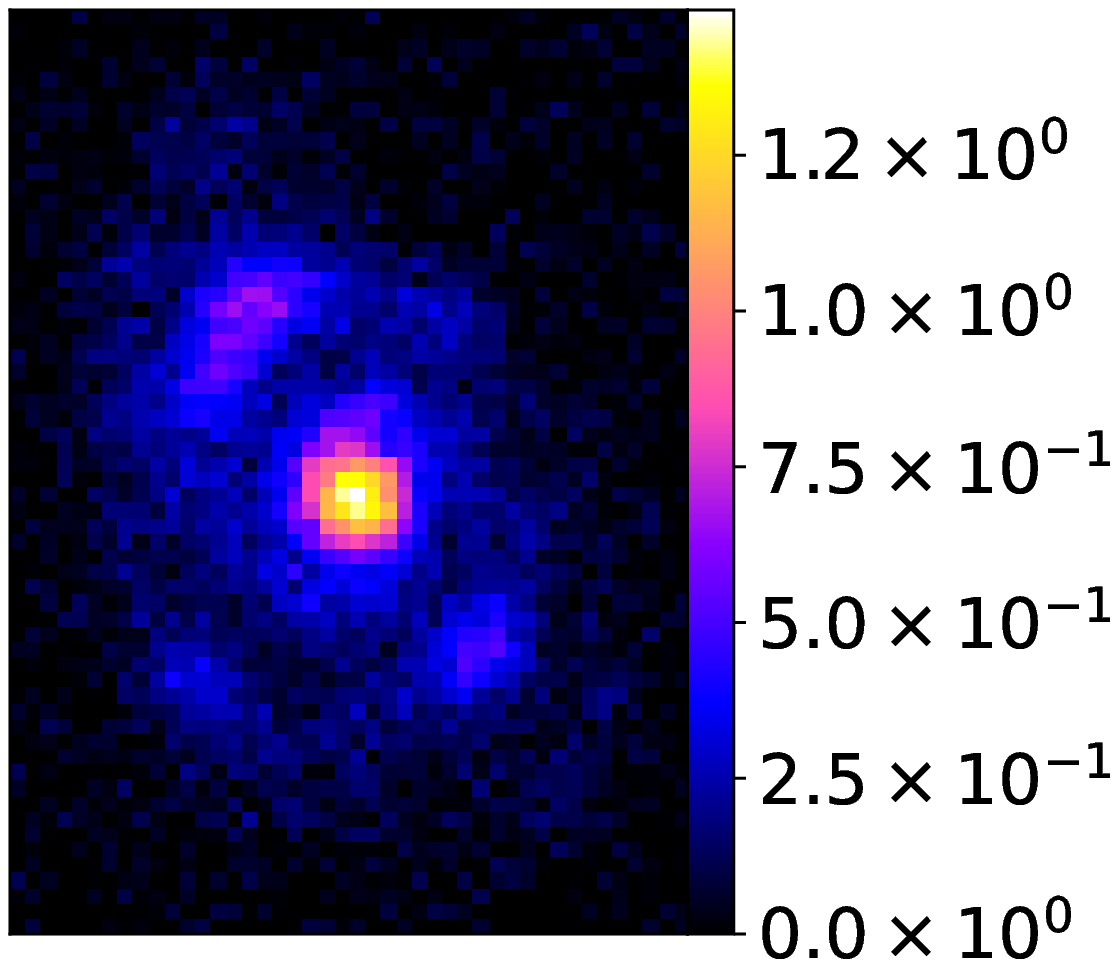}
    \includegraphics[width=3.2cm]
    {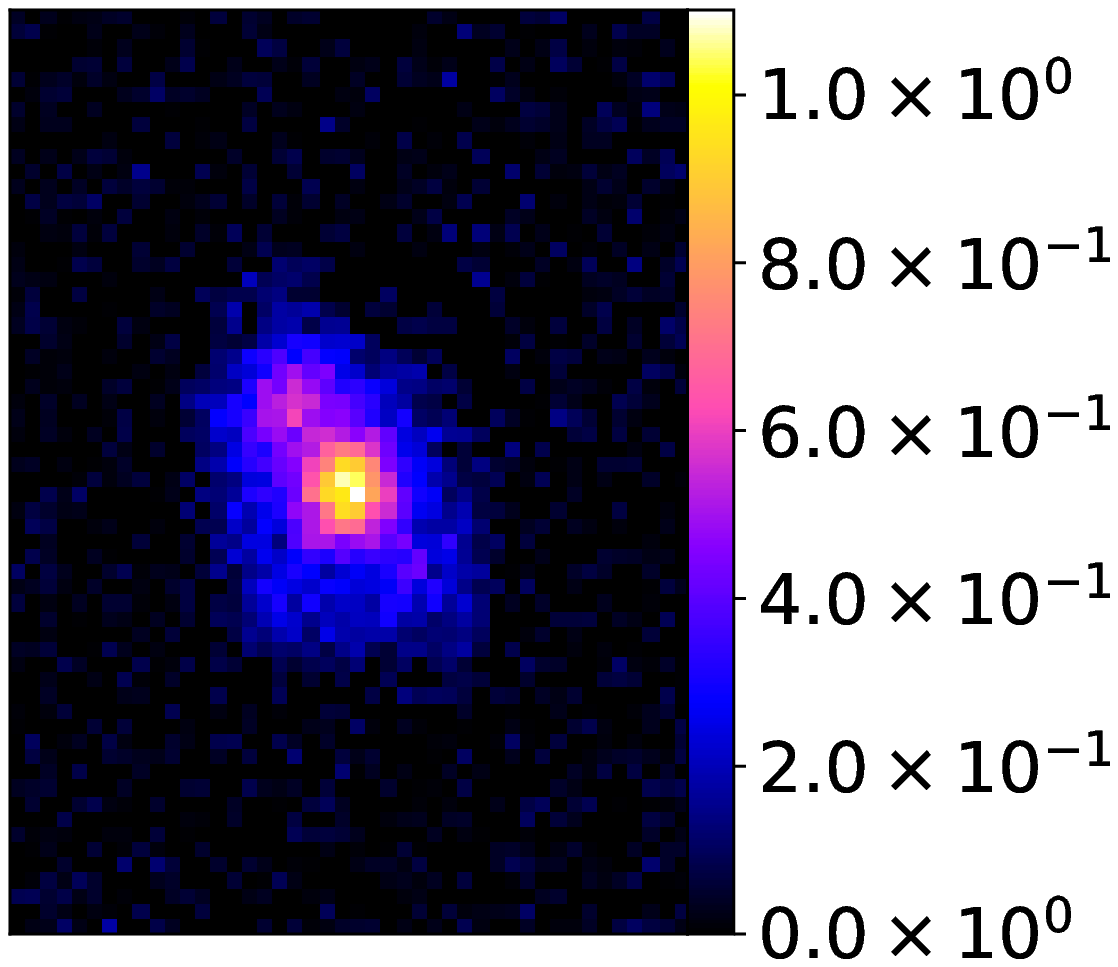}
    \includegraphics[width=3.2cm]
    {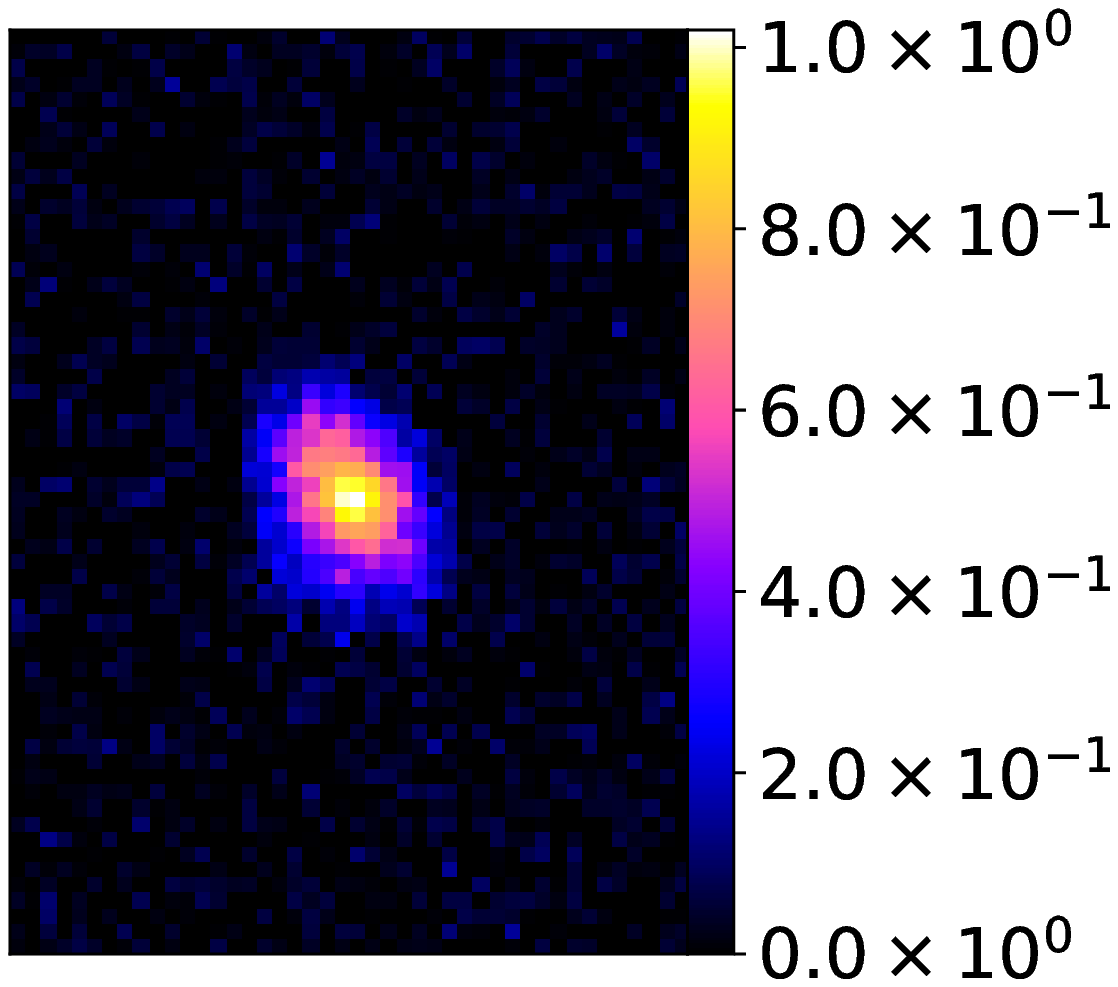}
     \includegraphics[width=3.2cm]
    {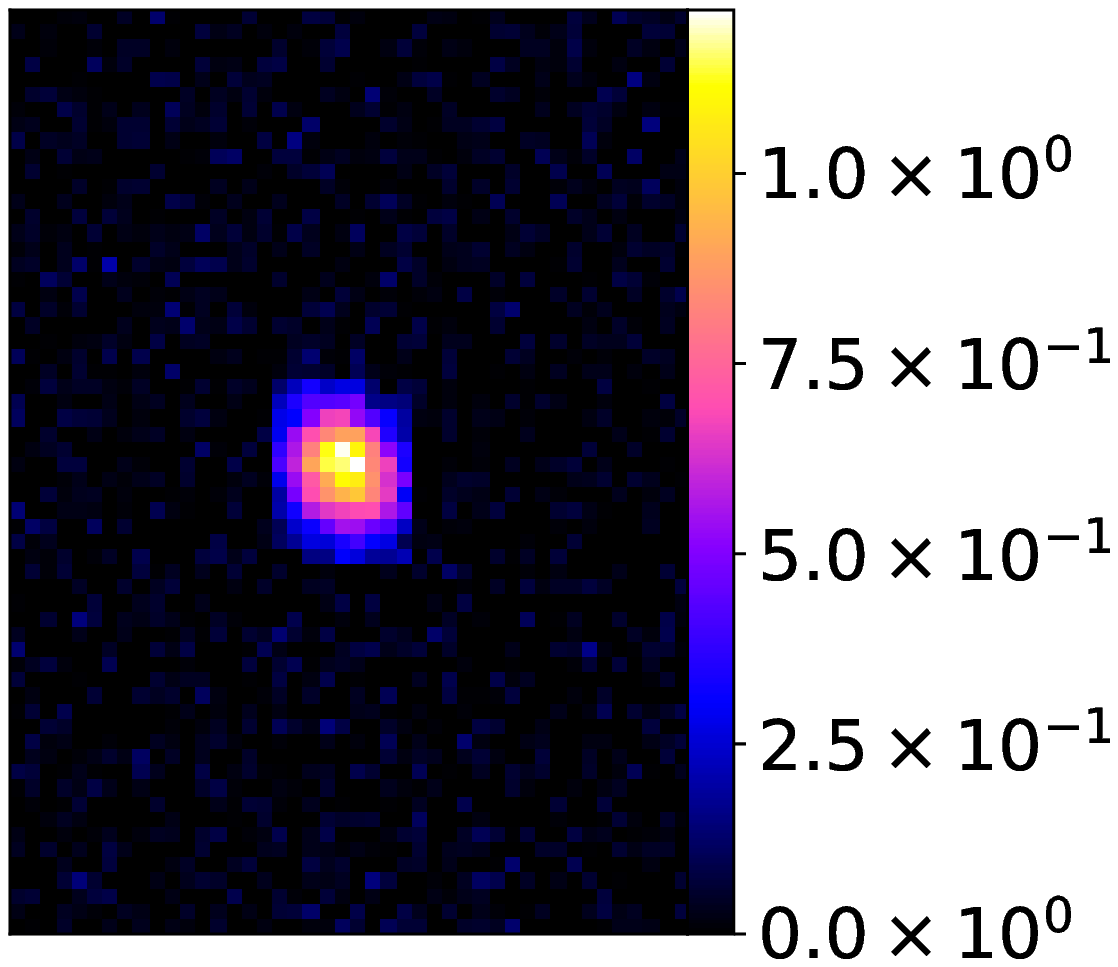}
     \includegraphics[width=3.2cm]
    {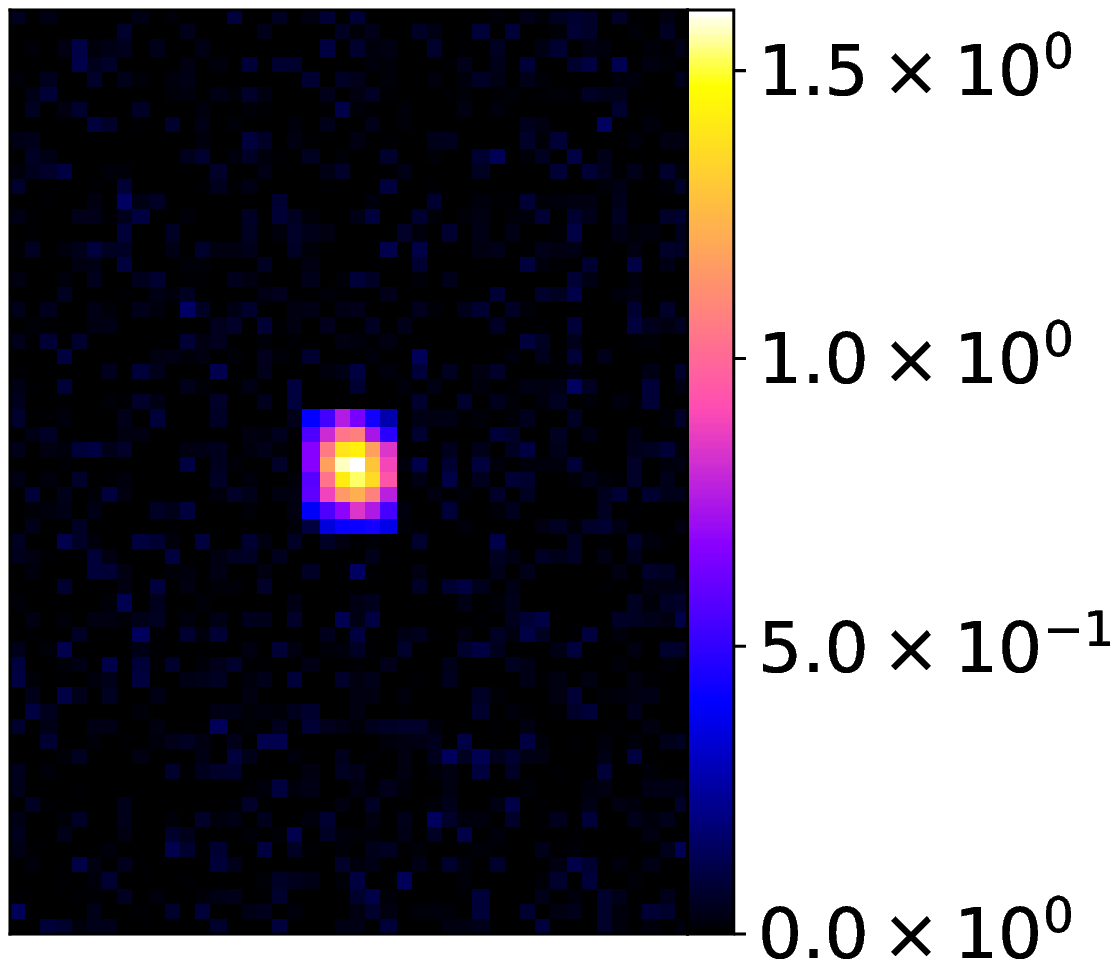}
    \caption{Thermal}   
\end{subfigure}
\hspace{1em}
\begin{subfigure}[t]{0.20\textwidth}
    \includegraphics[width=3.1cm]  
    {M51-nt-z0.eps}
    \includegraphics[width=3.1cm]
    {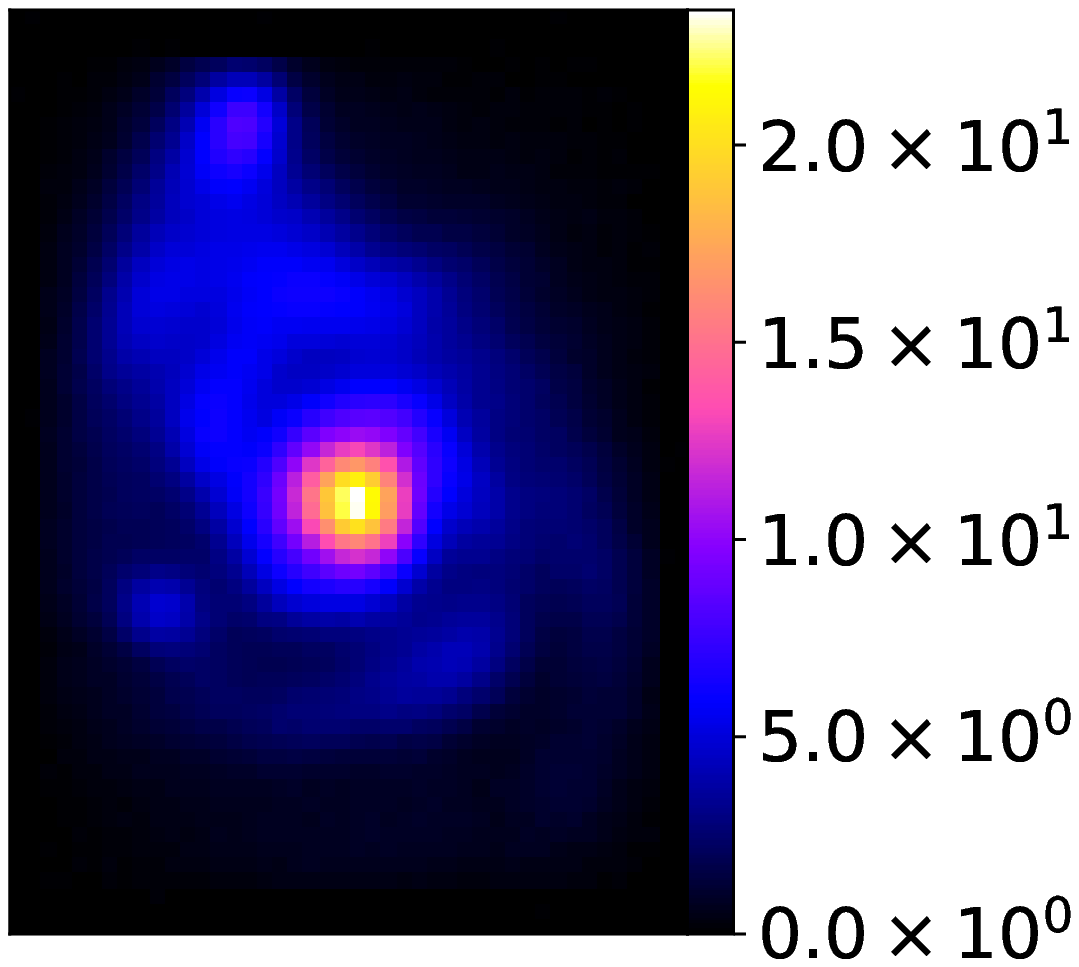}
    \includegraphics[width=3.1cm]
    {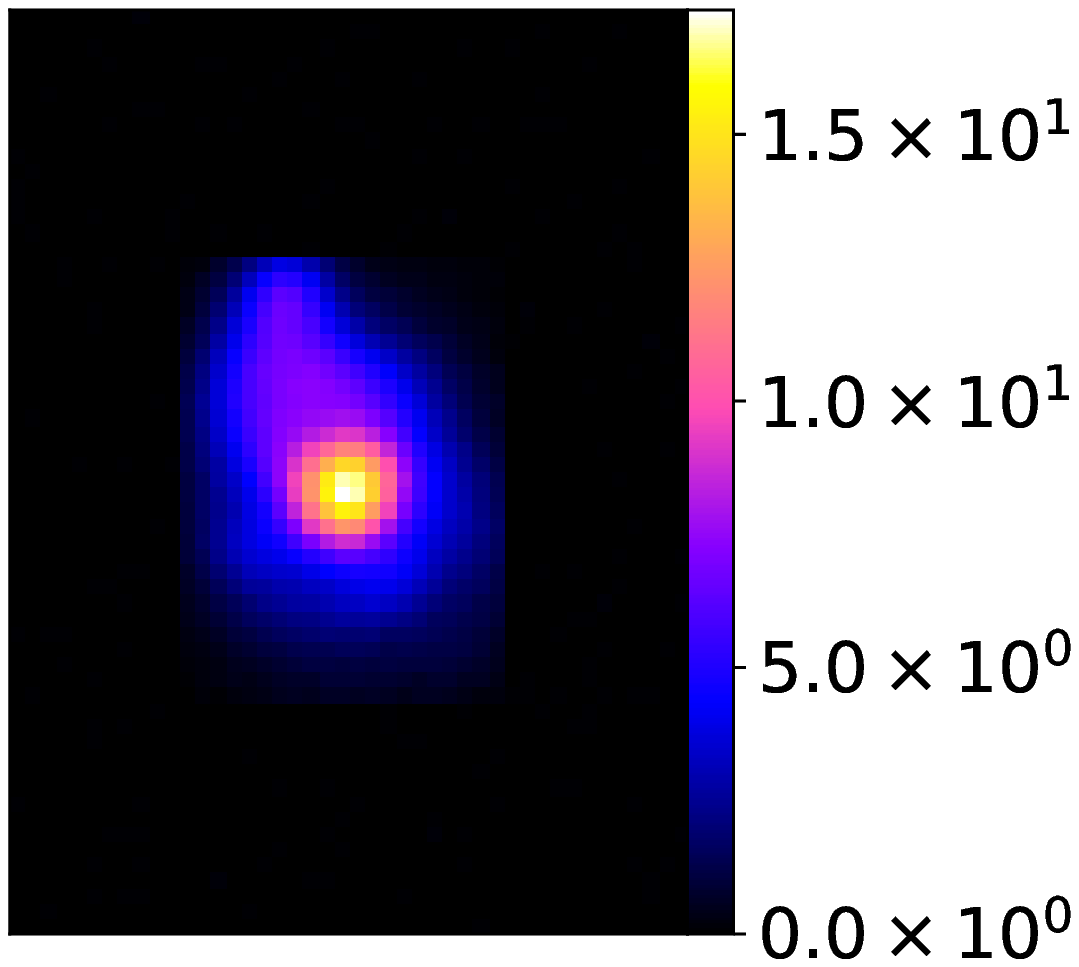}
    \includegraphics[width=3.1cm]
    {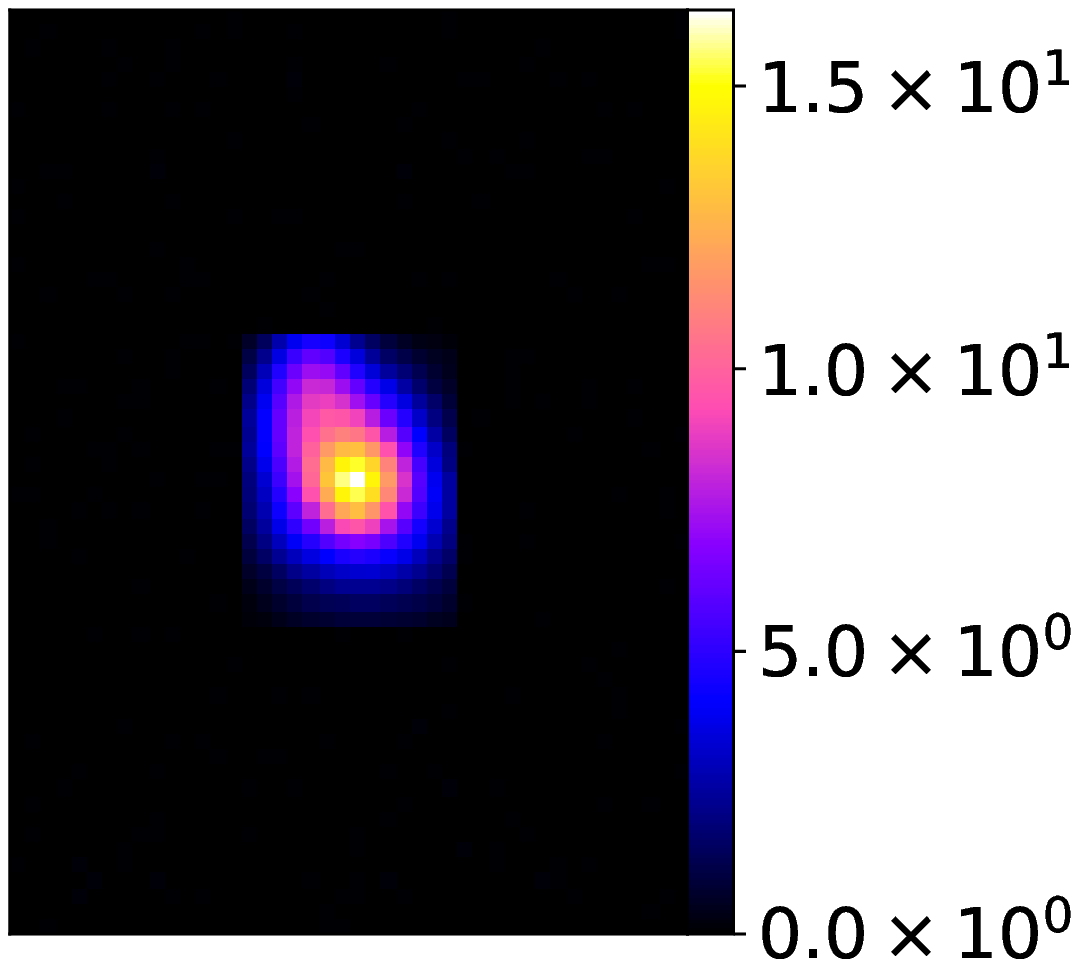}
    \includegraphics[width=3.1cm]
    {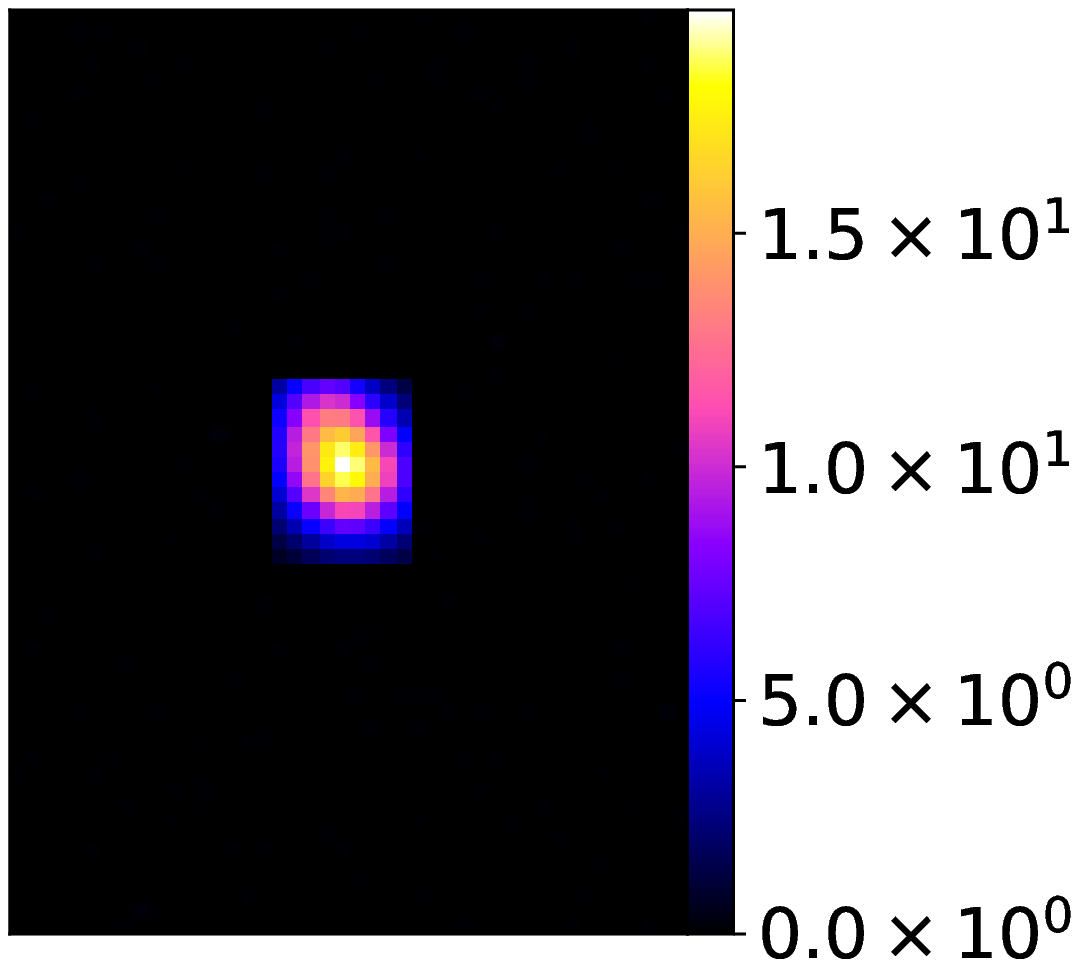}
    \includegraphics[width=3.1cm]
    {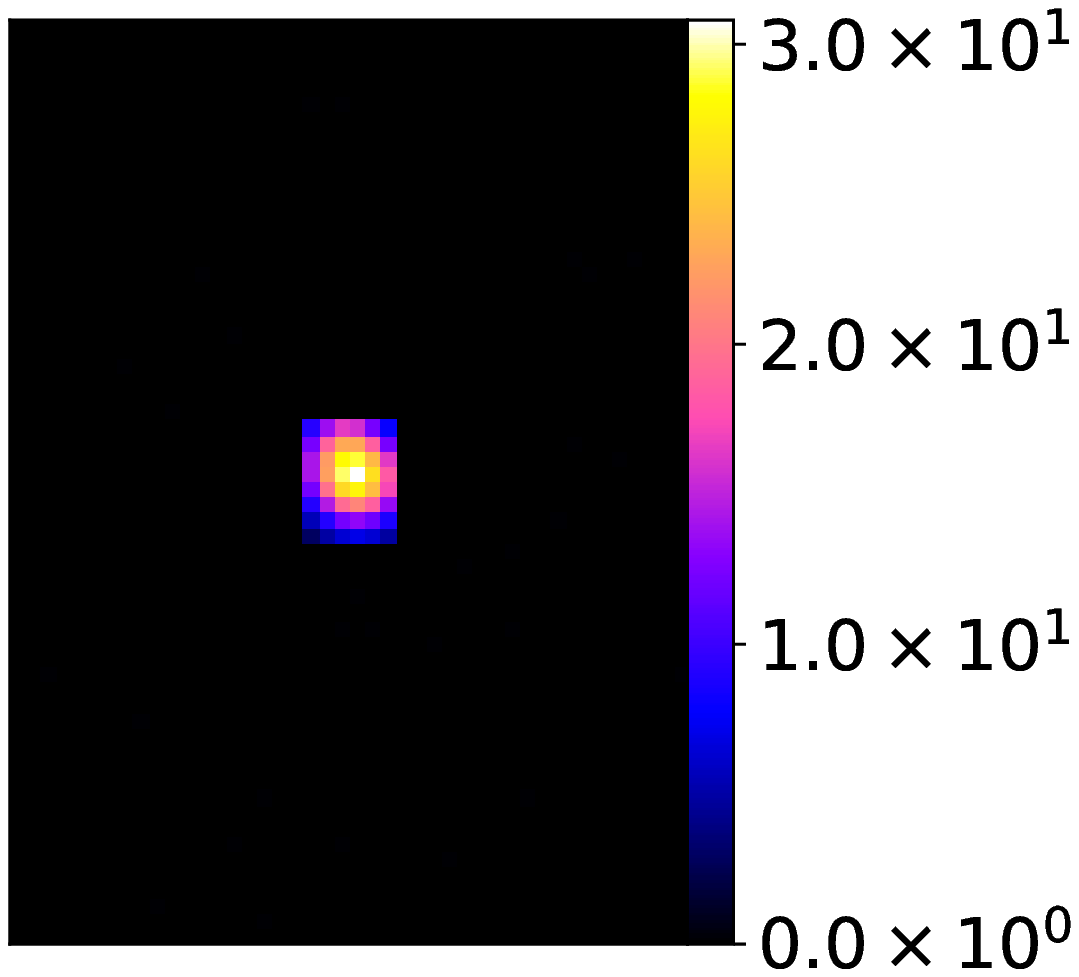}
    \caption{Nonthermal}
\end{subfigure}
\caption{ Simulated surface brightness maps for a M51--like galaxy { with the UDT-noise sky level} at the observed frequency of 1.4 GHz  showing the case(1)-thermal emission  ({\it first column}), case(1)-nonthermal emission ({\it second column}), case(2)-thermal emission ({\it third column}), and case(2)-nonthermal emission ({\it forth column}) at $z=$0.15, 0.3, 0.5, 1, and 2 from the second to last row, respectively.  First row shows the thermal and non-thermal maps of M51 separated  \citep[following][]{tab7}. Bars show the surface brightness in units of $\mu$Jy/beam.\label{mapM51c12}}
\end{figure*}

% Bars show the surface brightness in units of $\mu$Jy/beam. \cite{tab7}

\begin{figure*}
\centering
\textbf{case 1}
\hspace{6cm}
\textbf{case 2}\par\medskip
\begin{subfigure}[t]{0.20\textwidth}
    \makebox[0pt][r]{\makebox[30pt]{\raisebox{40pt}{\rotatebox[origin=c]{90}{Input}}}}%
    \includegraphics[width=3.5cm]
    {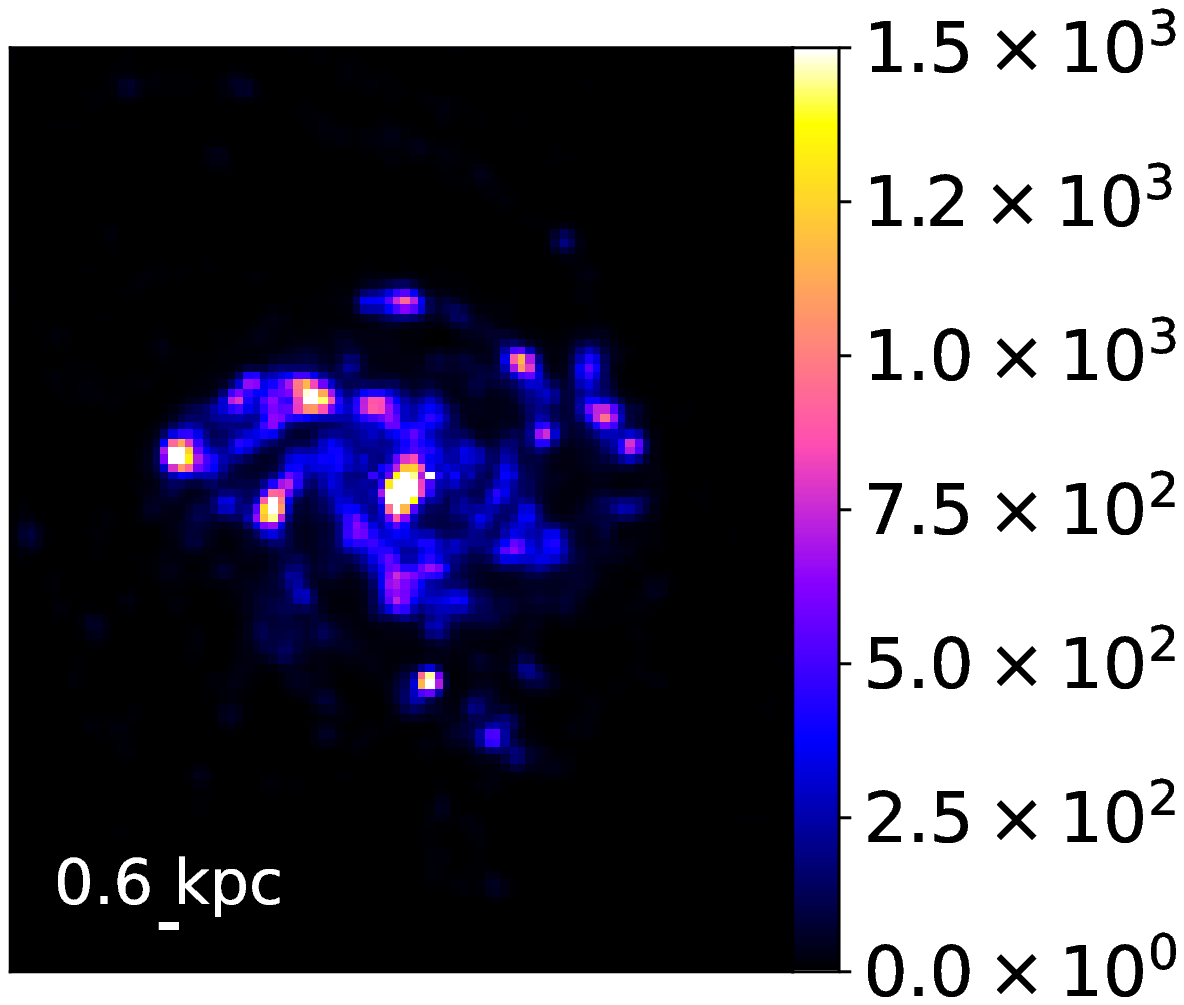}
    \makebox[0pt][r]{\makebox[30pt]{\raisebox{40pt}{\rotatebox[origin=c]{90}{$z=0.15$}}}}%
    \includegraphics[width=3.5cm]
    {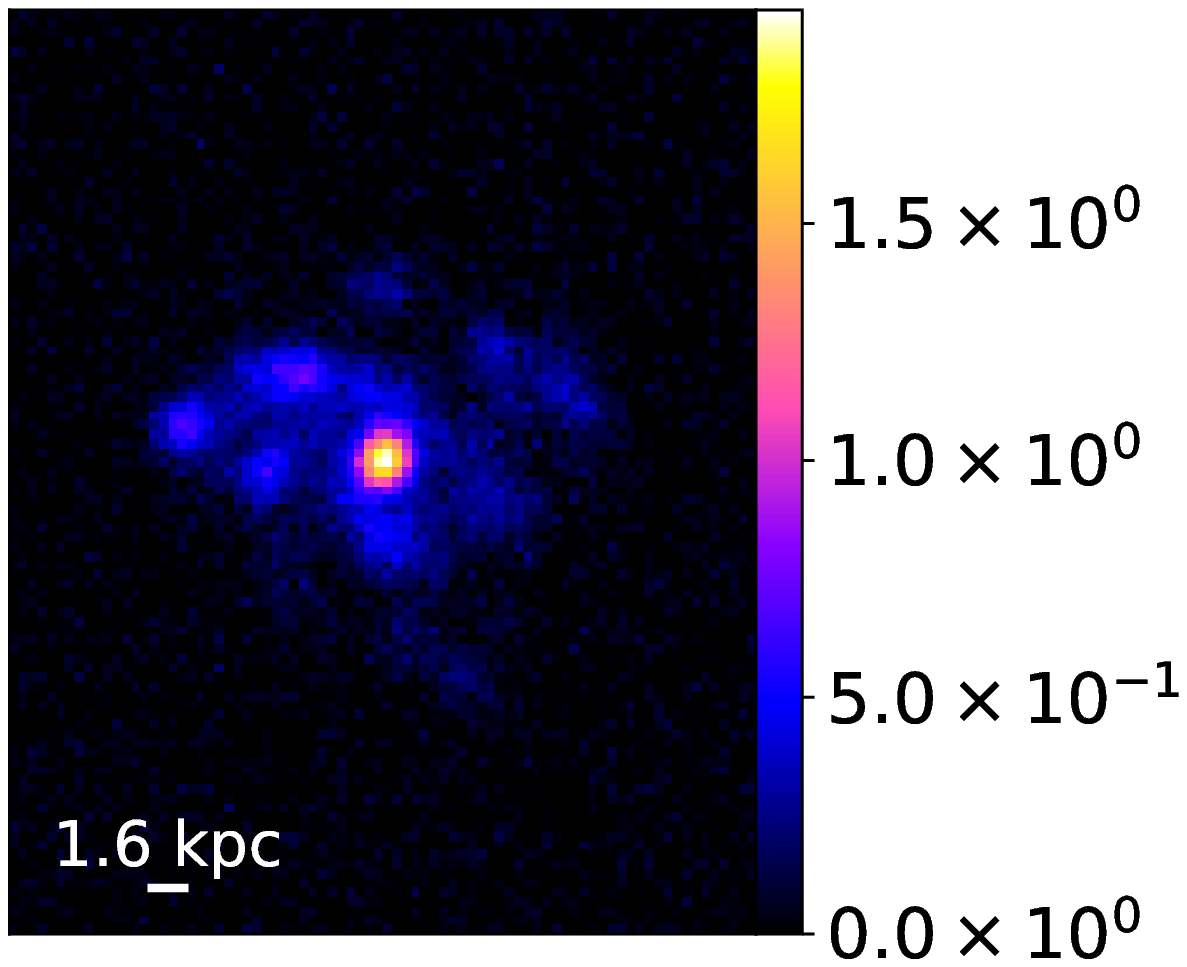}
    \makebox[0pt][r]{\makebox[30pt]{\raisebox{40pt}{\rotatebox[origin=c]{90}{$z=0.3$}}}}%
    \includegraphics[width=3.5cm]
    {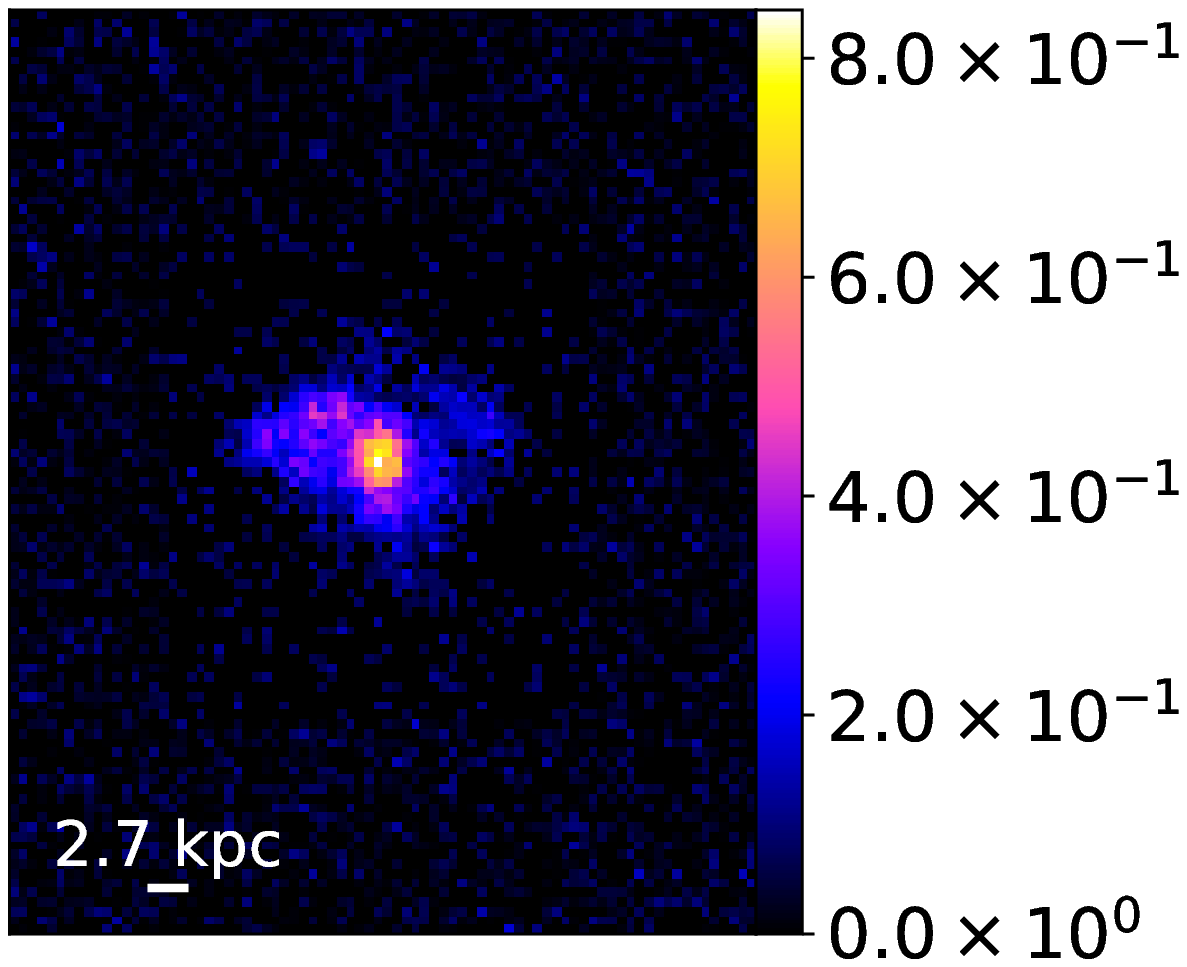}
    \makebox[0pt][r]{\makebox[30pt]{\raisebox{40pt}{\rotatebox[origin=c]{90}{$z=0.5$}}}}%
    \includegraphics[width=3.5cm]
    {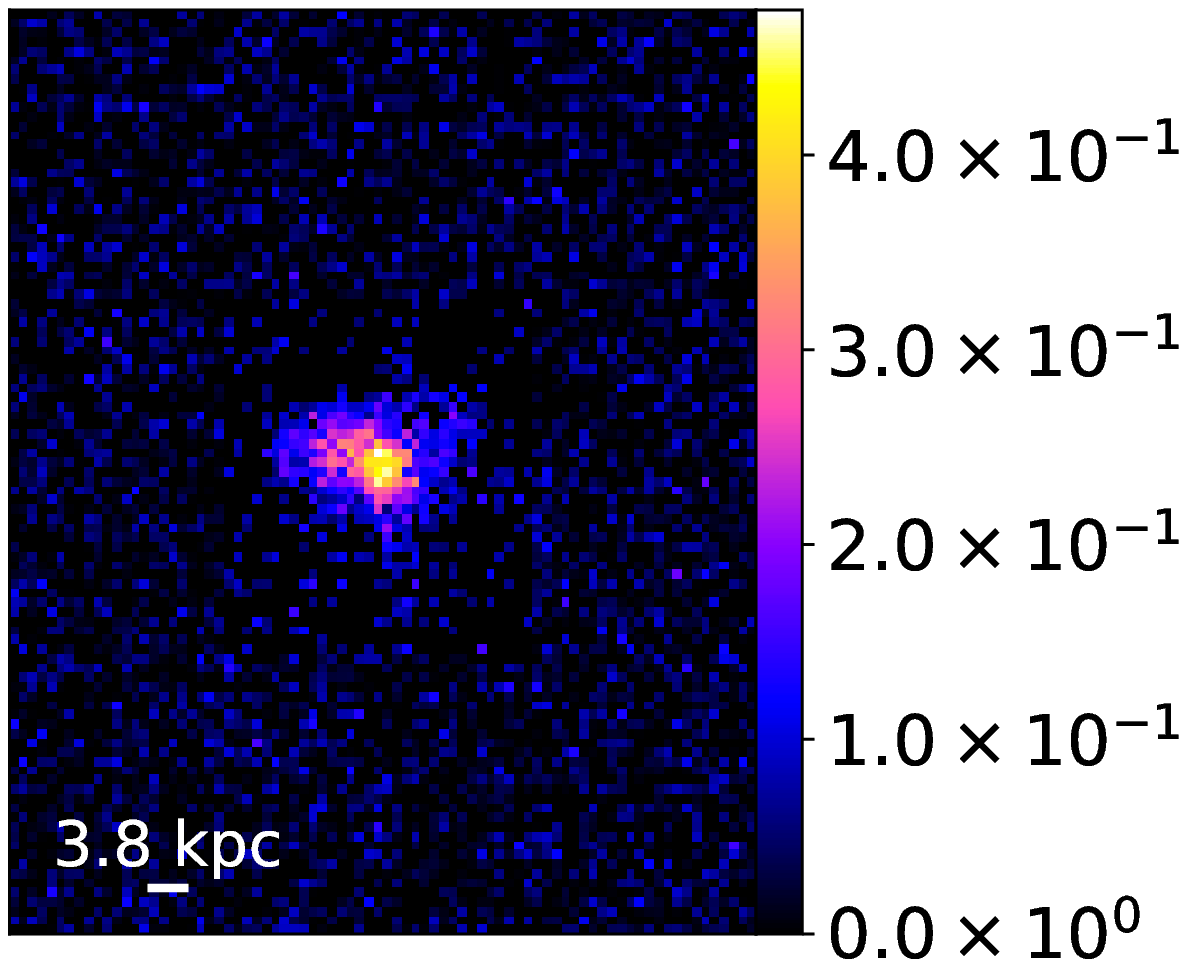}
    \makebox[0pt][r]{\makebox[30pt]{\raisebox{40pt}{\rotatebox[origin=c]{90}{$z=1$}}}}%
    \includegraphics[width=3.5cm]
    {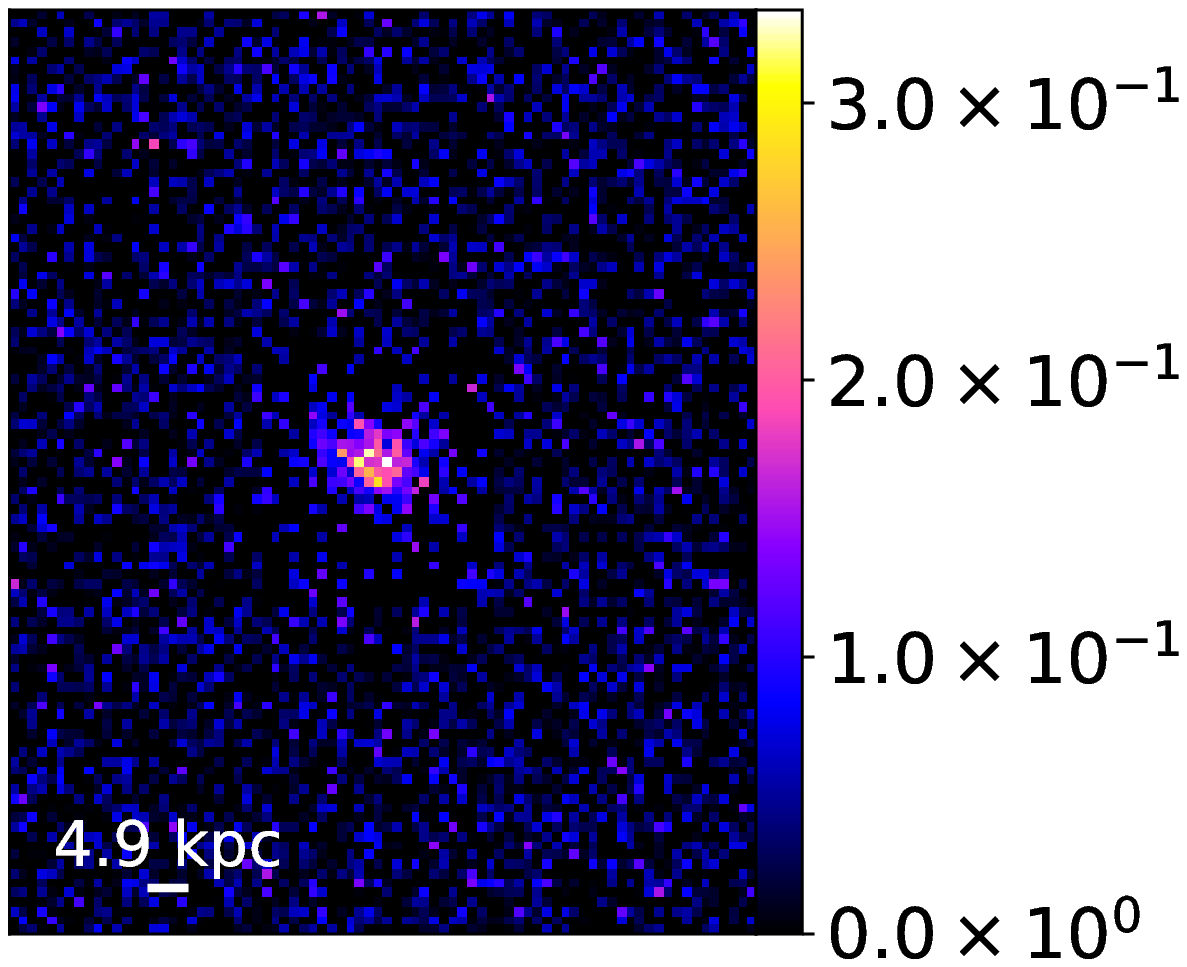}
     \makebox[0pt][r]{\makebox[30pt]{\raisebox{40pt}{\rotatebox[origin=c]{90}{$z=2$}}}}%
    \includegraphics[width=3.5cm]
    {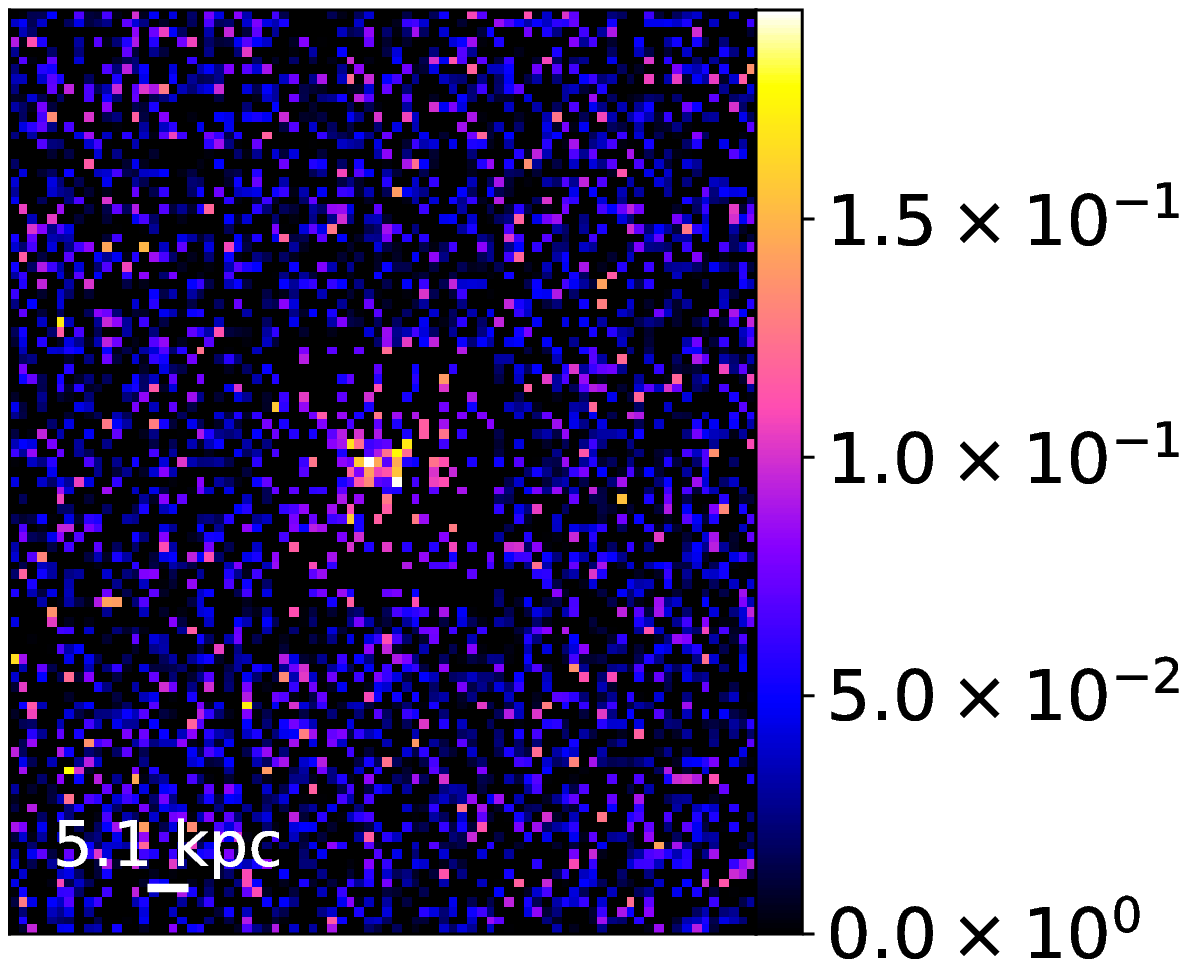}
    \caption{Thermal}
\end{subfigure}
\hspace{1em}
\begin{subfigure}[t]{0.20\textwidth}
    \includegraphics[width=3.4cm]  
    {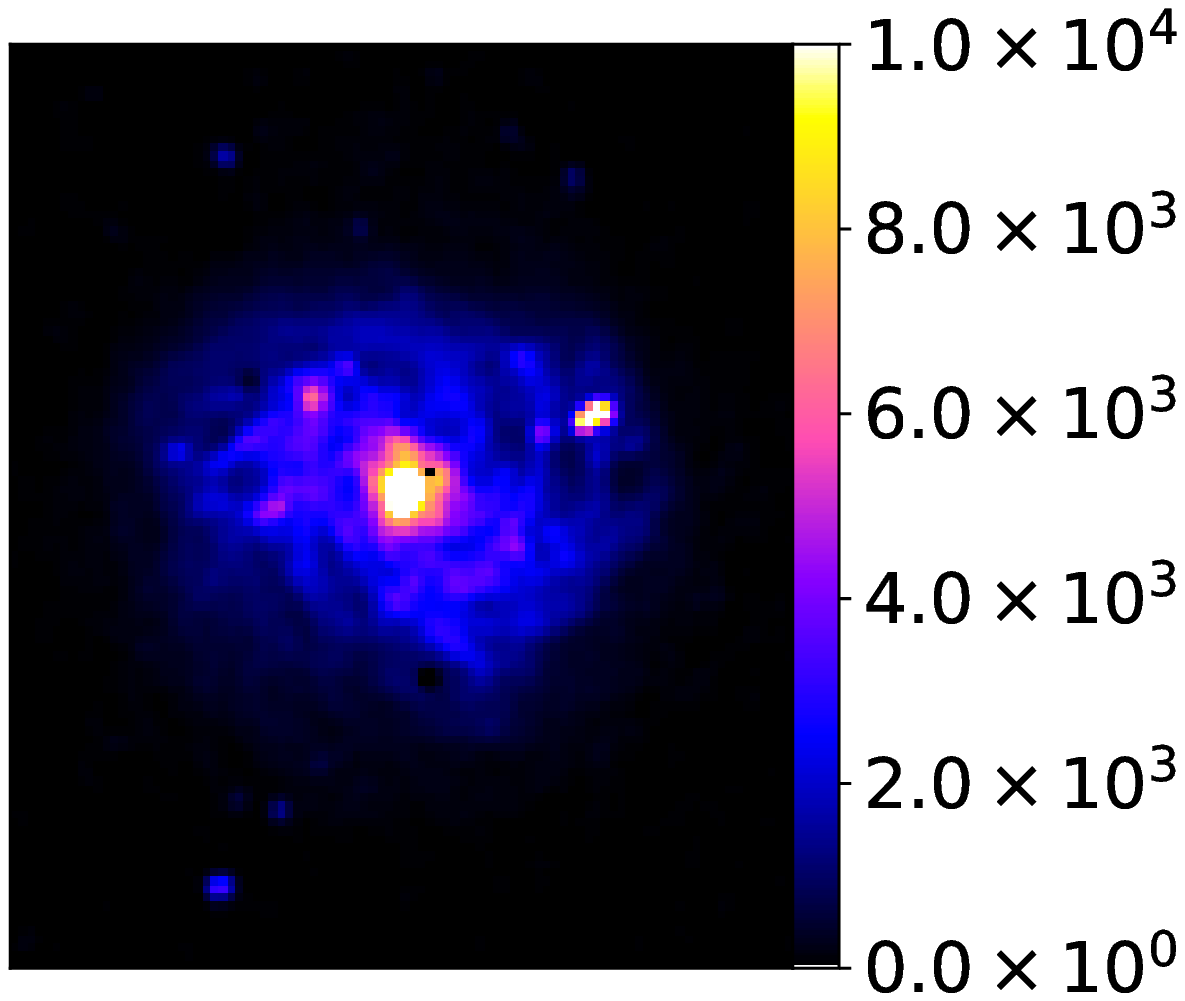}
    \includegraphics[width=3.4cm]
    {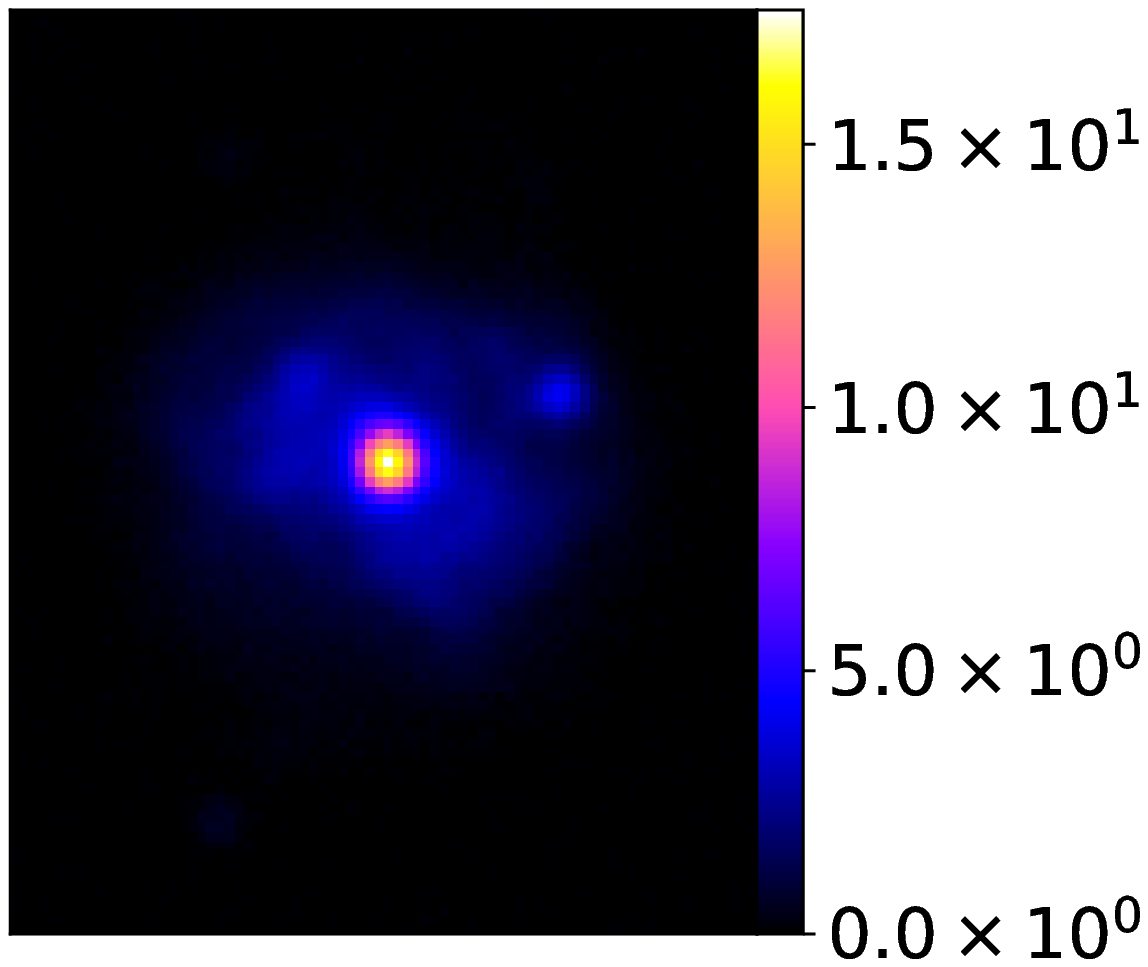}
    \includegraphics[width=3.4cm]
    {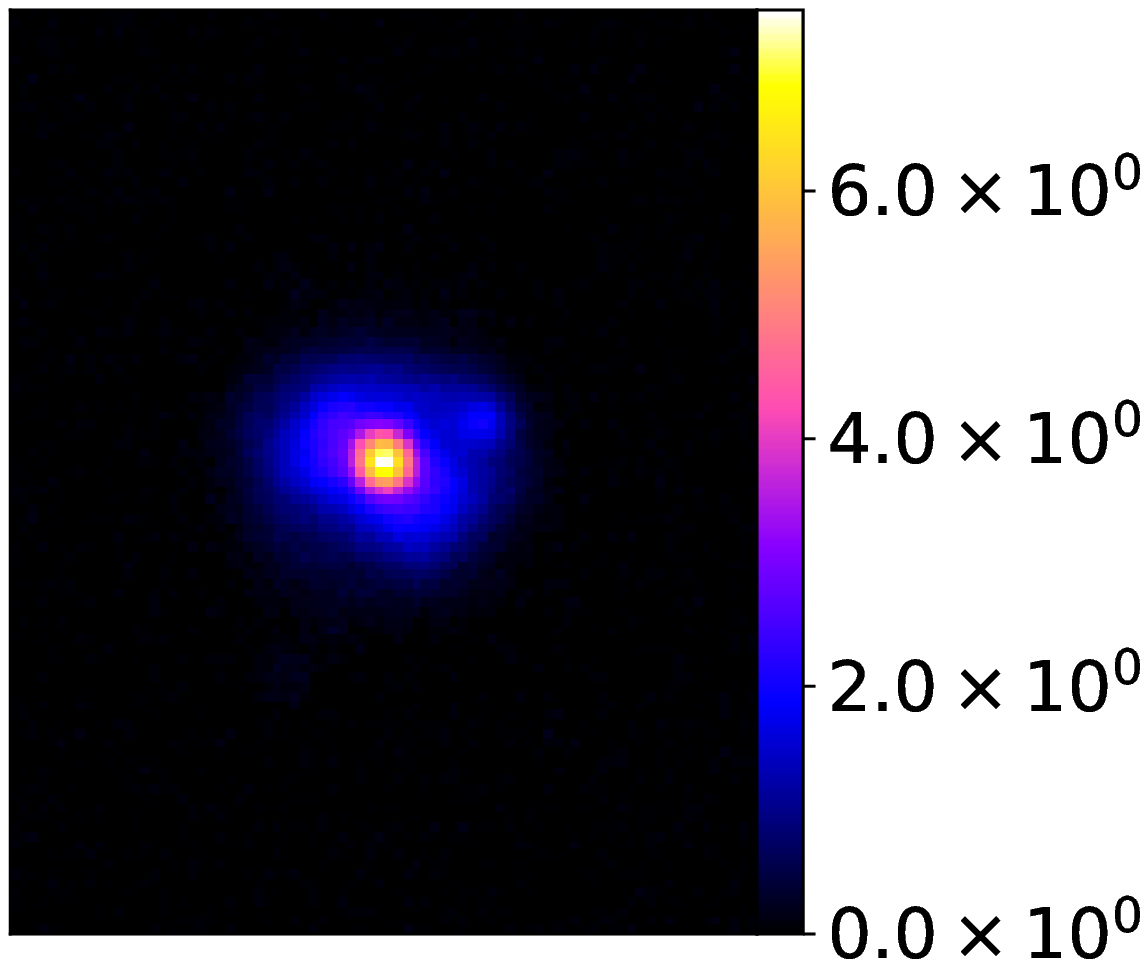}
    \includegraphics[width=3.4cm]
    {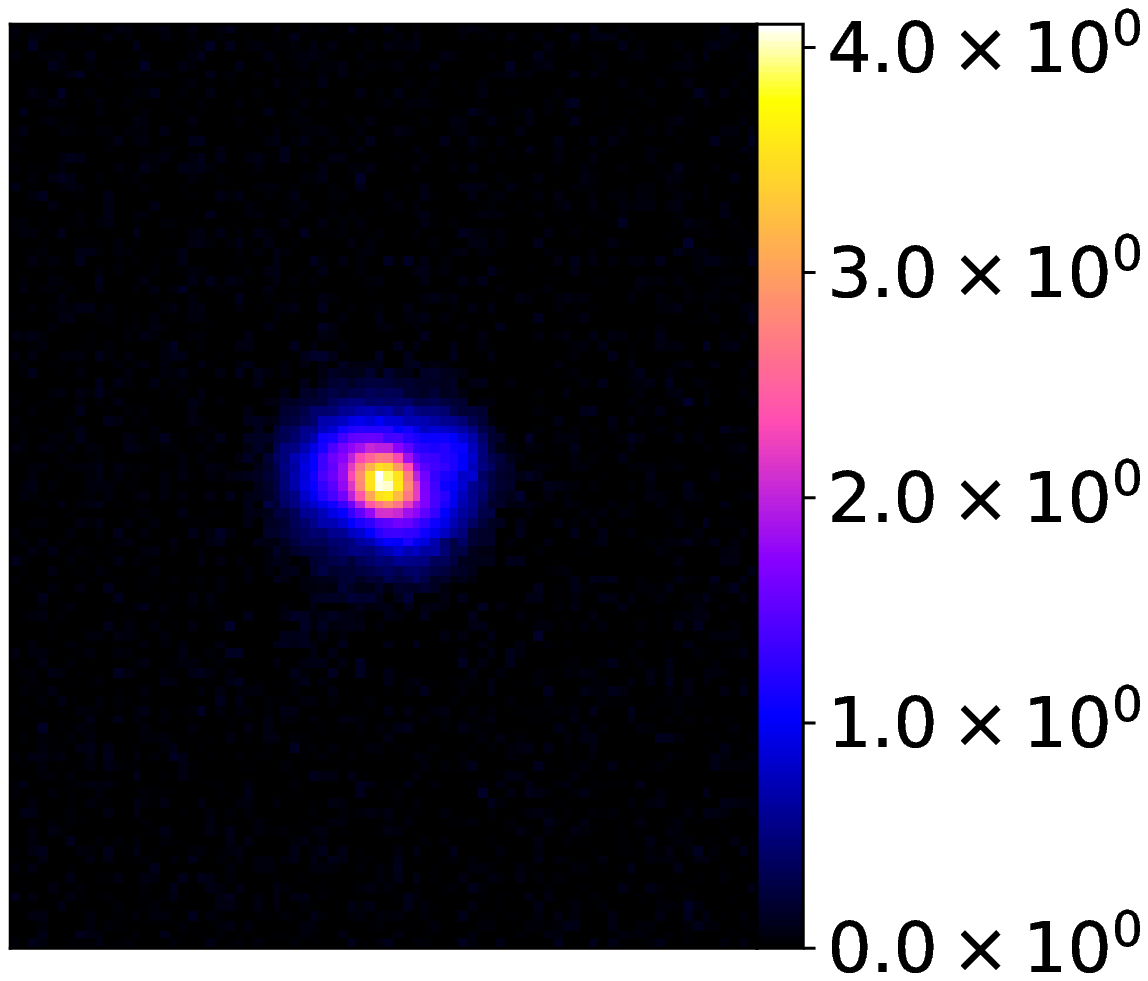}
    \includegraphics[width=3.4cm]
    {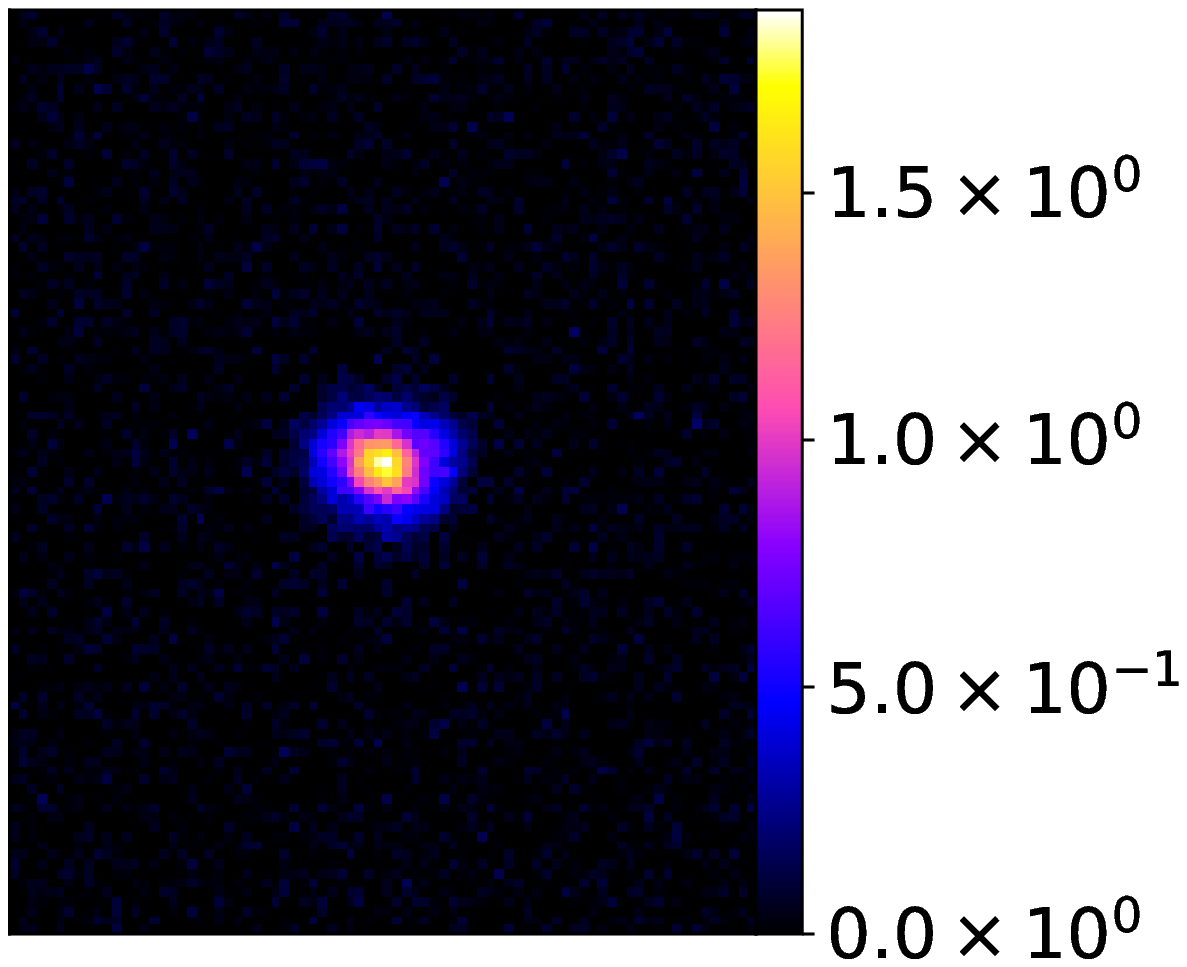}
     \includegraphics[width=3.4cm]
    {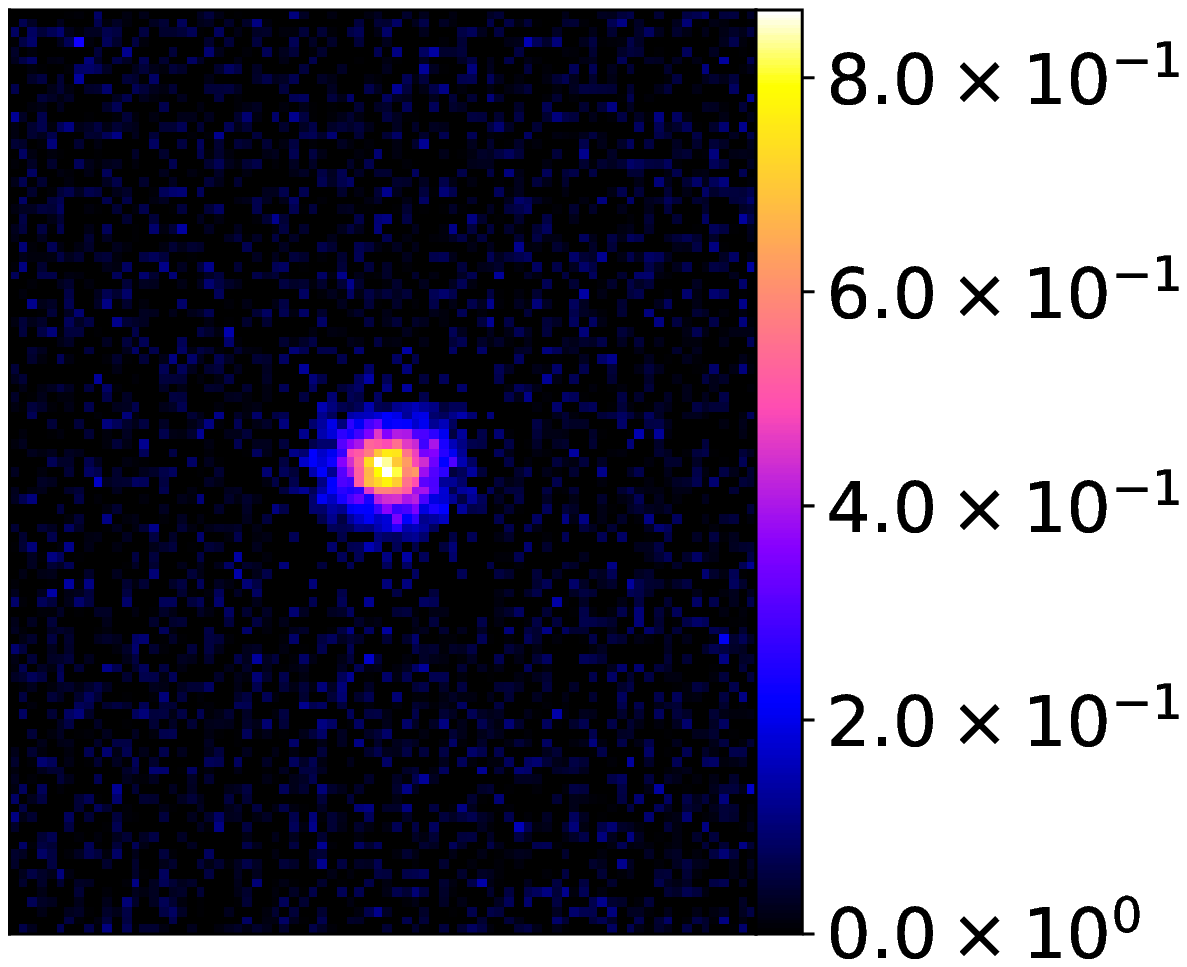}
    \caption{Nonthermal}
\end{subfigure}
\hspace{1em}
\begin{subfigure}[t]{0.20\textwidth}
    \includegraphics[width=3.5cm]  
    {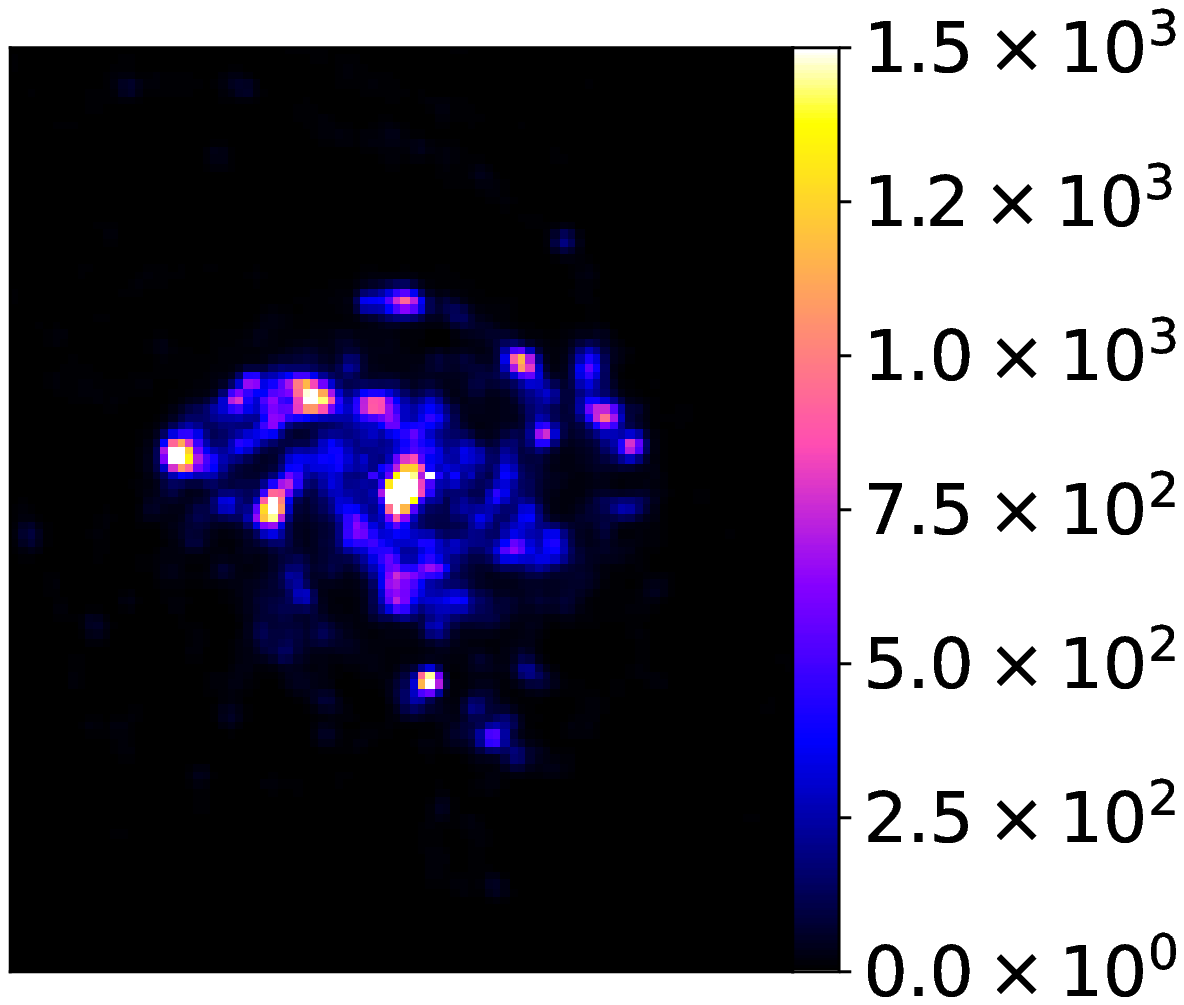}
    \includegraphics[width=3.5cm]
    {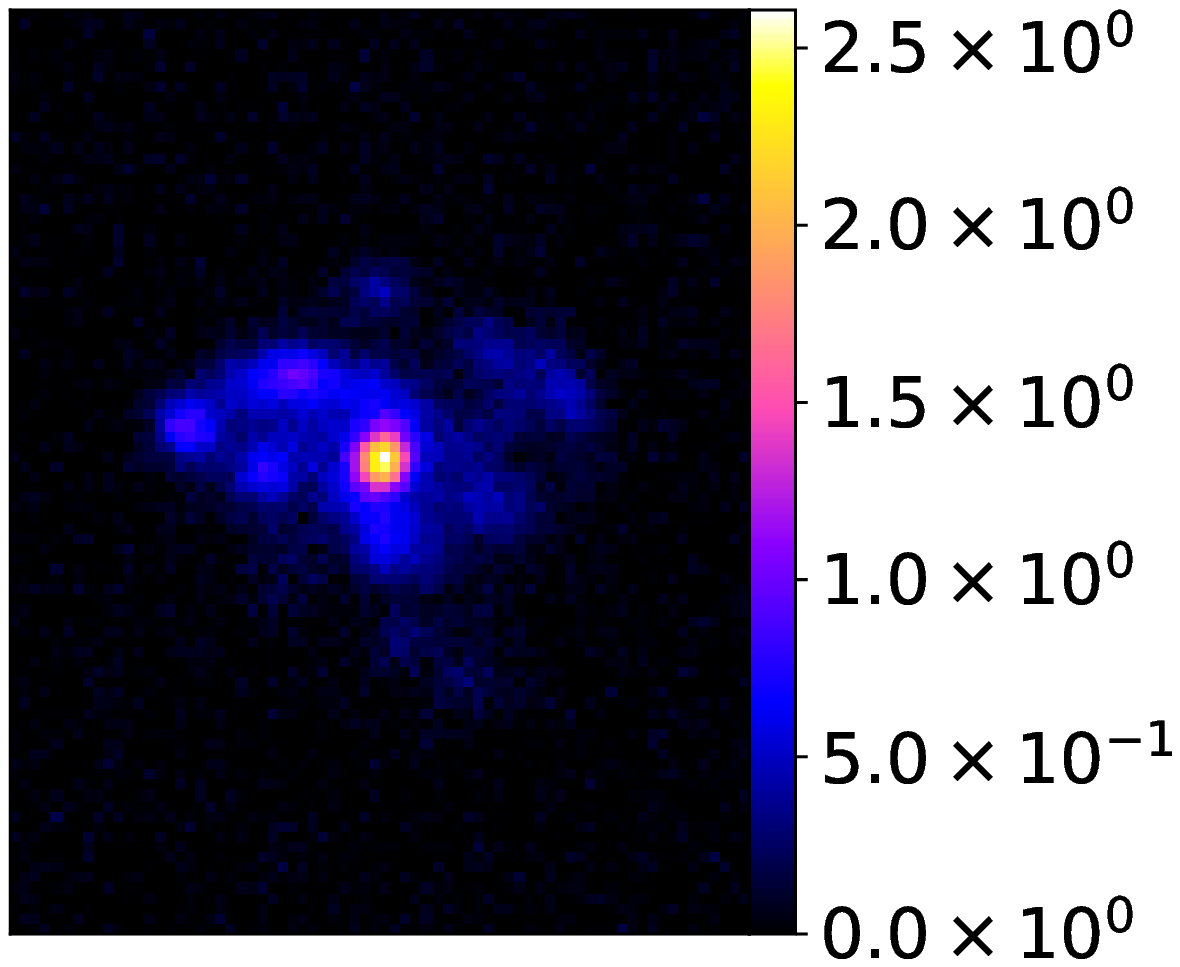}
    \includegraphics[width=3.5cm]
    {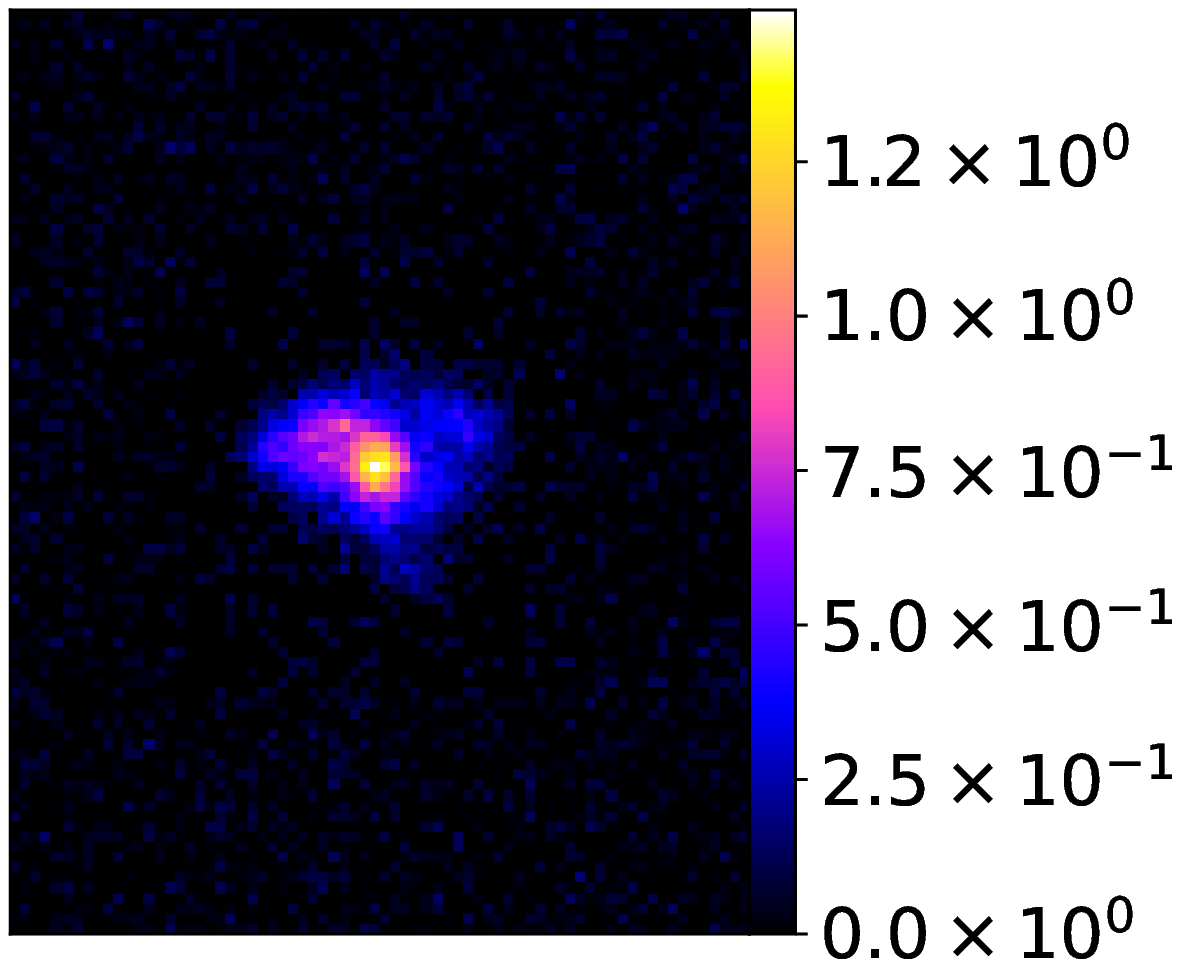}
    \includegraphics[width=3.5cm]
    {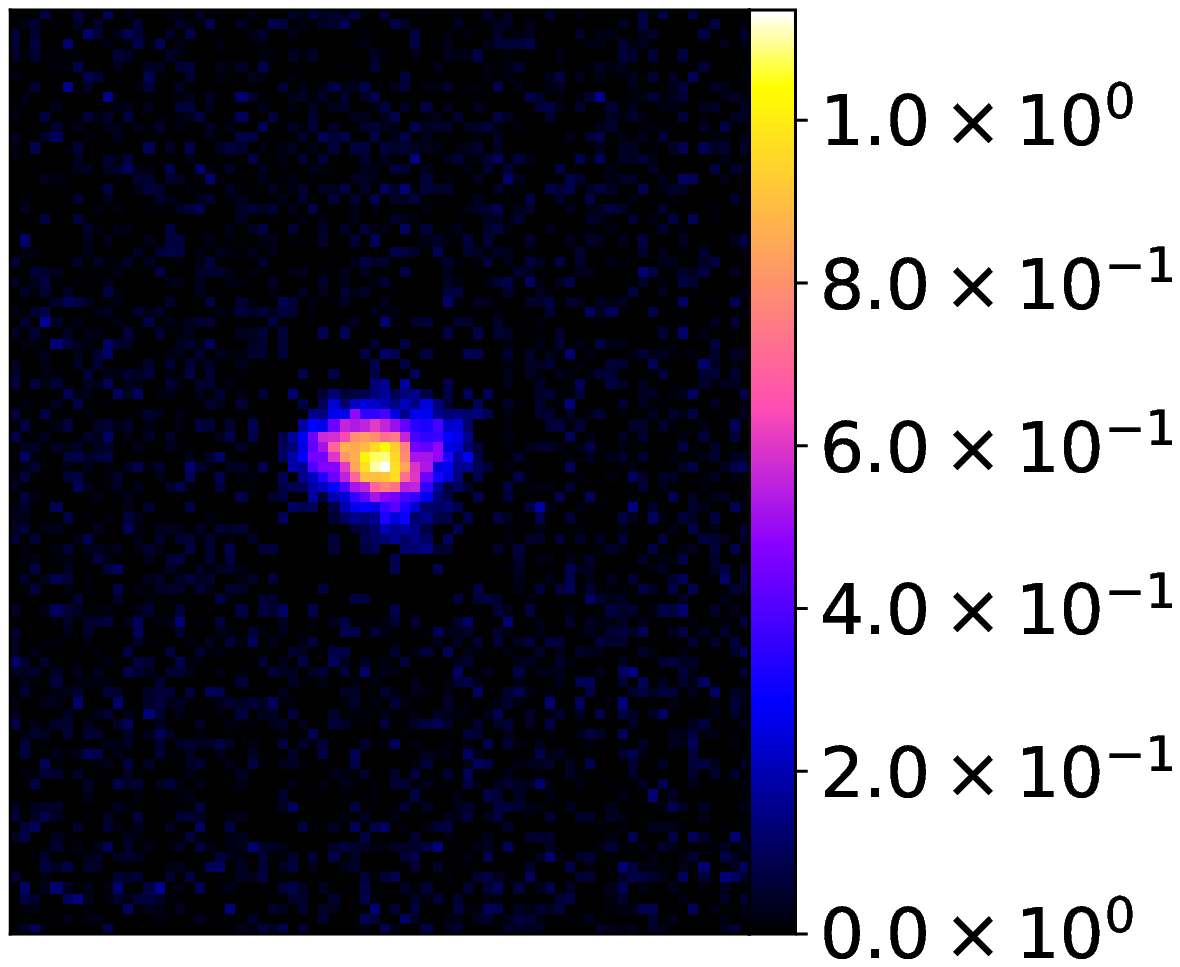}
     \includegraphics[width=3.5cm]
    {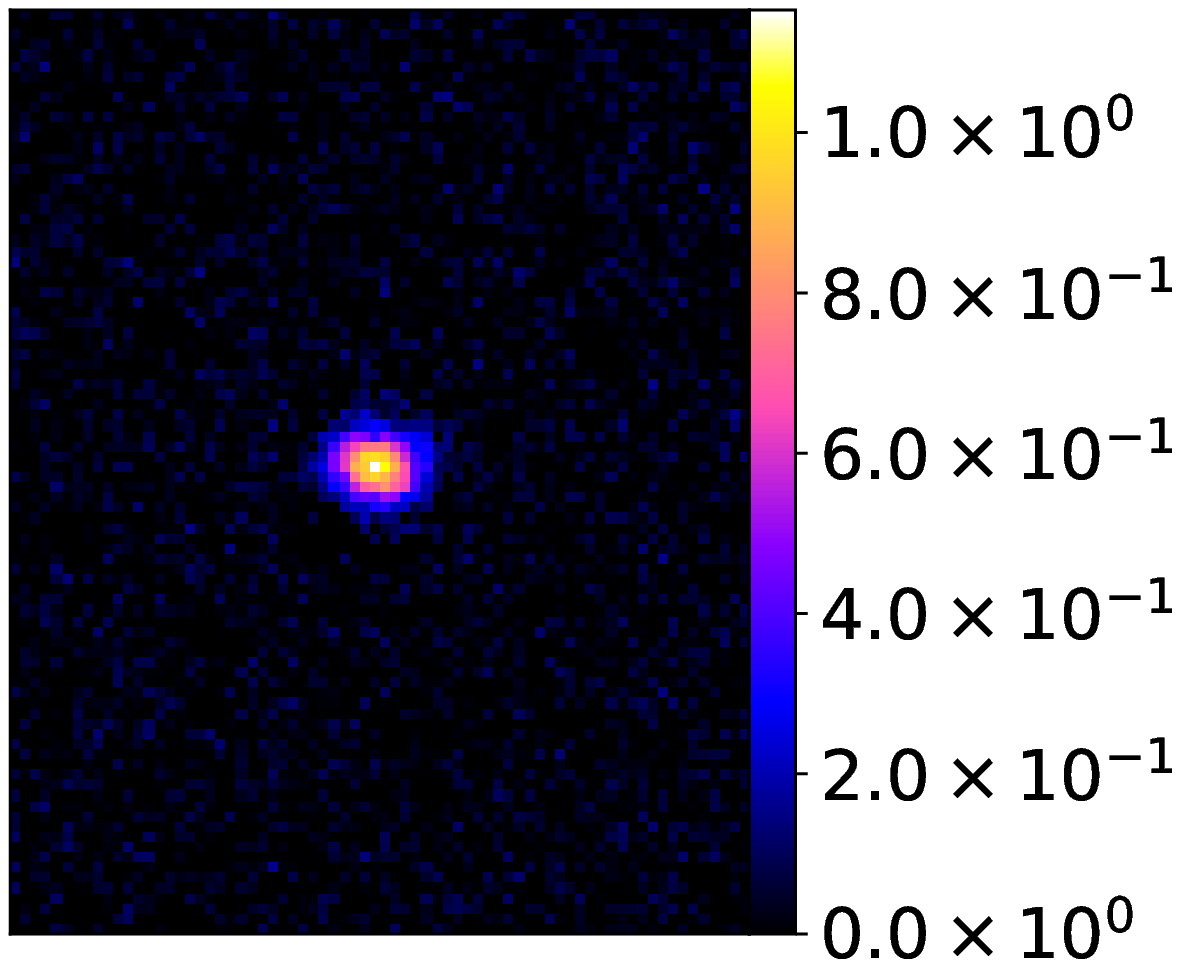}
     \includegraphics[width=3.5cm]
    {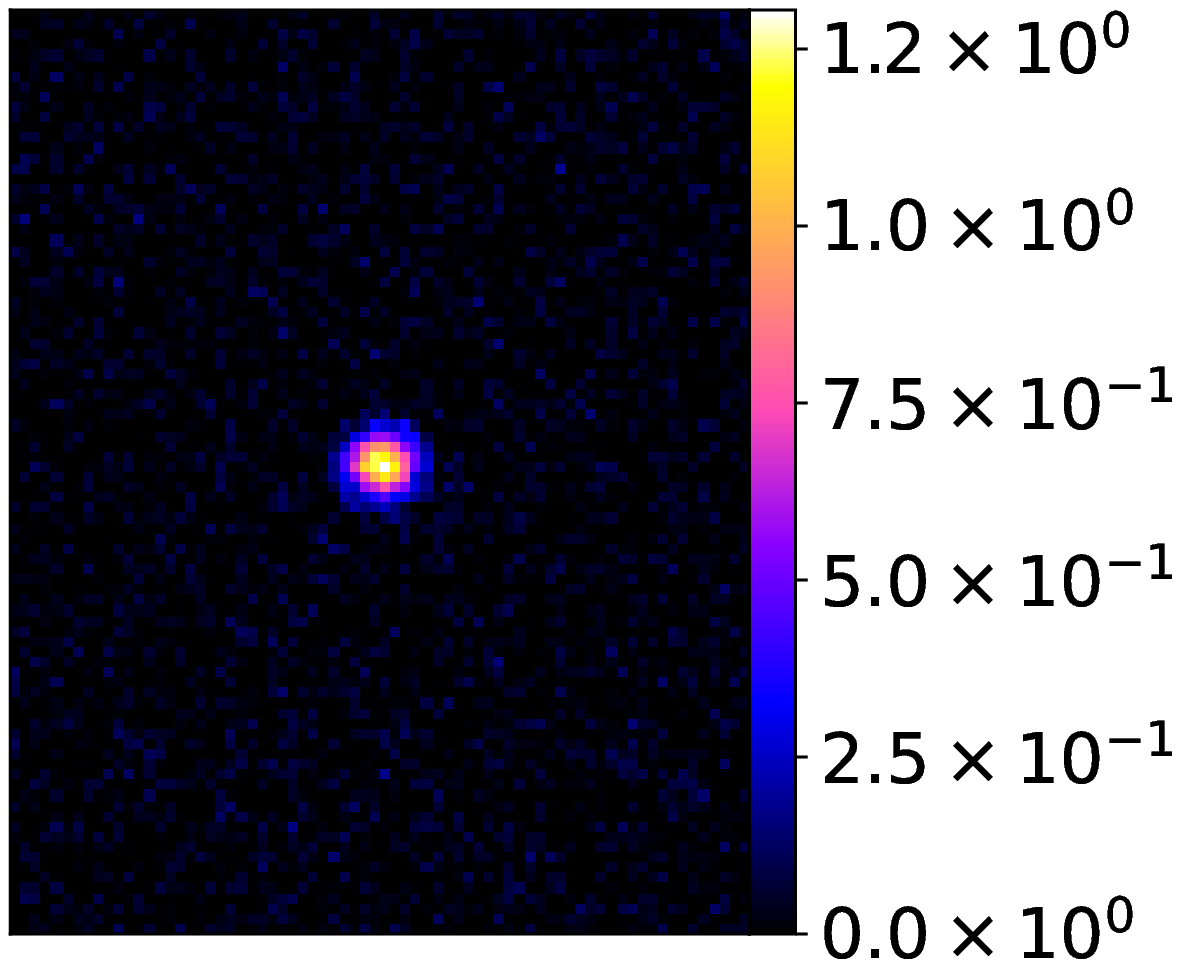}
    \caption{Thermal}   
\end{subfigure}
\hspace{1em}
\begin{subfigure}[t]{0.20\textwidth}
    \includegraphics[width=3.4cm]  
    {N6946-nt-z0.eps}
    \includegraphics[width=3.4cm]
    {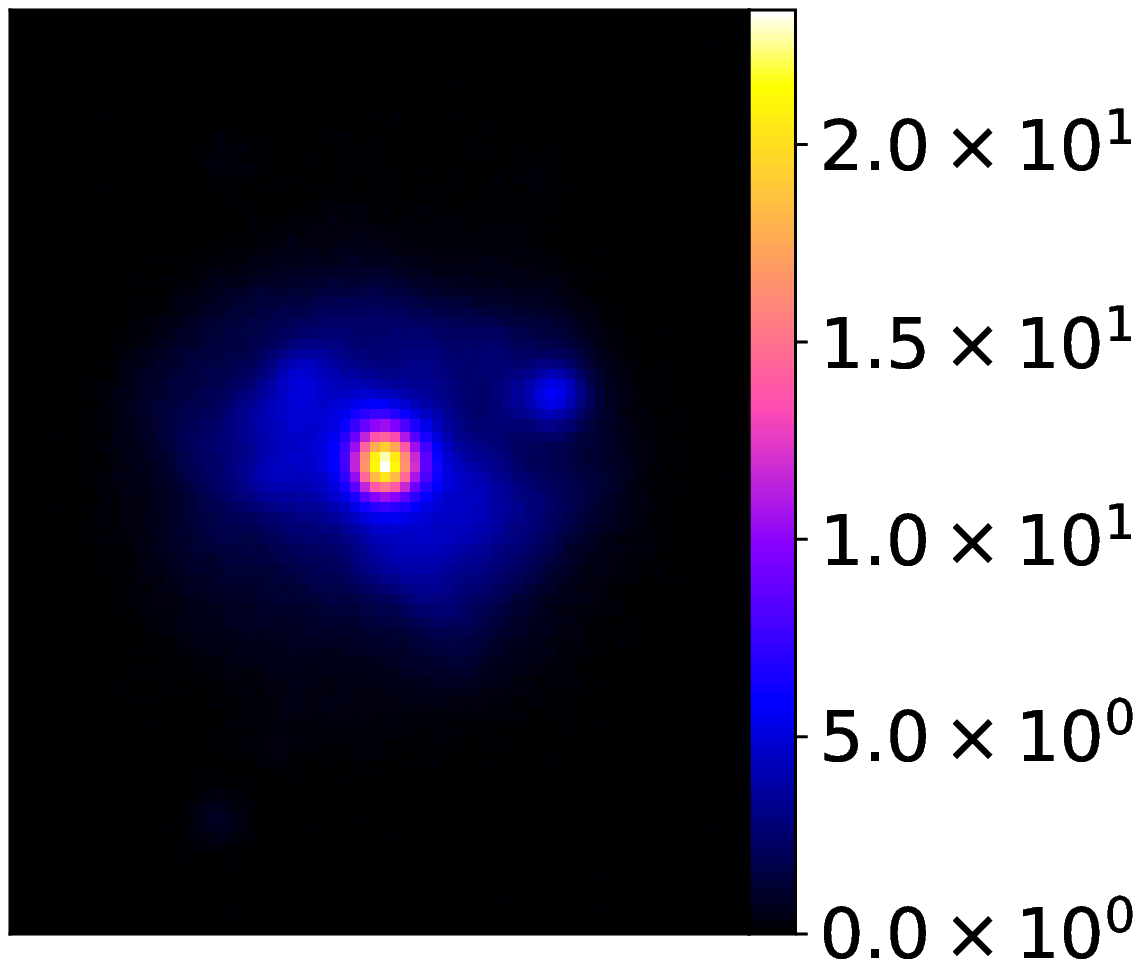}
    \includegraphics[width=3.4cm]
    {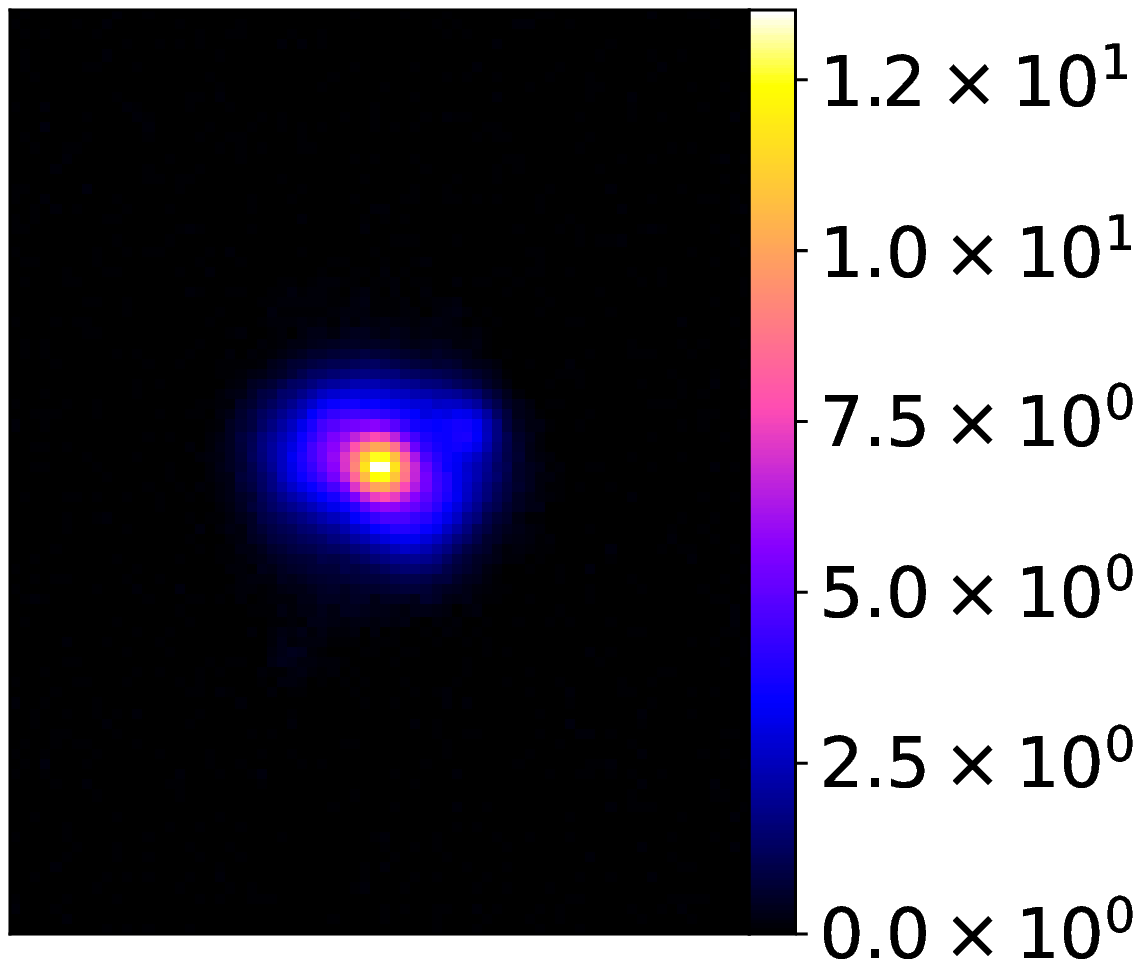}
    \includegraphics[width=3.4cm]
    {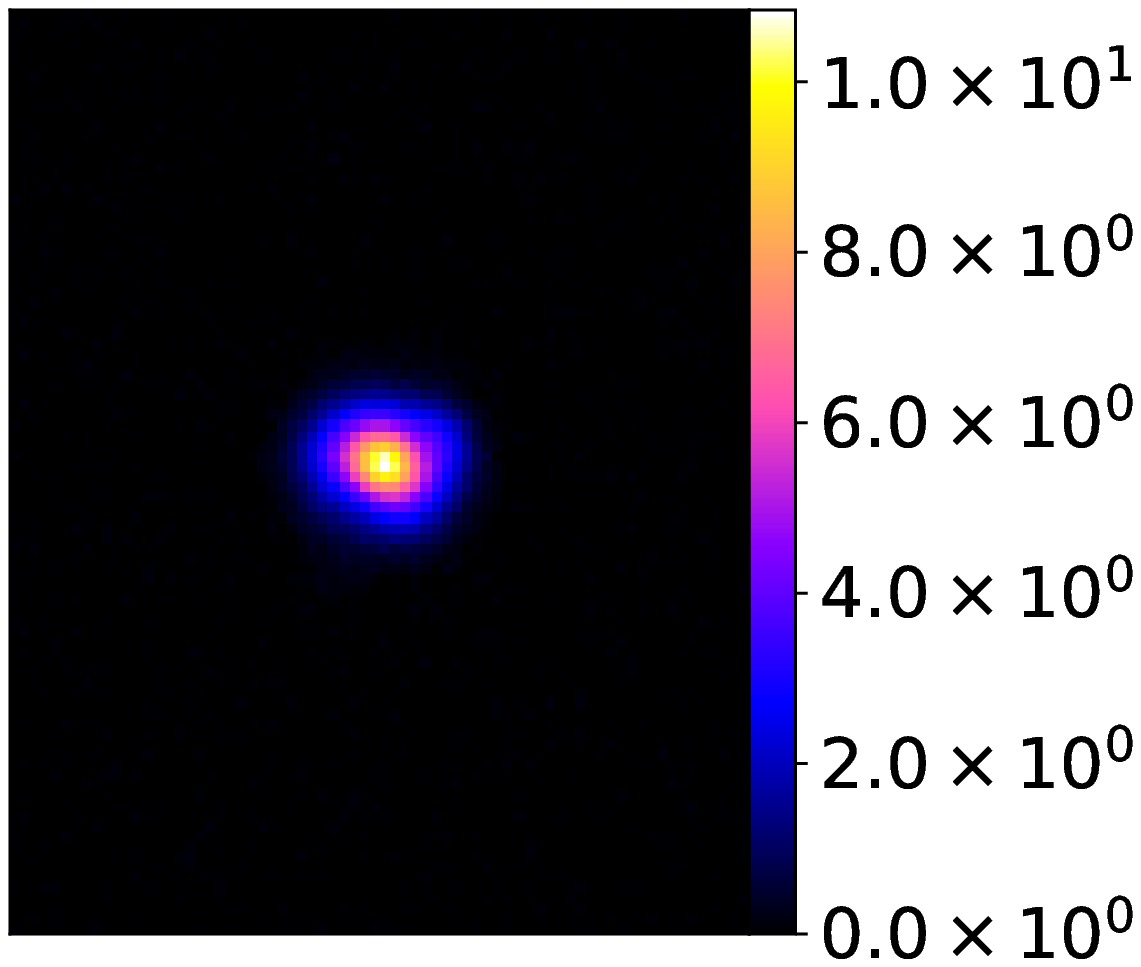}
    \includegraphics[width=3.4cm]
    {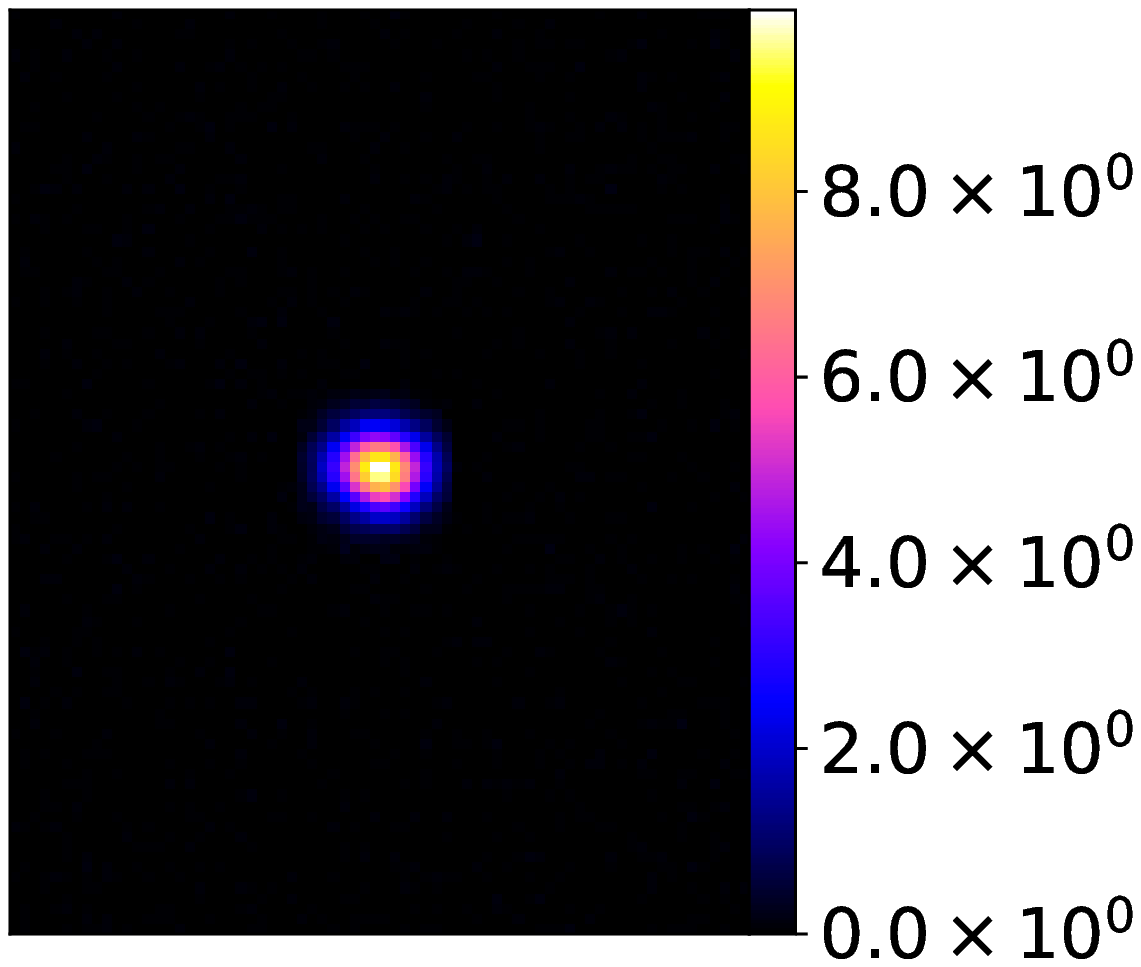}
    \includegraphics[width=3.4cm]
    {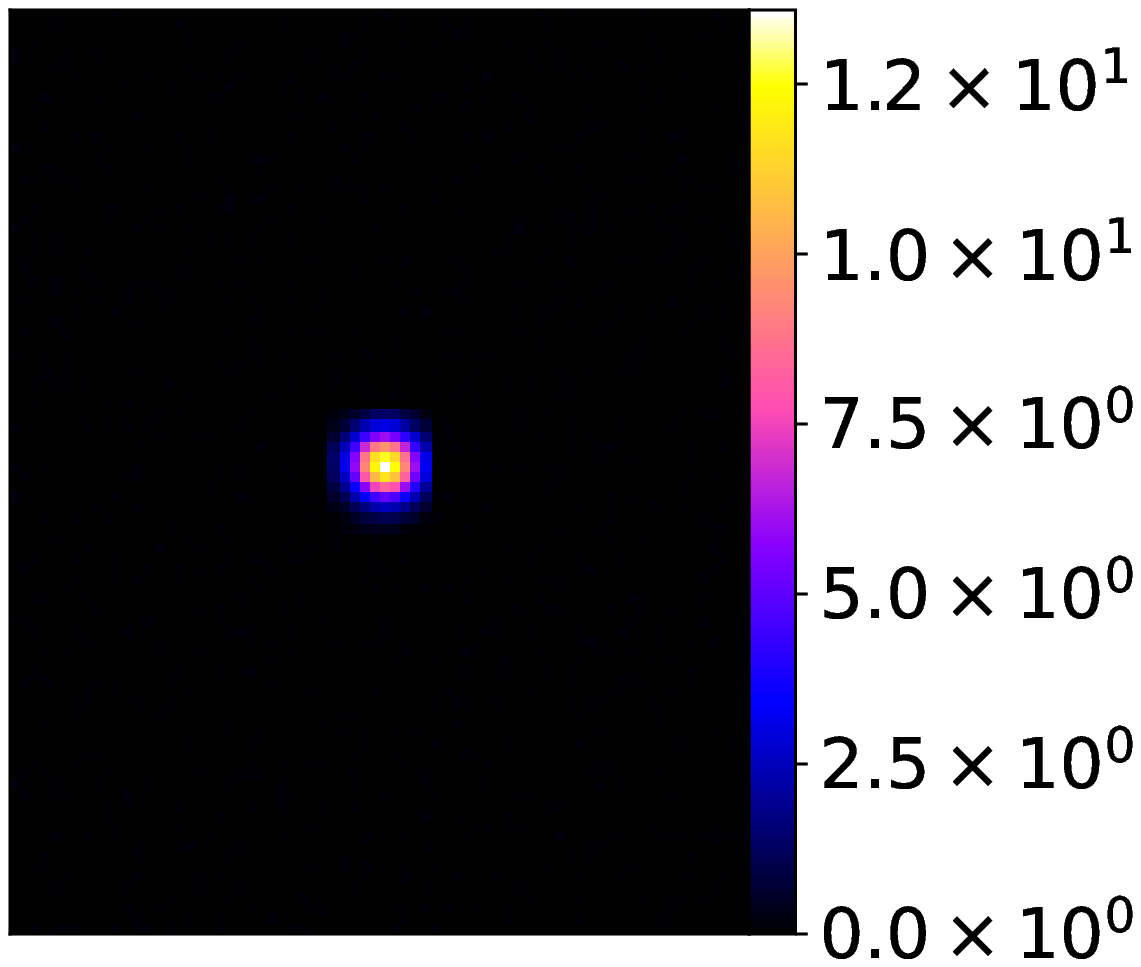}
    \caption{Nonthermal}
\end{subfigure}
\caption{ Same as Fig.~\ref{mapM51c12} for a NGC6946--like galaxy.  First row shows the thermal and non-thermal maps of NGC6946 \citep[from][]{tab_a}. Bars show the surface brightness in units of $\mu$Jy/beam.\label{mapN6946c12}}
\end{figure*}

\begin{figure*}
\centering
\textbf{case 1}
\hspace{6cm}
\textbf{case 2}\par\medskip
\begin{subfigure}[t]{0.20\textwidth}
    \makebox[0pt][r]{\makebox[30pt]{\raisebox{40pt}{\rotatebox[origin=c]{90}{Input}}}}%
    \includegraphics[width=3.5cm]
    {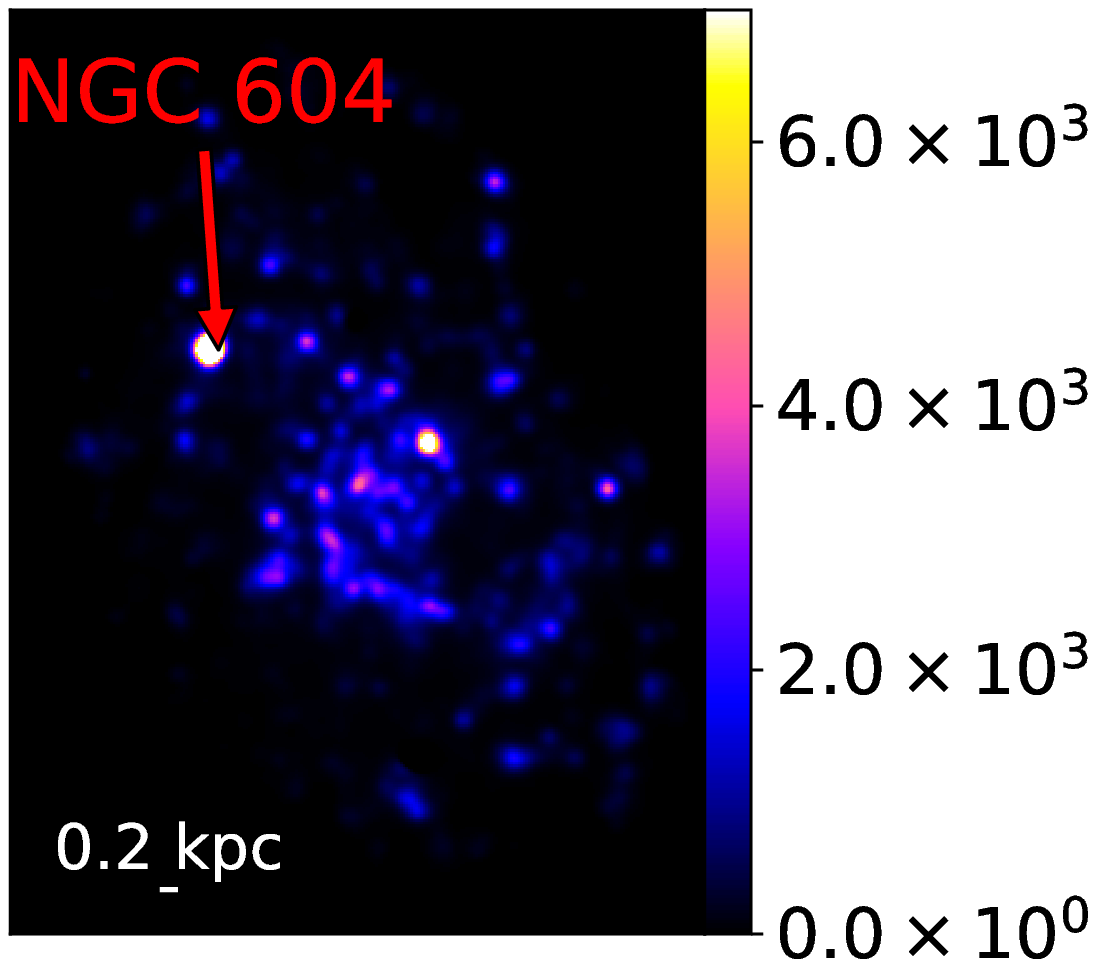}
    \makebox[0pt][r]{\makebox[30pt]{\raisebox{40pt}{\rotatebox[origin=c]{90}{$z=0.15$}}}}%
    \includegraphics[width=3.5cm]
    {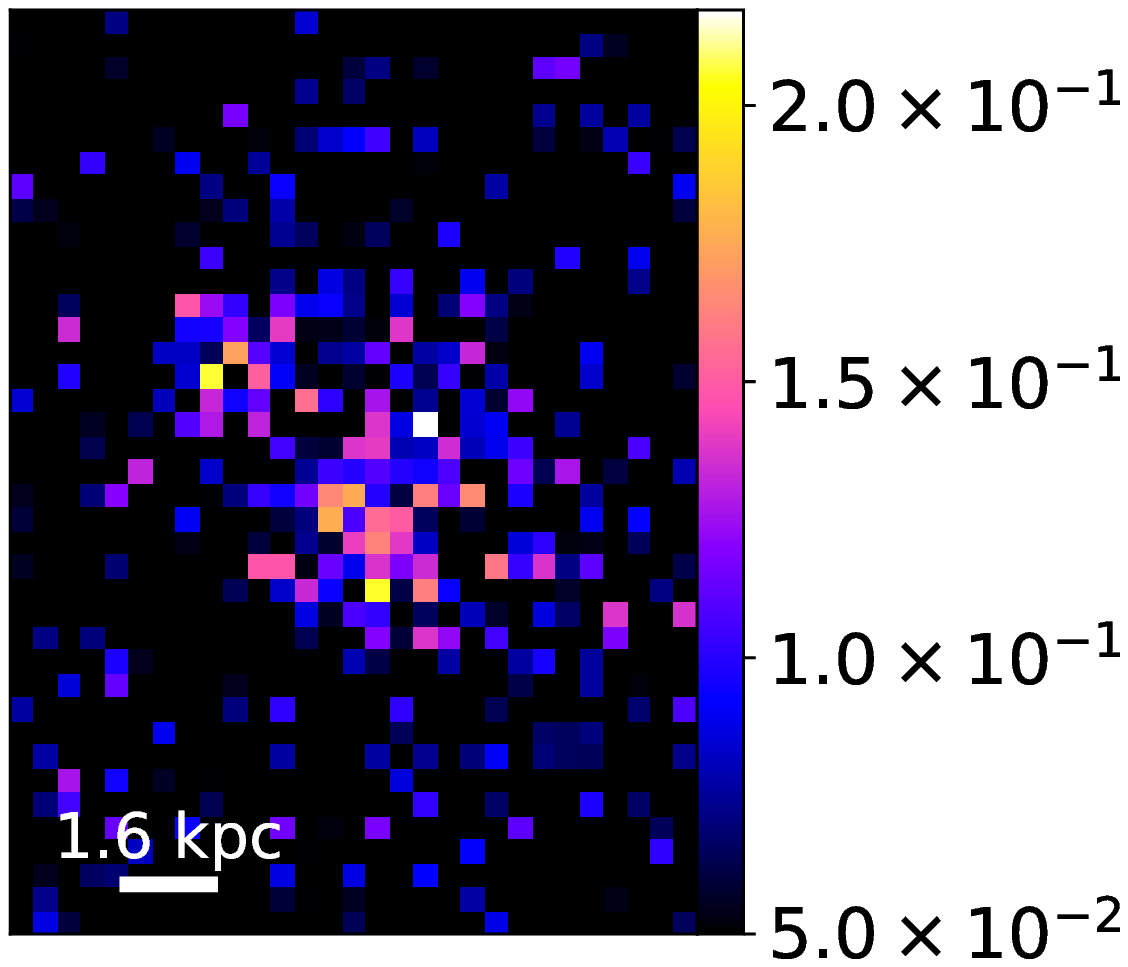}
    \makebox[0pt][r]{\makebox[30pt]{\raisebox{40pt}{\rotatebox[origin=c]{90}{$z=0.3$}}}}%
    \includegraphics[width=3.5cm]
    {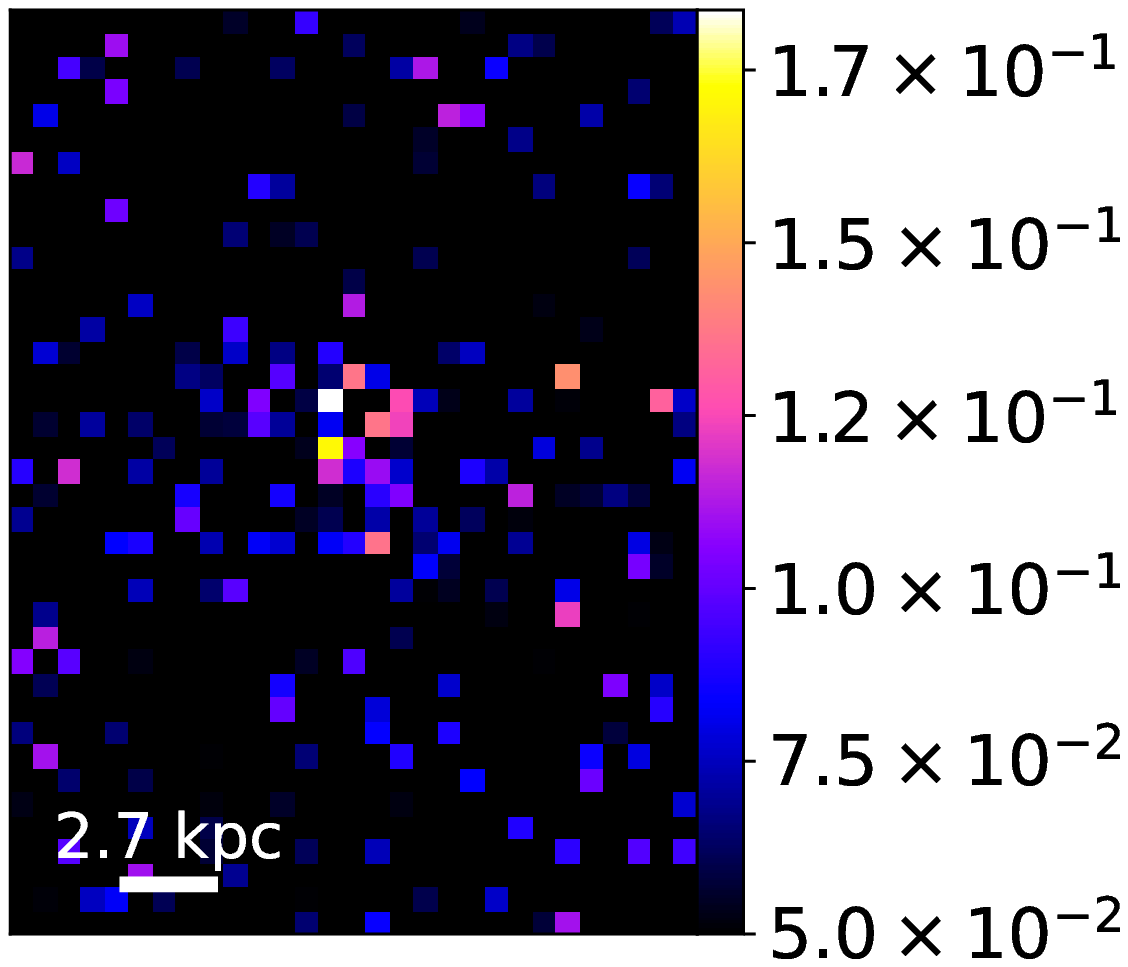}
    \makebox[0pt][r]{\makebox[30pt]{\raisebox{40pt}{\rotatebox[origin=c]{90}{$z=0.5$}}}}%
    \includegraphics[width=3.5cm]
    {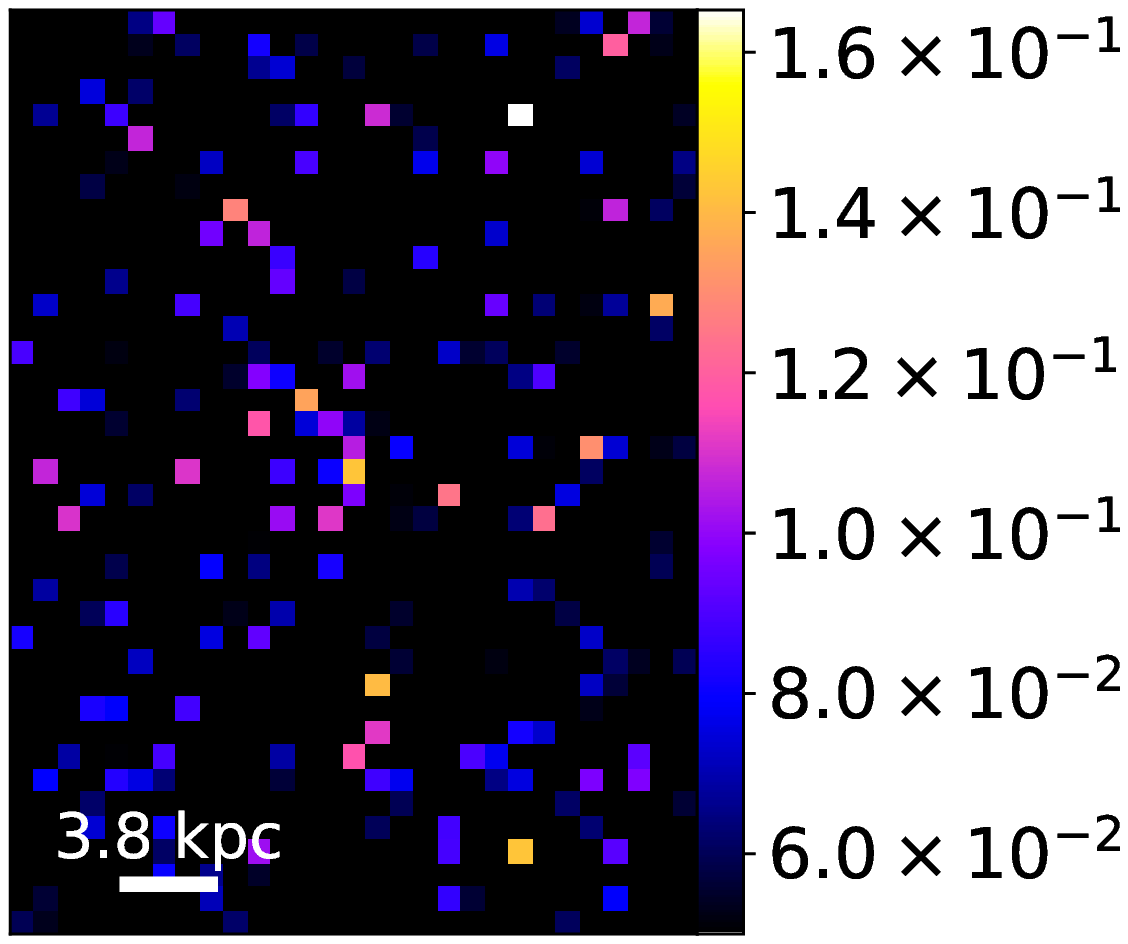}
    \makebox[0pt][r]{\makebox[30pt]{\raisebox{40pt}{\rotatebox[origin=c]{90}{$z=1$}}}}%
    \includegraphics[width=3.5cm]
    {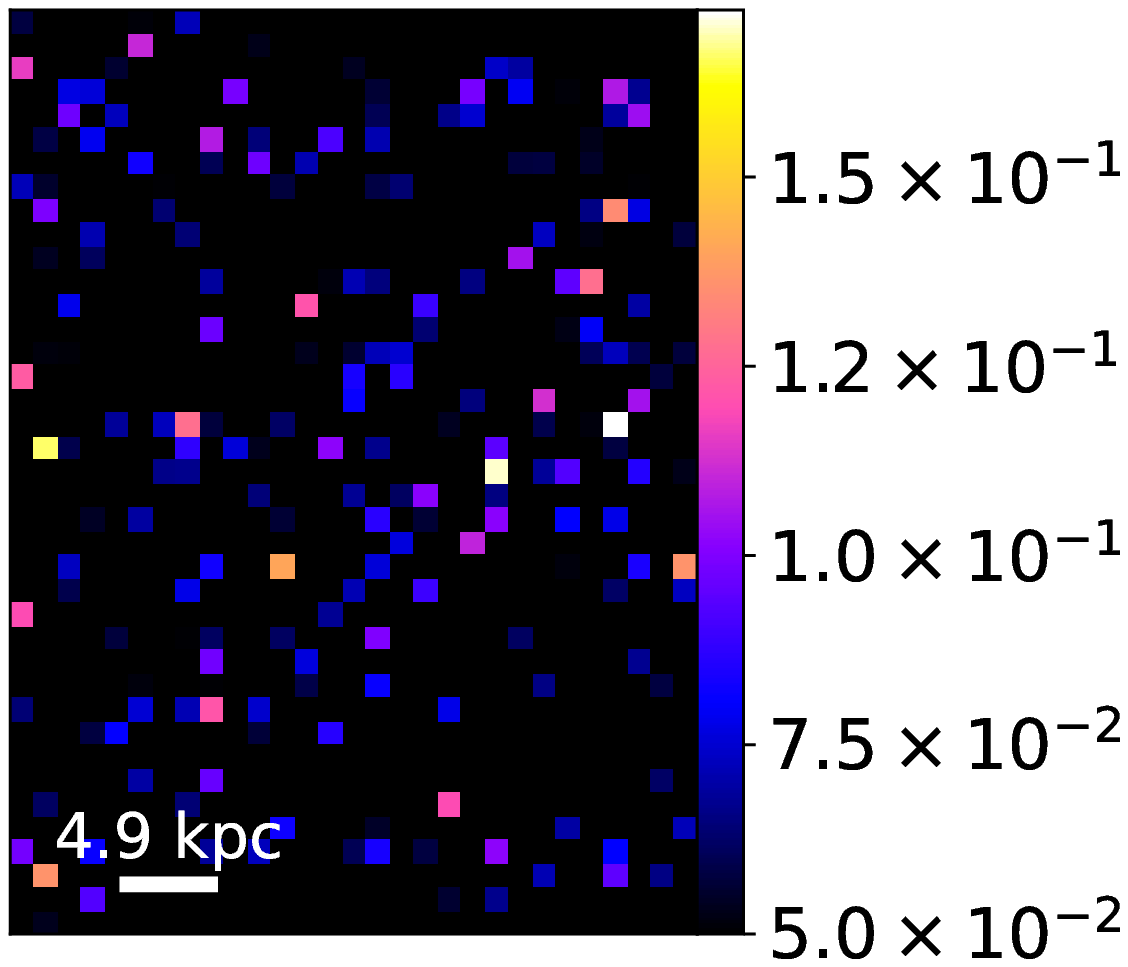}
     \makebox[0pt][r]{\makebox[30pt]{\raisebox{40pt}{\rotatebox[origin=c]{90}{$z=2$}}}}%
    \includegraphics[width=3.5cm]
    {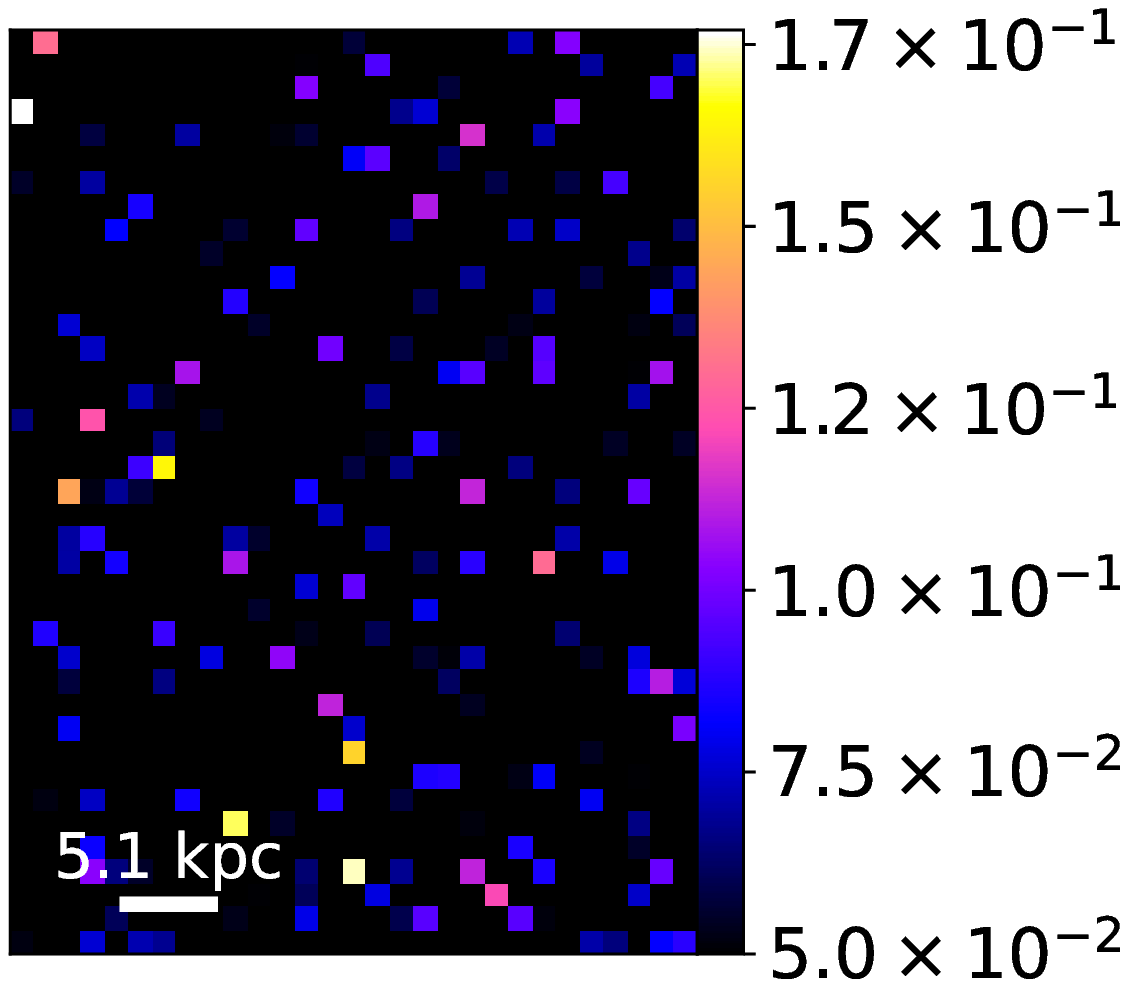}
    \caption{Thermal}
\end{subfigure}
\hspace{1em}
\begin{subfigure}[t]{0.20\textwidth}
    \includegraphics[width=3.45cm]  
    {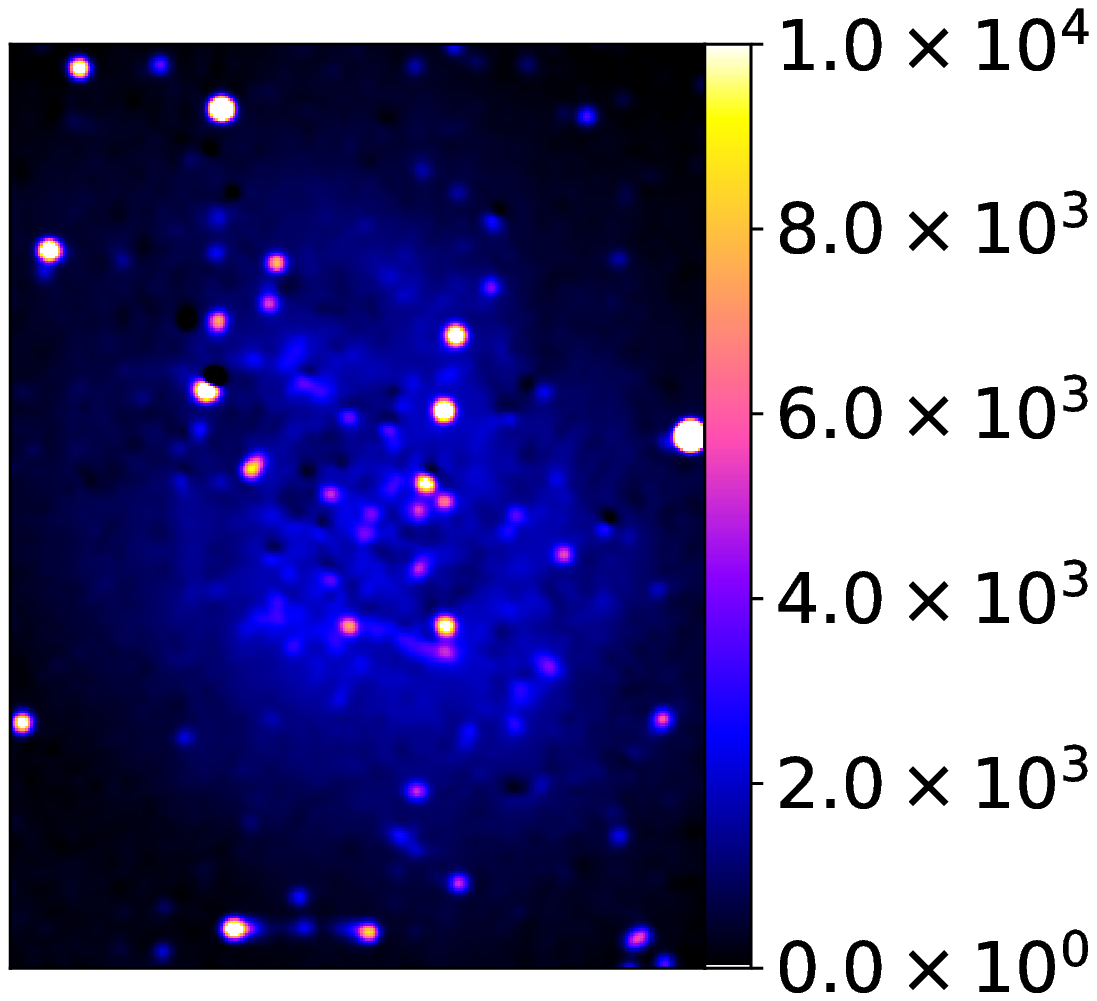}
    \includegraphics[width=3.45cm]
    {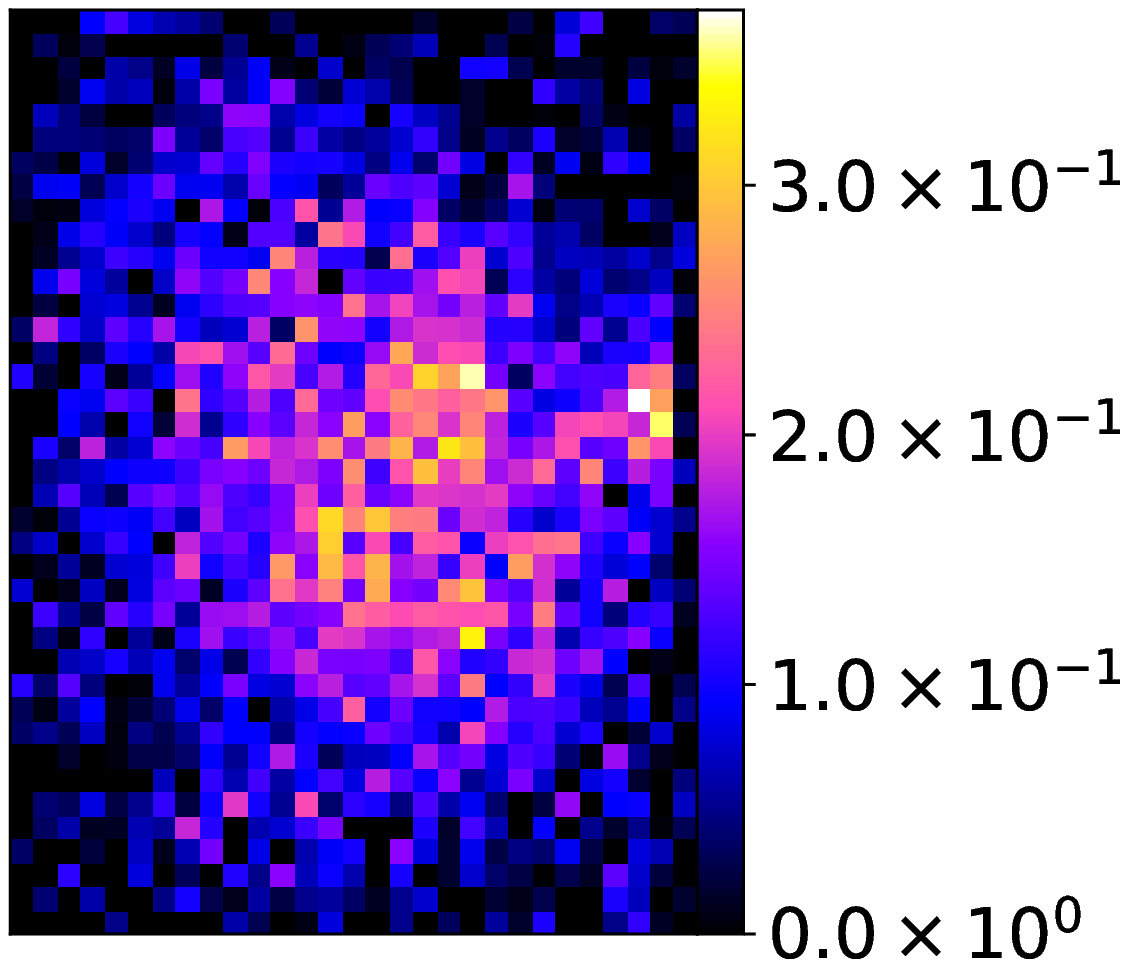}
    \includegraphics[width=3.45cm]
    {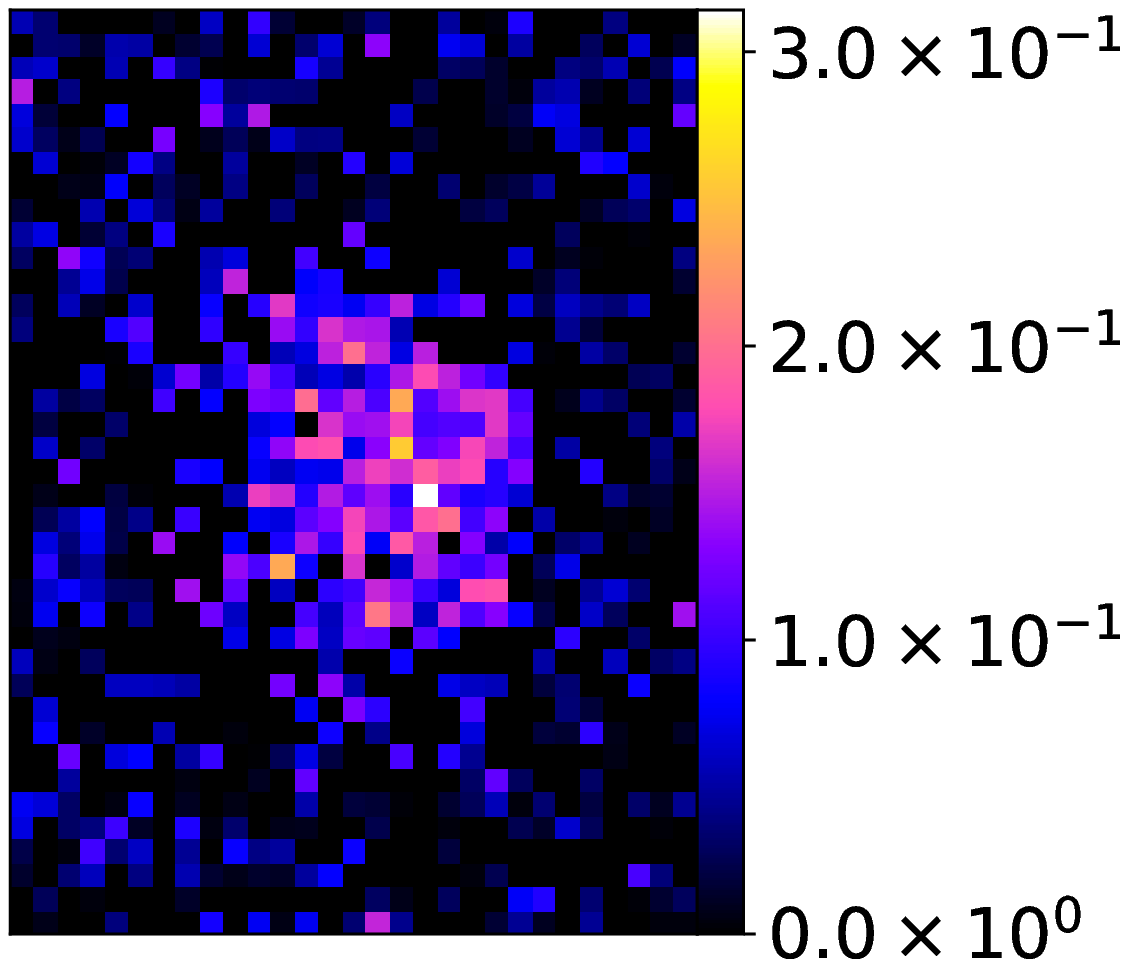}
    \includegraphics[width=3.45cm]
    {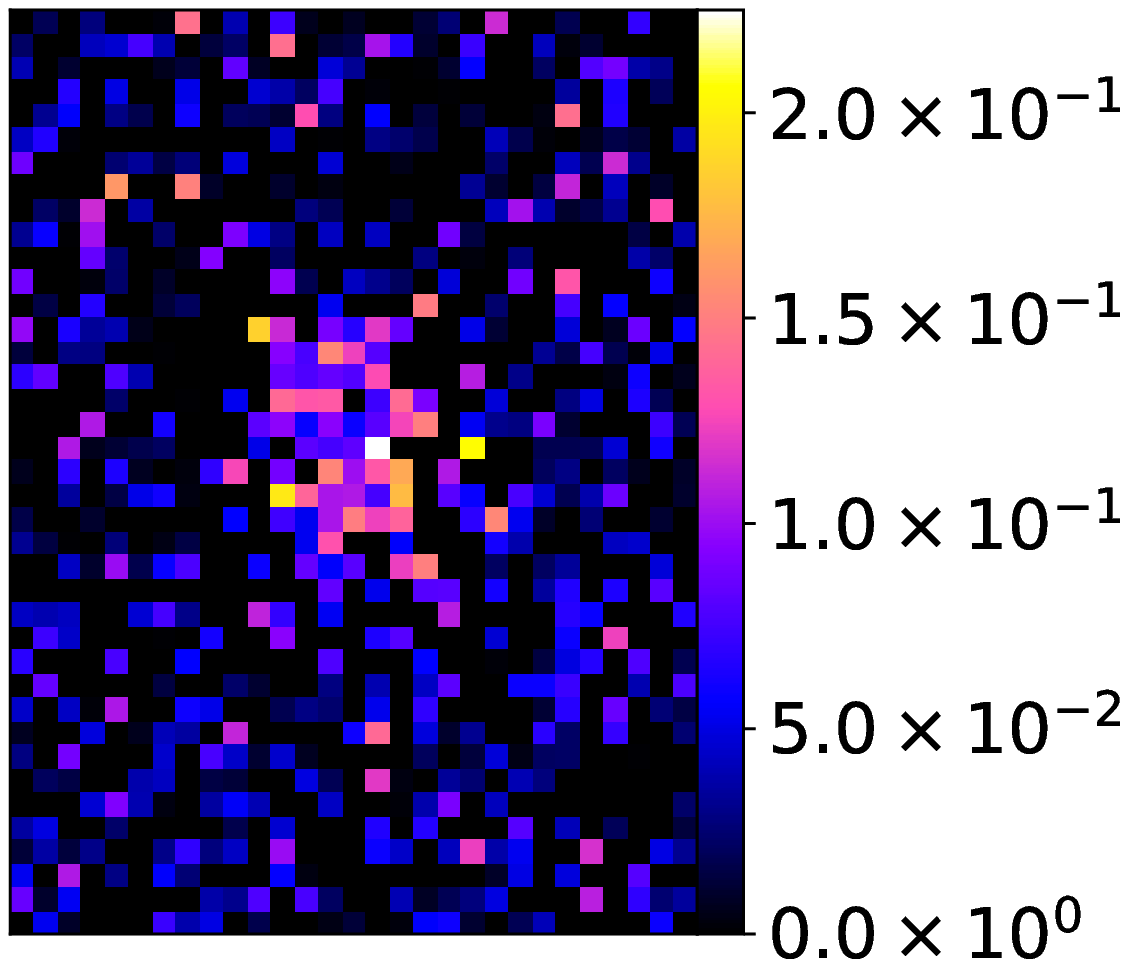}
    \includegraphics[width=3.45cm]
    {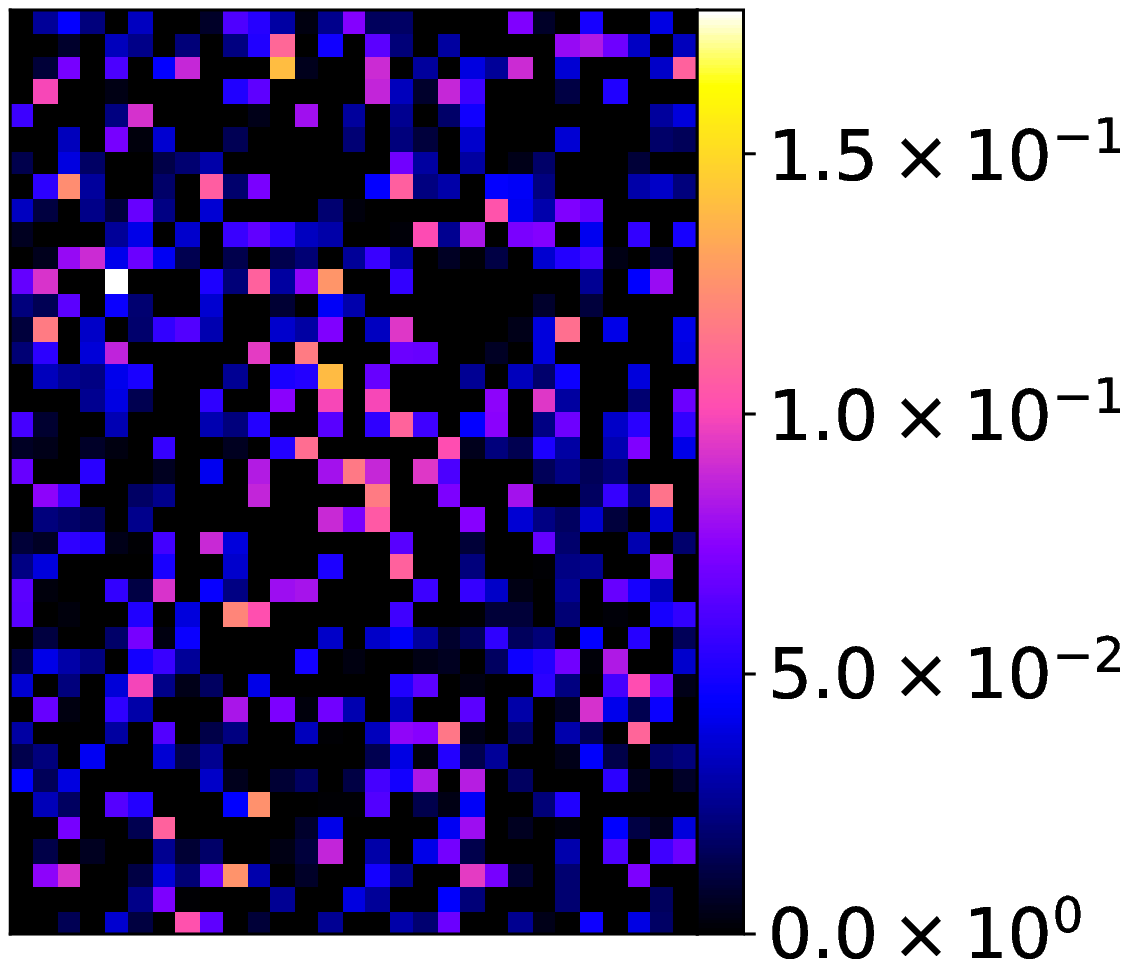}
     \includegraphics[width=3.45cm]
    {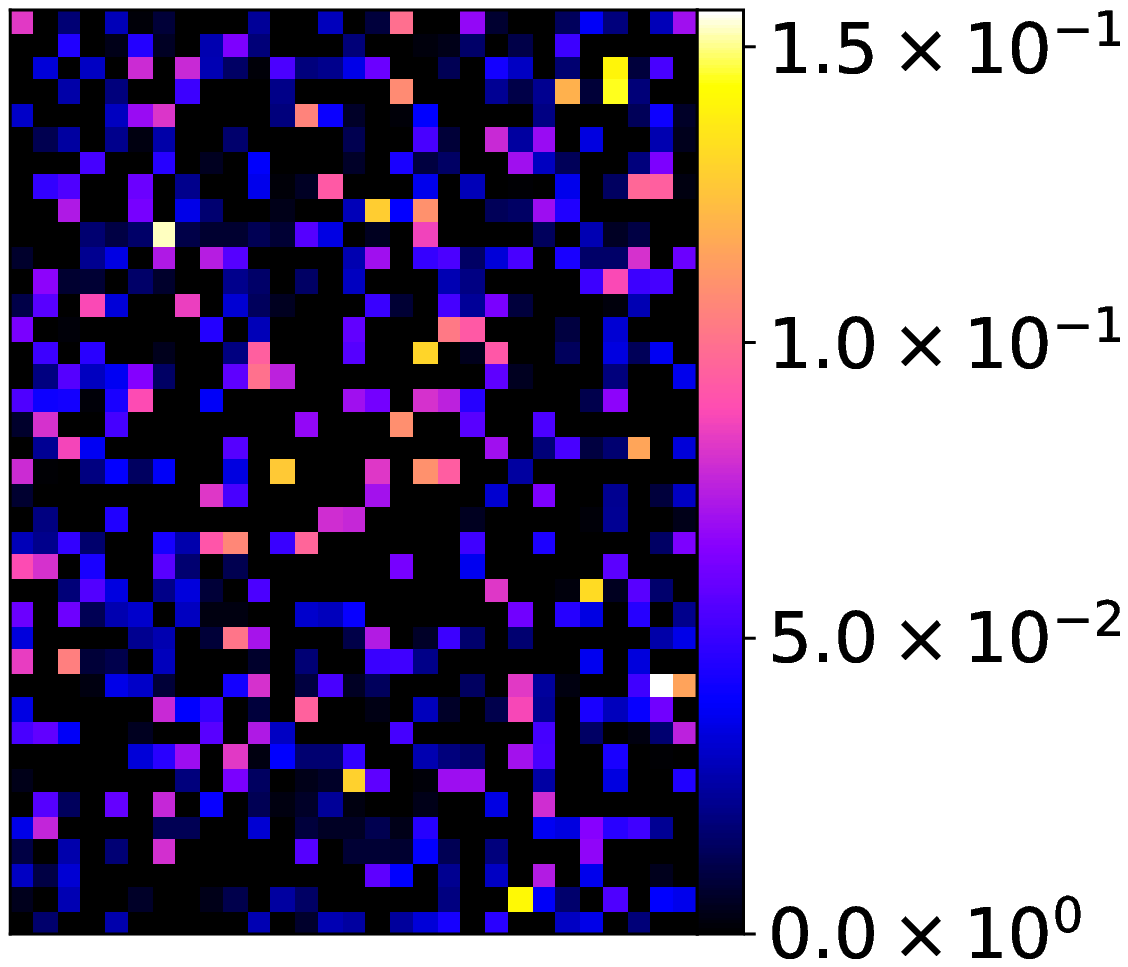}
    \caption{Nonthermal}
\end{subfigure}
\hspace{1em}
\begin{subfigure}[t]{0.20\textwidth}
    \includegraphics[width=3.52cm]  
    {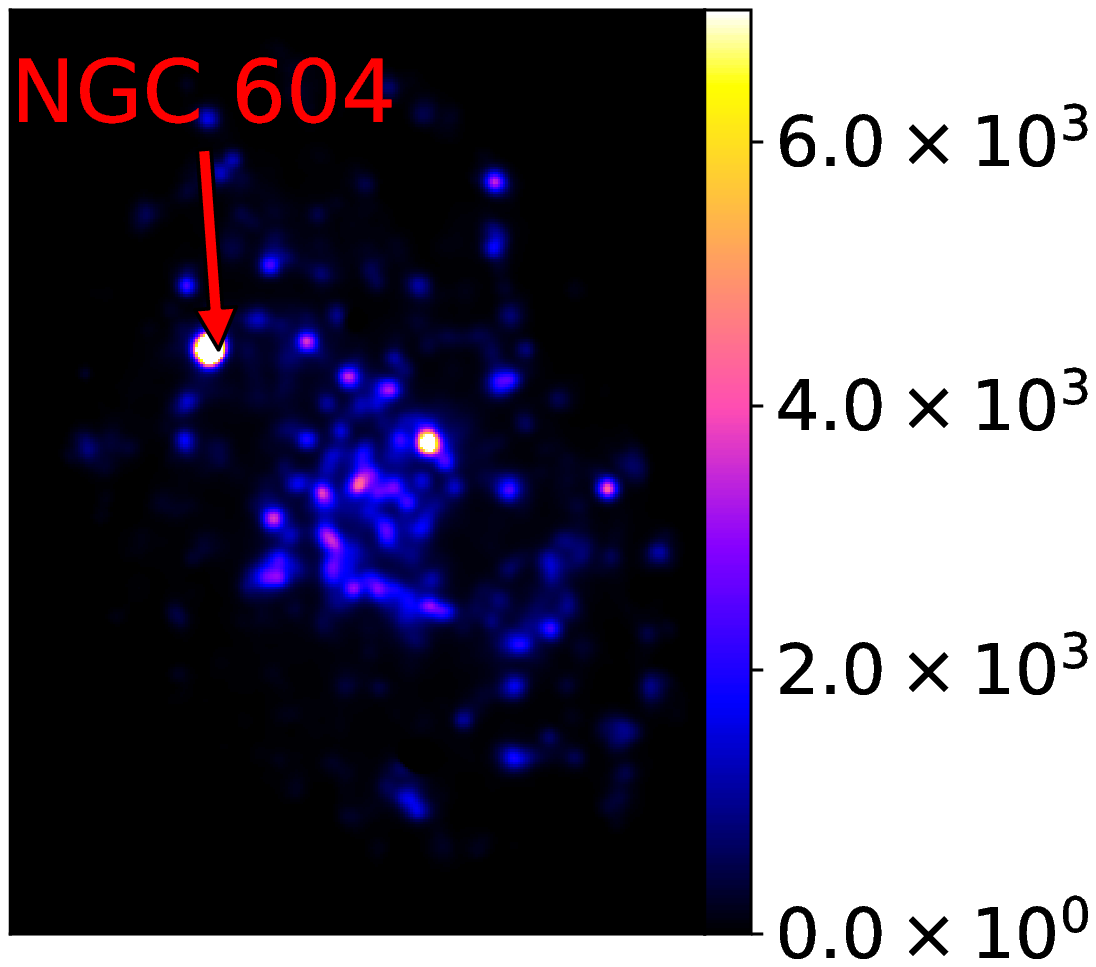}
    \includegraphics[width=3.52cm]
    {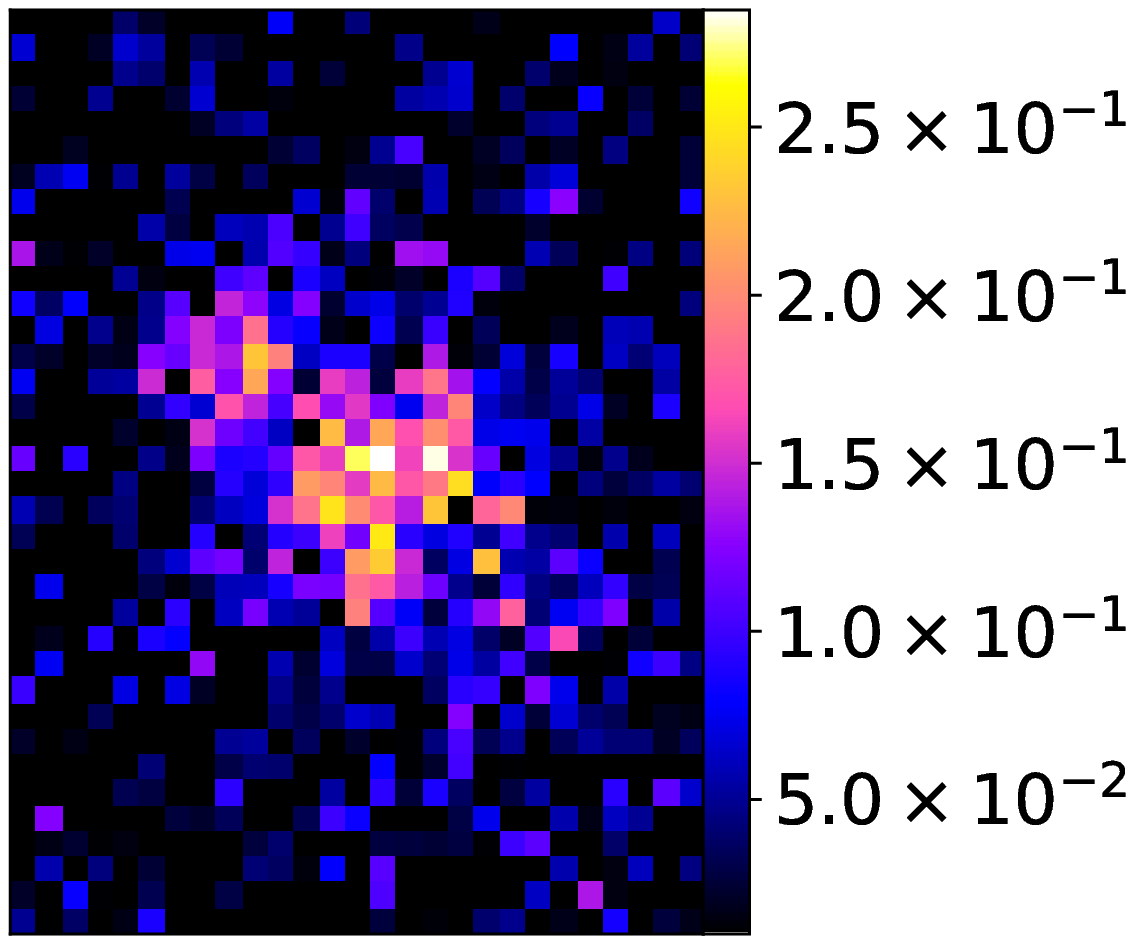}
    \includegraphics[width=3.52cm]
    {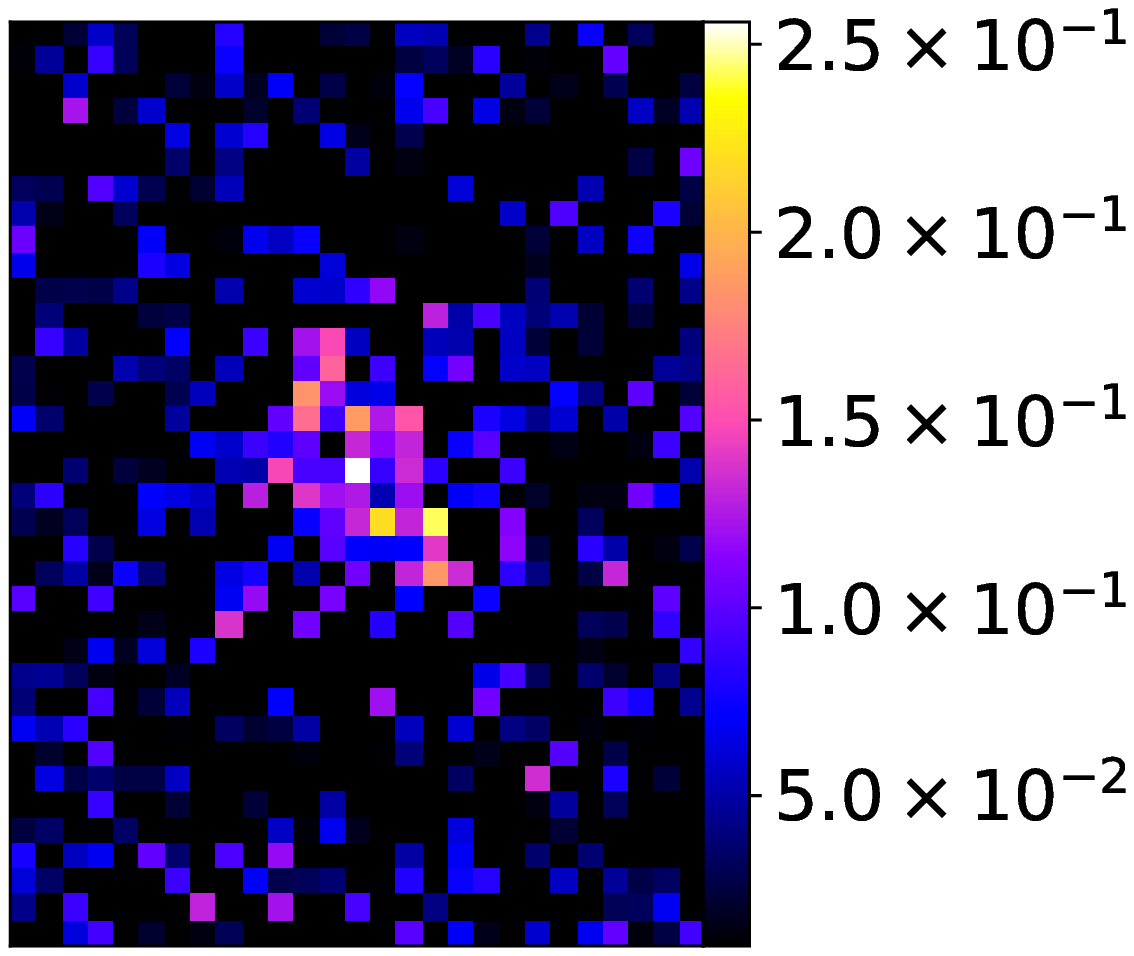}
    \includegraphics[width=3.52cm]
    {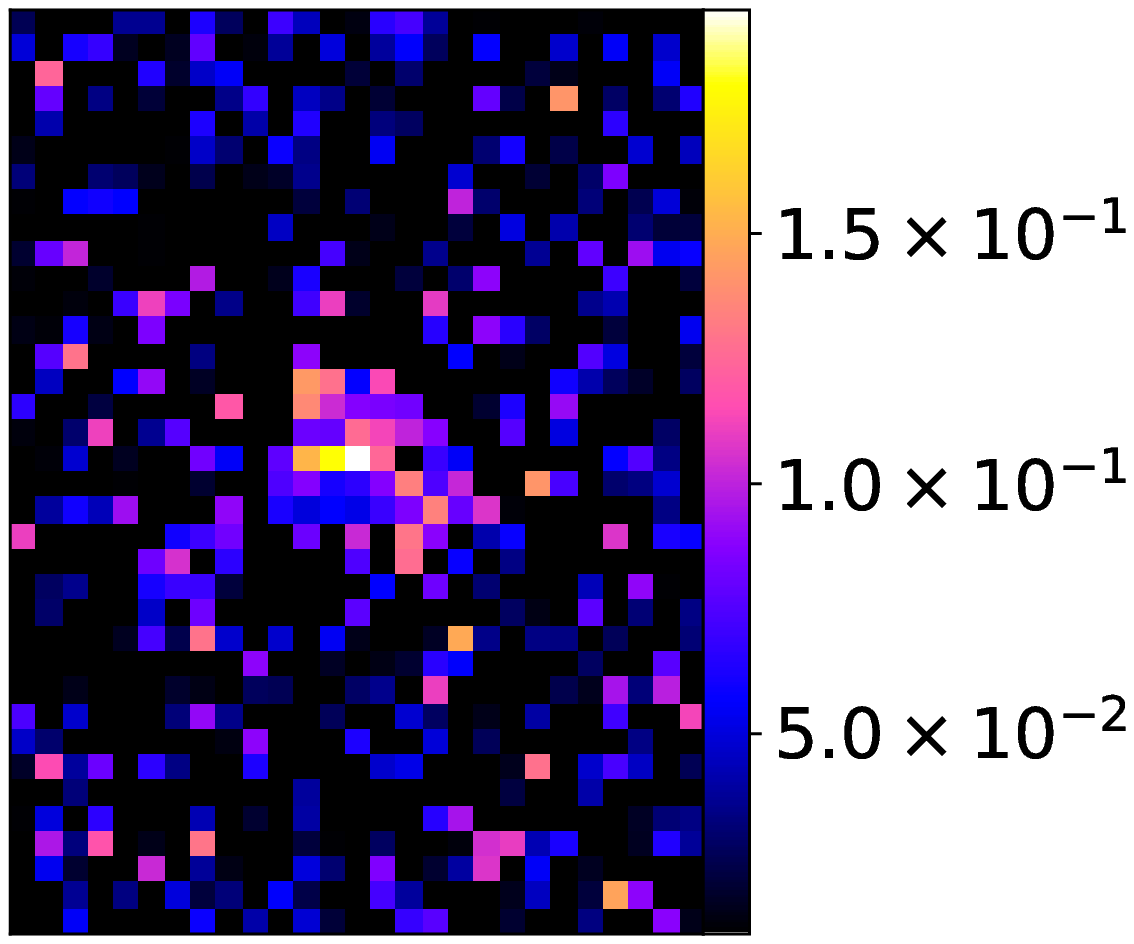}
     \includegraphics[width=3.52cm]
    {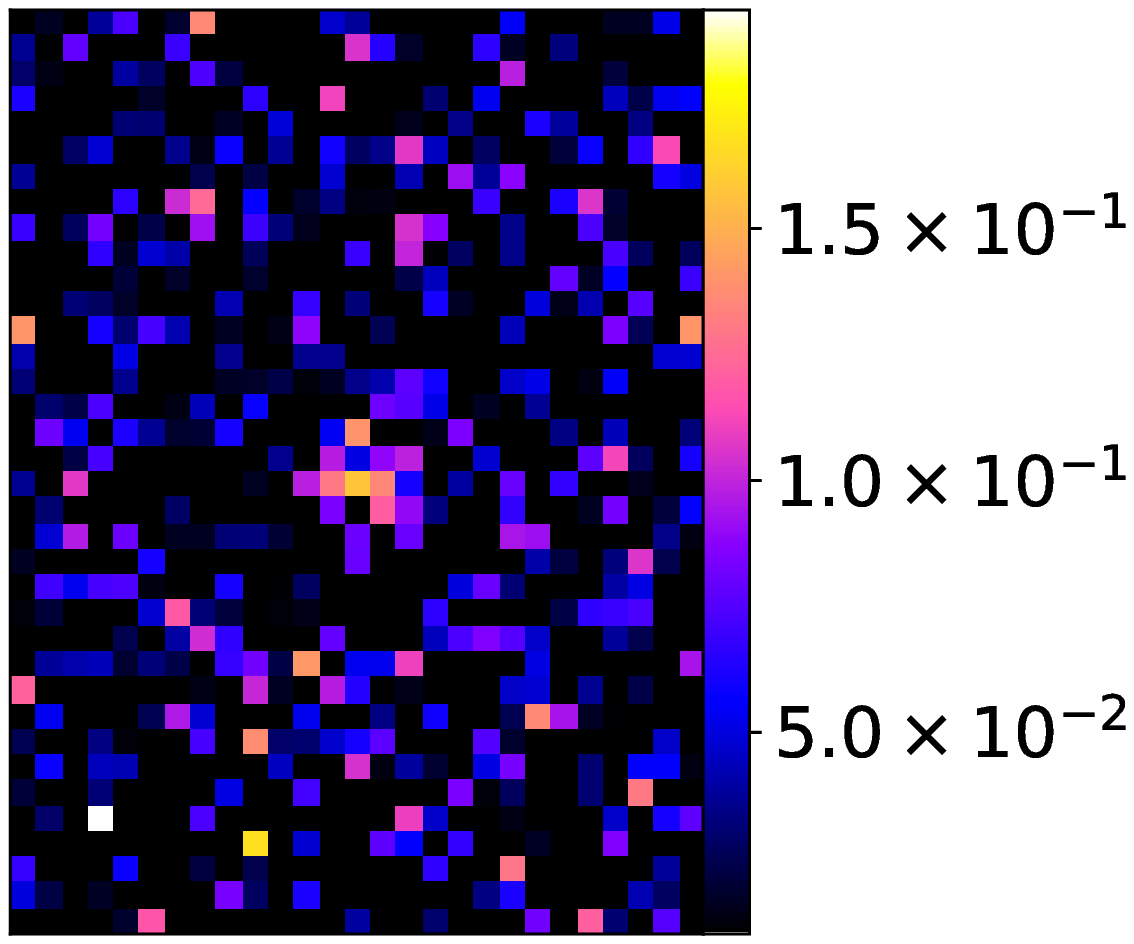}
     \includegraphics[width=3.52cm]
    {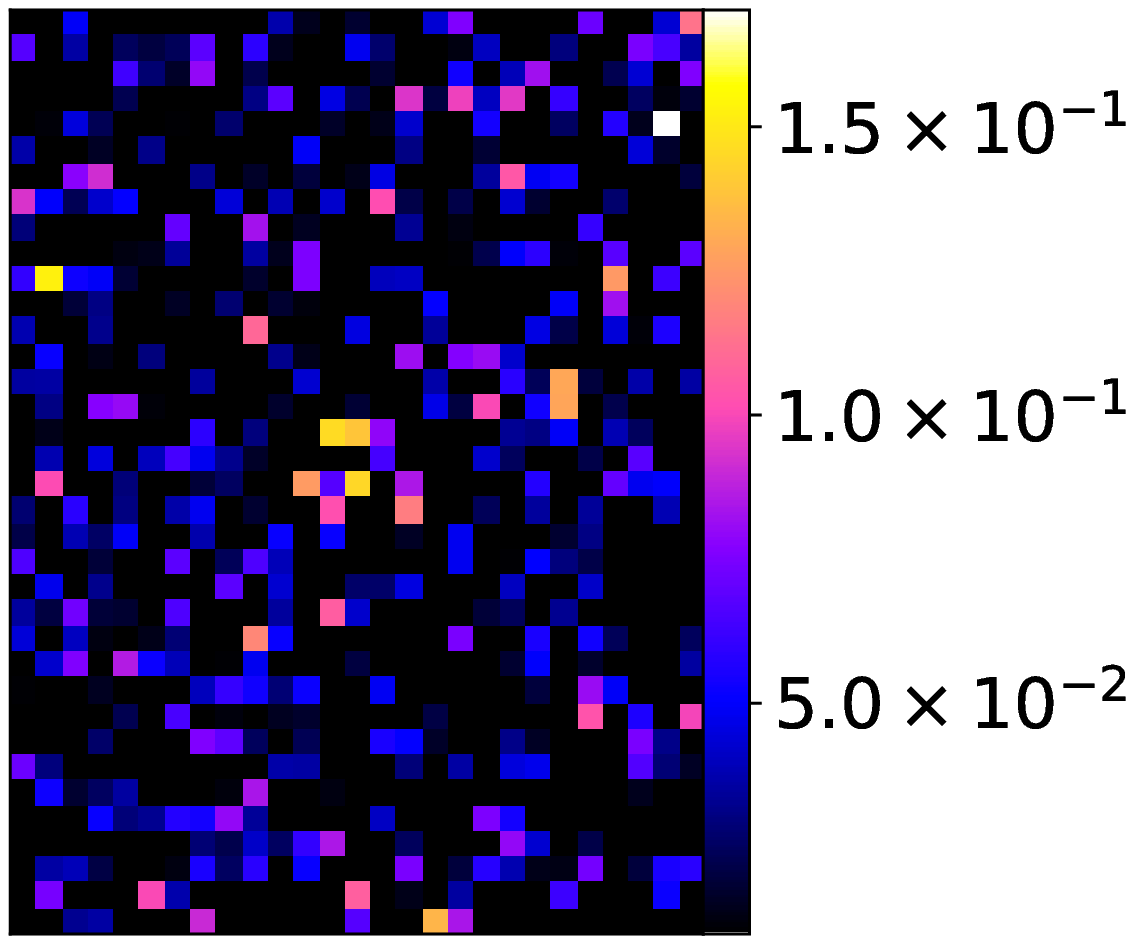}
    \caption{Thermal}   
\end{subfigure}
\hspace{1em}
\begin{subfigure}[t]{0.20\textwidth}
    \includegraphics[width=3.45cm]  
    {M33-nt-z0.eps}
    \includegraphics[width=3.45cm]
    {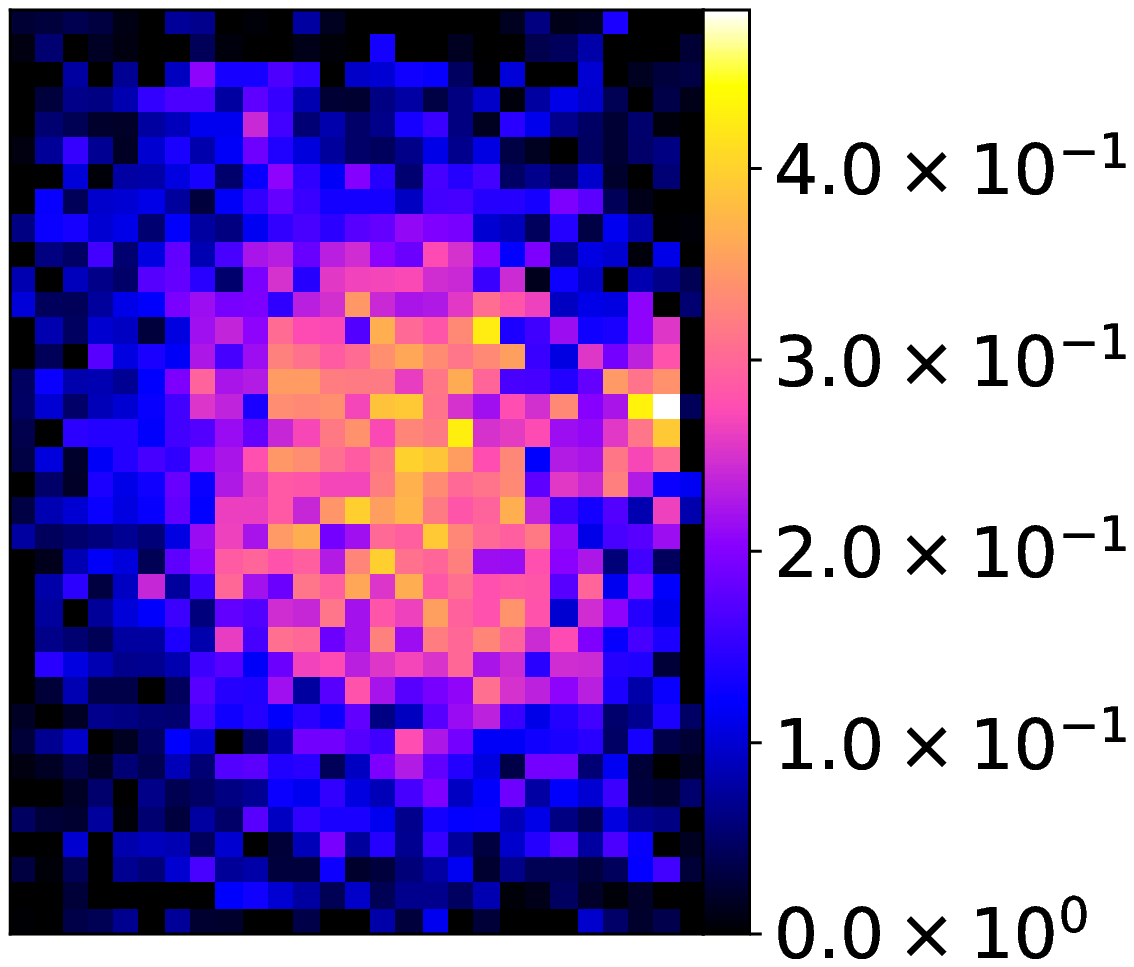}
    \includegraphics[width=3.45cm]
    {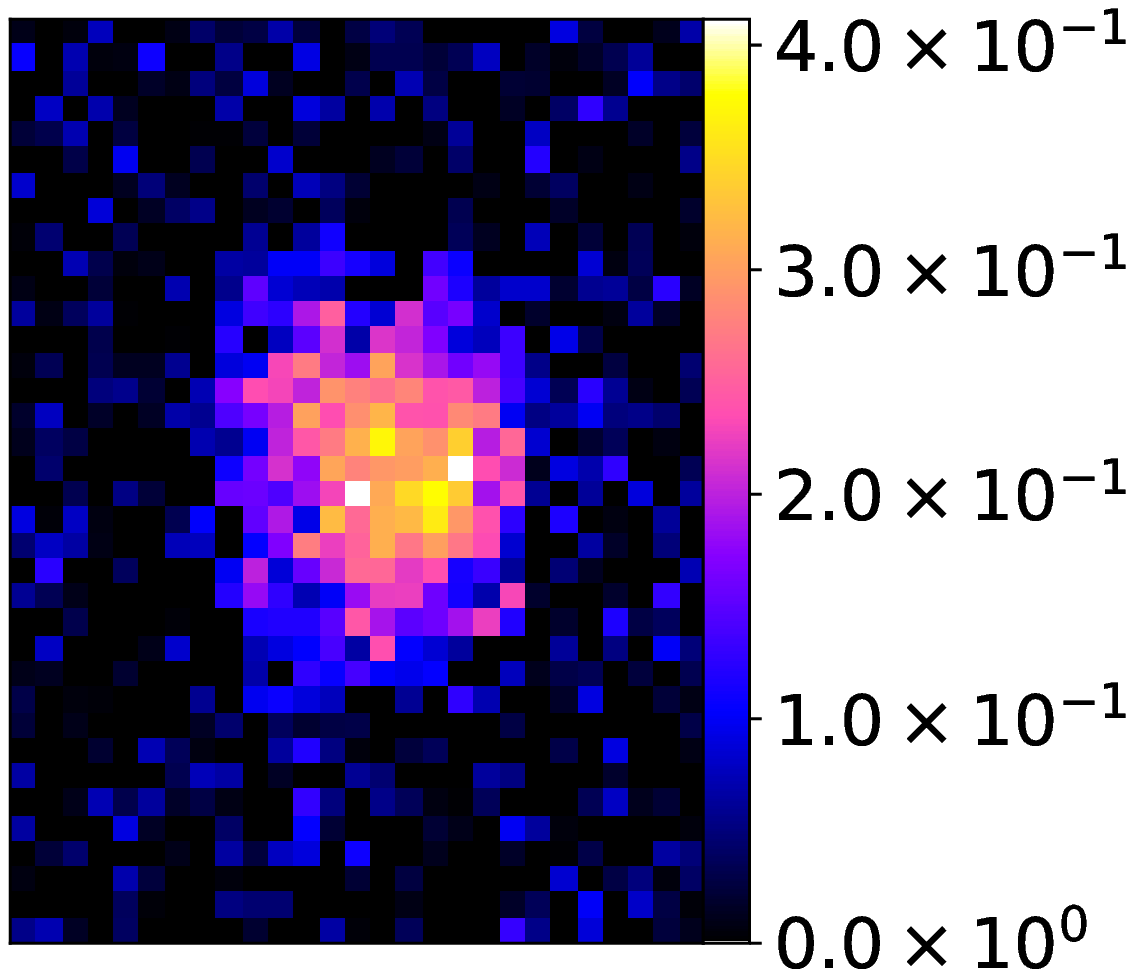}
    \includegraphics[width=3.45cm]
    {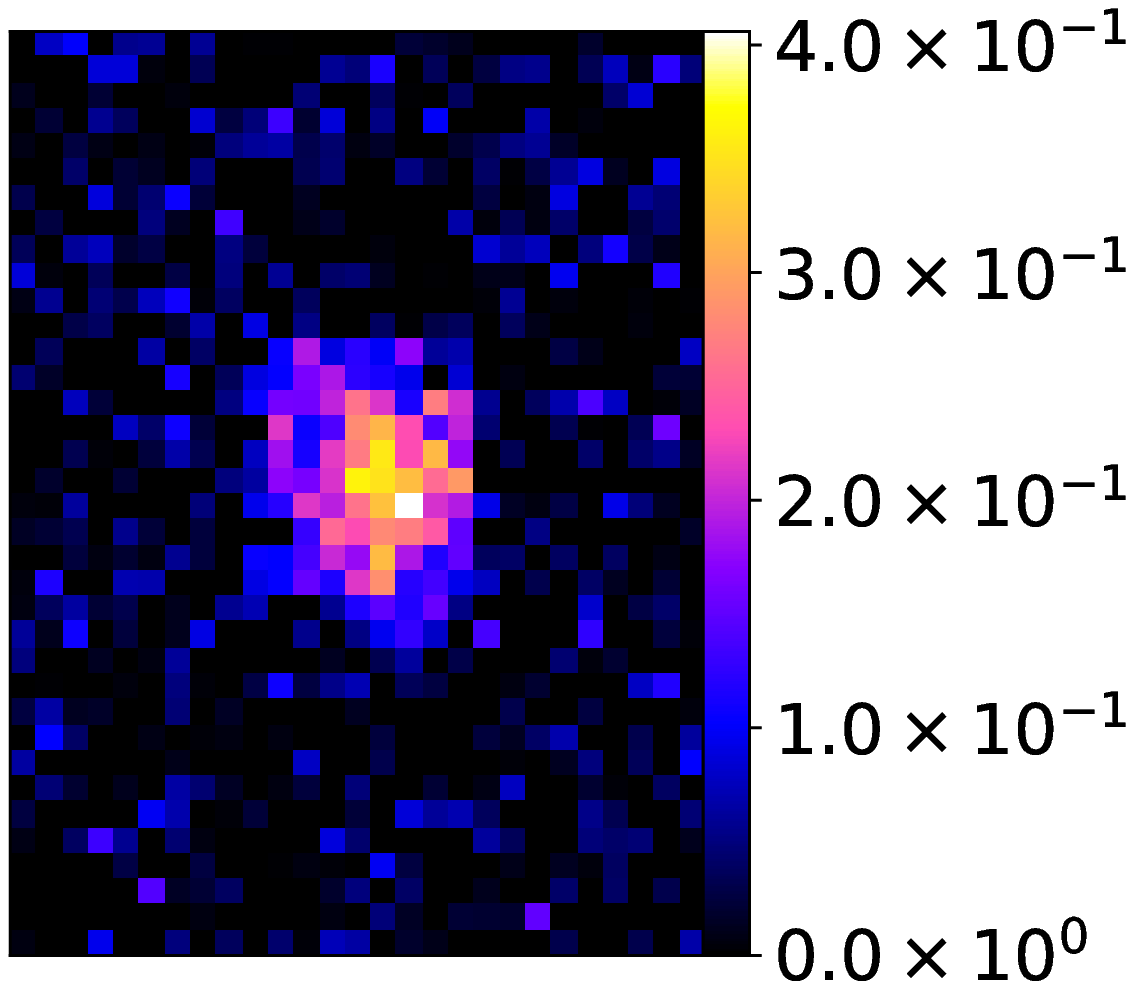}
    \includegraphics[width=3.45cm]
    {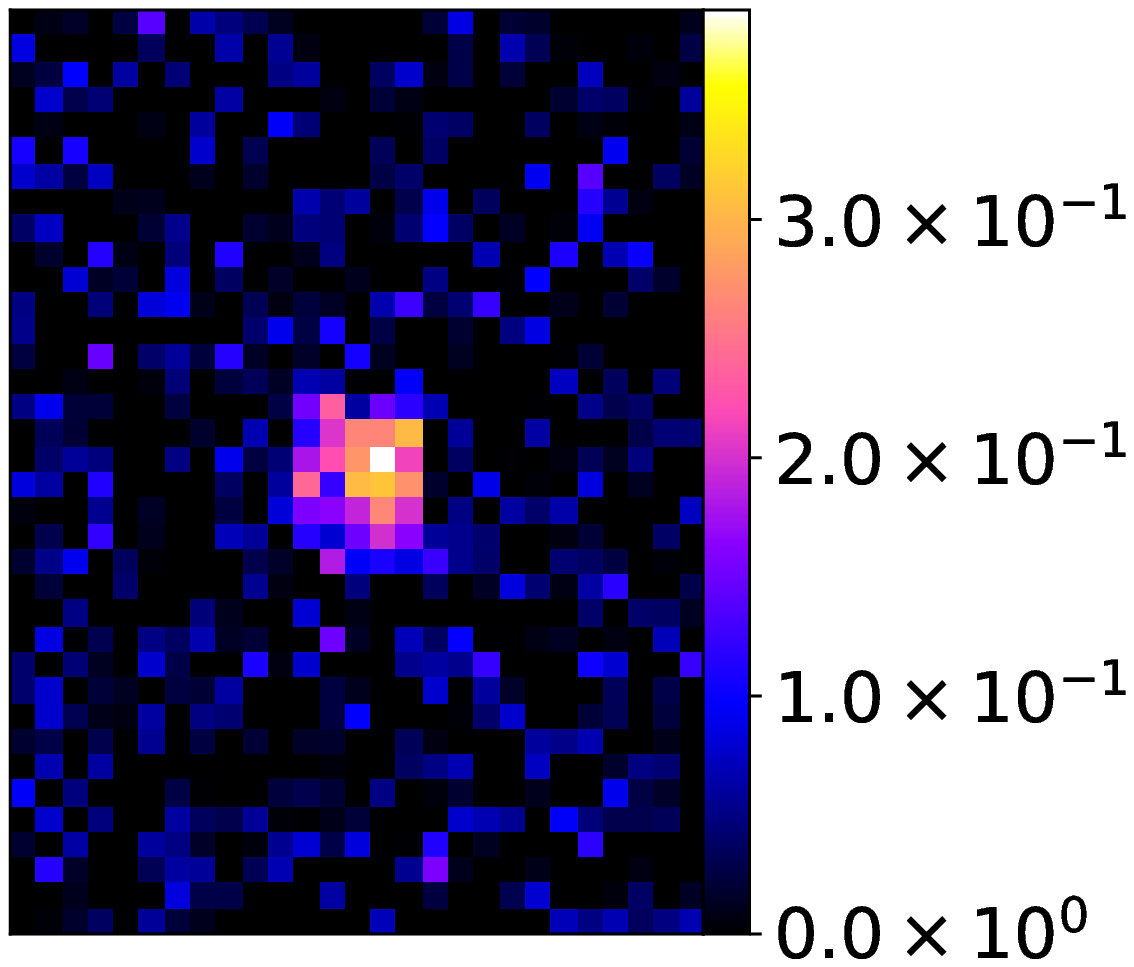}
    \includegraphics[width=3.45cm]
    {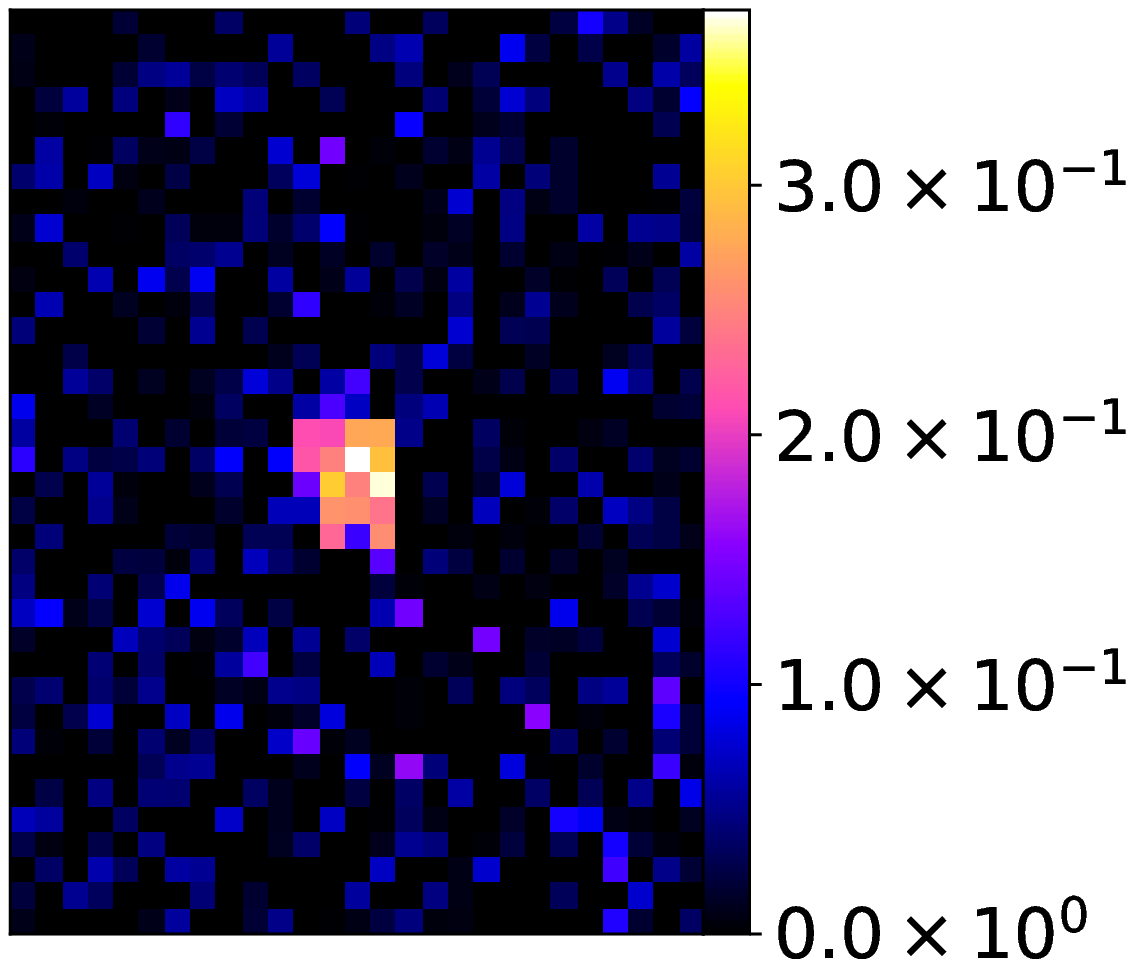}
    \caption{Nonthermal}
\end{subfigure}
\caption{ Same as Fig.~\ref{mapM51c12} for M33--like galaxy.  First row shows  the thermal and non-thermal maps of M33 \citep[from][]{tab_b}. Bars show the surface brightness in units of $\mu$Jy/beam.\label{mapM33c12}}
\end{figure*}

%%%%%%

\section{Results}\label{res}
Following Sect. \ref{theory}, the thermal and nonthermal surface brightness maps are generated at all redshifts in the range $0<z\leq 3$ with $\Delta z = 0.1$ at the observed frequency $\nu_2=1.4$\,GHz %\footnote{corresponding to the central frequency of the SKA Band~2}
for cases (1) and (2). {Our findings are presented} by first looking at variations in the galaxy mean values with $z$ and then describing structures observed at selected redshifts. {These results are mainly based on the maps with the UDT-noise sky level  but Tables~\ref{table2}~and~\ref{table3} also report signal-to-noise ratios for the DT and WT sky levels.}  %We note that all quantities in this section are reported in the observer rest frame.} 

%%%%%%
\subsection{Redshift evolution}\label{prof}
{ To investigate the redshift evolution of the mean surface brightnesses $\langle{I_{\nu_2}^{\rm th}}\rangle$ and $\langle{I_{\nu_2}^{\rm nt}}\rangle$, the thermal and nonthermal maps are averaged at each $z$. The resulting redshift evolution} of $\langle{I_{\nu_2}^{\rm th}}\rangle$ and $\langle{I_{\nu_2}^{\rm nt}}\rangle$ are then plotted  (Fig.~\ref{thernon}) and the corresponding values at selected redshifts $z=0.15, 0.3, 0.5, 1, 2$ are listed in Table~\ref{table2} for case~(1) { and Table~3 for case (2)}. The errors in the mean values reported are the statistical errors, i.e., the square root of the standard deviation around the mean divided by the number of pixels in each map.
{ To illustrate better the global evolutionary trends, we fitted functions of the form $a(1+z)^b$ to the distribution of the surface brightness vs. $z$ (curves in Fig.~\ref{thernon}) using a least-square regression with parameters listed in Table~\ref{param}. We find that, in case~(1), the nonthermal surface brightness $\langle{I_{\nu_2}^{\rm nt}}\rangle$ drops globally with $z$  for all 3 kinds of galaxies, although there are some fluctuations particularly at $z>1$ in low-mass M33-like galaxies. The thermal emission surface brightness $\langle{I_{\nu_2}^{\rm th}}\rangle$ exhibits more fluctuations than $\langle{I_{\nu_2}^{\rm nt}}\rangle$ but it also drops globally with $z$. This global change is faster in more massive galaxies. In case~(2), the trends should be reversed  because the sizes shrink with $z$. This is observed for $\langle{I_{\nu_2}^{\rm nt}}\rangle$ in all galaxies but only in M51-- and NGC6946--like galaxies for $\langle{I_{\nu_2}^{\rm th}}\rangle$. No certain global trend is found for $\langle{I_{\nu_2}^{\rm th}}\rangle$ in low-mass M33--like galaxies ($b^{(2)}=-0.09\pm0.10$).  We note that, in these galaxies, the thermal emission signal predicted is generally not robust because it is close to the sky level fluctuations injected (e.g., 0.05$\mu$Jy/beam {for UDT}, see also $\langle{I_{\nu_2}^{\rm th}}\rangle$ in Tables~\ref{table2}~and~\ref{table3}.}  

 { The mean thermal fraction defined as $\langle{f_{\nu_2}^{\rm th}}\rangle=\, \langle{I_{\nu_2}^{\rm th}}\rangle/ \langle{I_{\nu_2}^{\rm tot}}\rangle$, with 
 $\langle{I_{\nu_2}^{\rm tot}}\rangle= \langle{I_{\nu_2}^{\rm th}}\rangle + \langle{I_{\nu_2}^{\rm nt}}\rangle$ is also calculated and plotted against $z$ (Fig~\ref{thernon}). In case~(1),  $\langle{f_{\nu_2}^{\rm th}}\rangle$ follows a global increase. From $z=0.15$ to $z=2$,  $\langle{f_{\nu_2}^{\rm th}}\rangle$ increases by $\simeq$30\% in galaxies like M\,51. The increase is faster in NGC\,6946-- and M33--like galaxies by about a factor of two as shown by their fitted slope $b$ being larger than that of M\,51. Again, we caution that the evolution of $\langle{f_{\nu_2}^{\rm th}}\rangle$ in M\,33--like galaxies is not as certain due to low signal-to-noise ratios and large fluctuations particularly at  $z \gtrsim 0.7$.  
  
{In case~(2)}, the mean thermal fraction increases slightly from $z=0.15$ to $z\simeq 1$, then decreases by about the same rate from  $z=1$ to $z=3$ in M\,51--  and NGC\,6946--like galaxies.  {This occurs because the increase of the non-thermal emission is faster than that of} the thermal emission at $z>1$. Hence, the fitted single-slope function shows almost no evolution from $z=0$ to $z=3$.  In M33-like galaxies, the {global decrease of} $\langle{f_{\nu_2}^{\rm th}}\rangle$  occurs because of the increase in $\langle{I_{\nu_2}^{\rm nt}}\rangle$. However, this evolution inherit the large uncertainty in  $\langle{I_{\nu_2}^{\rm th}}\rangle$ at high redshifts. 
  
{Over the redshift range of $z=0.15-3$, a lower thermal fraction $\langle{f_{\nu_2}^{\rm th}}\rangle$ is found in more massive galaxies. These results are further discussed and compared with the literature in Sect.~4.4.}

\subsection{Maps at selected redshifts} \label{map}
%A galaxy emitting at a rest-frame frequency $\nu_1$ at redshift $z$ is observed at frequency $\nu_2=\nu_1/(1+z)$. {Hence, observing at a fixed frequency,  we can in principle receive information from objects at different redshifts emitting at frequencies $\nu_1 \geq \nu_2$.
 { Figures~\ref{mapM51c12} to \ref{mapM33c12} show the maps of the thermal and nonthermal surface brightness, $I_{\nu_2}^{th}(z)$ and $I_{\nu_2}^{nt}(z)$, for galaxies like M51, NGC6946, and M33 at selected redshifts of $z=$\,0.15, 0.3, 0.5, 1, and 2 observed at $\nu_2=$1.4\,GHz\footnote{Corresponding to the rest-frame frequencies $\nu_1=\nu_2\, (1+z)=1.6, 1.8, 2.8, 4.2$\,GHz, respectively.}. The  input maps of $I^{\rm th}(0)$ and $I^{\rm nt}(0)$  are also shown for comparison.} The resulting spatial distributions and morphologies of these galaxies change with redshift as follows: 
\begin{itemize}

	\item { M\,51--like galaxies:} The grand-design spiral structure {remains} visible both in the thermal and nonthermal maps at $z=0.15$. The maps are dominated by clumps of few kpc size at $z=0.3$. The spiral arms are not distinguishable at $z\geq1$. {The north-south asymmetry in the nonthermal emission is evident in case~(1) at all redshifts.} This structure becomes unresolved in case (2) for $z\geq1$. As shown in Fig. \ref{mapM51c12}, the structures remain more resolved in case (1) than in case (2), particularly at higher redshifts. However, the galaxy { has a higher surface brightness} in case (2) than case (1) at each redshift. The inner disk is always the most luminous part of the galaxy  in both cases. { The thermal emission from the disk becomes closer to the sky level with increasing $z$ in case~(1), while it is well above it ($\gtrsim 10\,\sigma$) in case~(2). 
		%The thermal fraction does not exceed the average in the center ($f_{\nu2}^{\rm th}<\langle{f_{\nu_2}^{\rm th}}\rangle \lesssim 6\%$) possibly due to M51's low-luminosity AGN that has a small radio jet~\citep{cecil,que}. Instead, $f_{\nu2}^{\rm th}$ is highest in star-forming regions particularly in the southern spiral arm reaching $>10\%$ at all redshifts in case (1) at $z=$0.15 and out to $z=$0.3 in case (2). 
	  }\\ 
		
	\item { NGC\,6946--like galaxies:} As shown in Fig.~\ref{mapN6946c12}, the thermal maps are dominated by bright clumps corresponding to giant HII complexes (including the one in its nucleus) {which remains visible at redshifts $z=0.15$ and $z=0.3$. In case (1), the spiral arms can also be traced up to $z=0.3$. The nonthermal map, however, exhibits a clumpy structure at $z=0.15$ which becomes }{ smoother at $z=0.3,~0.5$.} Only unresolved disks with a smooth decrease of brightness toward the outer disk are observable at $z=1$ and higher. { In case~(1), the nonthermal emission from the outer disk {gets below the detection threshold} at $z=2$ and the thermal emission can hardly be detected even at the center ($\lesssim 4\,\sigma$). } At this redshift, the information from the entire disk is preserved in case (2), although the disk has a smaller physical dimension than in the other case. { The thermal fraction in bright star-forming regions is above 15\% (reaching to 30\%) in case (1) but it falls below 12\% at $z=~2$ in case (2).  Similar to M51-like galaxies, $f_{\nu2}^{\rm th}$ is always below the mean galaxy value $\langle{f_{\nu_2}^{\rm th}}\rangle$ in the center due to a strong nonthermal component \citep[][]{tab_a}.  }\\

	\item { M\,33--like galaxies:} { The surface brightness of the RC emission is much weaker than that in other galaxy types, hence, the maps simulated are noisier generally (Fig.~\ref{mapM33c12}). The structures are severely affected by { noise} fluctuations and galaxies cannot be distinguished at $z=1,~2$ in case~(1).  By $z=0.15$, giant HII regions such as NGC\,604 can be traced in the thermal emission, while the nonthermal map exhibits an extended bright diffuse disk.  This disk remains visible out to $z=2$ in case~(2). Similar to their  more massive counterparts, these galaxies show a higher thermal fraction in star-forming regions than {in other regions of the ISM}  (e.g., at $z=0.15$, $f_{\nu2}^{\rm th}$ exceeds from 30\% in star-forming regions).  }
	
\end{itemize}
%To what extent these results are compatible with the standard SKAI-MID surveys is discussed in Sect.~\ref{discu}
%
%\subsection{Thermal fraction}

\begin{figure}
\vspace{0.4cm}
\begin{center}
\includegraphics[width=8.0cm]{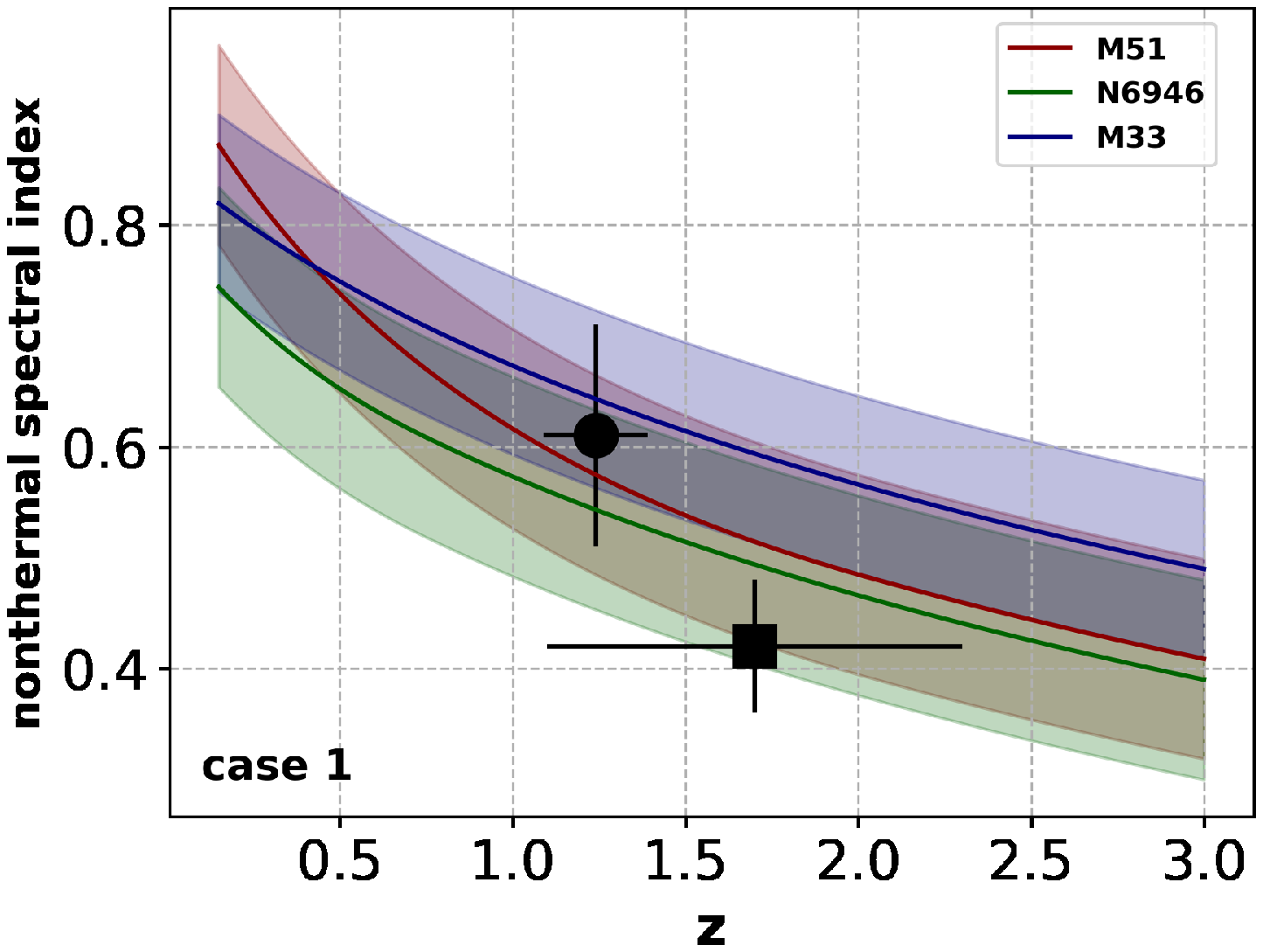}
\includegraphics[width=8.0cm]{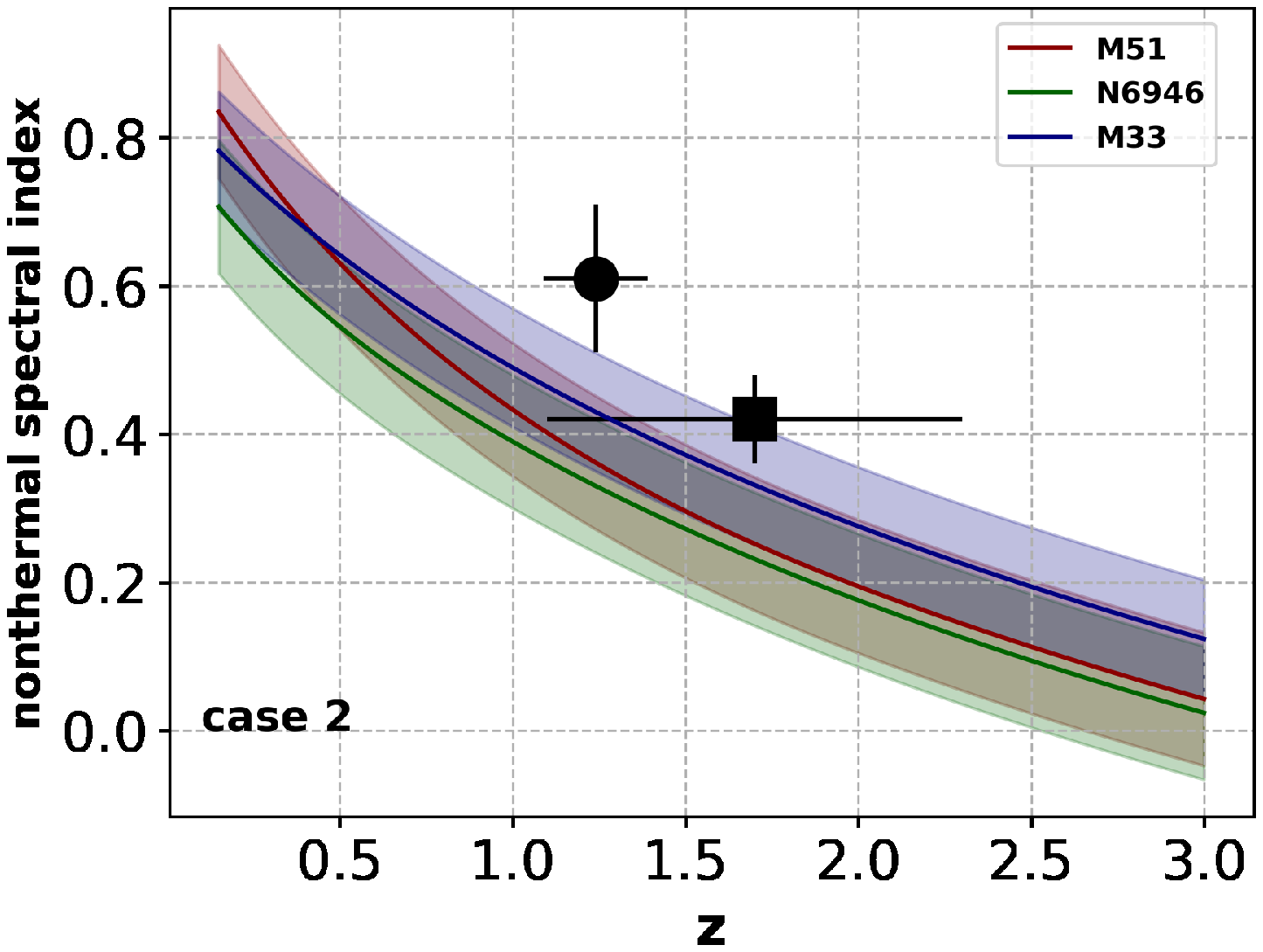}
\end{center}
\vspace{-0.3cm}
\caption{ Evolution of the spectral index $\alpha_{\rm nt}$ with redshift for case~1 ({\it top}) and case~2 ({\it bottom}). Shadowed bands show errors in the spectral indices given in Table~\ref{table}. Points indicate the measurements based on observations \citep[circle:][]{Murphy17} and \citep[square:][]{Tisanic} showing a better agreement with case 1(no radio size evolution). \label{alphaz}}
\end{figure}

%%%%%%%%%%%%%%%%%%%%%%%%%%%%%%%
\section{Discussion} \label{discu}
{{In this section, we discuss the redshift evolution of the nonthermal emission spectral index and the mid-radio SED}. We then address the detectability of kpc-scale emission from high-redshift objects similar to present-day galaxies with the planned SKAI-MID surveys. Finally, we compare our findings with observational results in the literature.  }

\subsection{Evolution of the nonthermal spectral index}
 As shown by \cite{cal2}, the nonthermal spectrum {at} mid-radio frequencies ($1\,<\nu<10$\,GHz) flattens with increasing star formation density in normal star-forming galaxies. This was explained by cosmic ray electrons being more energetic {in galaxies with higher star formation activity. In other words,  these particles are not much affected by  cooling because of the topology of the magnetic field: massive star formation produces more turbulent and tangled magnetic fields. }High-energy cosmic ray electrons injected in star-forming regions scatter off the very many pitch angles of the tangled field causing winds and outflows due to pressure gradients. {Thus, their energy spectrum remains globally flatter as winds can diffuse high energy CREs on timescale shorter than those of synchrotron cooling}. It is interesting to note that a flattening was also reported in highly star-forming galaxies (HSFGs) at high redshifts ($z \epsilon [0.3, 4], {\rm SFR} \geq 100 {\rm M_{\odot} yr^{-1}}$) in the VLA-COSMOS 3 GHz Large Project by \cite{Tisanic}. {These authors found a mean spectral index of $\simeq 0.42$ at $<z>=1.7\pm 0.6$ that is much flatter than that found for nearby normal star forming galaxies with SFR$\leq 10~{\rm M_{\odot} yr^{-1}}$ \citep[$\alpha_{\rm nt}\simeq 0.9$,][]{cal2}. 
 %As present-day normal star forming galaxies, like those considered here, have experienced higher star formation activities in the past, does they also show a flattening of $\alpha_{\rm nt}$ with $z$?   
 Did the present-day normal star forming galaxies have a flatter~$\alpha_{\rm nt}$ in the past when their SFR was higher?
 }

Following Eq.~\ref{alphant}, we obtain the evolution of the non-thermal spectral index $\alpha_{\rm nt}$ for galaxies like M51, NGC6946, and M33 for which the local values of $\alpha_{\rm nt}(0)$
% and $\Sigma$\,SFR(0) are 
are listed in Table~\ref{table}.{ As shown in Fig.~\ref{alphaz}, $\alpha_{\rm nt}$ drops with redshift  in both evolutionary cases (1) and (2), but with a different redshift dependence.  It drops from 0.8-0.9 to 0.5-0.6 from $z=0$ to $z=2$ in case (1), and even faster in case (2). The errors shown in Fig.~\ref{alphaz} represent measurement errors in $\alpha_{\rm nt}(0)$ (see Table~\ref{table}). However, we note that the main source of uncertainty in Eq.~\ref{alphant} is due to ${\rm SFR}(z)$ having a scatter of 0.3\,dex.} 
{As $\alpha_{\rm nt}$ reflects the energy index of cosmic ray electrons, the evolutionary trends obtained imply that these particles were more energetic at redshifts when galaxy star formation rates were higher and supernova explosions more frequent. 

{Besides the results based on the VLA-COSMOS observations of  HSFGs \citep{Tisanic},  the VLA observations of the $\mu$Jy radio sources  observed in the GOODS-N field also show a flatter spectrum at high $z$ \citep[$\simeq$0.61 at $<z>=1.24\pm 0.15$,][]{Murphy17}.  This sample has a mean SFR of $\sim 25{\rm M_{\odot} yr^{-1}}$ at $z\sim 1$ \citep{Murphy17}, similar to that of a M51-like galaxy at the same redshift (see Table~\ref{table}). Hence, a better agreement of our predictions is expected with the VLA GOODS-N than with the VLA-COSMOS survey. The mean values of the spectral indices measured by both surveys are overlaid in our simulations in Fig~\ref{alphaz}. The GOODS-N measurement \citep{Murphy17} agrees with the case~(1) simulation, particularly for M51-like galaxies as expected, while it is steeper than that of the case~(2) predictions. Due to their higher SFR, the VLA-COSMOS measurements \citep{Tisanic} should be flatter than the prediction of the M51-like galaxies at the same $z$. This is nicely seen in case~(1). Therefore, these observations are in favor of no (or a shallow) evolution of the radio size of galaxies.  } 
%	 The points in  show the spectral index measurements reported by \cite{Murphy17} and \cite{Tisanic} indicating  a better match with case~(1) than with case (2) of the $\Sigma$SFR evolution scenario. We note that highly star-forming galaxies presented by \cite{Tisanic} are expected to show flatter $\alpha_{\rm nt}$ than those galaxies simulated here that contradicts with the case~(2) results.
	
{We note that} observational studies of the evolution of the spectral index can be complicated by   $a$) contamination by thermal emission,  $b$) large observational uncertainties and inconsistent sensitivities at different frequencies, (noting that the uncertainty in measuring the spectral index is often $\simeq$\,25\%),  and $c$) the presence of obscured steep spectrum AGNs that are likely to be more abundant at higher redshifts \citep[e.g.][]{Magneli,Ivison,Bourne,An}. 
We note that AGNs can complicate any correlation between the spectral index and $z$ in unresolved studies.  The observed radio spectrum of AGNs can be flat at low redshifts $\alpha<0.5$ \citep[eg.,][]{Murphy2013} but steep at high redshifts $\alpha>\,1$ \citep[e.g.,][]{Huynh2007}. Radio galaxies also do not obey the $\alpha_{\rm nt}-z$ trend used here because their radio spectrum steepens at high $z$ \citep[e.g.,][]{Ishwara-Chandra,Roettgering,Singh}.   Overcoming all these limitations requires sensitive multi-frequency radio observations and hints on the important role of next generation radio telescopes such as the SKA.  }

\subsection{Evolution of the radio spectral energy distribution} \label{SED}
The evolution of the spectral index means that the rest-frame radio { SED} of galaxies changes with redshifts. This is investigated for the integrated mid-radio SED of a M51-like galaxy with known { integrated flux densities of the thermal $S_{\nu}^{\rm th}$ and the nonthermal $S_{\nu}^{\rm nt}$ emission at various frequencies 1.4, 2.7, 4.8, 8.4, 10.7\,GHz  at $z=0$  (Table~\ref{table4}).  The corresponding integrated flux densities at other redshifts observed at frequencies $\nu_2$ can be obtained following Eq.~\ref{S}. In rest-frame, we have $S_{\nu_1}(z)= S_{\nu_2}(z)\, (\nu_1/\nu_2)^{-\alpha}= S_{\nu_2}(z)\, (1+z)^{-\alpha}$, with $\nu_1$ the rest-frame frequency. Thus, the total integrated RC flux density ($S_{\nu_1}=S_{\nu_1}^{\rm th} + S_{\nu_1}^{\rm nt}$) evolves with redshift as } 
\be\label{sed}
S_{\nu_1}(z)&=&\left [S^{\rm th}_{\nu_1}(0)\,(1+z)^{-2\alpha_{\rm th}} + S^{\rm nt}_{\nu_1}(0)\,(1+z)^{-2\alpha_{\rm nt}} \right] \nonumber\\ 
&& \times  \frac{\rm SFR(z)}{\rm SFR(0)}\, (1+z)\, \frac{D^2}{D_L^2}, 
\ee
%\be\label{sed}
%S_{\nu}(z)&=&\left [S^{\rm th}_{\nu}(0)\,(1+z)^{1-\alpha_{\rm th}} + S^{\rm nt}_{\nu}(0)\,(1+z)^{1-\alpha_{\rm nt}} \right] \nonumber\\ 
%&& \times  \frac{\rm SFR_{\nu_1}(z)}{\rm SFR(0)}\, \frac{D^2}{D_L^2} 
%\ee
%
with $S^{\rm th}_{\nu_1}(0)$ and $S^{\rm nt}_{\nu_1}(0)$ the integrated thermal and nonthermal flux densities at $z=0$ (note that $\nu_1=\nu_2$ at $z=0$). The nonthermal spectral index $\alpha_{\rm nt}$ is given by Eq.~\ref{alphant} and $\alpha_{\rm th}=0.1$ (see Sect.~\ref{evothnon}). { Using Eq.~\ref{sed}  at different frequencies, the intrinsic flux densities can be obtained and the mid-radio (1-10\,GHz) SEDs can be constructed  at different redshifts. We model the SEDs following \cite{cal2},

\be\label{SEDeq}
I_{\nu} = A_1 \, \nu_0^{-0.1} \left(\frac{\nu}{\nu_0}\right)^{-0.1} + A_2 \,  \nu_0^{-\alpha_{nt}}  \left(\frac{\nu}{\nu_0}\right)^{-\alpha_{\rm nt}}
\ee
with $A_1$ and $A_2$ the scaling factors, $\nu_0$ the reference frequency, and $\alpha_{\rm nt}$ given by Eq.~\ref{alphant}. We selected $\nu_0=1.4$\,GHz at all redshifts. A non-linear least square fit is used to obtain $A_1$ and $A_2$ at each $z$ (by adopting the task ${\rm curve\_fit}$ in Python).}  Figure~\ref{SEDfig1} shows the resulting SEDs at selected redshifts of  $z=$0.15, 0.3, 0.5, 1, and 2. As expected, the SEDs become flatter towards higher $z$. As the flattening is due to the synchrotron emission it mainly occurs at lower frequencies leading to a more pronounced curvature at higher $z$. 

We note that these { mid-radio SEDs are shifted to lower frequencies in the observer-frame}. For instance, to study the intrinsic mid-radio SED of galaxies at $z<0.5$, observations with the SKA Bands 2, 3, 4, and 5 are required. Moving to higher redshifts, SKA Band 1 becomes important as well. At $z=2$,  the intrinsic mid-radio SED can be studied through Bands 1 to 4. The corresponding observed SEDs along with the SKA frequency windows are shown in Fig.~\ref{SEDfig1}.  The offset between the observed and the rest-frame SEDs increases with redshift.
\begin{table}
\begin{center}
\begin{tabular}{ l c c }
  \hline
Frequency & $S^{\rm th}_{\nu}(0)$ & $S^{\rm nt}_{\nu}(0)$  \\ 
(GHz) & (mJy)   & (mJy) \\ 
  \hline
 1.4   & $70\,\pm\,5$      &  $1330\,\pm\,109$ \\  
 2.7   & $66\,\pm\,4   $ & $714\,\pm\,66  $\\
 4.8   & $62\,\pm\,3 $& $358\,\pm\,25 $\\
 8.4   & $58\,\pm\,3$& $248\pm\,20$\\
 10.7 & $57\,\pm\, 4$& $178\,\pm\,18 $\\
  \hline
\end{tabular}
\caption{The integrated flux density of the thermal ($S^{\rm th}_{\nu}(0)$) and nonthermal  ($S^{\rm nt}_{\nu}(0)$) emission from M51 \citep[from][]{cal2}. \label{table4}}
\end{center}
\end{table}

%As we explained in section \ref{evothnon} radio continuum emission is often taken as power law function. For frequency range of $1<\nu<10$ GHz  can be written as  
%\be\label{SEDeq}
%I_{\nu} = A_1 \, \nu_0^{-0.1} \left(\frac{\nu}{\nu_0}\right)^{-0.1} + A_2 \,  \nu_0^{-\alpha_{nt}}  \left(\frac{\nu}{\nu_0}\right)^{-\alpha_{nt}}
%\ee
%where we consider $\nu_0$ as $1.4$ here. First, we fit the equation \ref{SEDeq} using data of \cite{taba2017} for M51 galaxy. We have $A_1 = 0.04$ and $A_2 = 1.87$ for local universe. Then, we evolve equation \ref{SEDeq} by equations \ref{intensityth} and \ref{intensitynth}. The results are shown in figure \ref{SEDfig}. The red points show our data. We show evolution for 5 different redshifts 0.15,0.3,0.5,1 and 2. 

\begin{figure}
\vspace{0.4cm}
\begin{center}
\includegraphics[width=9.0cm]{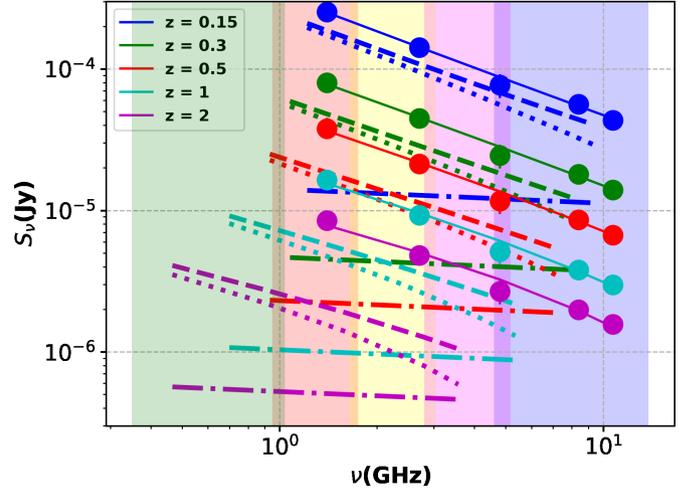}
\end{center}
\vspace{-0.3cm}
\caption{ The { fitted }rest-frame SEDs ({\it solid lines}) of a M51-like galaxy at 5 different redshifts. Points show the rest-frame integrated flux densities  at 1.4, 2.7, 4.8, 8.4, 10.7\,GHz obtained following Eq.~\ref{sed} at different redshifts.  The SEDs were fitted following Eqs.~\ref{SEDeq} and \ref{alphant} taking into account the errors in the fluxes (smaller than the symbols). Also shown are the corresponding SEDs in the observer-frame ({\it dashed lines}) and their thermal ({\it dashed-dotted lines}) and nonthermal ({\it dotted lines}) components. Vertical colored shades show the SKA frequency bands 1 (green),  2 (red), 3 (yellow), 4 (pink), and 5 (blue). \label{SEDfig1}}
\end{figure}

%The data points were fitted using non-linear least squares (${\rm curve_fit}$ in Python) taking into account the errors in the fluxes (smaller than the symbol size).

\subsection{Compatibility with SKA1-MID band 2 survey}\label{ska}
We further discuss to what extent the { proposed SKA1-MID band 2 reference survey (Sect.~\ref{srs})} can satisfy studies related to the RC emission from the ISM in selected galaxies. 
As the observed emission is the sum of the thermal and nonthermal components, the mean signal-to-noise ratio at the observer-frame is given by $\langle{S/N}\rangle = \langle{I_{\nu_2}^{\rm tot}}\rangle/N$, with $\langle{I_{\nu_2}^{\rm tot}}\rangle= \langle{I_{\nu_2}^{\rm th}}\rangle + \langle{I_{\nu_2}^{\rm nt}}\rangle$  and $N$ the sensitivity or the one $\sigma$ rms noise level of the survey. The mean signal-to-noise ratios obtained for the 3 tiers of UDT, DT, and WT are plotted against redshift (Fig.~\ref{snrexpandtol}) and the corresponding values at selected redshifts of 0.15, 0.3, 0.5, 1, 2 are listed in Tables~\ref{table2}~ and~3.
Generally, the signal-to-noise ratios decrease with redshift in case(1), while they increase in case(2) as expected from the redshift evolution of the mean surface brightnesses (Sect.~\ref{prof}).  
%the mean signal-to-noise ratio for the expected total radio continuum { surface brightness} obtained as  $\langle{I_{\nu_2}^{\rm tot}}\rangle= \langle{I_{\nu_2}^{\rm th}}\rangle + \langle{I_{\nu_2}^{\rm nt}}\rangle$ divided by $1 \sigma$ rms sensitivity of the ultra-deep tier (S/N$_{\rm UDT}$) 

From $z=0$ out to $z=3$, UDT can detect the mean RC emitting ISM above 8\,$\sigma$ and 4\,$\sigma$ levels  in galaxies like M51 and NGC6946, respectively, in case (1). In M33--like galaxies, {$\langle{S/N_{\rm UDT}}\rangle \gtrsim 3$ occurs only at $z\lesssim 0.3$.} DT can detect the mean ISM of the M51--like galaxies at $>3\sigma$  while a $3\sigma$ detection is only possible up to $z=0.6-0.7$ for NGC6946-like galaxies. WT detects none of these galaxies {  at $3 \sigma$ }level in case of no evolution in radio size.  

Based on the case~(2) results, the mean RC emitting ISM is detected in all galaxies above the $4\sigma$ level in UDT at all redshifts.  However, only M51-- and NGC6946--like galaxies are detected in DT with  $\langle{S/N_{\rm DT}}\rangle \geq 8$.  Such a detection is only possible for M51--like galaxies at $z\gtrsim 1$ in WT. 
%Taking into account the cosmological dimming  (Fig.~\ref{snrexpandtol}), the UDT can provide a $>3\sigma$ level detection of the mean ISM up to $z=1, 0.7, 0.2$ for the M51-, NGC6946- and M33-like galaxies, respectively, in case (1).  This level of detection is expected up to the redshifts as high as $z=2.7,1.6, \& 0.3$ for  the M51-, NGC6946- and M33-like galaxies, respectively, in case (2). 

{According to \cite{pran}, the scope of the { three-tiered} surveys { was} to study galaxies at different SFR levels at different redshifts (their Table~2).
Specifically, UDT was designed to probe SFR$\sim 10$ galaxies at $z \sim 3-4$, DT to study SFR$\sim 10$ galaxies at z$\sim 1-2$, and WT for detecting galaxies with $0.5 < {\rm SFR} < 10$ at $0<z<1$. As these SFR levels are similar to those of our sample  at corresponding redshifts  (see Table~\ref{table}), we further discuss 
%to what extent the above scopes
the extent to which the above scopes 
 can be met.  The ultimate goal of the UDT survey is to study M33-like galaxies (with SFR$\sim 10$ at z$\sim 3-4$) which can be achieved only in case~(2). Similarly, the DT goal can be reached only in case~(2), as the SFR$\sim 10$ galaxies at z$\sim 1-2$ are like NGC~6946 which are below $3\sigma$ detection limit  in case~(1) at those redshifts.  The WT goal is met only for M~51--like galaxies at $z\lesssim0.3$ in case~(2). At higher redshifts ($0.3<z<1$), galaxies have SFRs similar to M~33 and NGC~6946 ($0.5 < {\rm SFR} < 10$) which are undetected in both cases~(1)~and~(2). It is hence important to reconsider the Band~2 three tiered surveys particularly if there is no radio-size evolution.  }

\begin{figure}
	\vspace{0.4cm}
	\centering
	\textbf{case 1}
		\includegraphics[width=9.0cm]{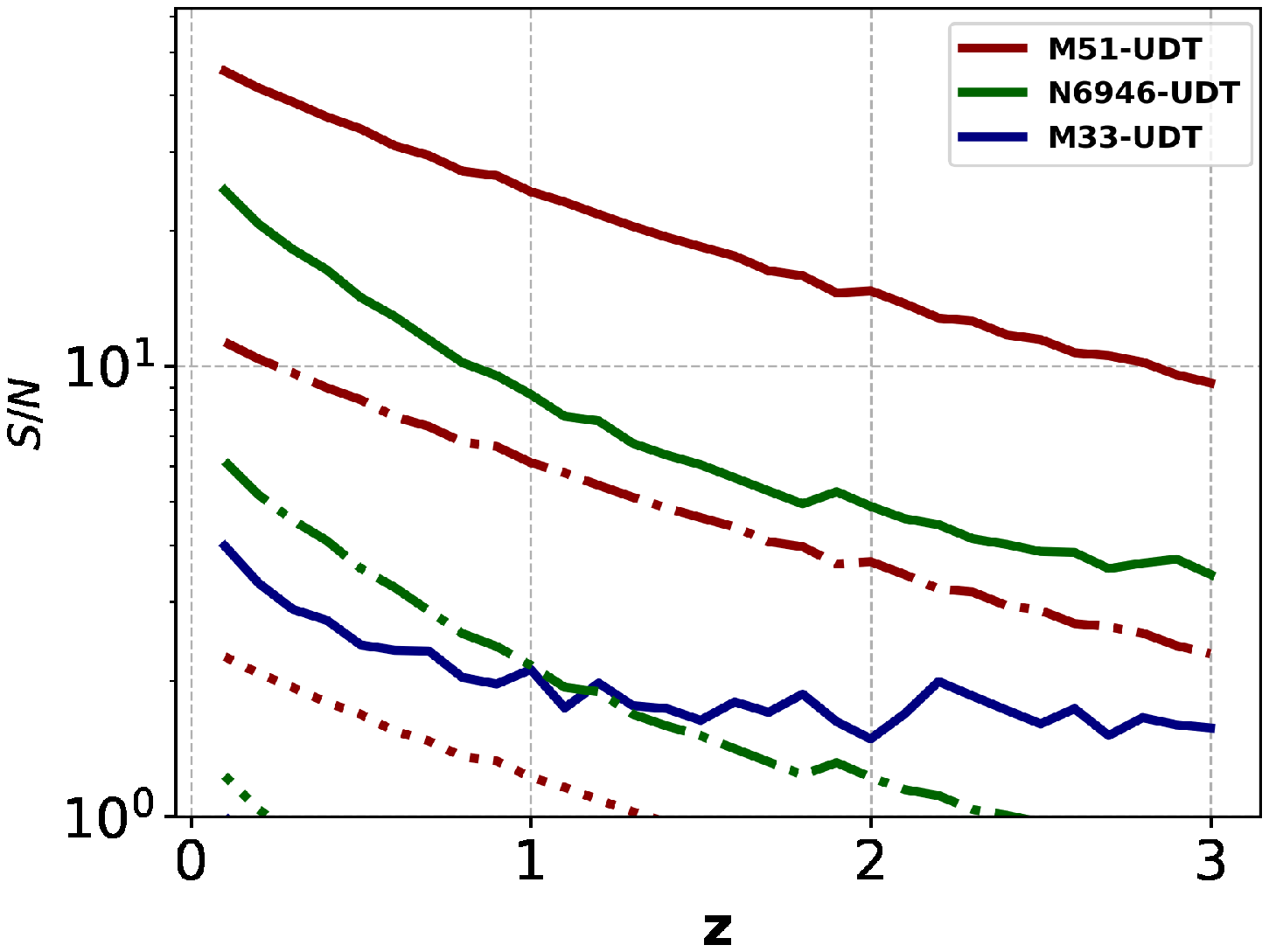}
		\textbf{case 2}\par\medskip
		\includegraphics[width=9.0cm]{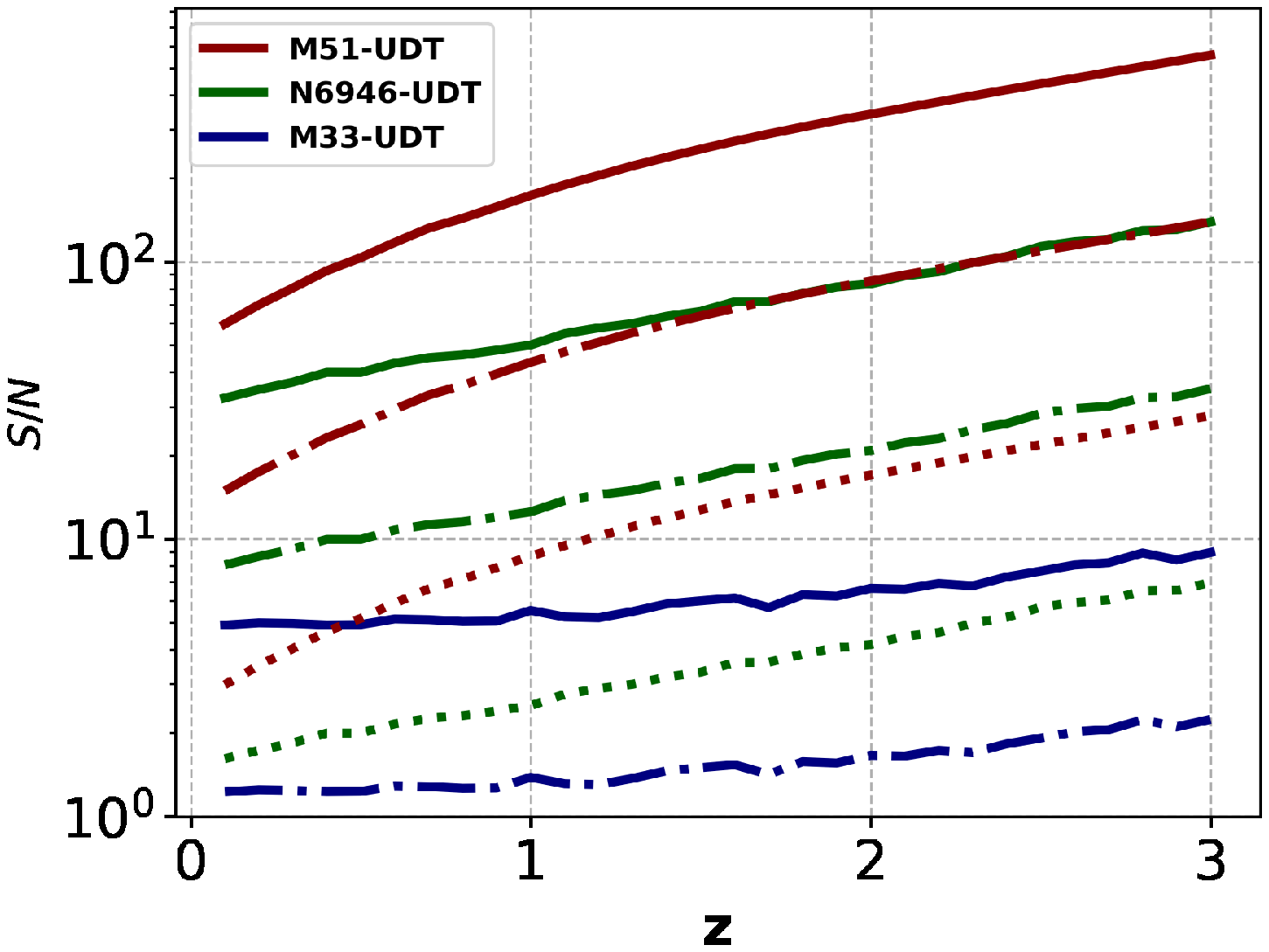}
	\vspace{-0.3cm}
	\caption{ The mean signal to noise ratio $\langle{\rm S/N}\rangle$ based on the SKA1-MID band~2 UDT (solid), DT (dashed-dotted), and WT (dotted) for case~1 ({\it top}) and case~2 ({\it bottom}) at the observed frequency of 1.4\,GHz (equivalent to a rest-frame frequency of 1.6--5.6\,GHz at $0.15\leq z \leq3$)).\label{snrexpandtol}}
\end{figure}

\subsubsection{Detecting galactic structures}\label{struc}
Moving from $z=1$ to $z=0.15$, galactic structures such as clumps in disks, spiral arms, and star-forming associations are potentially observable at $0.6\arcsec$ angular resolution  (see Figs.~\ref{mapM51c12}~to~\ref{mapM33c12}). The spiral arms can be resolved at scales of $\leq$1 kpc at $z=0.15$ for both cases (1 and 2). At $z=0.3$, the spiral arms can still be distinguished in case (1), while only a clumpy disk appears in case (2). Detecting structures is limited to spatial scales of few kpc at $z\geq 1$ in both cases (1) and (2). Hence, studies of the inner vs. outer disks as well as  structures at larger spatial scales in the intergalactic medium are in principle feasible.  { To assess the capability of the SKA1-MID band 2 survey in detecting these structures, we derive the maps of signal-to-noise ratio ({$S/N = {I_{\nu_2}^{\rm tot}}/N$}) by dividing the sum of the thermal and nonthermal maps simulated by the sensitivities of the three-tiered survey given in Sect.~\ref{srs}. These maps are shown in Fig.~\ref{mapM51sntol}  at the selected redshifts of $z=$0.15, 0.3, 0.5, 1, and 2  for UDT.
The structures detected in the three tiers are:  
\begin{itemize}
	\item { M51--like galaxies:} In case (1), all structures described in Sect.~\ref{map} can be detected  above 10$\sigma$ at all redshifts in UDT, but  the outer disk is not detected at $z\gtrsim 2$ in DT and at $z>0.3$ in WT. In the latter tier, only the very central part is detected at $z>1$. In case (2), all structures are detected in all three tiers.	
	
	\item { N6946--like galaxies:} The disk  is visible at S/N$>$5 and 50 in case~(1) and (2), respectively, in UDT. DT cannot detect the entire disk at $z=1$ and only the inner disk can be studies at $z=2$ for case (1).  This tier can detect all structures in case(2). WT can hardly detect the entire disk at $z\gtrsim 0.3$ in case (2). It  may only detect the central part at $0.3 \lesssim z\lesssim 1$ in case(1).
	%However, the structures are hardly detected at $z\geq1$ in case (1) and $z\geq2$ in case (2) due to cosmological dimming.

	\item { M33--like galaxies:}  In UDT,  the diffuse emission from the outer disk, spiral arms and star-forming regions can all be detected (S/N$>3$) up to $z$=0.15 in case (1). The sensitivity to detect the outer disk is reduced for $z \geq0.3$.  The disk is undetected for $z=1$ and higher. DT and WT cannot detect these galaxies. In case~(2), i.e., if radio sizes evolve with z, almost all  structures can be observed at all redshifts in UDT, but  only the central disk up to $z=0.3$ in DT and no detection in WT. 
	%
	%Considering the cosmological dimming,  the entire galaxy is seen only at $z=0.15$. At higher redshift of $z=0.3$, only the center and star forming regions can be observed with  S/N$>5$ in case (2).
	
	 \end{itemize}
 }
It is worth noting that the size evolution of galaxies such as that considered in case (2) is  established based on the rest-frame ultraviolet and optical observations by HST \citep[e.g., ][]{Ferguson,Mosleh,Ono,van,Shibuya}. However,  radio observations provide no clear consensus on the size evolution \citep[e.g., ][]{Murphy17,Guidetti,Cotton,Bondi}. One reason can be a possible dilution by the presence of radio AGNs \citep{Bondi}. Another reason is the different spatial { scales} on which synchrotron emission occurs at different frequencies. Assuming that the size evolution is shallower at longer wavelengths \citep[e.g., ][]{van} and that the case (2) evolution is more justified for a rest-frame optical emission, our simulation of the thermal radio emission traced using the H$\alpha$ emission is best represented through this case, and the nonthermal emission by an evolution that is closer to case (1). Hence, the galaxy size evolution observed in the radio (sum of the thermal and nonthermal emission) might occur in between the cases (1) and (2) as extreme scenario.  This agrees with \cite{Jimenez-Andrade} finding  a shallow evolution of $r_e\sim (1+z)^{(-0.26\pm0.08)}$ for the radio continuum size evolution of star-forming galaxies over the redshift range 0.35<z<2.25 using the VLA COSMOS 3GHz survey.   

%\iffalse

\begin{figure*}
\centering
\hspace{-1cm}
\textbf{case 1}
\hspace{1.4cm}
\textbf{case 2}
\hspace{1.7cm}
\textbf{case 1}
\hspace{1.4cm}
\textbf{case 2}
\hspace{1.7cm}
\textbf{case 1}
\hspace{1.4cm}
\textbf{case 2}
\par\medskip
\begin{subfigure}[t]{0.15\textwidth}
    \makebox[0pt][r]{\makebox[30pt]{\raisebox{40pt}{\rotatebox[origin=c]{90}{$z=0.15$}}}}%
    \includegraphics[width=2.5cm]
    {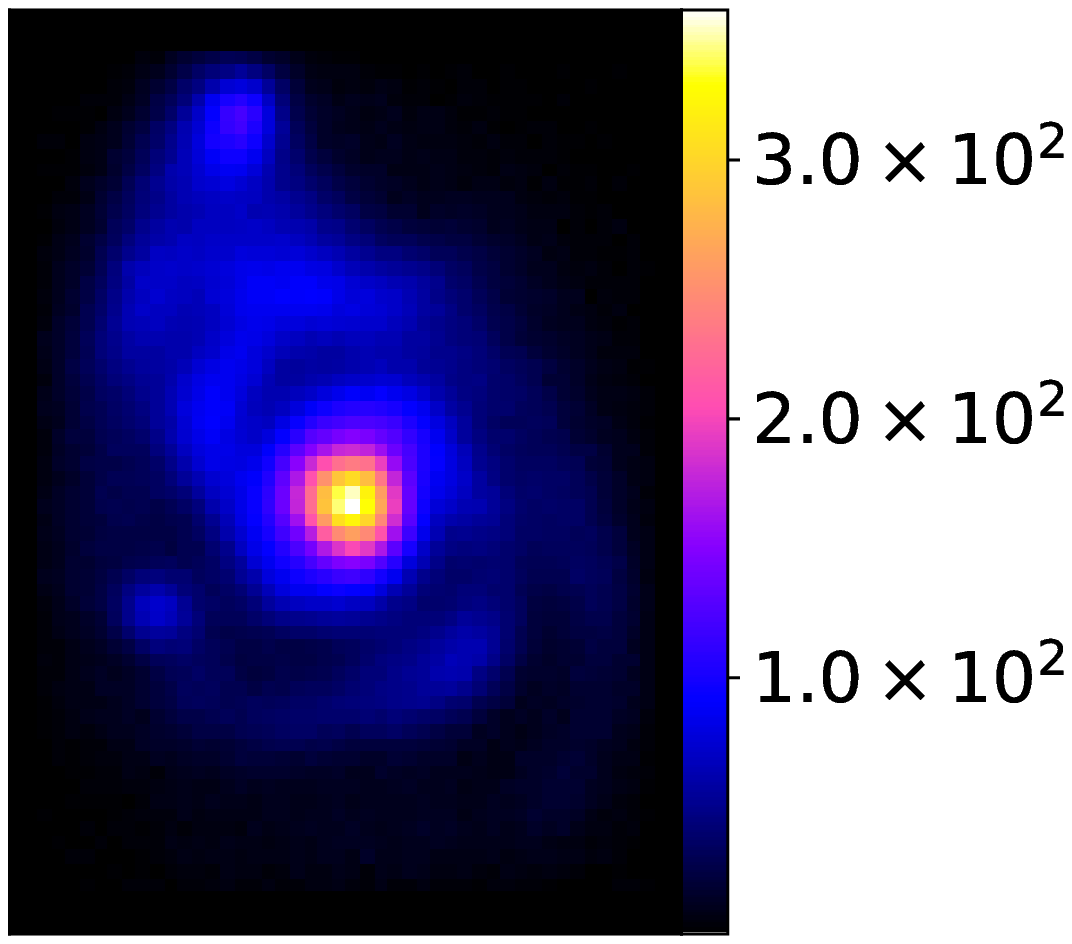}
    \makebox[0pt][r]{\makebox[30pt]{\raisebox{40pt}{\rotatebox[origin=c]{90}{$z=0.3$}}}}%
    \includegraphics[width=2.5cm]
    {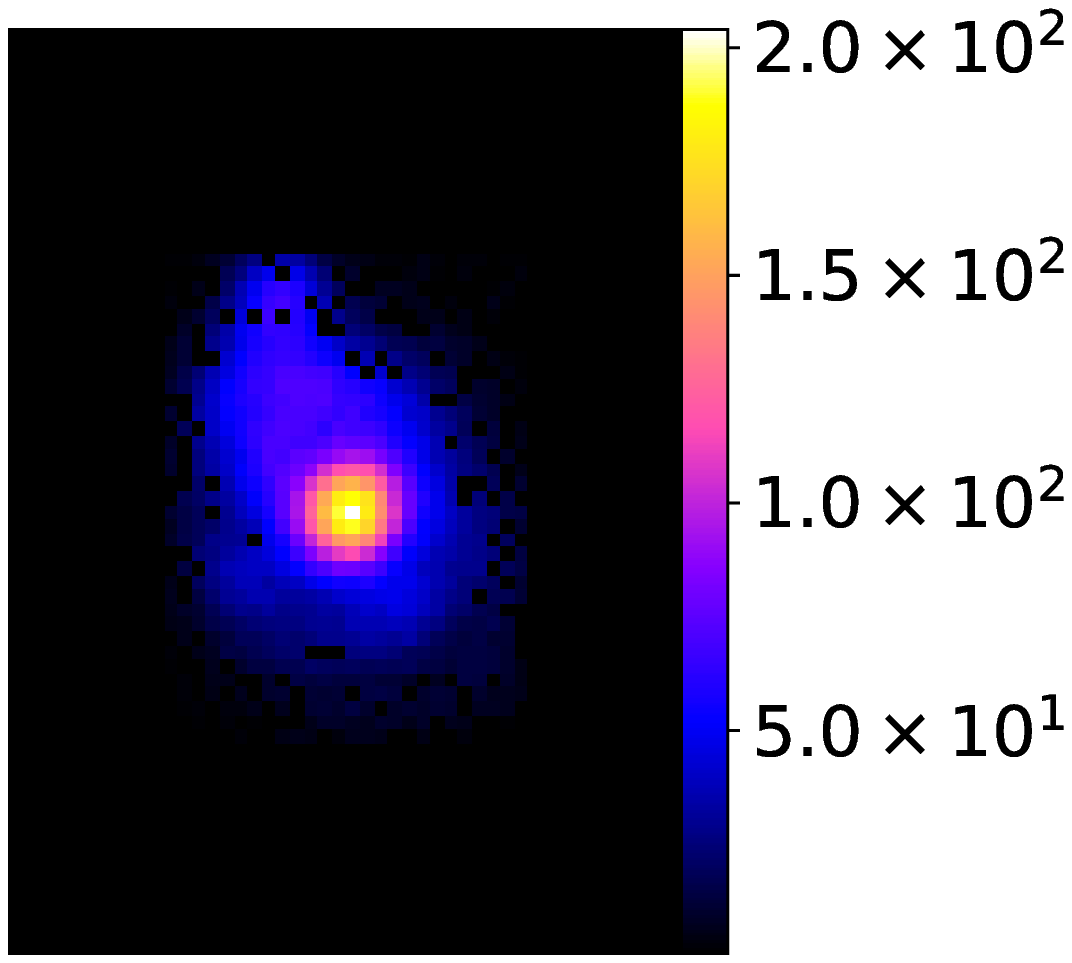}
    \makebox[0pt][r]{\makebox[30pt]{\raisebox{40pt}{\rotatebox[origin=c]{90}{$z=0.5$}}}}%
    \includegraphics[width=2.5cm]
    {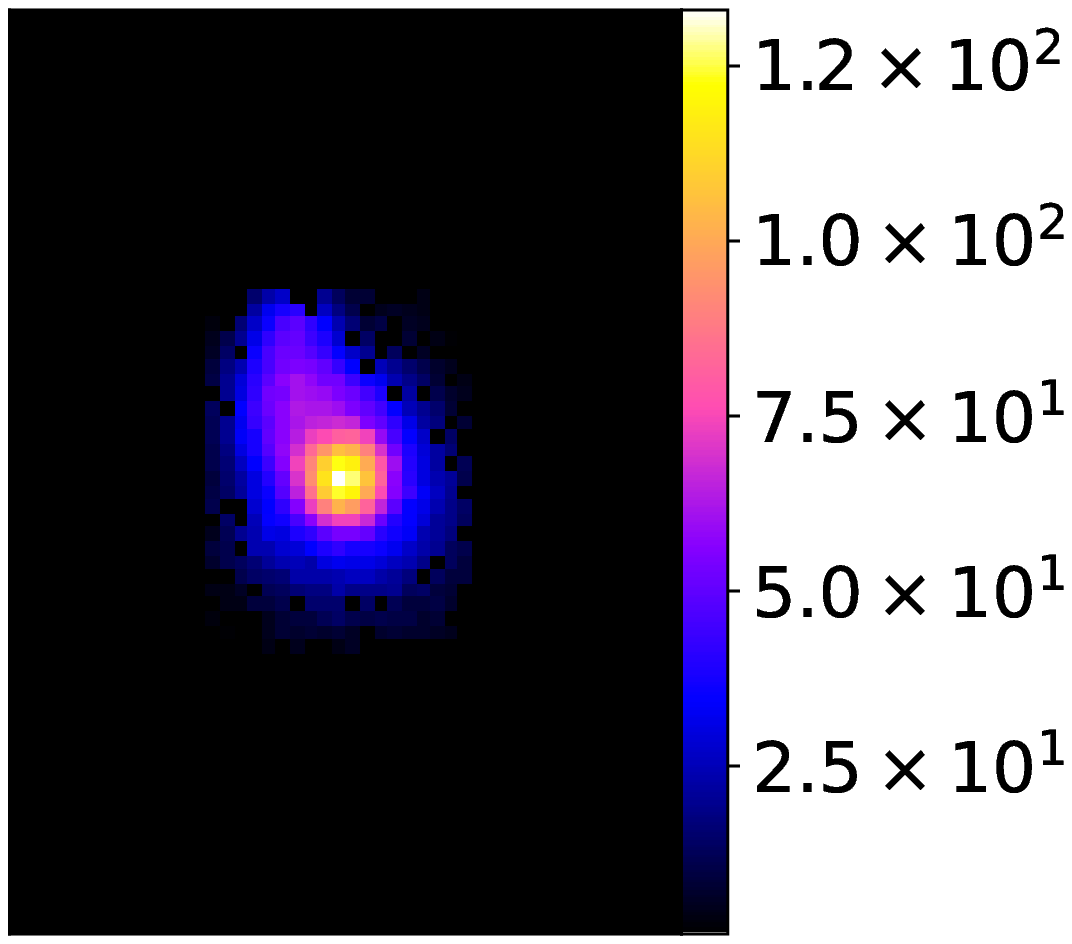}
    \makebox[0pt][r]{\makebox[30pt]{\raisebox{40pt}{\rotatebox[origin=c]{90}{$z=1$}}}}%
    \includegraphics[width=2.5cm]
    {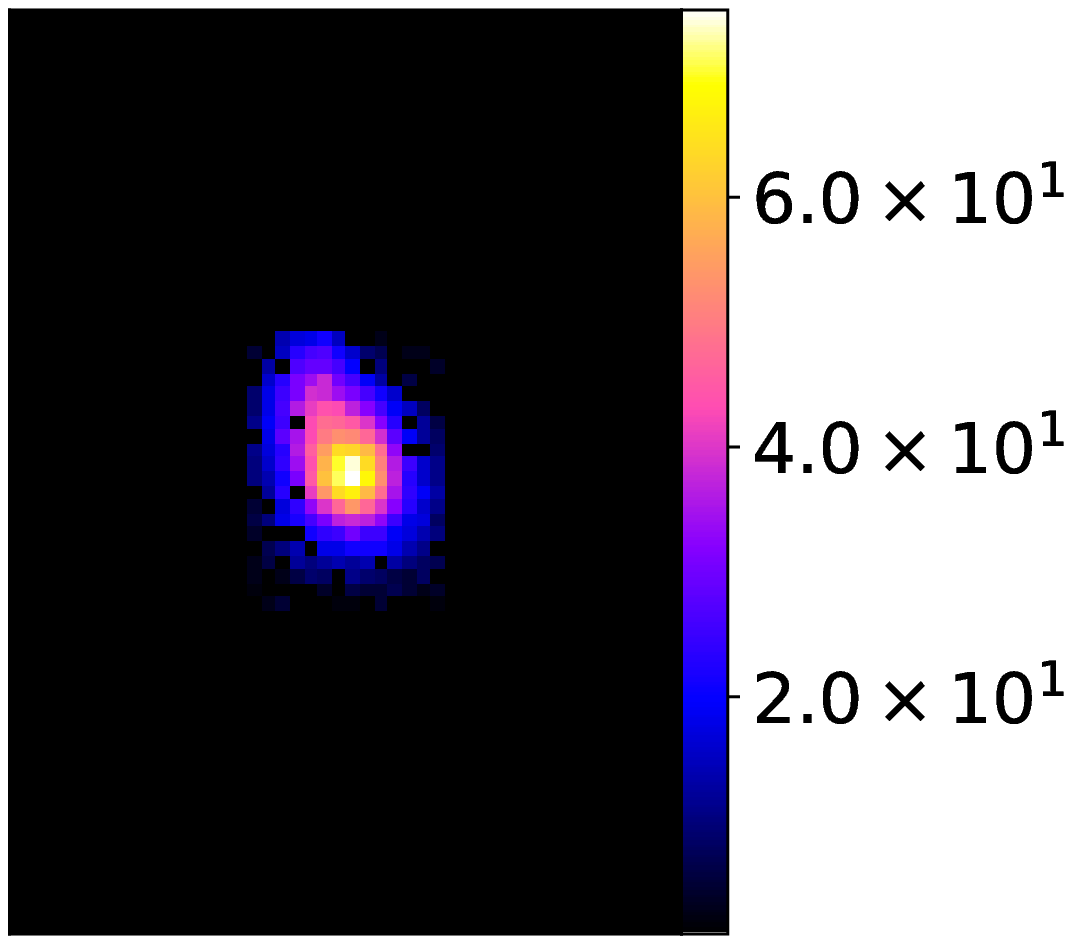}
     \makebox[0pt][r]{\makebox[30pt]{\raisebox{40pt}{\rotatebox[origin=c]{90}{$z=2$}}}}%
    \includegraphics[width=2.5cm]
    {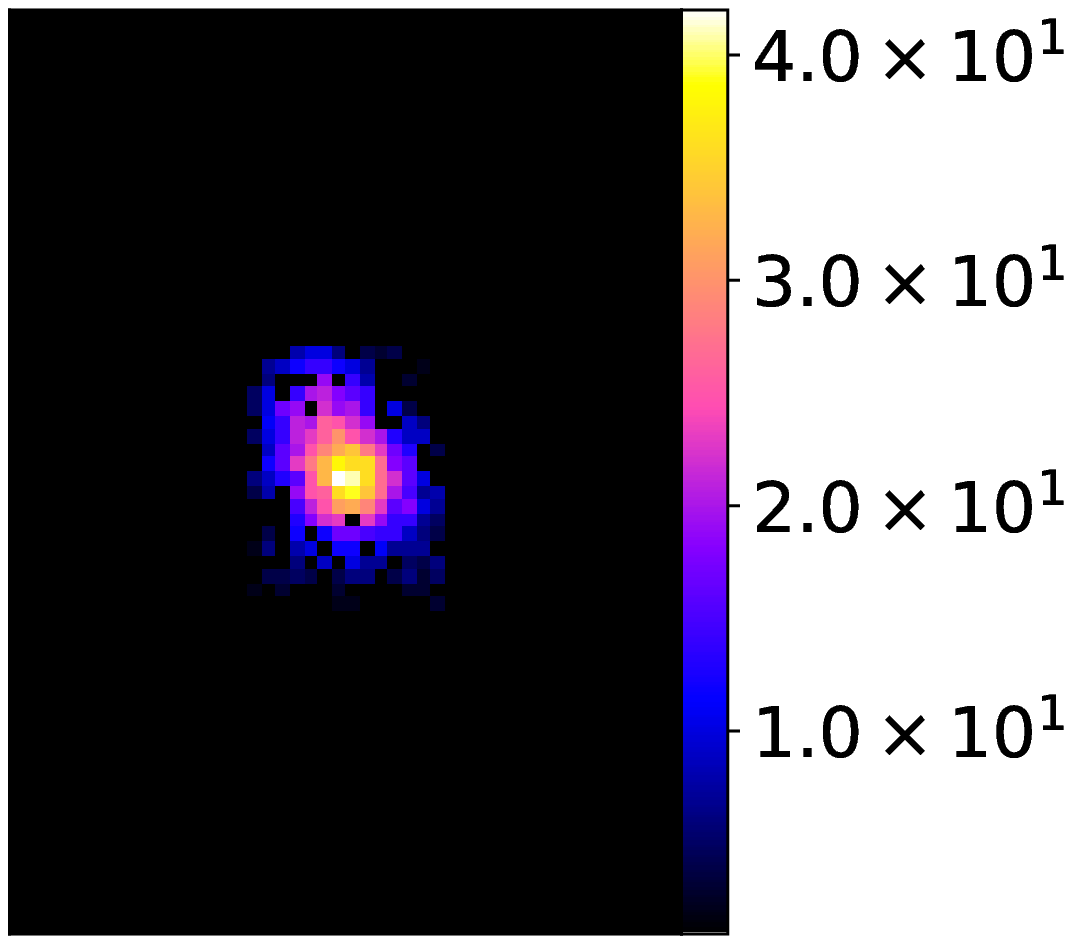}
    \caption{M51}
\end{subfigure}
\begin{subfigure}[t]{0.15\textwidth}
    \includegraphics[width=2.5cm]
    {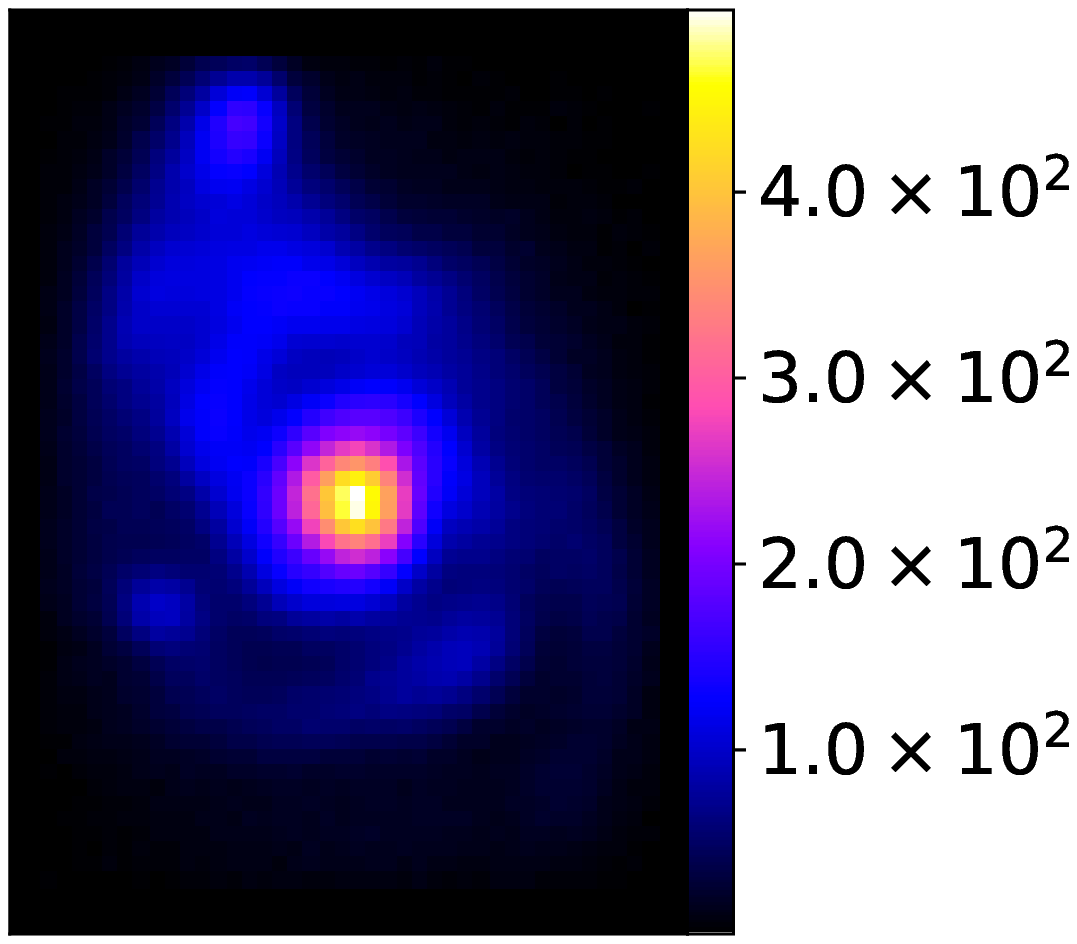}
    \includegraphics[width=2.5cm]
    {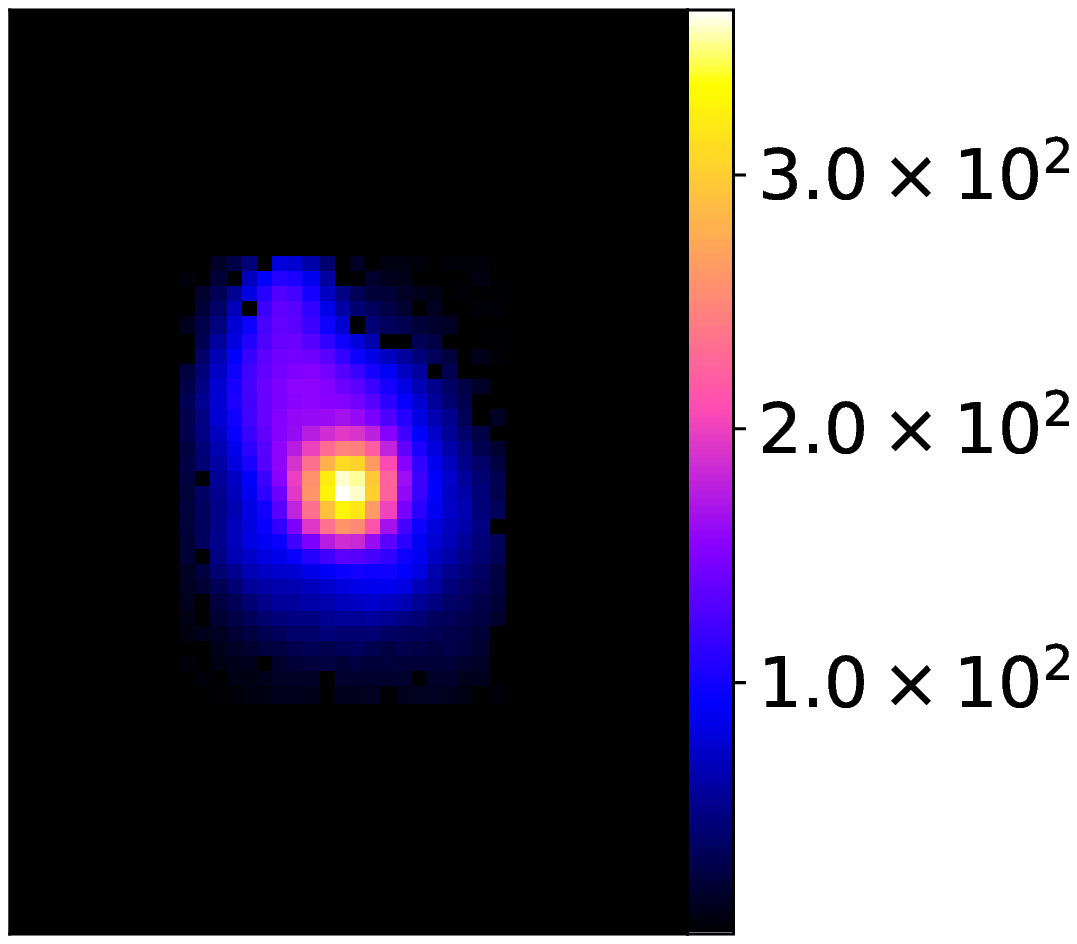}
    \includegraphics[width=2.5cm]
    {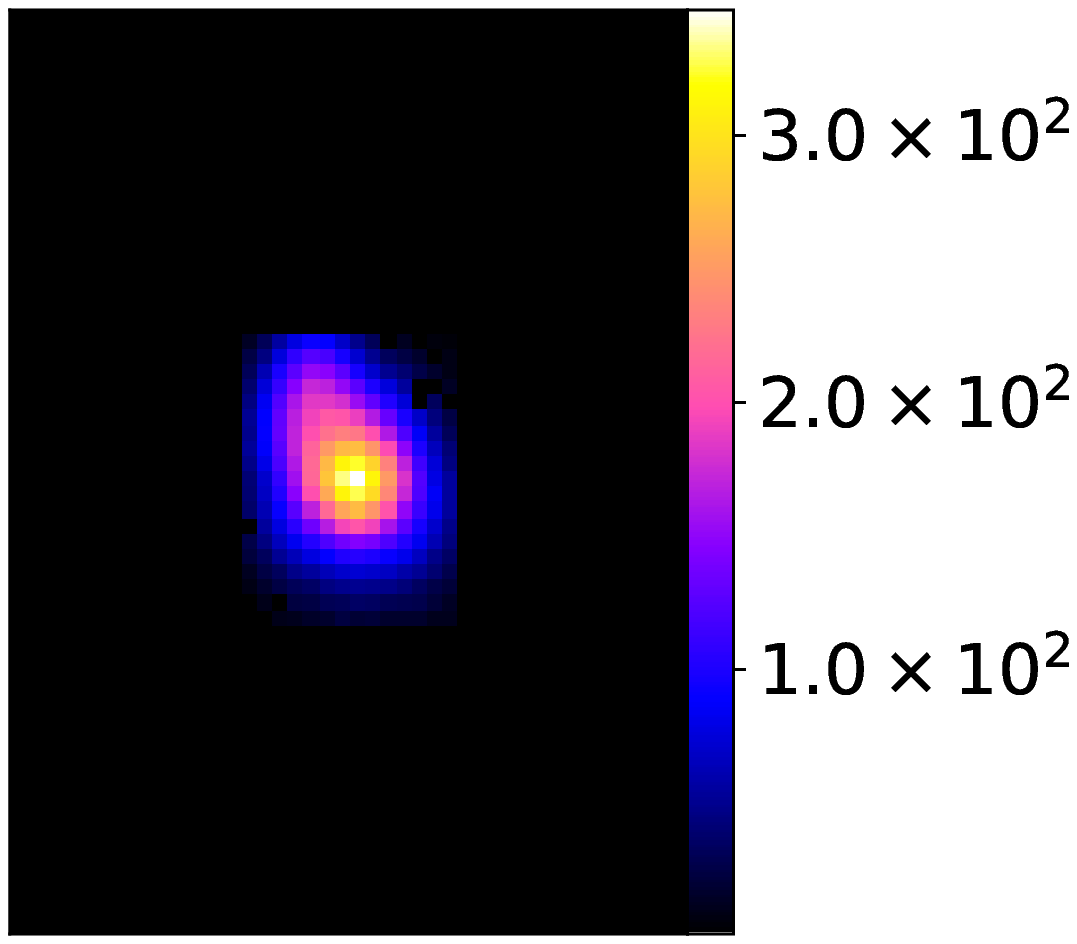}
    \includegraphics[width=2.5cm]
    {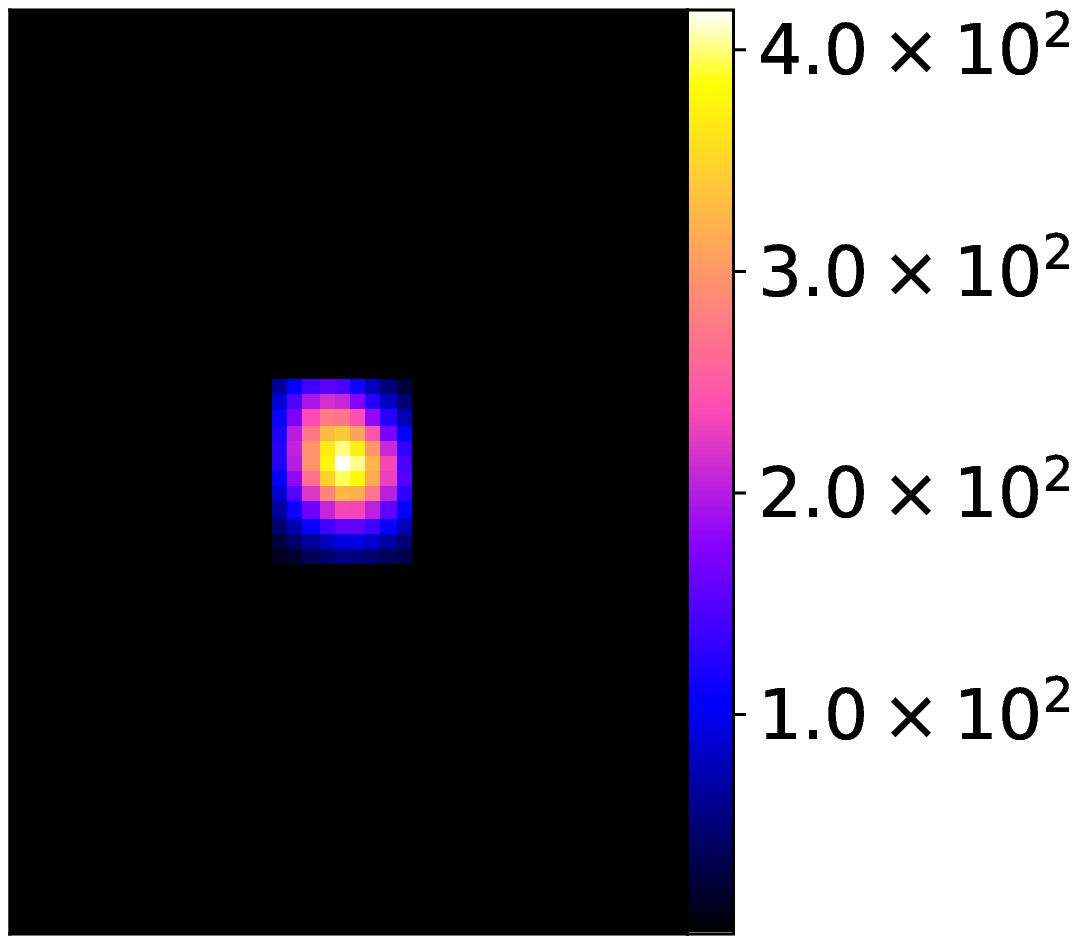}
     \includegraphics[width=2.5cm]
    {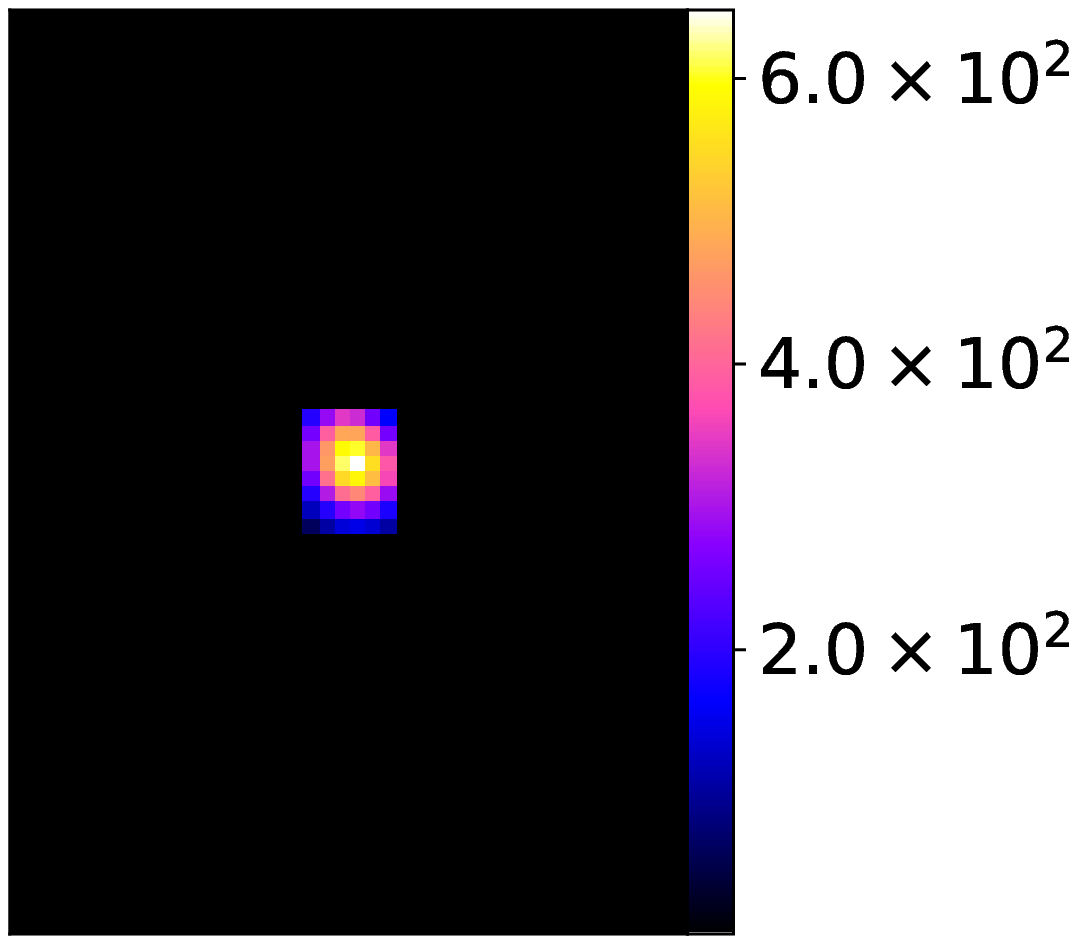}
    \caption{M51}
\end{subfigure}
\begin{subfigure}[t]{0.15\textwidth}
    \includegraphics[width=2.65cm]
    {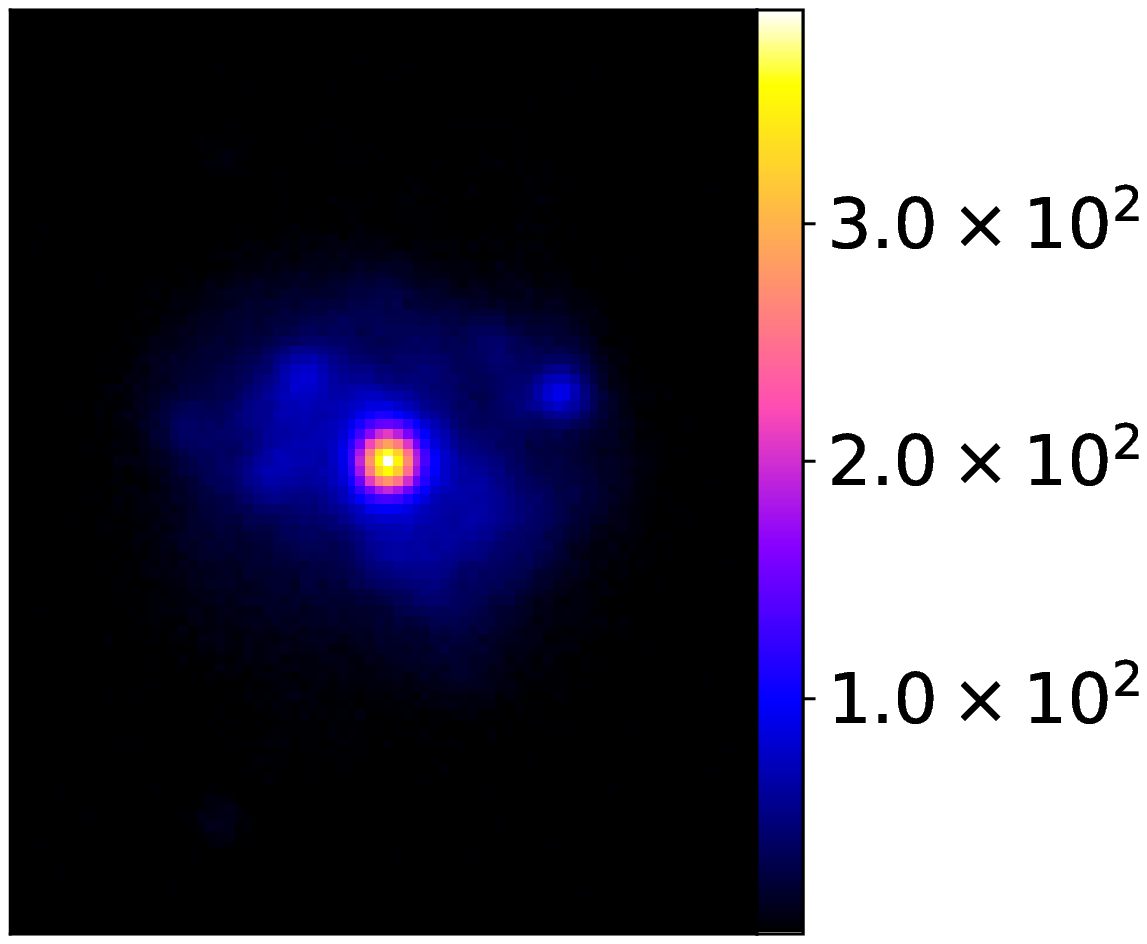}
    \includegraphics[width=2.65cm]
    {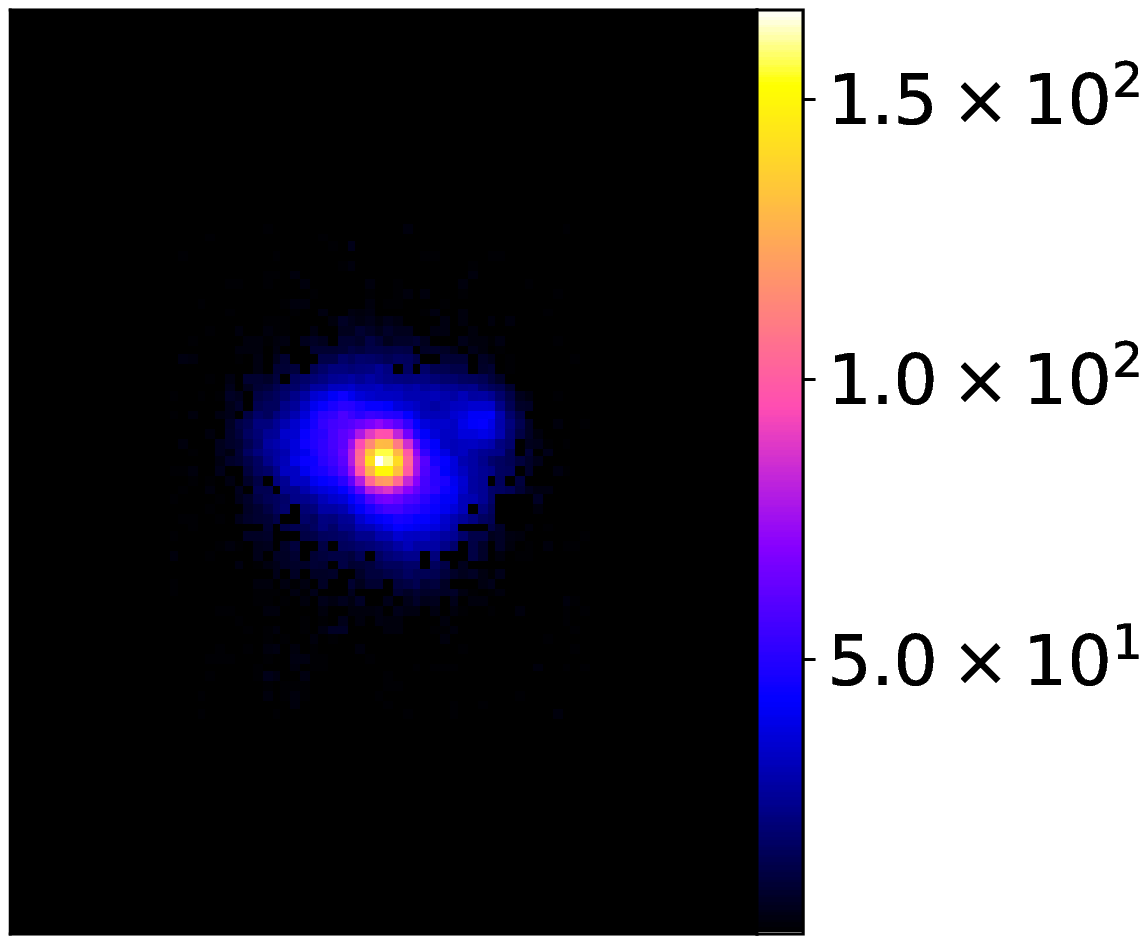}
    \includegraphics[width=2.65cm]
    {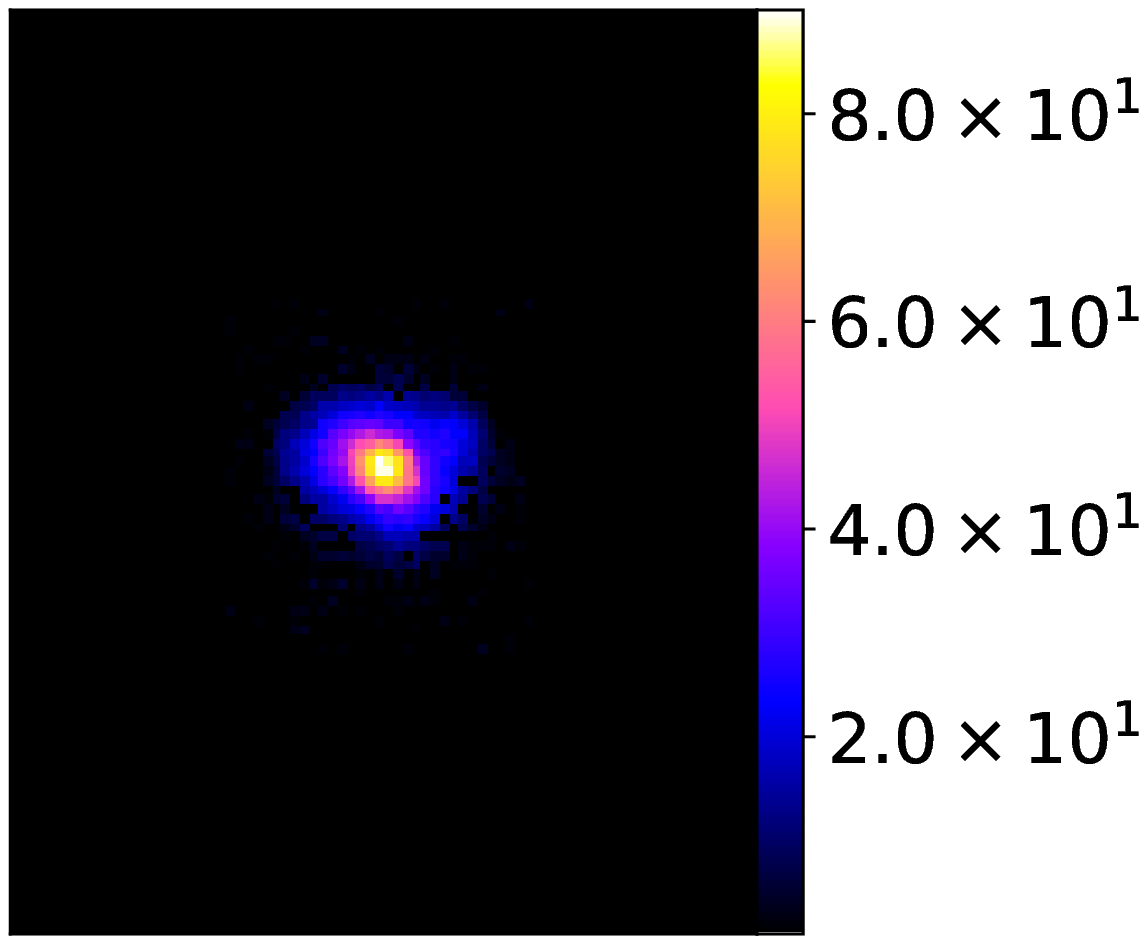}
    \includegraphics[width=2.65cm]
    {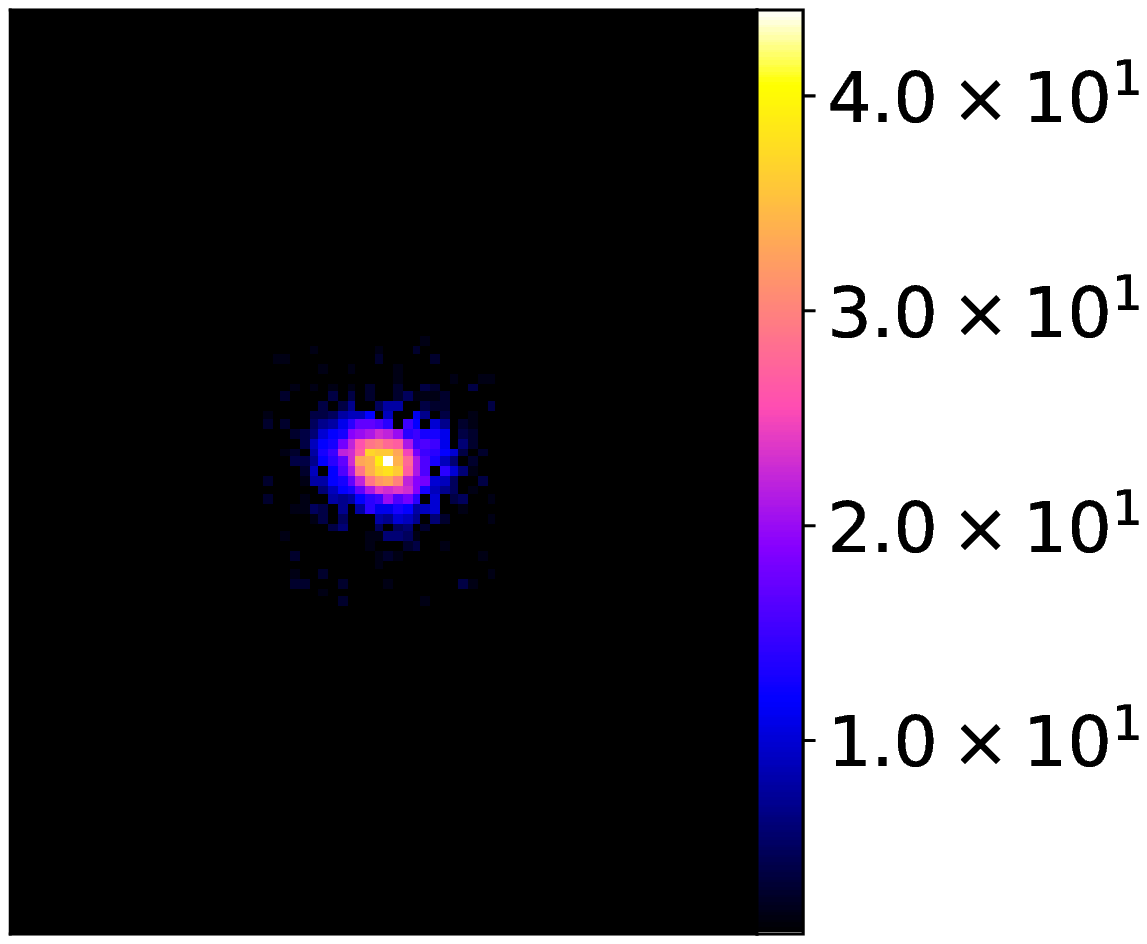}
     \includegraphics[width=2.65cm]
    {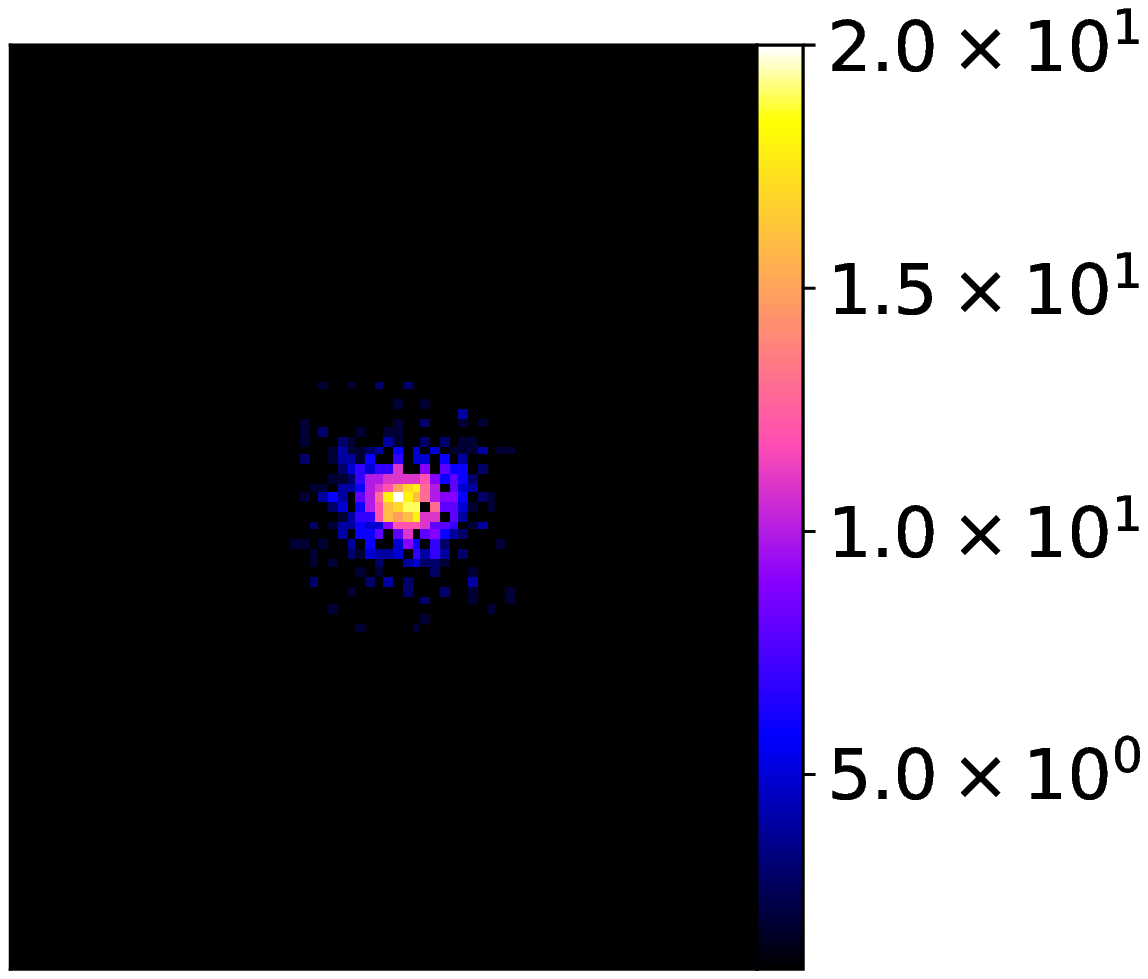}
    \caption{N6946}
\end{subfigure}
\begin{subfigure}[t]{0.15\textwidth}
    \includegraphics[width=2.65cm]
    {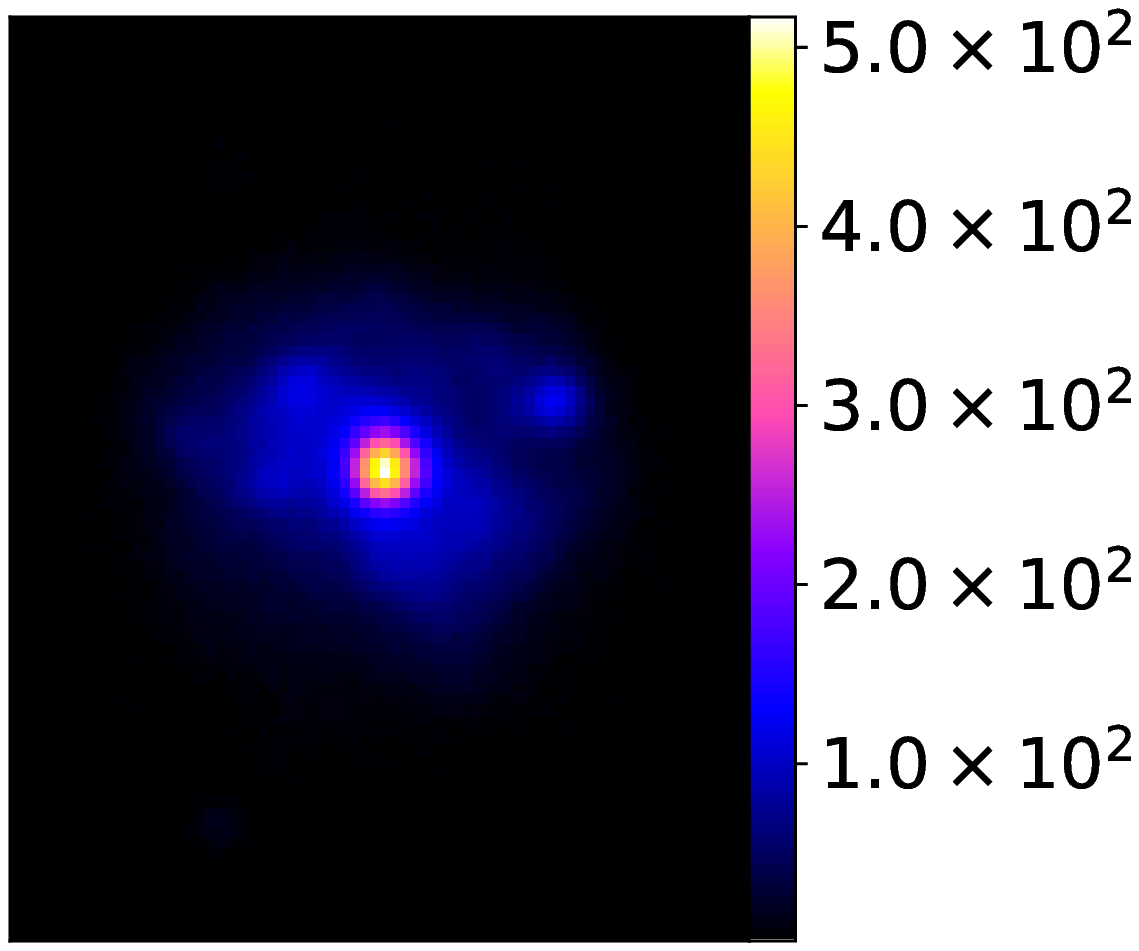}
    \includegraphics[width=2.65cm]
    {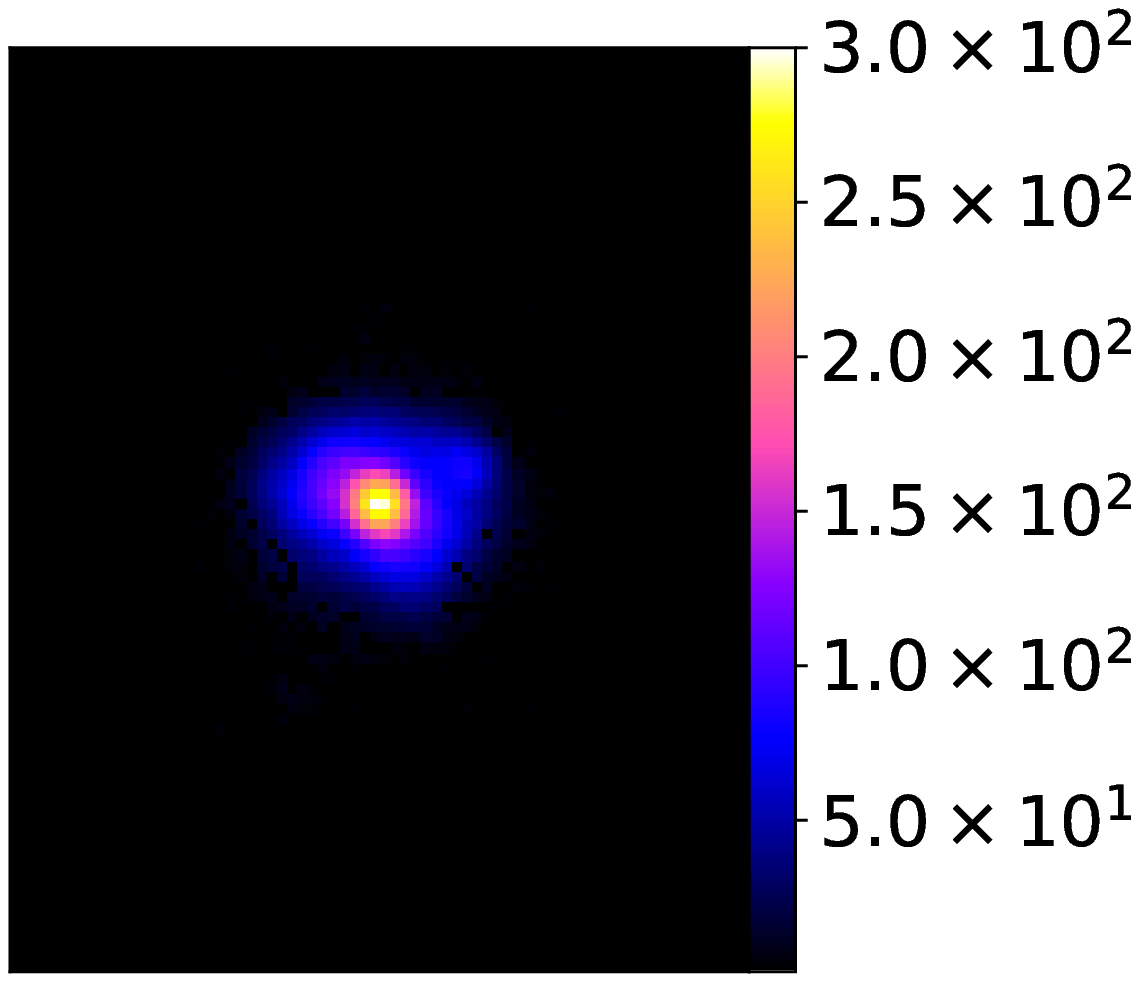}
    \includegraphics[width=2.65cm]
    {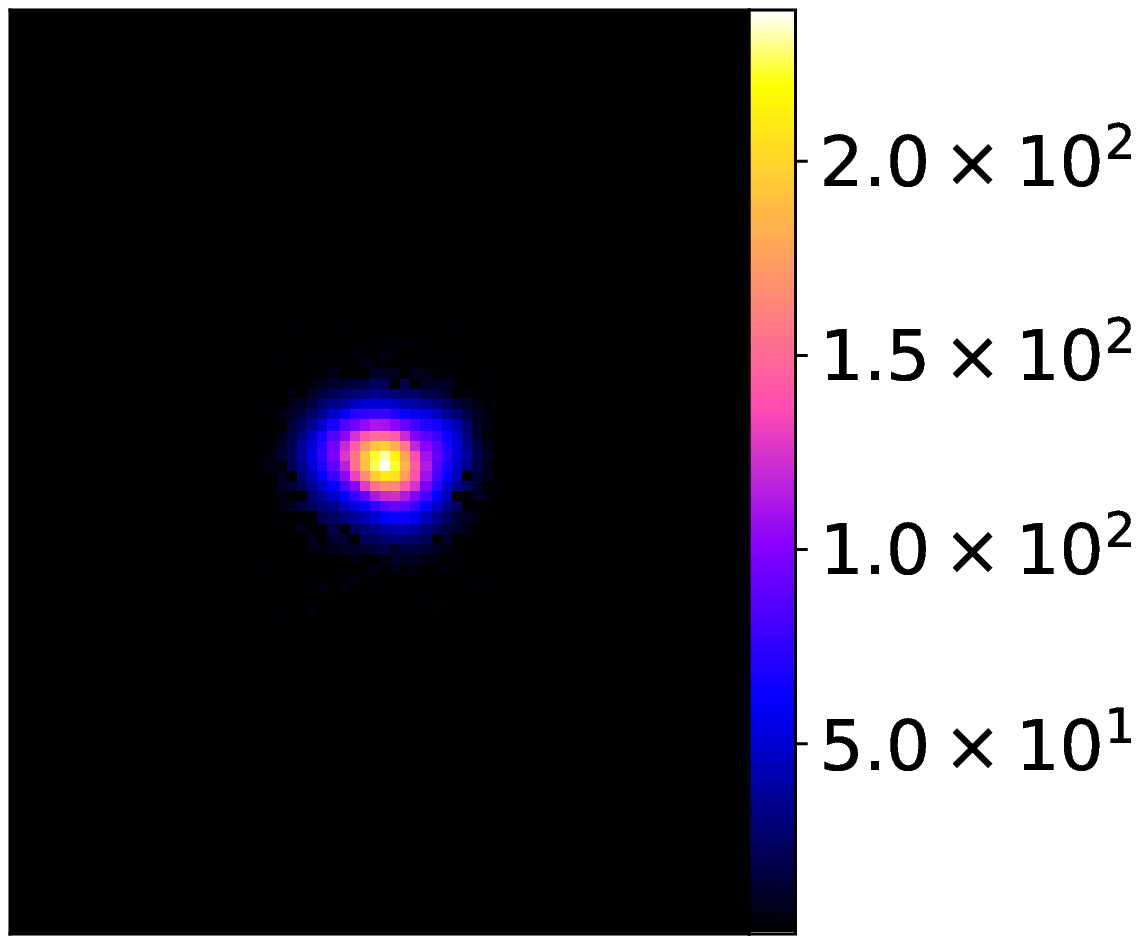}
    \includegraphics[width=2.65cm]
    {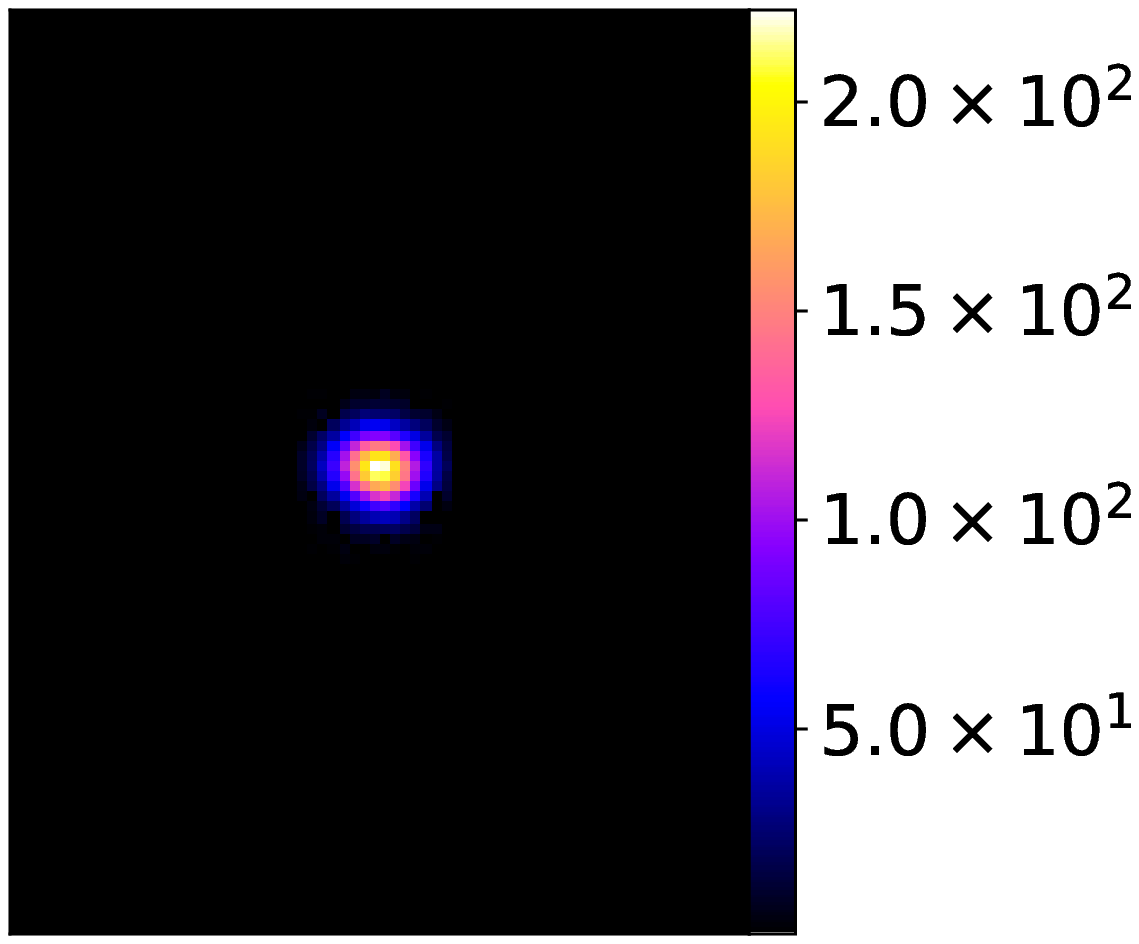}
     \includegraphics[width=2.65cm]
    {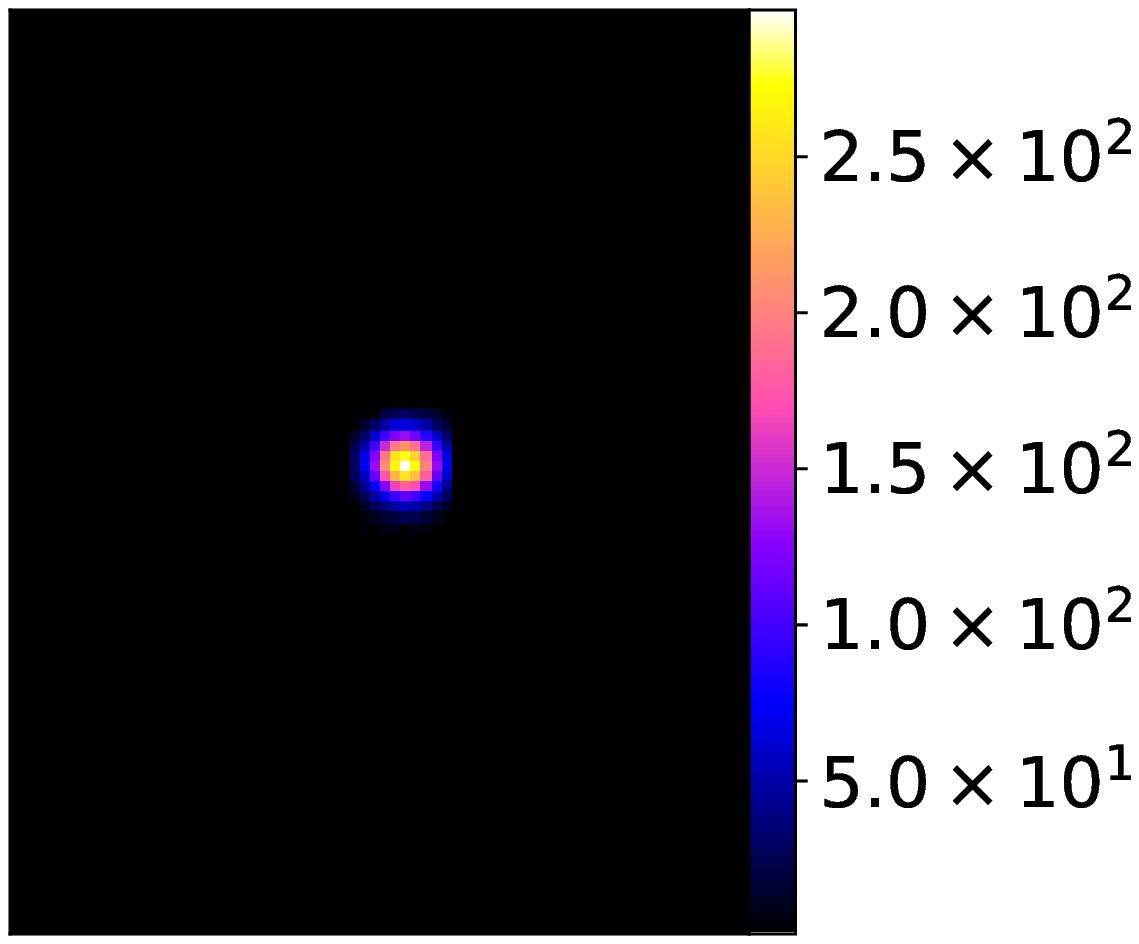}
    \caption{N6946}
\end{subfigure}
\begin{subfigure}[t]{0.15\textwidth}
    \includegraphics[width=2.45cm]
    {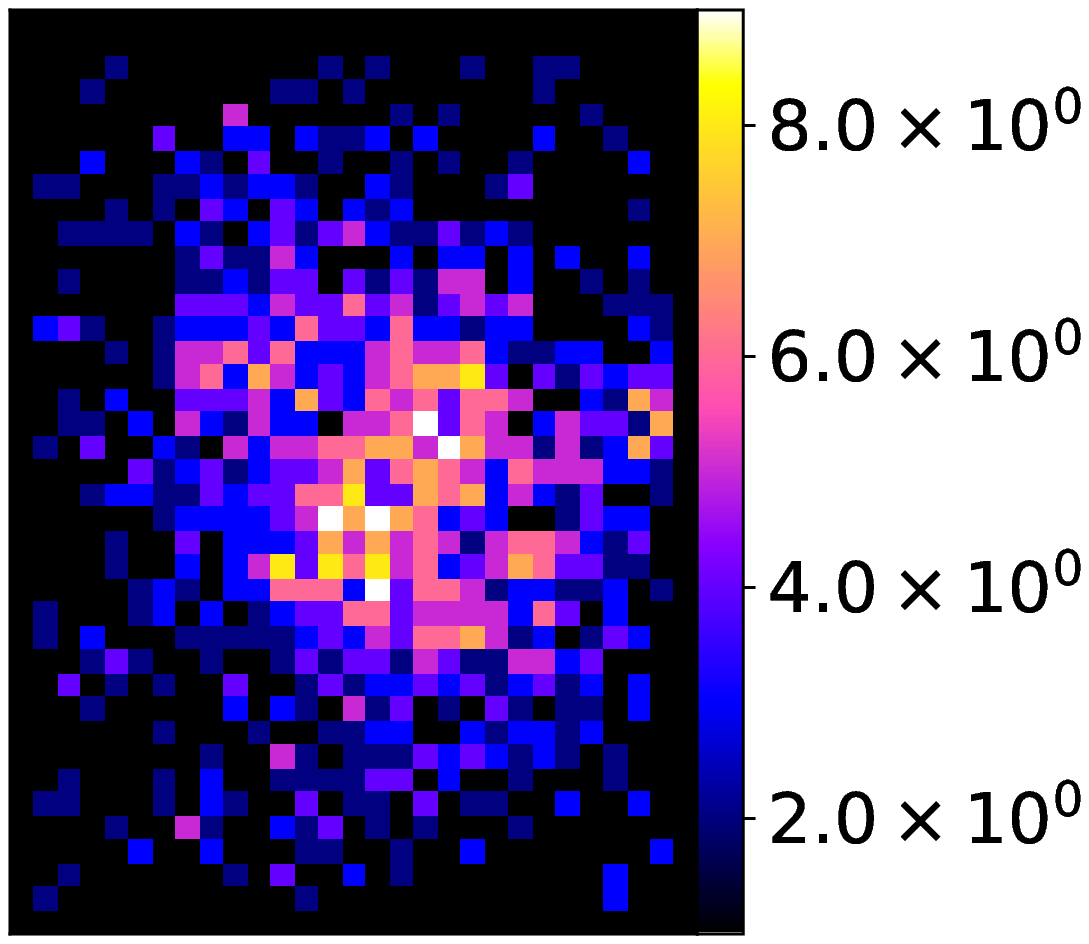}
    \includegraphics[width=2.45cm]
    {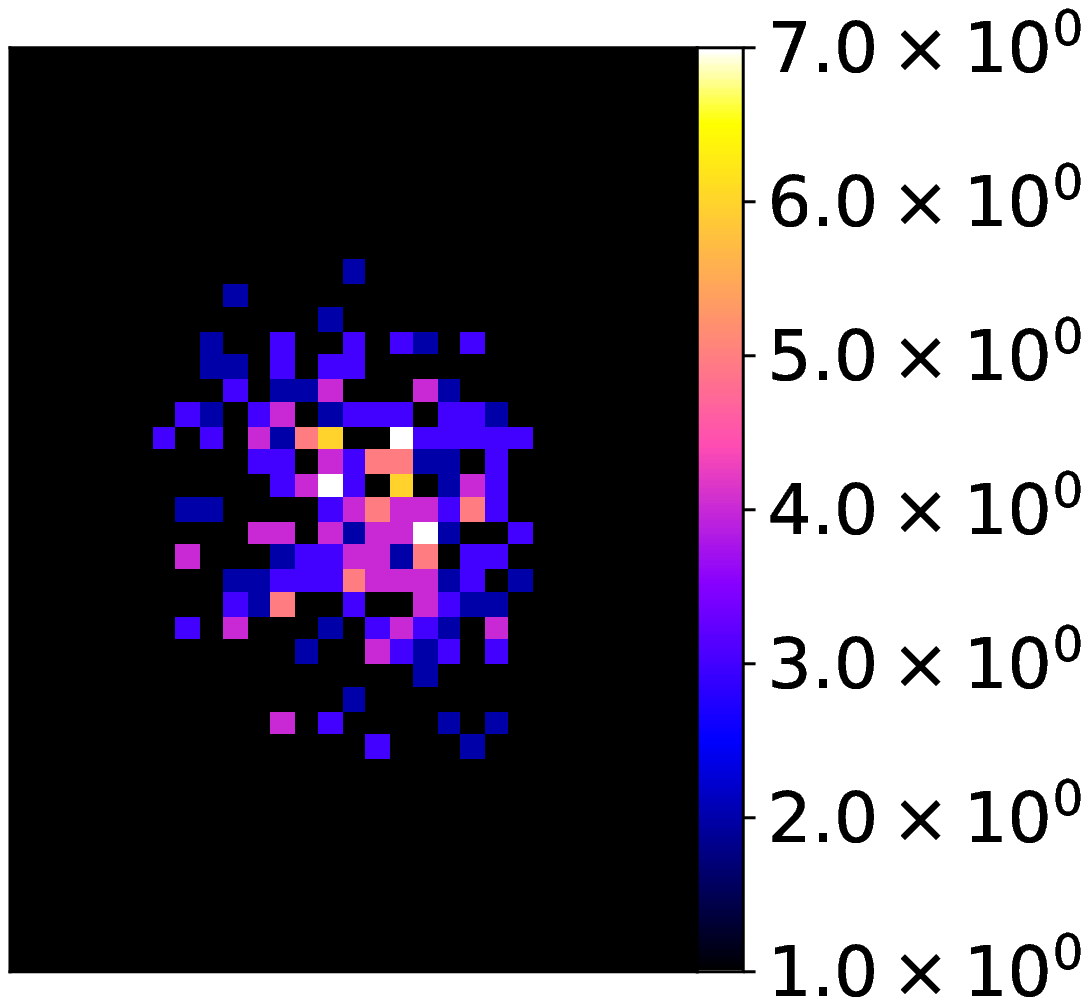}
    \includegraphics[width=2.45cm]
    {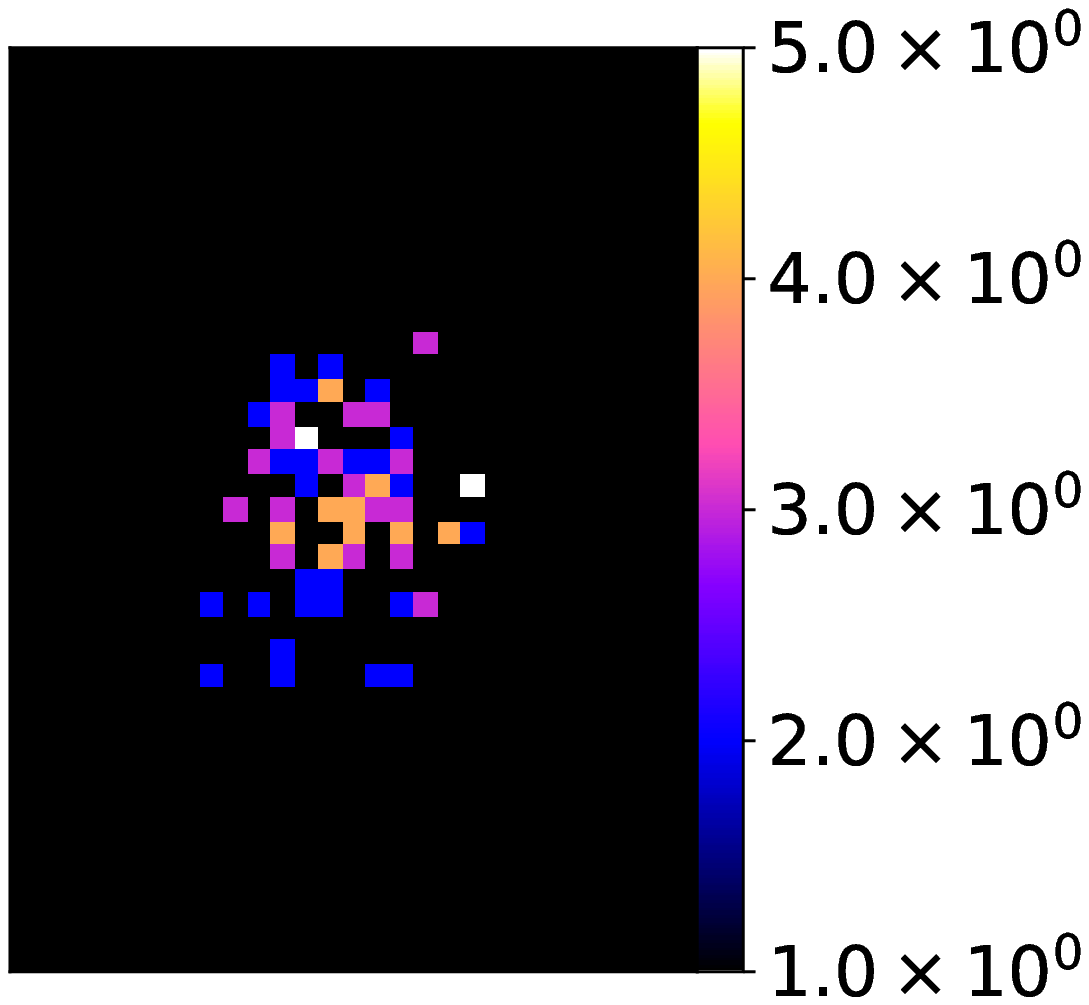}
    \includegraphics[width=2.45cm]
    {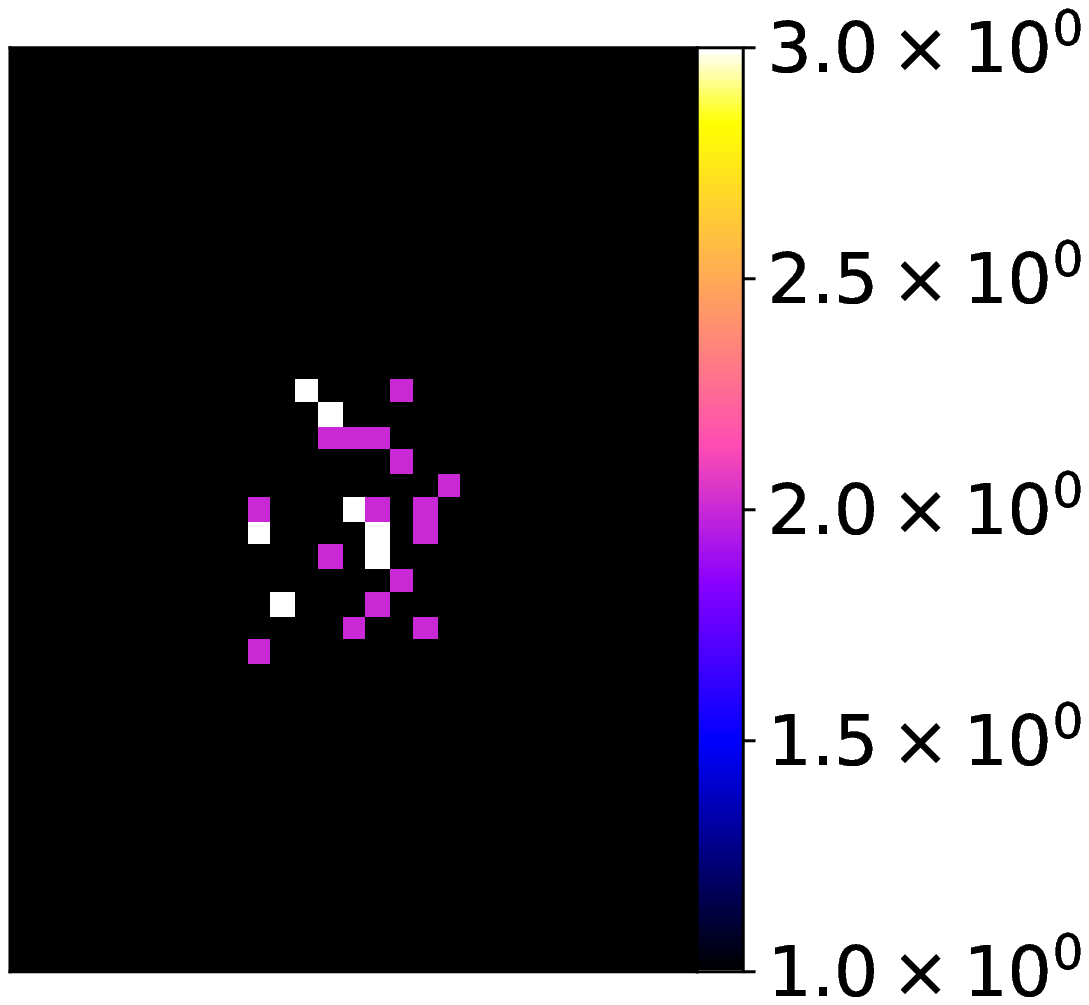}
     \includegraphics[width=2.45cm]
    {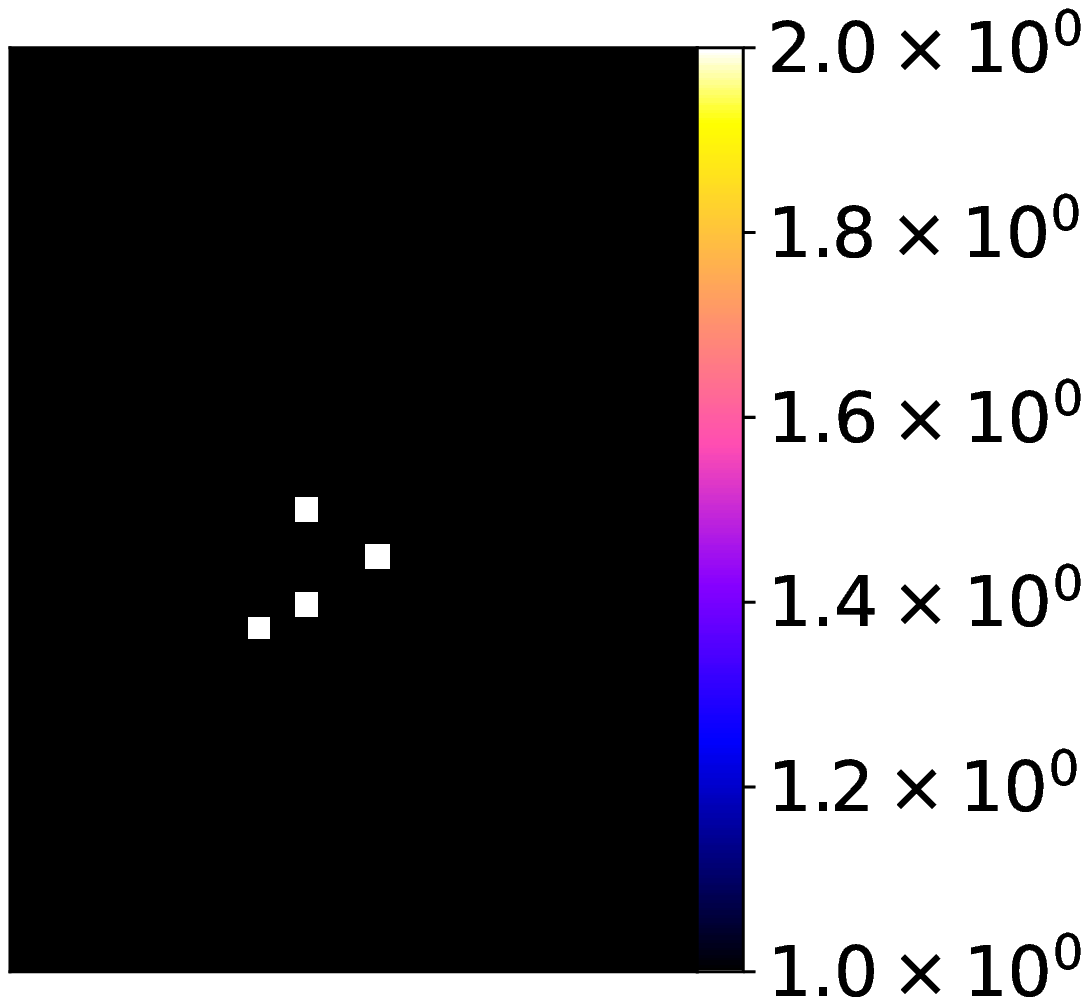}
    \caption{M33}
\end{subfigure}
\begin{subfigure}[t]{0.15\textwidth}
    \includegraphics[width=2.57cm]
    {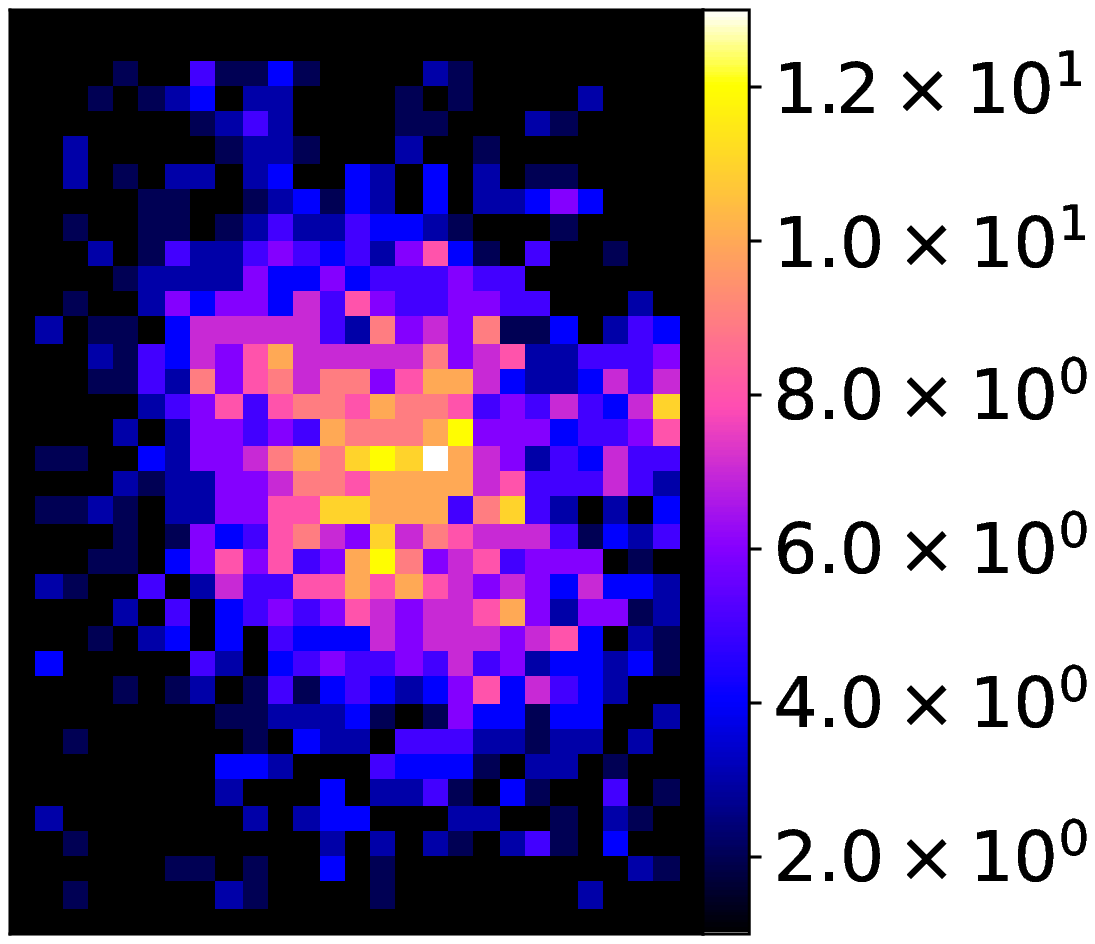}
    \includegraphics[width=2.57cm]
    {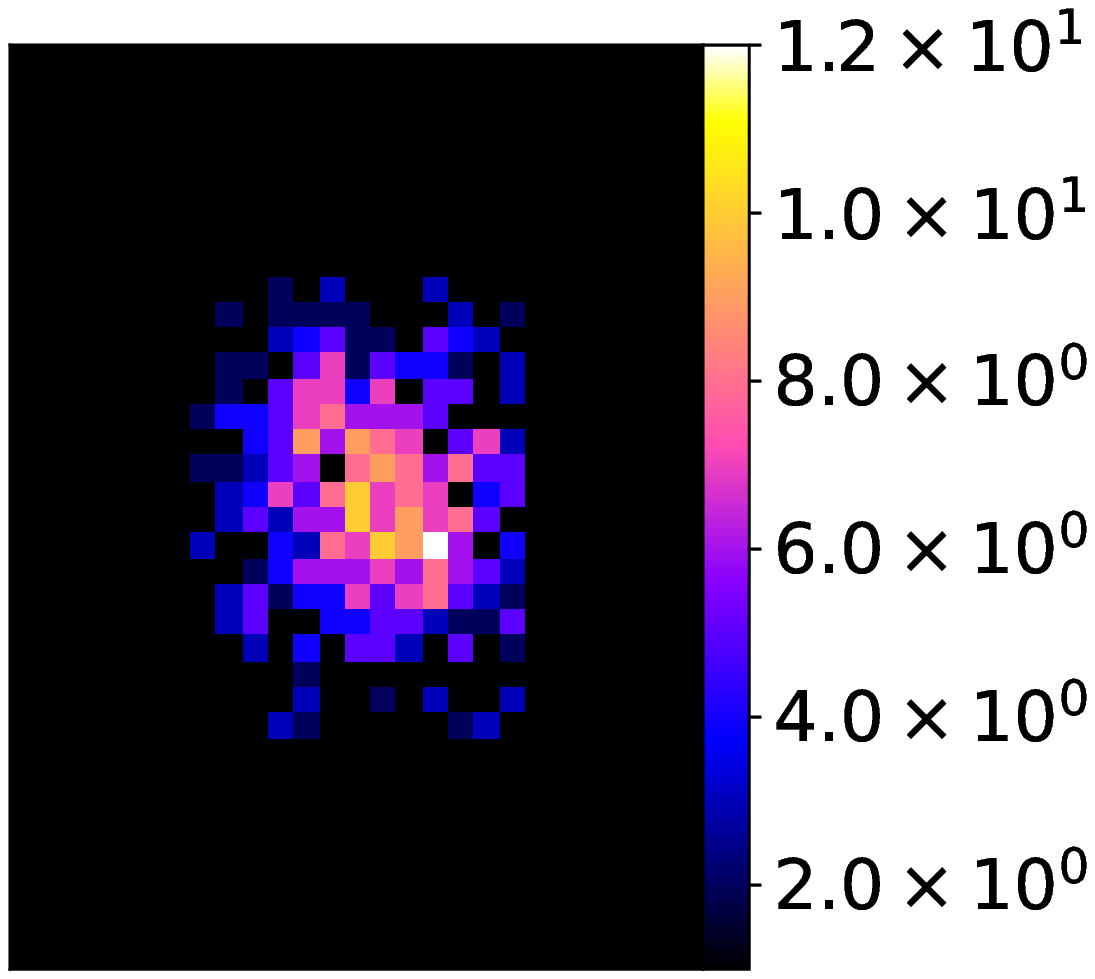}
    \includegraphics[width=2.57cm]
    {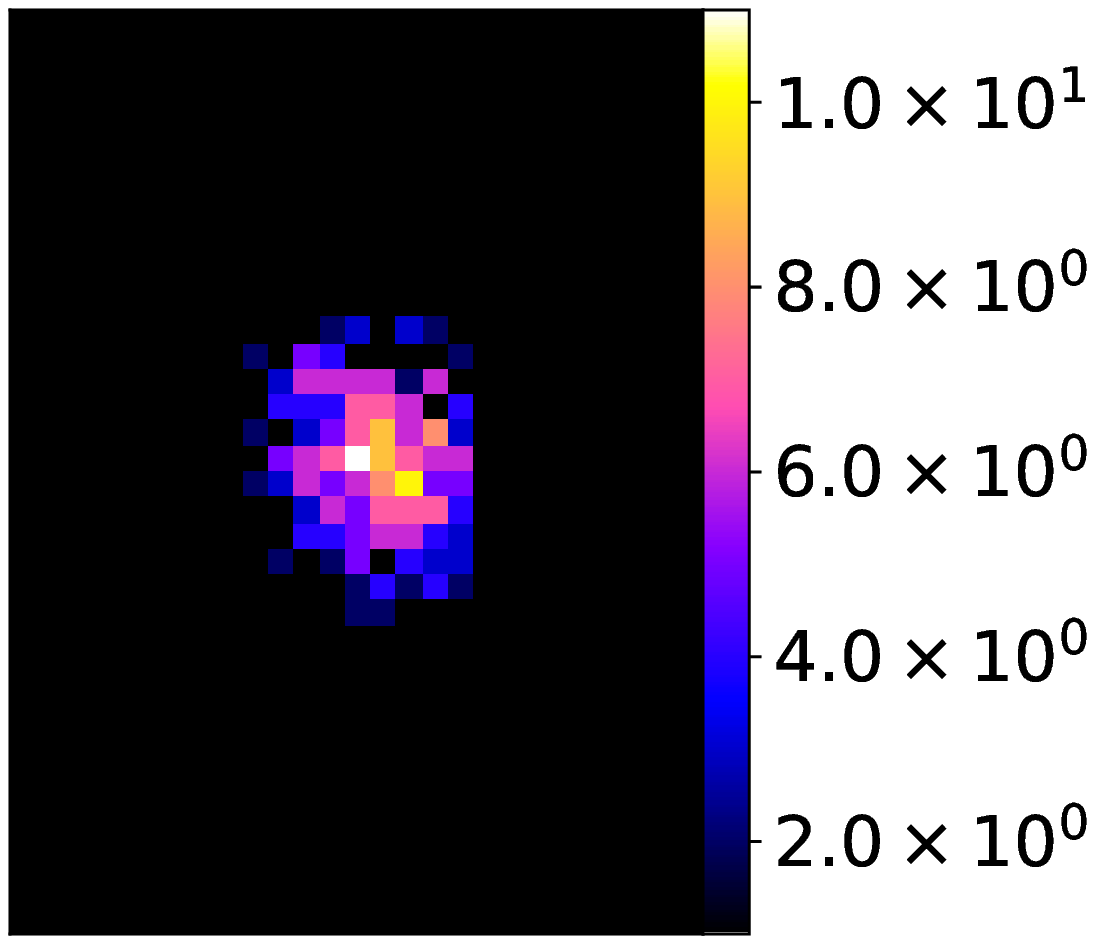}
    \includegraphics[width=2.57cm]
    {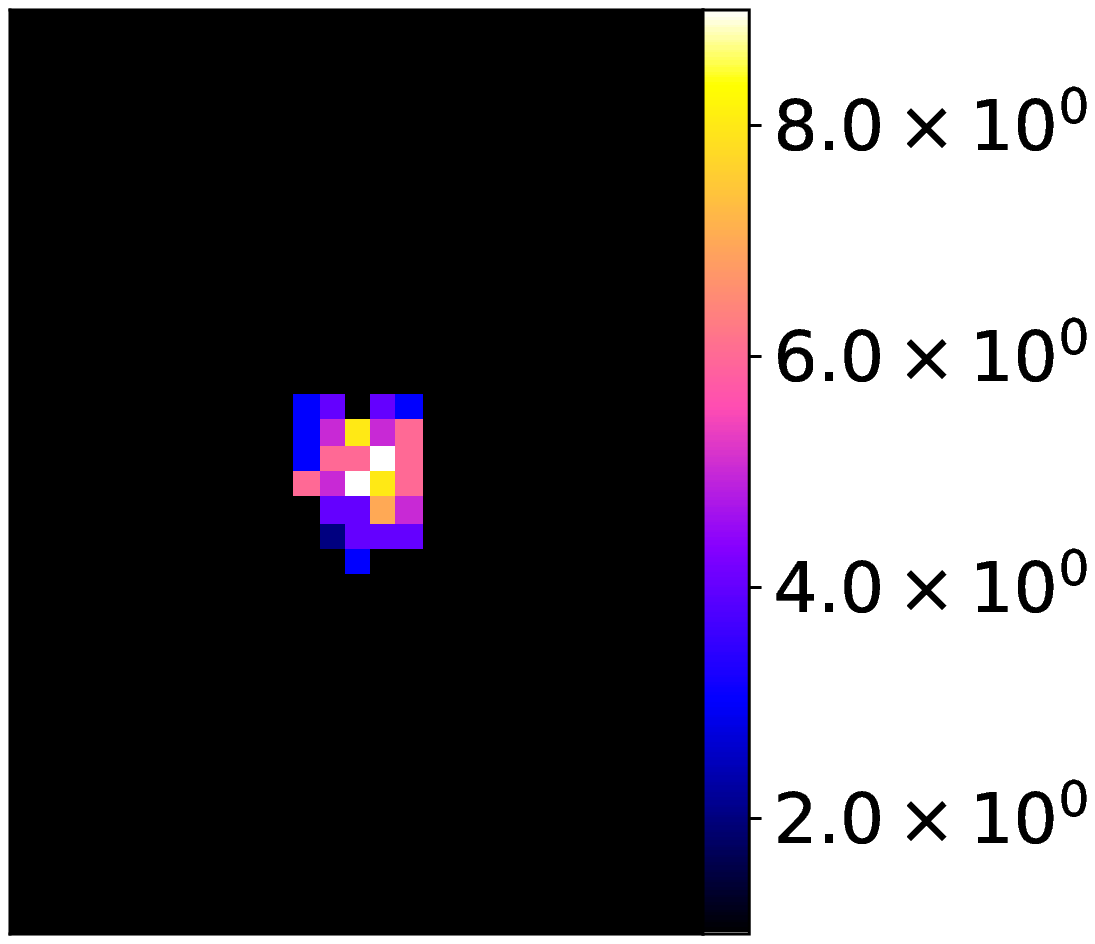}
     \includegraphics[width=2.57cm]
    {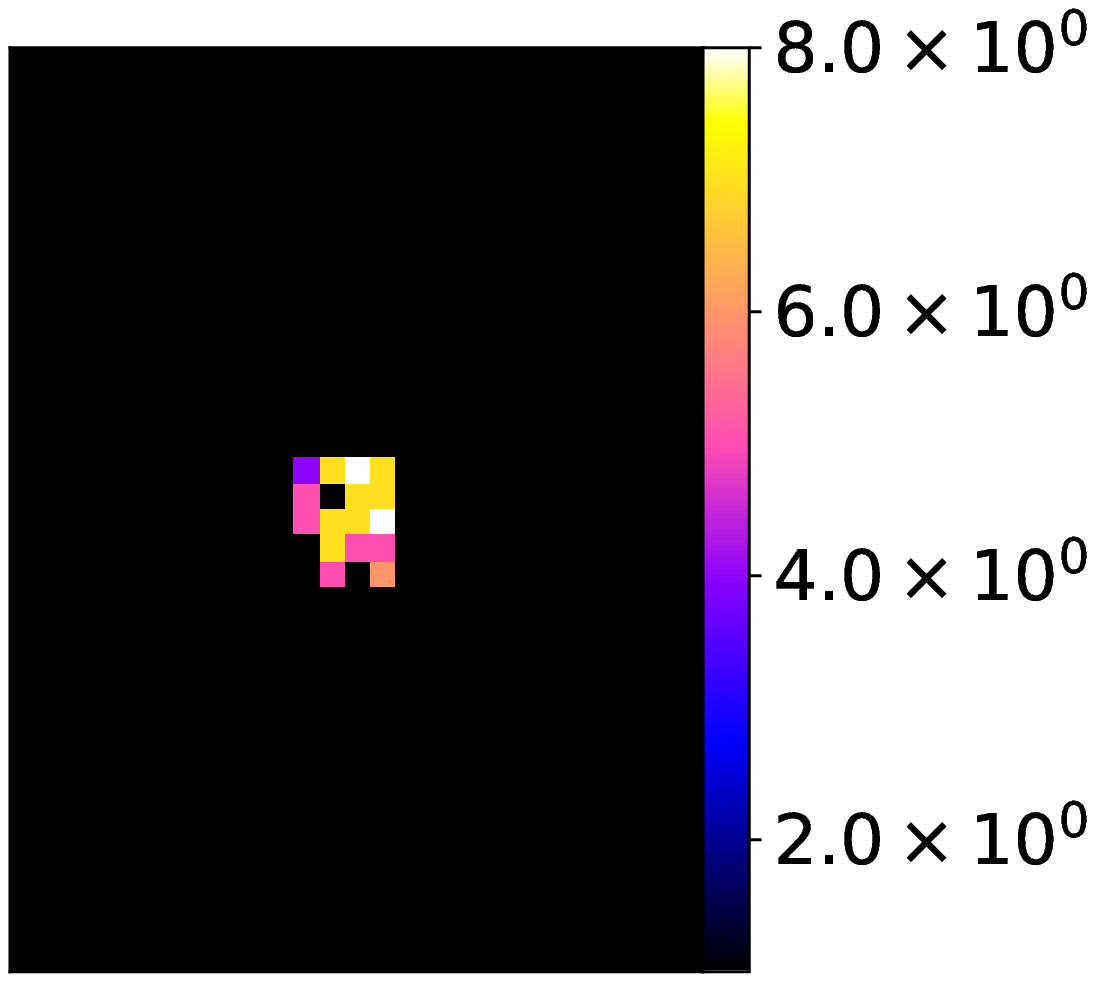}
    \caption{M33}
\end{subfigure}
\caption{ The signal-to-noise ratio S/N maps for a M51-like galaxy for case~1 ({\it first column}), case~2 ({\it second column}), NGC6946-like galaxy for case~1 ({\it third column}), case~2 ({\it fourth column}) and M33-like galaxy for case~1 ({\it fifth column}), case~2 ({\it sixth column}) based on the SKA1-MID band 2 UDT survey at the observed frequency of 1.4\,GHz. From top to bottom: $z=$ 0.15, 0.3, 0.5, 1, and 2.\label{mapM51sntol}}
\end{figure*}

%\fi

{ 
	\subsection{Comparisons with the literature studies}
	Observational studies are mostly limited to integrated properties up to now. Therefore, we discuss and compare the integrated flux densities, $S_{\nu_2}$,  with those in the literature. This should be instructive for unresolved studies in general.  The evolution in $S_{\nu_2}$ given by Eq.~\ref{S} is similar to that of the surface brightness $I_{\nu_2}$ in case (1) (Eq.~\ref{intensityth}). We are witnessing a drop in both the thermal and the nonthermal RC flux densities\footnote{However, we note that their corresponding luminosity increases with $z$ following the expected increase in SFR.}, $S_{\nu_2}^{\rm th}$ and $S_{\nu_2}^{\rm nt}$, with $z$ similar to what is shown in the two upper left panels in Fig.~\ref{thernon} at $\nu_2=1.4$\,GHz (equivalent to 1.6--5.6\,GHz rest-frame frequencies at $0.15\leq z \leq3$). As explained in Sect.~\ref{prof}, the global evolution is uncertain for  M33--like galaxies and, hence, here we compare the evolutionary trends for other galaxy types only. From $z=0.15$ to $z=2$,  $S_{\nu_2}^{\rm nt}$ decreases by a factor of $\simeq$3 and 4.5 in M51-- and NGC6946--like galaxies, respectively. At the same redshift range, $S_{\nu_2}^{\rm th}$ drops by a factor of $\simeq$ 1.6 and 2 for M51-- and NGC6946--like galaxies, respectively (see Table~\ref{table2}). Thus, the rate of this drop is faster for lower mass galaxies in general. As a matter of fact, a mass dependency is expected through the evolution of SFR (Eq.~\ref{sfrz}). It is also interesting to note that the nonthermal emission drops faster than the thermal emission. This is caused by the nonthermal emission dropping faster with increasing frequency (we remind that the rest-frame frequency probed increases with redshift).
 This effect can reduce the detection chance of high-z normal star-forming galaxies at observed frequencies higher than the mid-radio frequencies studied here. {Such a deficit in the nonthermal emission at {high frequencies} can also lead to an underestimation of the radio-based SFR at high-redshifts as noted by \cite{Algera} studying a sample of galaxies in the COSMOS  and GOODS-North fields at the rest-frame frequencies $\simeq$65-90\,GHz.}
	
%	Through a stacking analysis and SED modelling, \cite{Algera} obtained the thermal and nonthermal flux densities of a sample of galaxies in the COSMOS  and GOODS-North fields. They found that, at the rest-frame frequencies $\simeq$65-90\,GHz, the RC emission underestimates the SFR and interpret it as a deficit in high-frequency synchrotron emission in agreement with what we predict. However, they also saw a deficit in the thermal-based SFR linking it to their SED fitting routine leading to thermal fractions which are by a factor of 1.5-2 lower than that is expected at those high-frequencies. Apart from a possible role of a high-frequency break in the SED discussed by \cite{Algera}, their assumed prior (peaked at $\alpha_{\rm nt}=0.85$) or using a fixed nonthermal spectral index could cause such deviations as well. 
	
	In the MIGHTEE-COSMOS field, \cite{An} found that the ratio of the observed total flux densities observed at 1.3\,GHz-to-3\,GHz increases very slowly with stellar mass, though with a large scatter. Neglecting thermal contamination, this was explained by synchrotron losses in more massive galaxies. Here, we find a steeper 1-10\,GHz synchrotron spectrum for the most massive, M51-like galaxies compared  to the others only for $z<0.5$. We also caution that, although more massive galaxies are expected to have a larger magnetic field strength \citep{tab16}, the cooling of cosmic ray electrons does not depend only on the magnetic field strength in galaxies \cite[e.g.,][]{Longair}. Moreover, the structure of the magnetic field can also play a role: A more tangled field increases the chance of scattering and decoupling of high-energy particles from the magnetic field \cite[e.g.,][]{cal2}. This can explain the faster drop of $\alpha_{\rm nt}$ vs. $z$ in M51-like galaxies than in others due to a faster increase of SFR activities, following  Eq.~\ref{sfrz}, inducing more turbulent/tangled fields at high $z$. The result obtained by \cite{An} will be revisited after separating the thermal/nonthermal emission for the MIGHTEE-COSMOS galaxies (Tabatabaei et al in prep.).
	
}

\section{Summary} \label{summary}
{ 
The thermal and nonthermal components of the radio continuum (RC) emission are ideal tracers of energetic processes in galaxies, hence, studying them over cosmic time can shed light on the evolution of galaxies. This paper presents a study of the cosmic evolution of the thermal and nonthermal RC emission from normal star-forming galaxies by simulating the RC emission properties of present-day galaxies at earlier cosmic epoch between $0.15<z<3$. The goals are to investigate 1) the expected structure of thermal and nonthermal emission on $\gtrsim$ 1 kpc scales at different redshifts, 2) a possible evolution of the thermal fraction and  synchrotron spectral index at mid-radio frequencies (1-10 GHz), and 3) the potential of the upcoming SKA1-MID reference surveys to detect RC emitting structures that are needed to address the role of thermal/nonthermal processes in the evolution of galaxies. Surface brightness maps are created at the observed frequency of 1.4\,GHz at an angular resolution of 0.6\arcsec~considering two different scenarios: 1) no galaxy size evolution in the radio and 2) that radio sizes evolve with redshift similarly to optical size variation. The most important findings are summarized as follows:

\begin{itemize}
	\item { General trends:} The mean thermal and nonthermal surface brightnesses drop with $z$ in case (1) but they increase in case (2) because, at higher $z$, galaxy sizes become smaller. As a result of the size shrinking, the structures simulated are less resolved in {the} case (2) than in case (1).   We note that the information about the diffuse emission ({spatial }extent) from the outer disks is more redshift-dependent in { the} case~(1).
	
	\item { Evolution of the thermal fraction:}
	 As the synchrotron emission drops faster than the free-free emission at higher frequencies (due to the spectral dependencies, $\alpha_{\rm nt} > \alpha_{\rm th}$ ), the mean thermal fraction observed at 1.4\,GHz increases with redshift (by $\gtrsim$30\% from $z=0.15$ to $z=2$) in case (1). The trend is, however, more complex in case (2) as the surface brightness enhancement with $z$ is dominated by that of the nonthermal emission. This leads to a turnover in $\langle{f_{\nu_2}^{\rm th}}\rangle (z)$ at $z\simeq 1$. In lower mass galaxies, the thermal fraction is generally higher than in higher mass galaxies. %The thermal fraction in galaxy centers is less than average in all galaxies but M33.
	
	\item { Evolution of the nonthermal spectrum:} 
	We predict that the mid-radio (1-10\,GHz) nonthermal spectrum flattens with increasing redshift:
	$\alpha_{\rm nt}$ drops from 0.8-0.9 to 0.5-0.6 in case~(1) and even more for case~(2) moving back in time from  $z=0$ to $z=2$, indicating that cosmic ray electrons used to be more energetic due to higher star formation activities. Current observations match better with the case~(1) than the case~(2) predictions. In more massive galaxies like M51,  the spectral index evolves faster with $z$, { and }$ \alpha_{\rm nt}$ is steeper than in other galaxies only at $z<0.5$. The flattening of $\alpha_{\rm nt}$, causes a curvature in the observed mid-radio SED of galaxies at higher $z$.

	\item { Progress with SKA:} 
	Assuming that galaxy sizes evolve little in the radio - as suggested by our comparison of case (1) predictions and observational data, as well as recent literature \citep{Jimenez-Andrade} - the SKA1-MID band 2 survey can detect ISM structures in  M51-- and NGC6946--like galaxies at all redshifts studied here ($z\leq3$) in UDT. Depending on redshift, the disk of galaxies can still be partly detected with the two other tiers, DT and WT.  The mean RC emission from low-mass galaxies like M33 can only be detected { at $3 \sigma$} level up to $z=0.5$ in UDT. {These findings show that the scientific goals defined for the SKA1-MID band~2 surveys cannot be fully met if the radio-size of galaxies remains fixed with redshift.} At the selected resolution of 0.6\arcsec,~spiral arms can be distinguished out to $z<1$. These can be in principle detected at higher $z$ with SKA adopting higher angular resolutions but at a lower S/N.
\end{itemize}

Last but not the least, constraining the mid-radio SED of galaxies is fundamental 
%to separate the thermal and nonthermal emission and to calibrate 
{to separating the thermal and non-thermal emissions and calibrating}
SFR \cite[][]{cal2} independently from the RC--IR correlation at high $z$. However, {complete sampling of the mid-radio SED will be possible only up to $z=1$} with the currently proposed SKA1-MID surveys (see Sect.~\ref{srs}).  To address the evolution of galaxies,  studies of the epoch of maximum star formation (i.e., $z\simeq2$) are crucial.  This study suggests that band~1 must be included in the SKA1-MID surveys to constrain the mid-radio SED of galaxies at this epoch. This will further help to disentangle AGN/starburst candidates and to understand the evolution of the radio--infrared correlation in galaxies.
}

 \section*{Acknowledgements}

The authors thank Robert Braun and the SKA organization team for their help and useful discussions. We also thank the anonymous referee for his/her helpful comments. MG-N acknowledges support from Iran's National Elites Foundation (INEF).
MTS acknowledges support from a Scientific Exchanges visitor fellowship (IZSEZO$\_$202357) from the Swiss National Science Foundation.

%%%%%%%%%%%%%%%%%%%%%%%%%%%%%%%%%%%%%%%%%%%%%%%%%

 %------------------------S/N maps Tolman----------

\subsection*{Data Availability}
The simulated data will be made available on \href{https://figshare.com/}{figshare}.



%\bibliographystyle{mnras}


%%%%%%%%%%%%%%%%%%%%%%%%%%%%%%%%%%%%%%%%%%%%%%%%%%

%%%%%%%%%%%%%%%%% APPENDICES %%%%%%%%%%%%%%%%%%%%%
\appendix
{
\section*{Appendix}
\section*{Thermal/Nonthermal Separation Method}
The thermal and nonthermal components of the radio continuum emission are mapped using the TRT method developed by \cite{tab7} and \cite{tab_a} in which a de-reddened H$\alpha$ map is used to trace the thermal free-free emission in galaxies. As the thermal radio emission is expected to arise generally from ionized media irrespective from their ionization sources, a physically motivated thermal tracer should map the thermal radio emission from not only star-forming regions but also diffuse ionized gas in galaxies. Recombination lines, with the H$\alpha$ line being the brightest one, can ideally do the job as a physical balance is expected between ionization and recombination rates  in an ionized gas.  

At frequency $\nu$, the relation between the thermal free-free brightness temperature $T_b$ in Kelvin (K) and the H$\alpha$ surface brightness $I_{{\rm H}\alpha}$ in erg\,cm$^{-2}$\,s$^{-1}$\,sr$^{-1}$  is given by
\begin{equation}
\left \{ \begin{array}{ll}
{T_b=T_e(1-e^{-A\,I_{{\rm H}\alpha}})} ,  \\ 
A=3.763\,\nu_{GHz}^{-2.1}\, T_{e4}^{-0.3}\, 10^{\frac{0.029}{T_{e4}}},  %
\end{array} \right.
\end{equation}
where $T_e$ is the electron temperature in K and $T_{e4}=T_e/10^4 K$. In the above relation, the contribution from singly ionized He is also taken into account. The thermal free-free surface brightness $I^{\rm th}$ in mJy/beam is then obtained using 
\begin{equation}
I^{\rm th}_{\nu} = \frac{\theta_i \theta_j}{1224} \,\nu_{GHz}^2\, T_b,
\end{equation}
with $\theta_i$ and $\theta_j$ the beam width along the major and minor axes in arcsec.
Subtracting this emission from the observed surface brightness $I^{\rm obs}$, the nonthermal surface brightness is obtained ${I^{\rm nt}_{\nu}}= {I^{\rm obs}_{\nu}} - {I^{\rm th}_{\nu}}$. 

To de-reddening  the H$\alpha$ map, different techniques are used depending on the available data. Ideally, the ratio of two recombination lines can be used to correct the effect of dust extinction \citep{tab18}. Using dust mass maps also provide an efficient information thanks to Spitzer and Herschel infrared maps of nearby galaxies \citep[e.g.][]{Hassani}.

%For M51, the Herschel PACS data at 70$\mu$m and 160$\mu$m (?) were used to map the effective dust optical depth across M51 following \cite{tab7}. Then the H$\alpha$ line emission (Thilker et al. 2000) was de-reddened using  $I = I_0 \,\,e^{-\tau_{\rm eff}}$ (The H$\alpha$ map as well as the PACS maps were first convolved and normalized to the same resolution (15\arcsec) and geometry of the radio continuum map at 1.4\,GHz.)

%\section*{Thermal/Nonthermal Separation with SKA}
%The sensitivity of the SKA is supposed to be about 50 times larger than the current most sensitive radio telescopes. Hence, it is expected to produce informative maps of the thermal and nonthermal emission in nearby galaxies using multi-frequency observations at consistent sensitivities and through pixel-by-pixel radio SED analysis. However, at larger distances, the sensitivity drops,  more severely at higher frequencies, decreasing the success chance of this method in studying the resolved ISM. Hence, it is wise to explore the use of the TRT method with SKA as well, at least at high-z. Combining the SKA continuum observations with the data of recombination lines in optical and near-infrared will allow us to study the thermal and nonthermal energy balance in galaxies at high-z. This is feasible taking into account the already available Hubble Deep Field surveys and the forthcoming observations with the James Webb Telescope along with the TMT, E-VLT, and others. This should justify our selection of the noise added to the thermal radio maps simulated. 

}

%%%%%%%%%%%%%%%%%%%%%%%%%%%%%%%%%%%%%%%%%%%%%%%%%%

% Don't change these lines
\bsp	% typesetting comment
\label{lastpage}
\end{document}